\documentclass{jpp}

\usepackage[toc, xindy, acronym]{glossaries}
\usepackage{natbib}

\usepackage[title]{appendix}

\usepackage[setpagesize=false]{hyperref} 
\hypersetup{
    colorlinks,
    linkcolor={blue!80!black},
    citecolor={blue!80!black},
    urlcolor={blue!80!black}
}

\usepackage{zref-savepos}

\usepackage[makeroom]{cancel}
\usepackage{scrextend} 
\usepackage{bold-extra} 
\usepackage{color} 
\usepackage{fancyhdr}
\usepackage{ulem}
\usepackage{xcolor}
\usepackage{relsize}
\usepackage{textcomp}
\usepackage{lipsum}
\usepackage{setspace}
\usepackage{empheq}

\usepackage{float} 

\usepackage{epstopdf, epsfig}

\usepackage{pdfpages} 
\usepackage{graphicx} 

\usepackage{caption}
\usepackage{tabu} 
\usepackage{tabulary} 

\usepackage{framed} 
\usepackage{listings} 

\usepackage{amsmath} 
\usepackage{amssymb} 
\usepackage{mathrsfs} 

\usepackage{algorithm} 
\usepackage{algpseudocode} 

\usepackage{tensor}
\usepackage{upgreek}
\usepackage{physics}

\usepackage{tikz}
\usepackage{circuitikz}
\usetikzlibrary{calc}
\usetikzlibrary{patterns}
\usetikzlibrary{arrows}
\usetikzlibrary{shapes}
\usetikzlibrary{plotmarks}
\usetikzlibrary{decorations.markings}

\usepackage{pgfplots}
\usepgfplotslibrary{polar}
\usepgflibrary{shapes.geometric}
\usepackage[]{pstricks}

\newcommand{\figref}[1]{figure~\ref{#1}}

\newcommand{\Figref}[1]{Figure~\ref{#1}}

\newcommand{\Secref}[1]{Section~\ref{#1}}
\newcommand{\secref}[1]{section~\ref{#1}}

\newcommand{\apref}[1]{appendix~\ref{#1}}
\newcommand{\Apref}[1]{Appendix~\ref{#1}}

\newcommand{\gradd}{\nabla}

\newcommand{\curll}{\nabla\times}
\renewcommand{\vec}{\boldsymbol}

\newcommand{\spc}{\:\:\:\:}

\newcommand{\taubar}{\bar{\tau}}
\newcommand{\lambdae}{\lambda_e}

\newcommand{\df}{\delta \! f}

\newcommand{\dB}{\delta \! \vec{B}}
\newcommand{\dBpar}{\delta \! B_\parallel}
\newcommand{\dBperp}{\delta \! \vec{B}_\perp}
\newcommand{\dTe}{\delta T_{e}}
\newcommand{\dne}{\delta n_{e}}
\newcommand{\rmd}{\mathrm{d}}
\newcommand{\sgn}{\mathrm{sgn}}

\newcommand{\s}{s}
\newcommand{\constant}{a}

\newcommand{\bea}{\begin{eqnarray}}
\newcommand{\eea}{\end{eqnarray}}
\newcommand{\beq}{\begin{equation}}
\newcommand{\eeq}{\end{equation}}

\newcommand{\dperp}{\nabla_\perp}
\newcommand{\dpar}{\nabla_\parallel}

\newcommand{\kperp}{k_\perp}

\newcommand{\kpar}{k_\parallel}

\newcommand{\kparc}{k_{\parallel\mathrm{c}}}

\newcommand{\vperp}{v_\perp}
\newcommand{\vpar}{v_\parallel}

\newcommand{\vths}{v_{{\rm th}s}}
\newcommand{\vthi}{v_{{\rm th}i}}
\newcommand{\vthe}{v_{{\rm th}e}}

\newcommand{\vE}{\vec{E}}
\newcommand{\vB}{\vec{B}}

\newcommand{\dn}{\delta n}

\renewcommand{\tt}{\tilde t}

\newcommand{\amp}{\bar}

\newcounter{NoTableEntry}
\renewcommand*{\theNoTableEntry}{NTE-\the\value{NoTableEntry}}

\newcommand{\customlabel}[2]{%
   \protected@write \@auxout {}{\string \newlabel {#1}{{#2}{\thepage}{#2}{#1}{}} }%
   \hypertarget{#1}{#2}
}

\title[]{Electromagnetic instabilities and plasma turbulence driven by electron-temperature gradient}
\author[T.\ Adkins et al.] 
{
T.~Adkins$^{1,2,3}$\thanks{Email: toby.adkins@physics.ox.ac.uk},
A.~A.~Schekochihin$^{1,2}$,
P.~G.~Ivanov$^{3}$,
and
C.~M.~Roach$^{3}$
}

\affiliation{
$^1$Rudolf Peierls Centre for Theoretical Physics, University of Oxford,\\ 
Oxford, OX1 3PU, UK
\\[\affilskip]
$^2$Merton College, Oxford, OX1 4JD, UK
\\[\affilskip]
$^3$Culham Centre for Fusion Energy, United Kingdom Atomic Energy Authority,\\ Abingdon, OX14 3DB, UK
}

\begin{document}

\maketitle

\begin{abstract}
Electromagnetic (EM) instabilities and turbulence driven by the electron-temperature gradient are considered in a local slab model of a tokamak-like plasma. Derived in a low-beta asymptotic limit of gyrokinetics, the model describes perturbations at scales both larger and smaller than the electron inertial length $d_e$, but below the ion Larmor scale $\rho_i$, capturing both electrostatic and EM regimes of turbulence. The well-known electrostatic instabilities --- slab and curvature-mediated ETG --- are recovered, and a new instability is found in the EM regime, called the Thermo-Alfv\'enic instability (TAI). It exists in both a slab version (sTAI, destabilising kinetic Alfv\'en waves) and a curvature-mediated version (cTAI), which is a cousin of the (electron-scale) kinetic ballooning mode (KBM). The cTAI turns out to be dominant at the largest scales covered by the model (greater than $d_e$ but smaller than $\rho_i$), its physical mechanism hinging on the fast equalisation of the total temperature along perturbed magnetic field lines (in contrast to KBM, which is pressure balanced). A turbulent cascade theory is then constructed, with two energy-injection scales: $d_e$, where the drivers are slab ETG and sTAI, and a larger (parallel system size dependent) scale, where the driver is cTAI. The latter dominates the turbulent transport if the temperature gradient is greater than a certain critical value, which scales inversely with the electron beta. The resulting heat flux scales more steeply with the temperature gradient than that due to electrostatic ETG turbulence, giving rise to stiffer transport. This can be viewed as a physical argument in favour of near-marginal steady-state in electron-transport-controlled plasmas (e.g., the pedestal). While the model is simplistic, the new physics that is revealed by it should be of interest to those attempting to model the effect of EM turbulence in tokamak-relevant configurations with high beta and large electron temperature gradients.
\end{abstract}


\section{Introduction}
\label{sec:introduction}
An understanding of the heat transport properties of a magnetically confined plasma is crucial to the design of successful tokamak experiments. Since the characteristic correlation lengthscales associated with the turbulence are small in comparison to the scale of the device, one can usually assume that the turbulence depends only on local equilibrium quantities --- such as density, velocity, temperature, electromagnetic fields --- and their gradients (though there are cases where the global features of the equilibrium can become important: see, e.g., \citealt{hatch21}). Much of the focus of current research is on the turbulence consisting of unstable microscale perturbations, the most important of which are driven either by the ion-temperature gradient (ITG) (see, e.g., \citealt{waltz88,cowley91,kotschenreuther95}) or the electron-temperature gradient (ETG) (see,  e.g., \citealt{dorland00,jenko00}). These perturbations typically live on ion and electron scales, respectively. Strongly driven plasma turbulence --- i.e., plasma turbulence with temperature gradients far above the linear-instability thresholds --- is believed to saturate by reaching a `critically balanced' state (\citealt{barnes11}), where, by analogy with the \cite{K41} theory of hydrodynamic turbulence, free energy injected by linear instabilities is nonlinearly transferred (cascaded) to smaller scales, at which it is thermalised by collisions. If one can determine the turbulent state of the plasma at saturation, then it is, in principle, possible to determine how the turbulent heat fluxes carried by these perturbations depends on the temperature gradients. Knowing this relationship, one can invert it to find the heating power that needs to be provided to support a particular temperature gradient. In many cases, the heat transport found in this context is described as `stiff' (\citealt{wolf03}): the heat flux scales sharply with the temperature gradient, so a large increase in heating power does very little to increase the temperature gradient, making achieving temperature gradients far above marginal a difficult task.  

Though it has long been understood that ion-scale physics can play a significant role in plasma transport (see references above), there is evidence to suggest that instabilities driven by the ion-temperature gradient can be suppressed by strong $\vec{E}\times \vec{B}$ shear in steep-gradient regions of a tokamak (e.g., the pedestal), particularly in spherical or low-aspect-ratio configurations (see \citealt{roach05,roach09,ren17,guttenfelder13,guttenfelder21}, and references therein). This has the effect of reducing the ion contribution to the turbulent heat transport, which instead becomes dominated by the electron channel. This means that the characterisation of electron-scale instabilities is not only worthwhile, but indeed necessary for a complete understanding of the heat transport in such systems. 

Furthermore, a comprehensive understanding of electromagnetic effects on the microinstability properties of the plasma, and the resultant turbulence, is becoming increasingly important as experimental values of the plasma beta (the ratio of the thermal and magnetic pressures) and, therefore, electromagnetic fluctuations, will be higher in reactor-relevant tokamak scenarios; e.g., ITER is projected to have a plasma beta of up to $2.5\%$ (\citealt{shimomura01,sips05}), while this value could exceed $15\%$ in a recently proposed STEP equilibrium (\citealt{patel21}). Though the investigation of electromagnetic instabilitites and turbulence is of general importance within many different types of plasma systems (e.g., astrophysical plasmas, laser plasmas), much of the research in fusion has focused on two particular microinstability classes: micro-tearing modes (MTM) --- initially, in simplified models (\citealt{hazeltine75,drake77,drake80,hassam80a,hassam80b,zocco15,larakers20,larakers21}), later in tokamak geometry (\citealt{applegate07,guttenfelder12mt,dickinson13,predebon13,moraldi13,rafiq16}) --- and kinetic ballooning modes (KBM) (\citealt{tang80,snyder99thesis,snyder01,snyder01gf,pueschel08,pueschel10,waltz10,wan12,wan13,guttenfelder13,ishizawa13,ishizawa14,ishizawa19,terry15,aleynikova17}). Both of these are intrinsically electromagnetic, requiring the ability to perturb the magnetic field's direction and (sometimes) magnitude. Despite significant numerical progress in understanding the behaviour of such modes, however, there is still a certain lack of clarity about the fundamental physical processes that are responsible for them, owing to the complexity of these modes in the general tokamak geometry. Progress in distilling the essential physical ingredients behind electromagnetic destabilisation can be made by means of constructing minimal models.

To this end, in this paper, we consider electromagnetic instabilities and turbulence driven by the electron-temperature gradient in a local slab model of a tokamak-like plasma, with constant equilibrium gradients, including magnetic drifts but not magnetic shear. The inclusion of the finite gradient and curvature of the magnetic field --- in addition to the conventional slab geometry (see, e.g., \citealt{howes06}) --- is motivated by recent evidence (\citealt{abel18,parisi20}) that the modes mediated by these equilibrium quantities can often be the fastest-growing ones in steep-gradient regions of the plasma (e.g., the pedestal), and thus significant in determining its nonlinear saturated state. The governing equations are derived in the low-beta asymptotic limit of gyrokinetics (see, e.g., \citealt{abel13}), and describe perturbations on scales both larger and smaller than the electron inertial scale~$d_e$, at which flux unfreezes, capturing both electrostatic and electromagnetic regimes of turbulence. Formally, the electron beta is ordered as $\beta_e \sim m_e/m_i$ (electron-ion mass ratio), while perpendicular wavenumbers are ordered as $\rho_i^{-1} \lesssim k_\perp \sim d_e^{-1} \ll \rho_e^{-1}$ (sub-ion-Larmor scales). The ordering is discussed in detail in \apref{app:low_beta_ordering}.

At appropriately short perpendicular wavelengths (below the $d_e$ scale), we recover the well-known, electrostatic slab electron-temperature-gradient (sETG, \citealt{liu71,lee87}) and curvature-mediated ETG (cETG, \citealt{horton88}) instabilities. Turning our attention to longer perpendicular wavelengths (above the $d_e$ scale, but still smaller than the ion gyroradius), we demonstrate the existence of the novel Thermo-Alfv\'enic instability (TAI) that arises in the electromagnetic regime. We show that it exists in both a slab version (sTAI, destabilising kinetic Alfv\'en waves) and a curvature-mediated version (cTAI), the latter of which is related to the (electron-scale version of) the KBM. In particular, we find that cTAI is the dominant instability on the largest scales covered by the model, with a maximum growth rate that is greater than that of the cETG. This maximum growth rate occurs at a specific, finite parallel wavenumber, unlike cETG, which is two-dimensional. Its physical mechanism hinges on the fast equalisation of the total temperature along perturbed magnetic field lines (in contrast to the KBM, which is approximately pressure balanced; see, e.g., \citealt{snyder01,kotschenreuther19}) due to the dominance of either parallel streaming (in the collisionless limit) or thermal conduction (in the collisional one). We also show that the sTAI is stabilised at large parallel wavenumbers by compressional heating, and at large perpendicular wavenumbers by the effects of finite eletron inertia (in the collisionless limit) or finite resisitivty (in the collisional one). We then map out all of these instabilities in parallel and perpendicular wavenumber space.

Using a critical-balance phenomenology analogous to \cite{barnes11}, we then construct a turbulent-cascade theory for the free energy injected by these instabilities. Assuming the cascade to be local, the theory is shown to allow two injection scales: $d_e$, where the drivers are sETG and sTAI, and a larger scale dependent on the parallel size of the system (the connection length, in the case of a tokamak), where the principal driver is cTAI. We find that the latter dominates the turbulent transport if the temperature gradient is greater than a certain critical value, which scales inversely with the electron beta. Using constant-flux arguments, we then derive scaling estimates for the turbulent electron heat flux carried by fluctuations at these injection scales, finding that the heat flux due to electromagnetic cTAI turbulence scales more steeply with the temperature gradient than the heat flux due to electrostatic sETG turbulence in this regime, and thus gives rise to stiffer transport. Note that we do not engage with ion physics here, formally assuming that the scale of dominant energy injection for the turbulent cascade lies on sub-Larmor scales.

The rest of the paper is organised as follows. In \secref{sec:low_beta_equations}, we describe and physically motivate our low-beta model equations, in both the collisionless and collisional limits. \Secref{sec:electrostatic_regime_etg} recovers the well-known electrostatic instabilities --- sETG and cETG --- while \secref{sec:electromagnetic_regime_tai} is devoted to the characterisation of the TAI, including a detailed treatments both sTAI and cTAI. \Secref{sec:summary_of_wavenumber_space} is a summary of the asymptotic behaviour of these instabilities in wavenumber space, providing a graphical representation of the linear results of this paper. In \secref{sec:free_energy_and_turbulence}, we construct a cascade theory for the turbulence driven by these instabilities, and derive scaling estimates for the turbulent electron heat fluxes as functions of the electron temperature gradient, parallel system size and the electron beta. Finally, results are summarised and limitations, implications, and future directions are discussed in \secref{sec:discussion}. 

\section{Low-beta equations}
\label{sec:low_beta_equations}
We wish to describe dynamics at electron scales (below the ion Larmor scale) of a magnetised plasma, in the presence of electromagnetic perturbations. Our electron species will have an equilibrium temperature gradient, and will be advected by the magnetic drifts associated with a magnetic geometry of constant curvature. Our equations will be derived in a low-beta asymptotic limit of gyrokinetics; this allows us to order out compressive magnetic field perturbations while retaining Alfv\'enic ones. In this section, we present a summary of these equations and the physical motivation behind them; their detailed derivation can be found in \apref{app:derivation_of_low_beta_equations}. 

\subsection{Magnetic equilibrium and geometry}
\label{sec:magnetic_equilibrium_and_geometry}
\begin{figure}
    \centering
    \begin{tikzpicture}[scale=1, thick, every node/.style={scale=1.2}]
        
            \def\axeslength{2.5}
            \def\xoffset{0.25}
            \def\yoffset{2}
            \def\zoffset{0.1}
            
            \draw[-latex] (\xoffset,\yoffset,\zoffset) -- (\xoffset + \axeslength,\yoffset,\zoffset);
            \draw (\xoffset + \axeslength,\yoffset,\zoffset) node[anchor=north] {$\hat{\vec{x}}$};
            
            \draw[-latex] (\xoffset,\yoffset,\zoffset) -- (\xoffset,\yoffset + \axeslength,\zoffset);
            \draw (\xoffset,\yoffset +\axeslength ,\zoffset) node[anchor=east] {$\hat{\vec{y}}$};
            
            \draw[-latex] (\xoffset,\yoffset,\zoffset) -- (\xoffset,\yoffset,\zoffset + \axeslength);
            \draw (\xoffset,\yoffset,\zoffset + \axeslength) node[anchor=east] {$\hat{\vec{z}}$};
            
            \draw[-] (\xoffset + \axeslength/2,\yoffset + \axeslength/2 ,\zoffset) ellipse [x radius=0.3, y radius =0.3];
            \draw[fill=black] (\xoffset + \axeslength/2,\yoffset +\axeslength/2 ,\zoffset) ellipse [x radius=0.1, y radius =0.1];
            \draw (\xoffset + \axeslength/2 + 0.2,\yoffset + \axeslength/2 + 0.3,\zoffset) node[anchor=west] {$ B_0 \vec{b}_0$};
            
            \def\tscale{0.2}
            \draw[latex-] (\xoffset +\tscale*\axeslength, \yoffset, \zoffset + 0.5) -- (\xoffset + \axeslength - \tscale*\axeslength, \yoffset  , \zoffset +0.5);
            \draw (\xoffset + \axeslength/2, \yoffset , \zoffset +0.5) node[anchor = north] {$\gradd T_{0e}, \gradd B_0$};
            
            \def\currentoffset{-3}
            \def\currentlength{4}
            \def\xradius{3}
            \def\yradius{0.6}

            \begin{scope}
                \clip (\currentoffset-\xradius,-0.2,0) rectangle (\currentoffset + \xradius, 0.3 ,0);
                \draw[thick,-] (\currentoffset,0,0) -- (\currentoffset,\currentlength,0) 
                node[
                currarrow,
                pos=0, 
                xscale=-1,
                sloped,
                scale=1] {};
                \draw (\currentoffset-0.1,0*\currentlength/4,0) node[anchor=east] {$I$};
                                \clip (\currentoffset-\xradius,0.5,0) rectangle (\currentoffset + \xradius, \currentlength ,0);
            \end{scope}
            
            \begin{scope}
                 \clip (\currentoffset-\xradius,0.5,0) rectangle (\currentoffset + \xradius, 1.3 ,0);
                \draw[thick,-] (\currentoffset,0,0) -- (\currentoffset,\currentlength,0);
            \end{scope}
            
             \begin{scope}
                 \clip (\currentoffset-\xradius,1.5,0) rectangle (\currentoffset + \xradius, 2.3 ,0);
                \draw[thick,-] (\currentoffset,0,0) -- (\currentoffset,\currentlength,0);
            \end{scope}
            
            \begin{scope}
                 \clip (\currentoffset-\xradius,2.5,0) rectangle (\currentoffset + \xradius, \currentlength ,0);
                \draw[thick,-] (\currentoffset,0,0) -- (\currentoffset,\currentlength,0);
            \end{scope}
            
            \draw[dashed,|-|] (\currentoffset+0.2,0,0) -- (\currentoffset +\xradius,0,0);
            \draw (\currentoffset +0.2/2 +\xradius/2 ,0,0) node[anchor=north] {$R$};

            
            \begin{scope}
                \clip (\currentoffset-\xradius-0.1,0,0) rectangle (\currentoffset-\xradius+1,\currentlength,0);
            
            \draw[-] (\currentoffset, 2*\currentlength/4, 0) ellipse [x radius = \xradius, y radius = \yradius];
            \draw (\currentoffset-\xradius,2*\currentlength/4,0) node[draw, draw, regular polygon, regular polygon sides=3, fill=black, scale=0.25, rotate=-20] {};
            \end{scope}

             \begin{scope}
                \clip (\currentoffset-\xradius+1.6,0,0) rectangle (\currentoffset-0.1,\currentlength,0);

            \draw[-] (\currentoffset, 2*\currentlength/4, 0) ellipse [x radius = \xradius, y radius = \yradius];
            \draw (\currentoffset-0.2,2*\currentlength/4-\yradius,0) node[draw, draw, regular polygon, regular polygon sides=3, fill=black, scale=0.25, rotate=90] {};
            \end{scope}
            
            \begin{scope}
                \clip (\currentoffset+0.1,0,0) rectangle (\currentoffset+\xradius-1.6,\currentlength,0);

            \draw[-] (\currentoffset, 2*\currentlength/4, 0) ellipse [x radius = \xradius, y radius = \yradius];
            \draw (\currentoffset+0.2,2*\currentlength/4+\yradius,0) node[draw, draw, regular polygon, regular polygon sides=3, fill=black, scale=0.25, rotate=-90] {};
            \end{scope}
            
            \begin{scope}
                \clip (\currentoffset+\xradius-1,0,0) rectangle (\currentoffset+\xradius+0.1,\currentlength,0);

            \draw[-] (\currentoffset, 2*\currentlength/4, 0) ellipse [x radius = \xradius, y radius = \yradius];
            \draw (\currentoffset+\xradius,2*\currentlength/4,0) node[draw, draw, regular polygon, regular polygon sides=3, fill=black, scale=0.25, rotate=160] {};
            \end{scope}
            
             \begin{scope}
                \clip (\currentoffset-\xradius-0.1,0) rectangle (\currentoffset+\xradius+0.1,\currentlength/2,0);

            \draw[-] (\currentoffset, 2*\currentlength/4, 0) ellipse [x radius = \xradius, y radius = \yradius];
            \end{scope}

           \begin{scope}
                \clip (\currentoffset-\xradius,0,0) rectangle (\currentoffset-0.2,\currentlength,0);
                \draw[dotted] (\currentoffset, 3*\currentlength/4, 0) ellipse [x radius = \xradius, y radius = \yradius];
           \end{scope}
           
           \begin{scope}
                \clip (\currentoffset+0.2,0,0) rectangle (\currentoffset+\xradius,\currentlength,0);
                \draw[dotted] (\currentoffset, 3*\currentlength/4, 0) ellipse [x radius = \xradius, y radius = \yradius];
           \end{scope}
           
           \begin{scope}
                \clip (\currentoffset-\xradius,0,0) rectangle (\currentoffset+\xradius,3*\currentlength/4,0);
                \draw[dotted] (\currentoffset, 3*\currentlength/4, 0) ellipse [x radius = \xradius, y radius = \yradius];
           \end{scope}

             \begin{scope}
                \clip (\currentoffset-\xradius-0.2,0,0) rectangle (\currentoffset-\xradius+1,\currentlength,0);
                   \draw[dotted] (\currentoffset, 1*\currentlength/4, 0) ellipse [x radius = \xradius, y radius =\yradius];
            \end{scope}

             \begin{scope}
                \clip (\currentoffset-\xradius+1.6,0,0) rectangle (\currentoffset-0.2,\currentlength,0);
                   \draw[dotted] (\currentoffset, 1*\currentlength/4, 0) ellipse [x radius = \xradius, y radius =\yradius];
            \end{scope}
            
            \begin{scope}
                \clip (\currentoffset+0.2,0,0) rectangle (\currentoffset+\xradius-1.6,\currentlength,0);
                   \draw[dotted] (\currentoffset, 1*\currentlength/4, 0) ellipse [x radius = \xradius, y radius =\yradius];
            \end{scope}
            
            \begin{scope}
                \clip (\currentoffset+\xradius-1,0,0) rectangle (\currentoffset+\xradius+0.2,\currentlength,0);
                  \draw[dotted] (\currentoffset, 1*\currentlength/4, 0) ellipse [x radius = \xradius, y radius =\yradius];
            \end{scope}
            
             \begin{scope}
                \clip (\currentoffset-\xradius,0) rectangle (\currentoffset+\xradius,\currentlength/4,0);
             \draw[dotted] (\currentoffset, 1*\currentlength/4, 0) ellipse [x radius = \xradius, y radius =\yradius];
            \end{scope}

        \end{tikzpicture}
    \caption{Illustration of the constant-curvature geometry, showing the domain positioned at a distance $R$ from the current axis, with the $\hat{\vec{x}}$ and $\hat{\vec{y}}$ directions pointing radially outwards and parallel to this axis, respectively. The equilibrium magnetic field is in the $\vec{b}_0$ direction. Both the equilibrium temperature $T_{0e}$ and equilibrium magnetic field $B_0$ vary radially, with their scale lengths $L_T$ and $L_B$, respectively, assumed constant across the domain.}
    \label{fig:magnetic_geometry}
\end{figure}
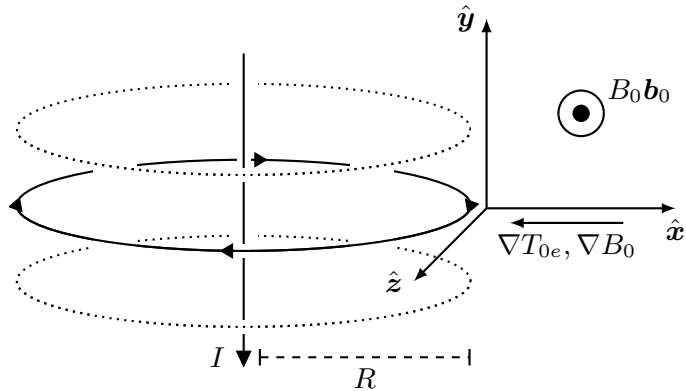

The magnetic geometry that we adopt is one of constant magnetic curvature, as this allows us to capture the effect of the magnetic drifts on our plasma while retaining most of the simplicity associated with conventional slab gyrokinetics (\citealt{howes06}, \citealt{ivanov20}). We consider a domain positioned in the magnetic field of a current line at a radial distance $R$ from the central axis, and define the $\hat{\vec{x}}$ and $\hat{\vec{y}}$ directions as pointing radially outwards and parallel to the central axis, respectively, as shown in \figref{fig:magnetic_geometry}. 
In the context of the outboard midplane in tokamak geometry, these are the `radial' and `poloidal' coordinates, respectively, terms that we shall adopt in our later discussions.
In such a geometry, the magnetic field consists of an equilibrium part that is oriented in the $\vec{b}_0 = \hat{\vec{z}}$ direction and varies radially, plus a time- and space-dependent fluctuating part:
\begin{equation}
    \vec{B}(\vec{r},t) = B_0(x) \vec{b}_0 + \dBperp(\vec{r},t).
    \label{eq:magnetic_field}
\end{equation}
In what follows, we shall express the perpendicular magnetic-field fluctuations in terms of the parallel component of the magnetic vector potential:
\begin{equation}
    \dBperp(\vec{r},t) =\curll \vec{A} =  -\vec{b}_0 \times \gradd A_\parallel.
\end{equation}
The component of the magnetic-field fluctuations parallel to the mean field is negligible in the limit of low beta [see \eqref{eq:ordering_dBpar}]. The electric field is related to the magnetic vector potential $\vec{A}$ and electrostatic potential $\phi$ by
\begin{equation}
    \vec{E} (\vec{r},t) = - \frac{1}{c} \frac{\partial \vec{A}}{\partial t} - \gradd\phi,
    \label{eq:electric_field}
\end{equation}
and is assumed to have no mean part. The equilibrium (mean) magnetic field has the scale length and radius of curvature
\begin{equation}
   L_B^{-1} =- \frac{1}{B_0}\frac{\rmd B_0}{\rmd x}, \quad  R^{-1}  = \left| \vec{b}_0 \cdot \gradd \vec{b}_0\right|,
    \label{eq:magnetic_field_scale_length}
\end{equation}
respectively, both of which are assumed to be constant across our domain. Note that for a low-beta plasma, $R = L_B$ (see \apref{app:equilibrium_and_fluctuations}). We assume that the background gradient of the temperature $T_{0e}$ associated with the equilibrium distribution of the electrons also varies radially, with scale length
\begin{equation}
    L_T^{-1} = - \frac{1}{T_{0e}} \frac{\rmd T_{0e}}{\rmd x},
    \label{eq:equilibrium_gradients}
\end{equation}
which, similarly, is assumed to be constant over the domain. The thermal speed of the electrons is then given by $\vthe = \sqrt{2T_{0e}/m_e}$, where $m_e$ is the electron mass.

With this local equilibrium, and adopting a low-beta ordering (see \apref{app:low_beta_ordering}), we derive evolution equations for the density ($\dne$), parallel velocity ($u_{\parallel e }$), parallel temperature ($\delta T_{\parallel e }$) and perpendicular temperature ($\delta T_{\perp e }$) perturbations of the electrons. 
These equations are presented in the following sections.
We assume everywhere that the electron Larmor radius $\rho_e$ is small, and so work in the drift-kinetic approximation for the electrons.

\subsection{Density perturbations}
\label{sec:density_perturbations}
The perturbed electron density satisfies the continuity equation:
\begin{align}
    \frac{\rmd}{\rmd t} \frac{\dne}{n_{0e}} + \gradd_\parallel u_{\parallel e} + \frac{\rho_e \vthe}{2L_{B}} \frac{\partial}{\partial y} \left( \frac{\delta T_{\parallel e}}{T_{0e}} + \frac{\delta T_{\perp e}}{T_{0e}} \right) = 0.
    \label{eq:density_moment_initial}
\end{align}
This says that the density perturbation is subject to three influences: (i) advection by the $\vec{E}\times \vec{B}$ motion of the electrons,
\begin{align}
    \frac{\rmd }{\rmd t}  = \frac{\partial}{\partial t} + \vec{v}_E \cdot \gradd_\perp , \quad \vec{v}_E = \frac{\rho_e \vthe}{2} \vec{b}_0 \times \gradd_\perp \varphi, \quad \varphi = \frac{e \phi}{T_{0e}},
    \label{eq:convective_derivative}
\end{align}
where $-e$ is the electron charge; (ii) compression or rarefaction due to the perturbed parallel electron flow $u_{\parallel e} \vec{b}$ along the exact magnetic field, including the perturbation of the magnetic field direction:
\begin{align}
     \gradd_\parallel = \vec{b} \cdot \gradd = \frac{\partial}{\partial z} + \frac{\dB_\perp}{B_0} \cdot \gradd_\perp , \quad \frac{\dB_\perp}{B_0} = - \rho_e \vec{b}_0 \times \gradd_\perp \mathcal{A}, \quad  \mathcal{A} = \frac{A_\parallel}{\rho_e B_0};
    \label{eq:parallel_derivative}
 \end{align}
(iii) the magnetic drifts due to the finite curvature and gradient of the magnetic field. The parallel and perpendicular temperature perturbations arise from the velocity dependence of the curvature and $\gradd B$ drifts in the gyrokinetic equation [see \eqref{eq:magnetic_drifts}]. The presence of these magnetic drifts is essential for the curvature-mediated instabilities that will be the focus of \secref{sec:curvature_ETG} and much of \secref{sec:electromagnetic_regime_tai}.

The continuity equation \eqref{eq:density_moment_initial} is \eqref{eq:g00} in \apref{app:electron_equations}, except in \eqref{eq:density_moment_initial} we have set the equilibrium density gradient to zero, and ignored the magnetic-drift terms proportional to $\dne/n_{0e}$ and $\varphi$, as they will always turn out to be smaller than the magnetic-drift terms proportional to the temperature perturbations in what follows. This is in a bid to make our equations as simple as possible, while retaining all of the relevant physics (see~\apref{app:strongly_driven_limit} for further details). We shall ignore similar terms in our other equations for the perturbations, for the same reason. Cautious readers may be reassured by the fact that all of the instabilities considered in sections \ref{sec:electrostatic_regime_etg} and \ref{sec:electromagnetic_regime_tai} are derived in a limit in which this is a valid approximation.

\subsection{Parallel velocity perturbations}
\label{sec:parallel_velocity_perturbations}
The parallel momentum equation associated with the electrons is [see \eqref{eq:g01}]
\begin{align}
    n_{0e} m_e  \frac{\rmd u_{\parallel e}}{\rmd t}  = - \gradd_\parallel p_{\parallel e} - en_e E_\parallel - \nu_{ei}m_e   u_{\parallel e},
    \label{eq:parallel_momentum}
\end{align}
The three forces appearing on the right-hand side are, from right to left: (i) the collisional drag against the ions (which are assumed motionless), where $\nu_{ei}$ is the electron-ion collision frequency, (ii) the parallel electric field 
\begin{align}
    E_\parallel  = \vec{b}\cdot \vec{E} = - \left( \frac{1}{c} \frac{\partial A_\parallel}{\partial t} + \gradd_\parallel \phi \right) = - \left( \frac{1}{c} \frac{\rmd A_\parallel}{\rmd t} + \frac{\partial \phi}{\partial z} \right),
    \label{eq:parallel_electric_field}
\end{align}
and (iii) the parallel pressure gradient, which consists of both the parallel gradient of the parallel-pressure perturbation and the projection of the equilibrium temperature gradient onto the perturbed magnetic field:
\begin{align}
    \gradd_\parallel p_{\parallel e} = \gradd_\parallel \delta p_{\parallel e} +  n_{0e} \frac{\delta B_x}{B_0} \frac{\rmd T_{0e}}{\rmd x} =  n_{0e}  T_{0e} \left[ \gradd_\parallel \left( \frac{\dne}{n_{0e}} + \frac{\delta T_{\parallel e}}{T_{0e}}\right) - \frac{\rho_e}{L_T} \frac{\partial \mathcal{A}}{\partial y} \right].
    \label{eq:pressure_gradient}
\end{align}
Since an electron flow uncompensated by an ion flow is a current, $u_{\parallel e}$ is related to $A_\parallel$ via Amp\`ere's law [see \eqref{eq:parallel_amperes_law_final}]:
\begin{align}
    - e n_{0e} u_{\parallel e} = j_{\parallel} = \frac{c}{4\pi} \vec{b}_0 \cdot \left( \gradd_\perp \times \dB_\perp \right) \quad \Rightarrow \quad u_{\parallel e} = \vthe d_e^2 \gradd_\perp^2 \mathcal{A}, 
    \label{eq:amperes_law_initial}
\end{align}
where $c$ is the speed of light, $d_e = c(m_e/4\pi e^2 n_{0e})^{1/2} = \rho_e /\sqrt{\beta_e}$ is the electron inertial scale, and $\beta_e = 8\pi  n_{0e} T_{0e}/B_0^2$. The electron inertial scale $d_e$ will be an important quantity throughout this work, as it demarcates the boundary between the electrostatic and electromagnetic regimes in the collisionless limit (see \secref{sec:flux_freezing}). In the collisional limit [see \eqref{eq:ordering_collisional_frequencies}], the frictional term on the right-hand side of \eqref{eq:parallel_momentum} dominates over the electron inertial term on the left-hand side, meaning that the electron inertia can be neglected. 

\subsection{Temperature perturbations}
\label{sec:temperature_perturbations}
The parallel temperature $T_{\parallel e} = T_{0e} + \delta T_{\parallel e}$ is advected by the local $\vec{E} \times \vec{B}$ flow and is locally increased (or decreased) by the compressional heating (or rarefaction cooling) due to $u_{\parallel e}$, as well as by the (appropriately normalised) perturbed parallel heat flux $\delta q_{\parallel e}$ [see \eqref{eq:g02}]:
\begin{align}
    \frac{\rmd  T_{\parallel e}}{\rmd t} = \frac{\rmd  \delta T_{\parallel e}}{\rmd  t} + \vec{v}_E \cdot \gradd_\perp T_{0e} = -  \gradd_\parallel \frac{\delta q_{\parallel e}}{n_{0e} } - 2 T_{0e} \gradd_\parallel u_{\parallel e} - \frac{4}{3} \nu_e\left(\delta T_{\parallel e} - \delta T_{\perp e} \right).
    \label{eq:parallel_temperature_initial}
\end{align}
The factor of $2$ in the compressional-heating term (the second on the right-hand side) is due to the fact that we only consider the parallel (1D) motion of the electrons [perpendicular motions are formally small within our ordering: see \eqref{eq:ordering_lengthscales}]. The last term on the right-hand side is a consequence of our choice of collision operator [see \eqref{eq:collision_operator} and the subsequent discussion], and is responsible for collisional temperature isotropisation, with $\nu_e = \nu_{ee} + \nu_{ei}$, and $\nu_{ee} = \nu_{ei}/Z$ the electron-electron collision frequency ($Ze$ is the ion charge). 

Similarly, the perpendicular temperature $T_{\perp e} = T_{0e} + \delta T_{\perp e}$ evolves according to [see~\eqref{eq:g10}]
\begin{align}
    \frac{\rmd  T_{\perp e}}{\rmd t} = \frac{\rmd\delta T_{\perp e}}{\rmd t} + \vec{v}_E \cdot \gradd_\perp T_{0e} = -  \gradd_\parallel \frac{\delta q_{\perp e}}{n_{0e}}  - \frac{2}{3}\nu_e\left(\delta T_{\perp e} - \delta T_{\parallel e} \right),
    \label{eq:perpendicular_temperature_initial}
\end{align}
where $\delta q_{\perp e}$ is the perturbed perpendicular heat flux. Note the absence of perpendicular compressional heating (perpendicular flows are incompressible).

The term expressing the seeding of both parallel and perpendicular temperature perturbations via the advection of the equilibrium temperature profile by the $\vE\times\vB$ flow becomes, after a straightforward manipulation, the familiar (electrostatic) linear drive responsible for extracting (`electrostatic') free energy from the equilibrium temperature gradient: 
\begin{align}
    \vec{v}_E \cdot \gradd_\perp T_{0e} = T_{0e} \frac{\rho_e \vthe}{2 L_T} \frac{\partial \varphi}{\partial y},
    \label{eq:linear_drive_phi}
\end{align}
where $L_T$ is defined in \eqref{eq:equilibrium_gradients}.
In order to determine the heat fluxes $\delta q_{\parallel e}$ and $\delta q_{\perp e}$, kinetic theory is needed, and so we must append to our emerging system of equations the drift-kinetic equation for electrons (see \apref{app:electron_equations}), of which \eqref{eq:density_moment_initial}, \eqref{eq:parallel_momentum}, \eqref{eq:parallel_temperature_initial} and \eqref{eq:perpendicular_temperature_initial} are four lowest-order moments. 

In the collisional limit [see \eqref{eq:ordering_collisional_frequencies}], the temperature isotropisation terms in \eqref{eq:parallel_temperature_initial} and \eqref{eq:perpendicular_temperature_initial} are dominant, enforcing $\delta T_{\parallel e} = \delta T_{\perp e} = \delta T_{e}$ to leading order. In this limit, therefore, we no longer distinguish between parallel and perpendicular temperature perturbations, and obtain an equation for $\delta T_{e}$ from the linear combination $(1/3)$\eqref{eq:parallel_temperature_initial}$+(2/3)$\eqref{eq:perpendicular_temperature_initial} [see \eqref{eq:col_temperature_linear_combination} and \eqref{eq:col_temperature_equation}]:
\begin{align}
     \frac{\rmd \dTe}{\rmd t} + \vec{v}_E \cdot \gradd_\perp T_{0e} = -  \frac{2}{3} \gradd_\parallel \frac{\delta q_e}{n_{0e} } - \frac{2}{3} T_{0e} \gradd_\parallel {u_{\parallel e}} , \label{eq:collisional_temperature_initial}
\end{align}
where the (collisional) heat flux $\delta q_e = \delta q_{\parallel e}/2 + \delta q_{\perp e}$ can be expressed in terms of the parallel gradient of the total temperature $T_e = T_{0e} + \dTe$ along the exact magnetic field direction [see \eqref{eq:col_combined_heat_flux}]:
\begin{align}
     \frac{\delta q_{e}}{n_{0e} T_{0e} } = - \frac{3}{2}\kappa \gradd_\parallel \log T_e, \quad \kappa = \frac{5\vthe^2}{18 \nu_e},
     \label{eq:collisional_heat_flux}
\end{align}
where $\kappa$ is the electron thermal diffusivity and
\begin{align}
    \gradd_\parallel \log T_e =  \gradd_\parallel \frac{\dTe}{T_{0e}} + \frac{\delta B_x }{B_0} \frac{1}{T_{0e}} \frac{\rmd T_{0e}}{\rmd x} =  \gradd_\parallel \frac{\dTe}{T_{0e}} - \frac{\rho_e}{L_T} \frac{\partial \mathcal{A}}{\partial y}
    \label{eq:logt_definition_intro}
\end{align}
is the parallel gradient of the \textit{total} electron temperature, which will prove a key quantity in what follows.

\subsection{Quasineutrality}
\label{sec:quasineutrality}
Finally, as usual, particle density is related to $\phi$ via quasineutrality, which is the route 
whereby ions contribute to dynamics. Since, at scales smaller than their 
Larmor radius $\sim \rho_i$, ions can be viewed as large motionless rings of charge, their density 
response is Boltzmann:
\begin{align}
    \frac{\dne}{n_{0e}}  = \frac{\delta n_i}{n_{0i}} = - \frac{Ze \phi}{T_{0i}}  = - \taubar^{-1} \varphi, \quad \taubar = \frac{\tau}{Z},
    \label{eq:quasineutrality_initial}
\end{align}
where $\tau = T_{0i}/T_{0e}$ is the ratio of the ion to electron equilibrium temperatures. The more general quasineutrality closure, for which $\taubar^{-1}$ is an operator and which includes scales comparable to the ion Larmor radius, is given in \apref{app:ion_equations}, but, since we shall focus on scales smaller than this in our discussions, \eqref{eq:quasineutrality_initial} is sufficient for our purposes. 

\subsection{Summary of equations}
\label{sec:summary_of_equations}
Assembling together all of the above, we end up with the following systems of equations,  in the collisionless limit [see \eqref{eq:ordering_timescales} with $\nu_{ei},\nu_{ee} \rightarrow 0$]:
\begin{align}
     &\frac{\rmd}{\rmd t} \frac{\dne}{n_{0e}} + \gradd_\parallel u_{\parallel e} + \frac{\rho_e \vthe}{2L_{B}} \frac{\partial}{\partial y} \left( \frac{\delta T_{\parallel e}}{T_{0e}} + \frac{\delta T_{\perp e}}{T_{0e}} \right) = 0, \label{eq:density_moment}\\
     & \frac{\rmd}{\rmd t} \left( \mathcal{A} - \frac{u_{\parallel e}}{\vthe} \right) = - \frac{\vthe}{2} \left[\frac{\partial \varphi}{\partial z} - \gradd_\parallel \left( \frac{\dne}{n_{0e}}+ \frac{\delta T_{\parallel e}}{T_{0e}} \right) + \frac{\rho_e}{L_T} \frac{\partial \mathcal{A}}{\partial y} \right] , \label{eq:velocity_moment} \\
     & \frac{\rmd }{\rmd t} \frac{\delta T_{\parallel e}}{T_{0e}} +  \gradd_\parallel \left(\frac{\delta q_{\parallel e}}{n_{0e} T_{0e} } + 2 u_{\parallel e} \right) + \frac{\rho_e \vthe}{2 L_T} \frac{\partial \varphi}{\partial y}  =0,\label{eq:tpar_moment}\\
     & \frac{\rmd}{\rmd t} \frac{\delta T_{\perp e}}{T_{0e}} +  \gradd_\parallel \frac{\delta q_{\perp e}}{n_{0e} T_{0e}} + \frac{\rho_e \vthe}{2 L_T} \frac{\partial \varphi}{\partial y} =0,\label{eq:tperp_moment}
\end{align}
or, in the collisional limit [see \eqref{eq:ordering_collisional_frequencies}],
\begin{align}
     &\frac{\rmd}{\rmd t} \frac{\dne}{n_{0e}} + \gradd_\parallel u_{\parallel e} + \frac{\rho_e \vthe}{L_{B}} \frac{\partial}{\partial y} \frac{\delta T_{ e}}{T_{0e}}  = 0, \label{eq:density_moment_collisional}\\
     & \frac{\rmd \mathcal{A}}{\rmd t} +\frac{\vthe}{2} \frac{\partial \varphi}{\partial z} =  \frac{\vthe}{2} \left( \gradd_\parallel  \frac{\dne}{n_{0e}} + \gradd_\parallel \log T_e \right) + \nu_{ei} \frac{u_{\parallel e}}{\vthe}, \label{eq:velocity_moment_collisional} \\
     & \frac{\rmd}{\rmd t} \frac{\dTe}{T_{0e}}  -   \kappa \gradd_\parallel^2 \log T_e +  \frac{2}{3} \gradd_\parallel u_{\parallel e}  +  \frac{\rho_e \vthe}{2 L_{T}} \frac{\partial \varphi}{\partial y} =0 \label{eq:t_moment_collisional},
\end{align}
to which we append the field equations:
\begin{align}
    \frac{\dne}{n_{0e}} = - \taubar^{-1} \varphi, \quad \frac{u_{\parallel e}}{\vthe} = d_e^2 \gradd_\perp^2 \mathcal{A}. \label{eq:field_equations}
\end{align}
This system is a minimal model for describing low-beta electromagnetic plasma dynamics --- whether collisionless or collisional --- driven by a background electron temperature gradient, and in the presence of magnetic drifts. 

\subsection{Flux-freezing}
\label{sec:flux_freezing}
These equations describe two broad classes of physical phenomena: electrostatic and electromagnetic, distinguished by whether the magnetic field lines are frozen into the electron flow or not. We shall refer to the perpendicular scale at which the transition between these two regimes occurs as the `flux-freezing scale'.
In the collisionless limit, this scale is given by the balance between the electron inertia and the inductive parallel electric field on the left-hand side of \eqref{eq:velocity_moment}, viz.,
\begin{align}
    k_\perp d_e \sim 1.
    \label{eq:flux_freezing_scale}
\end{align}
In the collisional limit, the analogous balance involves, instead of electron inertia, the resistive term --- the last on the right-hand side of \eqref{eq:velocity_moment_collisional}. However, in this limit, we shall always deal with perturbations for which the term in \eqref{eq:velocity_moment_collisional} that contains the projection of the equilibrium temperature gradient onto the perturbed magnetic field [the second part of $\gradd_\parallel \log T_e$ written in \eqref{eq:logt_definition_intro}] is larger than $\partial \mathcal{A}/\partial t$. Therefore, it is with this term that the effect of resistivity will be usefully compared:
\begin{align}
    \omega \lesssim \omega_{*e} \equiv \frac{k_y \rho_e \vthe}{2 L_T} \sim  k_\perp^2 d_e^2 \nu_{ei} ,
    \label{eq:flux_freezing_scale_col_comparison}
\end{align}
where $\omega_{*e}$ is the drift frequency associated with the electron-temperature gradient. For modes with $k_y \sim k_\perp$, the balance \eqref{eq:flux_freezing_scale_col_comparison} can be written as 
\begin{align}
    k_\perp d_e \sim \frac{\rho_e}{d_e} \frac{\vthe}{ L_T \nu_{ei}} = \sqrt{\beta_e} \frac{\lambdae}{L_T} \equiv \chi^{-1},
    \label{eq:flux_freezing_scale_col}
\end{align}
where $\lambdae = \vthe/\nu_e$ is the electron mean free path. It is the scale at which $k_\perp d_e \chi \sim 1$ that will effectively play the role of the flux-freezing scale in the collisional limit. Note that $\chi^{-1}\ll 1$ [see \eqref{eq:ordering_of_chi}], meaning that the flux-freezing scale occurs at much longer perpendicular wavelengths than in the collisionless limit.

We shall refer to scales below the flux-freezing scale \eqref{eq:flux_freezing_scale} or \eqref{eq:flux_freezing_scale_col} as electrostatic scales (on which electrons are free to flow across field lines without perturbing them), and to scales above the flux-freezing scale as electromagnetic scales (on which the magnetic field is frozen into the electron flow). In \apref{app:magnetic_flux_conservation}, we show that the electron flow into which the magnetic field lines are frozen on electromagnetic scales (while still remaining below the ion Larmor scale) is given by
\begin{align}
     \vec{u}_\text{eff} = \vec{v}_E - \frac{\rho_e \vthe}{2} \frac{\vec{b}_0 \times \gradd p_{\parallel e}}{n_{0e} T_{0e}},
    \label{eq:flux_freezing_velocity}
\end{align}
where $p_{\parallel e} = n_e T_{\parallel e}$, $n_e = n_{0e} + \delta n_e$, and $T_{\parallel e} = T_{0e} + \delta T_{\parallel e}$ are the total parallel pressure, density, and parallel temperature, respectively; in the collisional limit, $\delta T_{\parallel e} \rightarrow \delta  T_e$, as in \secref{sec:temperature_perturbations}. The flow \eqref{eq:flux_freezing_velocity} is simply the part of the electron flow velocity perpendicular to the total magnetic field $\vec{B}$, comprised of the usual $\vec{E}\times \vec{B}$ drift velocity $\vec{v}_E$ [see \eqref{eq:convective_derivative}], and a `diamagnetic' contribution coming from the electron (parallel) pressure gradient, manifest in the right-hand side of \eqref{eq:velocity_moment} or \eqref{eq:velocity_moment_collisional}. This is distinct from the MHD limit (above the ion Larmor scale), in which the magnetic field is only frozen into $\vec{v}_E$ due to the dynamics being pressure balanced, a distinction that will prove important in our considerations of electromagnetic instabilities in \secref{sec:electromagnetic_regime_tai}. 

In what follows, all orderings introduced should be considered subsidiary to the orderings that define the collisionless and collisional limits [\eqref{eq:ordering_timescales}, with $\nu_{ee},\nu_{ei} \rightarrow 0$, and \eqref{eq:ordering_collisional_frequencies}, respectively], and the resultant reduced equations thus to be particular limits of the collisionless [\eqref{eq:density_moment}-\eqref{eq:tperp_moment}] or collisional [\eqref{eq:density_moment_collisional}-\eqref{eq:t_moment_collisional}] equations. 

\section{Electrostatic regime: electron temperature gradient instability}
\label{sec:electrostatic_regime_etg}
Let us begin by examining the more familiar instabilities that occur at electrostatic scales, before considering what happens at electromagnetic ones.

\subsection{Collisionless slab ETG}
\label{sec:setg}
As explained above, the electrostatic limit corresponds to perpendicular scales $k_\perp d_e \gg 1$. If we strengthen this condition to [cf. \eqref{eq:flux_freezing_scale_col_comparison}]
\begin{align}
    k_\perp^2 d_e^2 \gg  \frac{\omega_{*e}}{\omega} \gtrsim 1,
    \label{eq:setg_de}
\end{align}
then both $\mathcal{A}$ terms in \eqref{eq:velocity_moment} can be neglected in comparison with the electron inertia. Furthermore, we would like to consider the slab approximation, in which the magnetic drifts are negligible in comparison with parallel compressions. In terms of wavenumbers, this means that we assume
\begin{equation}
    k_\parallel \vthe \gg \omega_{de} \left( \frac{L_B}{L_T} \right)^{1/4}, \quad \omega_{de} = \frac{k_y \rho_e \vthe}{2 L_B},
    \label{eq:setg_kpar}
\end{equation}
where $\omega_{de}$ is the magnetic drift frequency. Though not immediately obvious, it shall turn out that the limit \eqref{eq:setg_kpar} allows us to neglect the magnetic drifts in \eqref{eq:density_moment} [this follows from comparing the sizes of the last two terms in \eqref{eq:lin_es_dispersion_relation} under the ordering \eqref{eq:lin_electrostatic_limit_ordering}]. Then, the perpendicular temperature perturbation \eqref{eq:tperp_moment} becomes decoupled from the remaining equations, leaving us with an electrostatic three-field ($\dne$, $u_{\parallel e}$ and $\delta T_{\parallel e}$) system of the kind traditionally used to describe temperature-gradient instabilities in a slab (\citealt{cowley91}). The slab electron-temperature-gradient (sETG) instability (\citealt{liu71,lee87}) in its most explicit, fluid form is obtained if one further assumes [see \eqref{eq:lin_electrostatic_limit_ordering}]
\begin{align}
    k_\parallel \vthe \ll \omega \ll \omega_{*e}.
    \label{eq:setg_fluid_approx}
\end{align}
Then \eqref{eq:density_moment}-\eqref{eq:tpar_moment} reduce to, approximately,
\begin{align}
    \frac{\rmd }{\rmd t} \taubar^{-1} \varphi = \gradd_\parallel u_{\parallel e}, \quad \frac{\rmd u_{\parallel e}}{\rmd t} = - \frac{\vthe^2}{2} \gradd_\parallel \frac{\delta T_{\parallel e}}{T_{0e}}, \quad \frac{\rmd }{\rmd t} \frac{\delta T_{\parallel e}}{T_{0e}} = - \frac{\rho_e \vthe}{2L_T} \frac{\partial \varphi}{\partial y}.
    \label{eq:setg_equations}
\end{align}
Linearising and Fourier-transforming, we find the familiar dispersion relation [see \eqref{eq:lin_setg}]
\begin{align}
    \omega^3 = - \frac{k_\parallel^2 \vthe^2 \omega_{*e} \taubar}{2} \quad \Rightarrow \quad \omega  = \sgn(k_y) \left( -1, \frac{1}{2} \pm i \frac{\sqrt{3}}{2} \right) \left( \frac{k_\parallel^2 \vthe^2 |\omega_{*e}| \taubar}{2}\right)^{1/3}.
    \label{eq:setg_gamma}
\end{align}
The unstable root is the collisionless sETG. 

\begin{figure}
    \centering
    
\input{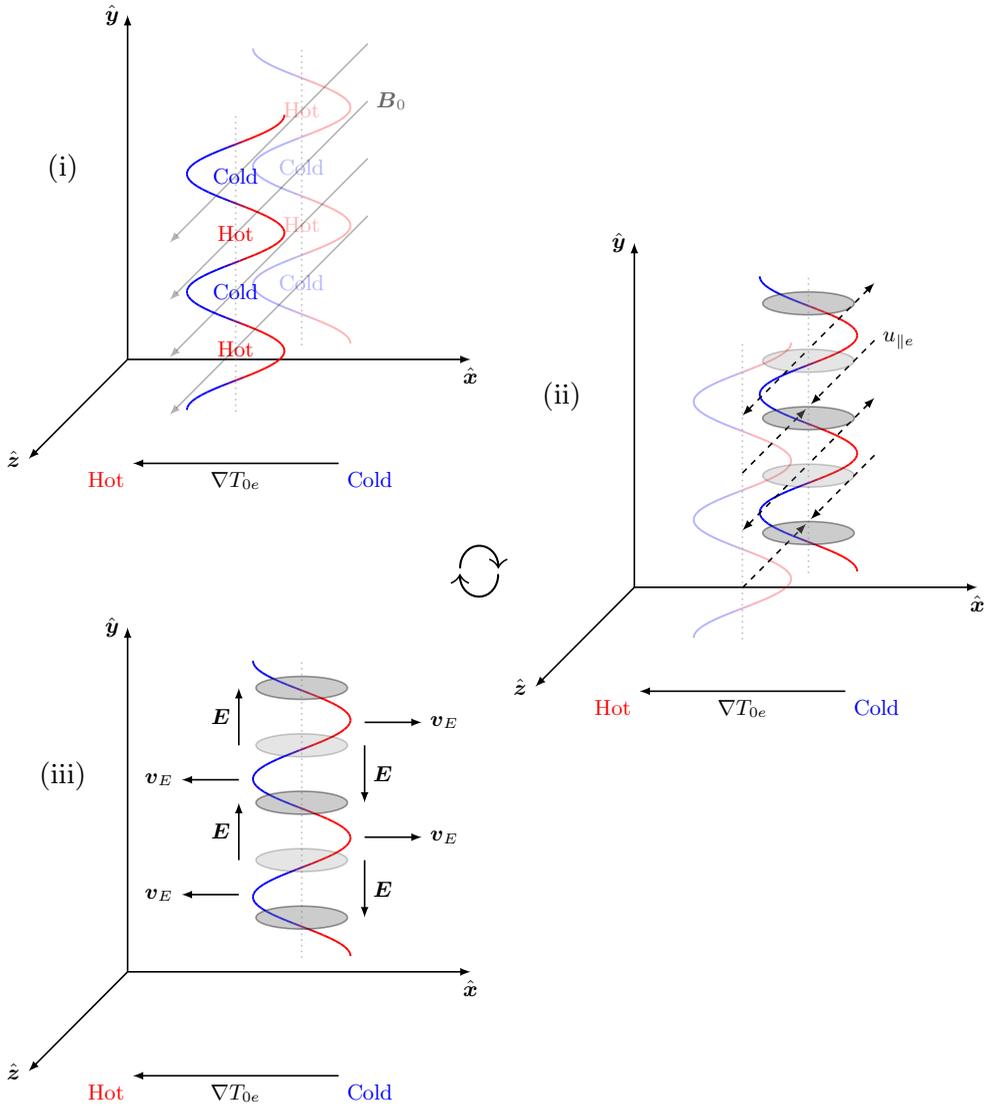}

    \caption{A cartoon illustrating the feedback mechanism of the (collisionless) sETG instability. (i) An electron temperature perturbation with $k_y \neq 0$ and $\kpar \neq 0$ (red-and-blue curves) has alternating hold and cold regions along the (unperturbed) magnetic field (grey arrows), and also along $\hat{\vec{y}}$. (ii) The resulting perturbed temperature gradients drive parallel electron flows $u_{\parallel e}$ (dashed arrows) from the hot regions into the cold regions, giving rise to increased electron density in the cold regions (over- and under- densities are indicated by the dark- and light-grey ellipses, respectively). (iii) By quasineutrality, the electron-density perturbation gives rise to an exactly equal ion-density perturbation, and that, via Boltzmann-ion response, creates alternating electric fields $\vec{E}$ in the perpendicular plane (vertical black arrows). This produces an $\vec{E}\times \vec{B}$ drift $\vec{v}_E$ (horizontal black arrows), which pushes hotter particles further into the colder region, and vice versa, reinforcing the initial temperature perturbation and thus completing the positive feedback loop required for the instability. In this cartoon, for the sake of simplicity, we have chosen not to include the phase information between the various perturbations involved; a reader seeking such information will find it in figure 1 of \cite{cowley91} (the equivalent picture for ITG).}
    \label{fig:setg}
\end{figure}

In this limit, the instability works as follows. Suppose that a small perturbation to the parallel electron temperature is created with $k_y \neq 0$ and $k_\parallel \neq 0$, bringing the plasma from regions with higher $T_{0e}$ to those with lower $T_{0e}$ ($\delta T_{\parallel e} >0 $), and vice versa ($\delta T_{\parallel e} < 0 $). This temperature perturbation produces alternating hot and cold regions along the (unperturbed) magnetic field. The resulting perturbed temperature gradients drive electron flows from the hot regions to the cold regions [second equation in \eqref{eq:setg_equations}], giving rise to increased electron density in the cold regions [first equation in \eqref{eq:setg_equations}]. By quasineutrality, the electron density perturbation is instantly replicated in the ion density perturbation, and that, via Boltzmann-ion response, creates an electric field that produces a radial $\vec{E} \times \vec{B}$ drift that pushes hotter particles further into the colder region, and vice versa [third equation in \eqref{eq:setg_equations}], reinforcing the initial temperature perturbation and thus completing the positive feedback loop required for the instability. This is illustrated in \figref{fig:setg}.

The `fluid' limit \eqref{eq:setg_fluid_approx} is physically transparent and easy to handle, primarily because the heat flux (and thus all kinetic effects, such as Landau damping, \citealt{landau46}) can, in this limit, be neglected in \eqref{eq:tpar_moment}. However, the approximation contains the seed of its own destruction: according to \eqref{eq:setg_gamma}, perturbations with a larger $k_\parallel$ grow faster, and can only be checked by Landau damping when [see \eqref{eq:lin_stab_setg}]
\begin{align}
    k_\parallel \vthe \sim  \omega \sim \omega_{*e}.
    \label{eq:setg_max}
\end{align}
Thus, the fastest-growing collisionless sETG modes are expected to sit in this latter, kinetic regime.

\subsection{Collisional slab ETG}
\label{sec:col_setg}
An important difference between the collisionless and collisional limits, exemplified by the form of the collisonal heat flux \eqref{eq:collisional_heat_flux}, is the replacement of the parallel streaming rate of electrons $k_\parallel \vthe$ with the parallel conduction rate $(k_\parallel \vthe)^2/\nu_{ei}$. The collisional analogues of \eqref{eq:setg_de}, \eqref{eq:setg_kpar} and \eqref{eq:setg_fluid_approx} are thus [see \eqref{eq:lin_col_es_short_wavelength_ordering}]
\begin{align}
     k_\perp^2  d_e^2  \gg  \frac{\omega_{*e}}{\nu_{ei}}, \quad \frac{(k_\parallel \vthe)^2}{ \nu_{ei}} \gg \omega_{de}, \quad \frac{(k_\parallel \vthe)^2}{ \nu_{ei}} \ll \omega \ll \omega_{*e},
         \label{eq:col_setg_fluid_approx}
\end{align}
respectively, for which \eqref{eq:density_moment_collisional}, \eqref{eq:velocity_moment_collisional} and \eqref{eq:t_moment_collisional} reduce to, approximately, 
\begin{align}
    \frac{\rmd }{\rmd t} \taubar^{-1} \varphi = \gradd_\parallel u_{\parallel e}, \quad \nu_{ei} u_{\parallel e} = - \frac{\vthe^2}{2 } \gradd_\parallel \frac{\delta T_{ e}}{T_{0e}}, \quad \frac{\rmd }{\rmd t} \frac{\delta T_{e}}{T_{0e}} = - \frac{\rho_e \vthe}{2L_T} \frac{\partial \varphi}{\partial y}.
    \label{eq:col_setg_equations}
\end{align}
These equations are similar to \eqref{eq:setg_equations}, except the parallel temperature gradient is now balanced by the electron-ion frictional force, rather than by electron inertia. The dispersion relation is [see \eqref{eq:lin_col_es_setg}]
\begin{align}
    \omega^2 = i \frac{k_\parallel^2 \vthe^2\omega_{*e} \taubar}{2\nu_{ei}} \quad \Rightarrow  \quad \omega = \pm \frac{1-i\sgn(k_y)}{\sqrt{2}}\left(\frac{k_\parallel^2 \vthe^2|\omega_{*e}| \taubar}{2\nu_{ei}} \right)^{1/2},
    \label{eq:col_setg_gamma}
\end{align}
where the unstable root is the collisional sETG. The physical mechanism of the instability is analogous to that for the collisionless sETG, except the parallel electron flow is now determined instantaneously by the parallel temperature gradient. Similarly to the collisionless sETG, the point of maximum growth of the instability occurs when 
\begin{align}
     \frac{(k_\parallel \vthe)^2}{ \nu_{ei}} \sim \omega \sim \omega_{*e},
     \label{eq:col_setg_max}
\end{align}
which is a balance between dissipation (through conduction, rather than Landau damping) and energy injection due to the background temperature gradient [see \eqref{eq:lin_col_es_stability_boundary} and the following discussion].

\subsection{Curvature-mediated ETG}
\label{sec:curvature_ETG}
Both the collisionless and collisional sETG instabilities were derived assuming that the parallel wavelengths were sufficiently short for the compressional terms in \eqref{eq:density_moment} and \eqref{eq:density_moment_collisional} to be dominant in comparison to the magnetic-drift terms, while still satisfying \eqref{eq:setg_fluid_approx} and \eqref{eq:col_setg_fluid_approx}. We now consider very long parallel wavelengths for which this is no longer true, ordering our frequencies as
\begin{align}
    k_\parallel \vthe \ll \omega_{de} \ll \omega \ll \omega_{*e}, \quad \frac{(k_\parallel \vthe)^2}{\nu_{ei}} \ll \omega_{de} \ll \omega \ll \omega_{*e}
    \label{eq:cetg_ordering}
\end{align}
in the collisionless and collisional regimes, respectively. This, in fact, amounts to setting $k_\parallel =0$ everywhere, i.e., we are considering purely two-dimensional modes (see appendices~\ref{app:lin_two_dimensional_perturbations} and \ref{app:lin_col_two_dimesional_perturbations}). In both regimes, our equations reduce to 
\begin{align}
    \frac{\rmd }{\rmd t} \taubar^{-1} \varphi = \frac{\rho_e \vthe}{L_B} \frac{\partial}{\partial y} \frac{\delta T_{\parallel e}}{T_{0e}} , \quad \frac{\rmd}{\rmd t} \frac{\delta T_{\parallel e}}{T_{0e}} = - \frac{\rho_e \vthe}{2L_T} \frac{\partial \varphi}{\partial y}, \quad \frac{\delta T_{\parallel e}}{T_{0e}}  = \frac{\delta T_{\perp e}}{T_{0e}}.
    \label{eq:cetg_equations}
\end{align}
The equality between the perpendicular and parallel temperature perturbations arises in the collisionless regime because the dominant balance in both \eqref{eq:tpar_moment} and \eqref{eq:tperp_moment} is between the time derivative and the ETG injection term, which is also true in the collisional limit and with strengthened isotropisation from collisions. The dispersion relation is [see \eqref{eq:lin_cetg} or \eqref{eq:lin_col_es_cetg}]
\begin{align}
    \omega^2 = - 2 \omega_{de} \omega_{*e}\taubar \quad \Rightarrow \quad \omega = \pm  i \left(2 \omega_{de} \omega_{*e} \taubar \right)^{1/2},
    \label{eq:cetg_gamma}
\end{align}
which is the familiar growth rate of the curvature-mediated ETG (cETG) instability (\citealt{horton88}). 
\begin{figure}
    \centering
    
\input{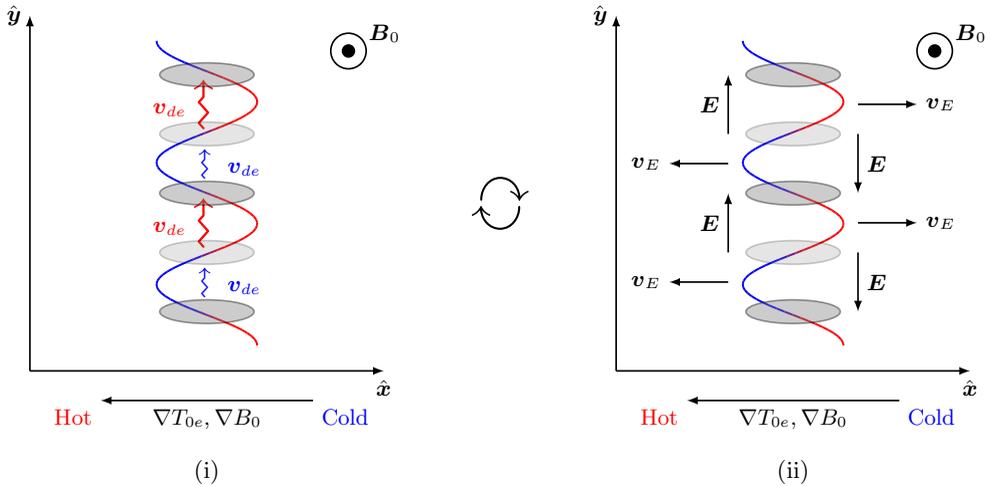}

    \caption{A cartoon illustrating the feedback mechanism of the (2D) cETG instability. (i) A ~temperature perturbation with $k_y \neq 0$ (red-and-blue curve) has alternating hot and cold regions along $\hat{\vec{y}}$. Due to the temperature dependence of the magnetic drifts $\vec{v}_{de}$, electrons in the hot regions will drift faster than those in the cold regions (red and blue arrows), creating an electron density perturbation (over- and under- densities are indicated by the dark- and light-grey ellipses, respectively). (ii) By quasineutrality, the electron-density perturbation gives rise to an exactly equal ion-density perturbation, and that, via Boltzmann-ion response, creates alternating electric fields $\vec{E}$ in the perpendicular plane (vertical black arrows). This produces an $\vec{E}\times \vec{B}$ drift $\vec{v}_E$ (horizontal black arrows), which pushes hotter particles further into the colder region, and vice versa, reinforcing the initial temperature perturbation and thus completing the positive feedback loop required for the instability.}
    \label{fig:cetg}
\end{figure} 
Physically, this arises due to the fact that the magnitude of the magnetic-drift velocity for a particle is proportional to its kinetic energy, and thus temperature. The presence of some temperature perturbation will cause an electron-density perturbation, as electrons in the hotter regions will drift faster than those in colder regions [first equation in \eqref{eq:cetg_equations}]. By quasineutrality, the electron density perturbation gives rise to an exactly equal ion density perturbation, and that, via Boltzmann-ion response, creates an electric field that produces an $\vec{E} \times \vec{B}$ drift which pushes hotter particles further into the colder region, and vice versa [second equation in \eqref{eq:cetg_equations}], completing the feedback loop required for the instability, as illustrated in \figref{fig:cetg}. This mechanism is unaffected by collisionality, hence the cETG instability is obtained in both the collisionless and collisional limits.

Given that both the collisionless and collisional sETG instabilities have maximum growth rates $\gamma_\text{max} \sim \omega_{*e}$ [see \eqref{eq:setg_max} and \eqref{eq:col_setg_max}, respectively], the growth rate of the cETG instability is always small in comparison, at least for large temperature gradients, viz.,
\begin{align}
    \frac{\gamma_\text{max}}{\sqrt{2\omega_{de} \omega_{*e}}} \sim \left( \frac{L_B}{L_T} \right)^{1/2} \gg 1.
    \label{eq:setg_vs_cetg_gamma}
\end{align}
However, the sETG instabilities only exist at perpendicular scales below the flux-freezing scales \eqref{eq:flux_freezing_scale} or \eqref{eq:flux_freezing_scale_col}, as they require electrons to be able to flow across field lines without perturbing them. While sETG is stabilised by magnetic tension above the flux-freezing scale, cETG is unaffected by flux freezing as it is an interchange ($k_\parallel = 0$) mode and does not trigger perpendicular magnetic field perturbations $\dBperp$. This means that it will happily survive in the electromagnetic regime.

\section{Electromagnetic regime: thermo-Alfv{\'e}nic instability}
\label{sec:electromagnetic_regime_tai}
In the electromagnetic regime [i.e. at perpendicular scales above the flux-freezing scales \eqref{eq:flux_freezing_scale} or \eqref{eq:flux_freezing_scale_col}], the magnetic field becomes frozen into the electron flow \eqref{eq:flux_freezing_velocity}, meaning that perpendicular magnetic field perturbations $\dBperp$ are created as electrons move across field lines and drag the latter along. This has two important physical consequences that make electrostatic and electromagnetic phenomena distinct: (i) perturbed magnetic fields give rise to currents that, being electron flows, oppose electron density perturbations [this is the sub-ion-scale version of Lorentz tension, manifest in the second term in \eqref{eq:density_moment_initial}], and (ii) the radial equilibrium temperature gradient now has a component along the exact field line, viz., its projection on to the radial perturbation of the magnetic field. As discussed above, the first effect stabilises the sETG instabilities at the flux-freezing scale [see \apref{app:electromagnetic_stabilisation_of_setg}], as well as giving rise to other electromagnetic phenomena to which we shall return shortly. It is the second effect, however, that will turn out to be crucial in the physics of instabilities in the electromagnetic regime.
\\

Throughout this section, we will focus on the collisional limit --- equations \eqref{eq:density_moment_collisional}, \eqref{eq:velocity_moment_collisional} and \eqref{eq:t_moment_collisional} --- as this will allow us to discuss all of the physics characteristic to the electromagnetic regime without being hampered by the technical detail associated with the full kinetic system. The physical similarly between the instability mechanisms in the collisionless and collisional limits means that we can just signpost the differences between these two limits where appropriate. 

Recalling the definition of the parallel derivative \eqref{eq:parallel_derivative}, we consider the parallel gradient of the total temperature \eqref{eq:logt_definition_intro}.
The first term is the familiar parallel gradient of the temperature perturbation that is present also in the electrostatic regime, the second is the projection of the equilibrium temperature gradient onto the perturbed magnetic field line that arises \textit{only} in the electromagnetic regime. This is the familiar magnetic-flutter drive (\citealt{callen77,manheimer78}). Like the electrostatic linear drive term~\eqref{eq:linear_drive_phi}, this term can also be responsible for extracting free energy from the equilibrium temperature gradient.

To aid our discussion, let us derive an evolution equation for $\gradd_\parallel \log T_e$. A useful result is that, for any field $\psi$,
\begin{align}
    \gradd_\parallel \frac{\rmd \psi}{\rmd t} -\frac{\rmd}{\rmd t} \gradd_\parallel \psi  = -\frac{c}{B_0} \left\{ E_\parallel ,\psi \right\} =  \rho_e  \left\{ \frac{\rmd \mathcal{A}}{\rmd t} + \frac{\vthe}{2} \frac{\partial \varphi}{\partial z},\psi \right\}.
    \label{eq:communting_identity}
\end{align}
The first equality follows by writing the nonlinear operators $\rmd/\rmd t$ and $\gradd_\parallel$, which we defined in \eqref{eq:convective_derivative} and \eqref{eq:parallel_derivative}, respectively, in terms of the Poisson bracket
\begin{align}
    \{ f,g \} = \vec{b}_0 \cdot (\gradd f \times \gradd g) = \frac{\partial f}{\partial x} \frac{\partial g}{\partial y} - \frac{\partial f}{\partial y} \frac{\partial g}{\partial x}
    \label{eq:poisson_bracket}
\end{align}
as follows:
\begin{align}
    \frac{\rmd}{\rmd t} = \frac{\partial}{\partial t} + \frac{\rho_e \vthe}{2} \left\{ \varphi, \dots \right\}, \quad \gradd_\parallel = \frac{\partial}{\partial z} - \rho_e \left\{ \mathcal{A}, \dots \right\},
    \label{eq:nonlinearities}
\end{align}
and noticing that the Poisson bracket satisfies the Jacobi identity:
\begin{align}
    \left\{ \mathcal{A}, \left\{ \varphi , \psi \right\} \right\} + \left\{ \varphi, \left\{ \psi, \mathcal{A} \right\} \right\} + \left\{ \psi, \left\{ \mathcal{A} , \varphi \right\} \right\} = 0.
    \label{eq:jacobi_identity}
\end{align}

Therefore, taking $\gradd_\parallel$ of \eqref{eq:t_moment_collisional}, we find
\begin{align}
    \frac{\rmd}{\rmd t}\gradd_\parallel \frac{\dTe}{T_{0e}} + \frac{\rho_e \vthe}{2L_T} \gradd_\parallel \frac{\partial \varphi}{\partial y} + \rho_e \left\{ \frac{\rmd \mathcal{A}}{\rmd t} + \frac{\vthe}{2} \frac{\partial \varphi}{\partial z}, \frac{\delta T_{ e}}{T_{0e}}  \right\} \nonumber \\
    -   \kappa \gradd_\parallel^3 \log T_e + \frac{2}{3} \gradd_\parallel^2 u_{\parallel e} = 0.
    \label{eq:tai_temperature_moment_derivative}
 \end{align}
Now taking $\partial/\partial y$ of \eqref{eq:velocity_moment_collisional} 
we find
\begin{align}
   \frac{\partial}{\partial y} \left(\frac{\rmd \mathcal{A}}{\rmd t} + \frac{\vthe}{2} \frac{\partial \varphi}{\partial z} \right) =  \frac{\rmd}{\rmd t} \frac{\partial \mathcal{A}}{\partial y} + \frac{\vthe}{2} \gradd_\parallel \frac{\partial \varphi}{\partial y} =   \frac{\vthe}{2}\frac{\partial}{\partial y} \gradd_\parallel \log p_e + \nu_{ei} \frac{\partial}{\partial y} \frac{u_{\parallel e}}{\vthe} ,
    \label{eq:tai_velocity_moment_derivative}
\end{align}
where we have recognised the first two terms on the right-hand side for what they are --- the parallel gradient of the \textit{total} electron pressure:
\begin{align}
    \gradd_\parallel \log p_e = \gradd_\parallel \frac{\dne}{n_{0e}}+ \gradd_\parallel \log T_e .
    \label{eq:log_p_definition}
\end{align}
Subtracting $(\rho_e/L_T)\cdot$\eqref{eq:tai_velocity_moment_derivative} from \eqref{eq:tai_temperature_moment_derivative}, we arrive at 
\begin{align}
    \frac{\rmd}{\rmd t} \gradd_\parallel \log T_e  + \rho_e \left\{ \frac{\rmd \mathcal{A}}{\rmd t} + \frac{\vthe}{2} \frac{\partial \varphi}{\partial z}, \frac{\delta T_{ e}}{T_{0e}}  \right\} + \frac{2}{3}\gradd_\parallel^2 u_{\parallel e} + \nu_{ei} \frac{\rho_e }{L_T} \frac{\partial}{\partial y} \frac{u_{\parallel e}}{\vthe} \nonumber \\
    = \kappa \gradd_\parallel^3 \log T_e - \frac{\rho_e \vthe}{2 L_{T}} \frac{\partial}{\partial y} \gradd_\parallel \log p_e. \label{eq:tai_logt_initial}
\end{align}

Let us now consider the regime
\begin{align}
    (k_\perp d_e)^2 \nu_{ei} \sim \omega_{de} \ll \omega \ll \omega_{*e} \sim \kappa k_{\parallel}^2.
    \label{eq:tai_ordering_collisional}
\end{align}
The ordering of the resistive rate and the magnetic drift frequency is such that we can retain perturbations of similar frequencies to the cETG, by analogy to \eqref{eq:cetg_ordering}. However, this time, we assume the parallel wavelength of the perturbations to be short enough, or, equivalently, their frequency to be low enough, for thermal conduction along the field lines to be rapid in comparison to the mode frequency. Then, in the limit \eqref{eq:tai_ordering_collisional}, the left-hand side of \eqref{eq:tai_logt_initial} is negligible in its entirety (being smaller than the right-hand side by at least a factor of $\omega/\omega_{*e}$), while the outcome of the competition between the two terms on the right-hand side is controlled by the ratio of the perpendicular drift-wave frequency to the parallel conduction rate:
\begin{align}
    \xi_{*} =\frac{\omega_{*e}}{\kappa k_{\parallel}^2}.
    \label{eq:xi_definition_collisional}
\end{align}
This divides our electromagnetic modes into two physically distinct classes: isothermal ($\xi_* \ll 1$) and isobaric ($\xi_* \gg 1$), the former of which is the focus of the next section, and the latter will be discussed in \secref{sec:isobaric_limit}. 

\subsection{Isothermal curvature-mediated TAI}
\label{sec:isothermal_ctai}
Previous studies of electromagnetic phenomena driven by an electron-temperature gradient have often assumed the electrons to be isothermal along the perturbed field line (e.g., \citealt{sch09}, \citealt{sch19}, \citealt{abel13b}, \citealt{zielinski17}). In our system, this assumption is valid if the thermal-conduction time dominates over all other timescales, viz., in addition to \eqref{eq:tai_ordering_collisional},
\begin{align}
    \xi_* \ll 1.
    \label{eq:isothermal_limit}
\end{align}
In the electrostatic regime, without the ability to have perturbations of the magnetic-field direction, adopting such a limit would simply lead to erasure of the temperature perturbation due to Landau damping or thermal conduction [see \eqref{eq:lin_setg} or \eqref{eq:lin_col_es_setg} and the following discussions], suppressing both the collisionless and collisional sETG, respectively. 

The isothermal limit allows the system more leeway in the electromagnetic regime. Given \eqref{eq:isothermal_limit}, the dominant term in \eqref{eq:tai_logt_initial} is the first term on the right-hand side, meaning that, to leading order,
\begin{align}
    \gradd_\parallel \log T_e = \gradd_\parallel \frac{\dTe}{T_{0e}} - \frac{\rho_e}{L_T} \frac{\partial \mathcal{A}}{\partial y} = 0,
    \label{eq:isothermal_limit_logt}
\end{align}
i.e., the temperature perturbations, rather than being zero, will always adjust to cancel the variation of the equilibrium temperature along the perturbed field line. At the next order in $\xi_*$,
\begin{align}
    \kappa \gradd_\parallel^3 \log T_e = \frac{\rho_e \vthe}{2 L_T} \frac{\partial}{\partial y} \gradd_\parallel \frac{\delta n_e}{n_{0e}} \quad \Rightarrow \quad \frac{|\gradd_\parallel \log T_e |}{|\gradd_\parallel \delta n_e/n_{0e} |} \sim \xi_* \ll 1,
    \label{eq:isothermal_limit_next_order}
\end{align}
meaning that we can neglect $\gradd_\parallel \log T_e$ in \eqref{eq:velocity_moment_collisional}. The $\gradd_\parallel u_{\parallel e}$ term in \eqref{eq:density_moment_collisional} is also negligible, as can be confirmed \textit{a posteriori}. Our system \eqref{eq:density_moment_collisional}-\eqref{eq:t_moment_collisional} therefore becomes
\begin{align}
    \frac{\rmd }{\rmd t} \frac{\dne}{n_e} = -\frac{\rho_e \vthe}{L_B} \frac{\partial}{\partial y} \frac{\dTe}{T_{0e}}, \quad \frac{\rmd  \mathcal{A}}{\rmd t} + \frac{\vthe}{2} \frac{\partial \varphi}{\partial z} = \frac{\vthe}{2} \gradd_\parallel \frac{\dne}{n_e}, 
     \quad \gradd_\parallel \frac{\dTe}{T_{0e}} = \frac{\rho_e}{L_T } \frac{\partial \mathcal{A}}{\partial y},
    \label{eq:isothermal_tai_equations}
\end{align}
where, by \eqref{eq:quasineutrality_initial}, $\varphi = - \taubar \dne/n_e$. The associated dispersion relation is
\begin{align}
    \omega^2 = - 2 \omega_{de} \omega_{*e} (1+ \taubar) \quad \Rightarrow \quad \omega = \pm i \left[ 2 \omega_{de} \omega_{*e} ( 1 + \taubar ) \right]^{1/2},
    \label{eq:isothermal_tai_dispersion_relation}
\end{align}
which looks like the familiar cETG growth rate \eqref{eq:cetg_gamma}, but enhanced by an extra order-unity contribution. In fact, this is a physically different and (as far as we know) new\footnote{\cite{zielinski17} proposed a fluid mechanism for the destabilisation of KAW (see \secref{sec:isothermal_kaw}) via their interaction with the cETG mode (see \secref{sec:curvature_ETG}), adopting a purely isothermal limit $\xi_* =0$ and thus neglecting any finite-heat-flux contributions. Under the ordering~\eqref{eq:tai_ordering_collisional}, neglecting equilibrium density gradients and electron finite-Larmor radius contributions, their dispersion relation (23) becomes, in our notation, $\omega^2 = -(2\omega_{de} \omega_{*e} - \omega_\text{KAW}^2) (1+\taubar)$. This the same as \eqref{eq:tai_dispersion_relation_general} to lowest order in $\xi_* \ll 1$. Obviously, it does not match the cETG growth rate \eqref{eq:cetg_gamma} at $k_{\parallel} = 0$, because isothermal limit cannot be valid as $k_{\parallel} \rightarrow 0$. Their dispersion relation displays behaviour qualitatively similar to ours in the isobaric limit for $k_\parallel < k_{\parallel c}$ (see \secref{sec:isobaric_limit}), in that they capture the stabilising effect of the KAW restoring force at $k_\parallel >0$, but miss the fact that the peak growth rate \eqref{eq:isothermal_tai_dispersion_relation} is achieved at a finite $k_\parallel$ (see \secref{sec:general_tai_dispersion_relation}). Their dispersion relation also does not contain the slab TAI mode (see \secref{sec:isothermal_kaw}) or any isobaric physics (\secref{sec:isobaric_limit}).} instability, which we shall refer to as the curvature-mediated \textit{thermo-Alfv\'enic instability} (cTAI). 

\begin{figure}
    \centering
    
   \input{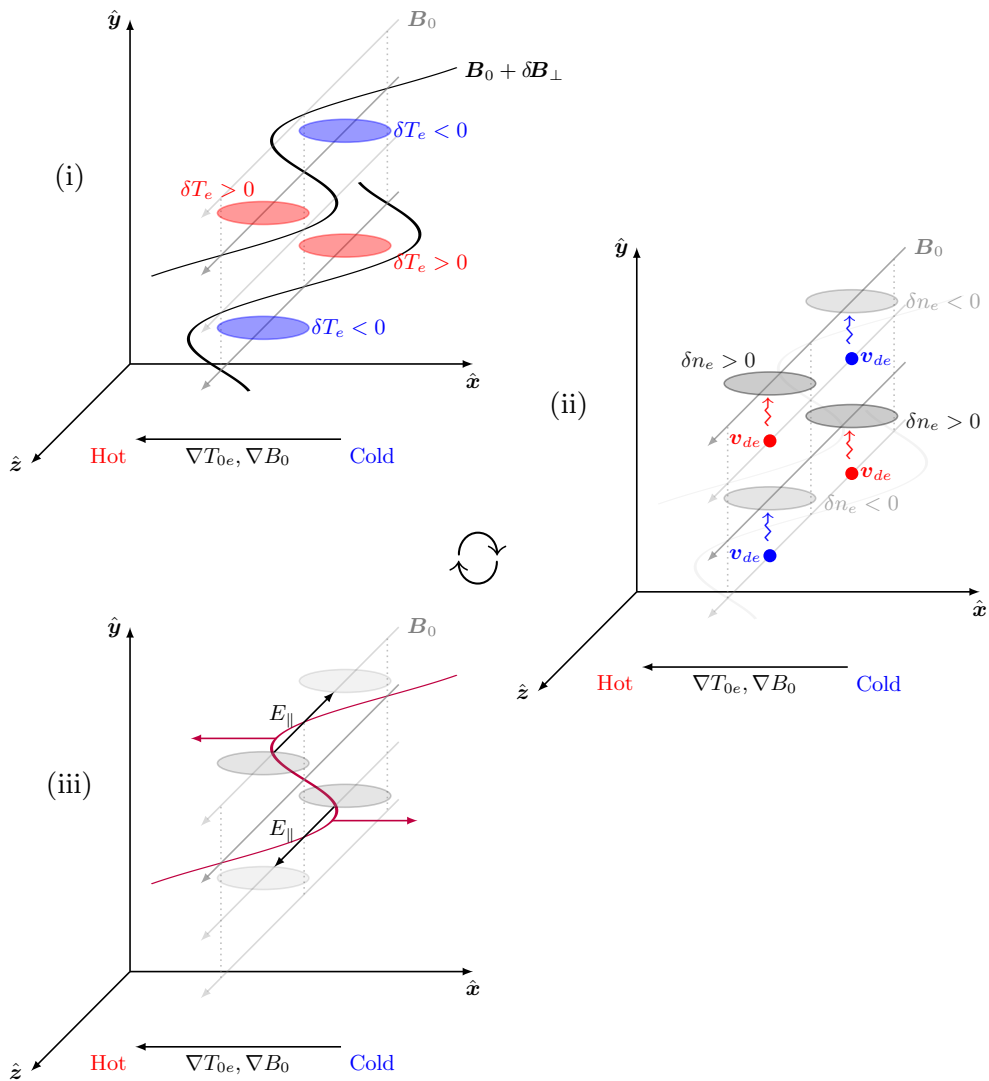}

    \caption{A cartoon illustrating the feedback mechanism of the (isothermal) cTAI. (i)~A~perturbation $\delta \! B_x = B_0 \rho_e \partial_y \mathcal{A}$ (solid black lines) to the equilibrium magnetic field (grey arrows, darker grey corresponding to the plane of constant $y$ containing $\delta \! B_x$, or the relevant perturbation in subsequent diagrams) is created with $k_y \neq 0$ and $k_\parallel \neq 0$ (we show half a wavelength of the mode along both $\hat{\vec{y}}$ and $\hat{\vec{z}}$). Due to the presence of the equilibrium temperature gradient, this will set up a (parallel) variation of the total temperature along the perturbed field line, as the field line makes excursions into hot and cold regions (on the left and right, respectively). However, rapid thermal conduction along the field line instantaneously creates a temperature perturbation that compensates for this temperature variation (red and blue ovals, located in the same planes of constant $y$ as $\delta \! B_x$). (ii) This temperature perturbation will cause a parallel density gradient (over- and under- densities are indicated by the dark and light grey ellipses, respectively, lying in the planes of constant $y$ a quarter of a wavelength above those of $\delta B_x$), as electrons in hotter regions will curvature-drift faster than those in colder regions ($\vec{v}_{de}$, red and blue arrows). (iii) The parallel density gradient must be balanced by a parallel electric field (black arrows, in the same planes of constant $y$ as the density perturbations), whose inductive part leads to an increase in the perturbation of the magnetic field (maroon arrows), deforming the field line further into the hot and cold regions, and in so doing completing the feedback loop for the instability. 
    Note that the maximal rate of change of $\delta \! B_x$ occurs where the $y$-gradient of $E_\parallel$ is at a maximum, as shown.
    }
    \label{fig:ctai}
\end{figure}

Physically, cTAI proceeds as follows. Suppose that a perturbation $\delta \! B_x = B_0 \rho_e \partial_y \mathcal{A}$ of the magnetic field is created, with $k_y \neq 0$ and $k_\parallel \neq 0$. According to the second equation in~\eqref{eq:isothermal_tai_equations}, such a perturbation is brought about by a radial displacement of the electron fluid associated with the velocity \eqref{eq:flux_freezing_velocity}, which, recalling the isothermal condition \eqref{eq:isothermal_limit_logt}, can be written~as
\begin{align}
   \vec{u}_\text{eff} = \vec{v}_E - \frac{\rho_e \vthe}{2} \vec{b}_0 \times \gradd \frac{\delta n_e}{n_{0e}} = - \frac{\rho_e \vthe}{2} \vec{b}_0 \times \gradd \left( 1 +\taubar \right) \frac{\delta n_e}{n_{0e}}.
    \label{eq:flux_freezing_velocity_isothermal}
\end{align}    
Due to the presence of the equilibrium temperature gradient, this magnetic-field perturbation will set up an apparent (parallel) variation of the equilibrium temperature along the perturbed field line, as the field line makes excursions into hot and cold regions. However, rapid thermal conduction along the field line instantaneously creates a temperature perturbation that compensates for this temperature variation, in order to enforce isothermality \eqref{eq:isothermal_limit_logt} [last equation in \eqref{eq:isothermal_tai_equations}]. This temperature perturbation will cause a parallel density gradient, as electrons in the hotter regions will curvature-drift faster than those in colder regions [first equation in \eqref{eq:isothermal_tai_equations}]. The resulting parallel pressure gradient must be balanced by a parallel electric field [second equation in \eqref{eq:isothermal_tai_equations}], whose inductive part leads to an increase in the perturbation of the magnetic field, deforming the field line further into the hot and cold regions, and in doing so completing the feedback loop required for the instability\footnote{Physically, this feedback loop is perhaps reminiscent of some MHD-like instabilities, such as kinetic ballooning modes (KBMs; see references in \secref{sec:introduction}). However, as is evident from the second equation in \eqref{eq:isothermal_tai_equations}, the isothermal cTAI does not satisfy the MHD constraint that $E_\parallel = 0$ typical of such modes. Indeed, in the isothermal regime, the magnetic field lines are not frozen into the $\vec{E}\times \vec{B}$ flow, as they would be in MHD, but instead into the electron flow velocity \eqref{eq:flux_freezing_velocity_isothermal}. We therefore consider that the isothermal cTAI can be regarded as a separate instability, rather than a sub-species of KBM --- unlike its isobaric counterpart discussed in \secref{sec:isobaric_limit}.}. This is illustrated in \figref{fig:ctai}. 

The physical distinction between cTAI and cETG can be made obvious by the following two observations. First, unlike cETG, cTAI relies vitally on $k_{\parallel} \neq 0$ and, indeed, on $k_\parallel$ being large enough for the condition \eqref{eq:isothermal_limit} to be satisfied --- even though the growth rate \eqref{eq:isothermal_tai_dispersion_relation} ends up being independent of $k_{\parallel}$. In \secref{sec:general_tai_dispersion_relation}, we shall show that this is the peak growth rate of the instability and that it is achieved at a finite $k_{\parallel}$, while at $k_{\parallel }= 0$, the cETG growth rate \eqref{eq:cetg_gamma} is recovered. Secondly, perturbations described by~\eqref{eq:isothermal_tai_equations} can be unstable without the need for them to contain any $\vec{E}\times \vec{B}$ flows (i.e., any electrostatic potential $\varphi$) --- this becomes obvious in the formal limit $\varphi =  - \taubar \delta n_e/n_{0e} \rightarrow 0$ as $\taubar \rightarrow 0$ (cold ions). In contrast, the cETG growth rate \eqref{eq:cetg_gamma} disappears in this limit. This is because cTAI extracts energy from the background temperature gradient not via $\vec{E} \times \vec{B}$ advection of said equilibrium gradient but via thermal conduction of it along the perturbed field lines. In order to complete the instability loop and reinforce the magnetic perturbation $\delta \! B_x$ required for this mechanism to work, the system only needs a perturbed density gradient. This is due to the fact that, as we discussed in \secref{sec:summary_of_equations}, below the ion Larmor scale, the magnetic field lines are frozen not into the $\vec{E}\times \vec{B}$ flow but in the electron flow \eqref{eq:flux_freezing_velocity}, which involves also a `diamagnetic' contribution from the electron pressure gradient --- which, in the isothermal limit, consists just of the perturbed density gradient, as in \eqref{eq:flux_freezing_velocity_isothermal}. It is because of the presence of this distinct destabilisation mechanism that the cTAI growth rate \eqref{eq:isothermal_tai_dispersion_relation} is always strictly greater than the cETG one \eqref{eq:cetg_gamma}. Thus, cTAI is not simply an `electromagnetic correction' to cETG, but rather the main effect at scales above the flux-freezing scale \eqref{eq:flux_freezing_scale_col} [or \eqref{eq:flux_freezing_scale} in the collisionless limit, where, as we shall see shortly, the same instability is present]. This suggests that a purely electrostatic description of these scales is inadequate.

\subsection{General TAI dispersion relation}
\label{sec:general_tai_dispersion_relation}
As we have noted above, despite cTAI relying on parallel dynamics, the dispersion relation \eqref{eq:isothermal_tai_dispersion_relation} is itself independent of $k_\parallel$. This is because we have thus far only captured the leading-order behaviour in our analysis, and further diligence is required in order to determine the details associated with the parallel dynamics. Let us give this problem the diligence that it is due, and adopt the ordering \eqref{eq:tai_ordering_collisional} but, for now, $\xi_{*} \sim 1$. Neglecting both the resistive term in \eqref{eq:velocity_moment_collisional} and the compressional term in \eqref{eq:t_moment_collisional} --- since both are small under \eqref{eq:tai_ordering_collisional} --- and determining $\gradd_\parallel \log T_e$ in \eqref{eq:velocity_moment_collisional} from the balance of the two terms on the right-hand side of \eqref{eq:tai_logt_initial}, viz.,
\begin{align}
   \left(\frac{\rho_e \vthe}{2 L_{T}} \frac{\partial}{\partial y}  - \kappa \gradd_\parallel^2\right)\gradd_\parallel \log T_e = -\frac{\rho_e \vthe}{2 L_{T}} \frac{\partial}{\partial y} \gradd_\parallel \frac{\dne}{n_{0e}},
   \label{eq:tai_logt_final_explicit}
\end{align}
we arrive at the following dispersion relation:
\begin{align}
    \omega^2 = -  \left(2 \omega_{de} \omega_{*e} - \omega_\text{KAW}^2 \right) \left( \taubar + \frac{1}{1+i\xi_*} \right), 
    \label{eq:tai_dispersion_relation_general}
\end{align}
where $\omega_\text{KAW} = k_\parallel \vthe k_\perp d_e /\sqrt{2}$ is the kinetic-Alfv\'en-wave (KAW) frequency, the physical origin of which will be discussed in \secref{sec:isothermal_kaw}. The cTAI growth rate is manifest in this expression; adopting the isothermal limit \eqref{eq:isothermal_limit} and neglecting $\omega_\text{KAW}$, we re-obtain \eqref{eq:isothermal_tai_dispersion_relation} to lowest order in $\xi_*$. 

Though we have thus far focussed on the collisional limit, it turns out that much of what we have done is directly applicable to the collisionless limit if we simply replace the parallel conduction rate with the parallel streaming rate [see \eqref{eq:lin_tai}], viz., \eqref{eq:tai_dispersion_relation_general} remains valid but with
\begin{align}
    \xi_* = \frac{\sqrt{\pi}}{2} \frac{\omega_{*e}}{k_{\parallel} \vthe}.
    \label{eq:xi_definition_collisionless}
\end{align}
The equivalent of the ordering \eqref{eq:tai_ordering_collisional} in the collisionless regime is [see \eqref{eq:lin_electromagnetic_ordering}]
\begin{align}
    \omega_{de} \ll \omega \ll \omega_{*e} \sim k_\parallel \vthe,
    \label{eq:tai_ordering_collisionless}
\end{align}
and the equations \eqref{eq:isothermal_tai_equations} are the same; note that in this regime, $\delta T_{\parallel e}=\delta T_{\perp e} = \dTe$ because both the parallel and perpendicular temperature are constant along the field line to leading order in $\omega/\omega_{*e}$. Furthermore, it is possible to show that \eqref{eq:tai_logt_initial} is also valid in the collisionless limit if one replaces $\dTe \rightarrow \delta T_{\parallel e}, \: -\kappa \gradd_\parallel \log T_e \rightarrow \delta q_{\parallel e}/n_{0e} T_{0e}, \: (2/3)\gradd_\parallel ^2 u_{\parallel e} \rightarrow 2 \gradd_\parallel^2 u_{\parallel e}, \: \nu_{ei} u_{\parallel e} \rightarrow \rmd u_{\parallel e}/\rmd t $, and the heat flux must now be determined kinetically [see \eqref{eq:tai_logt_parallel_initial}]. The effect is still to enforce isothermality along the field lines, but by means of parallel particle streaming, rather than collisional conduction. This means that cTAI, while being a `fluid' instability, is not an intrinsically collisional one, occuring also in the collisionless, kinetic limit. Its physical picture in the collisionless limit is exactly the same as in the collisional one.

\begin{figure}

\begin{tabular}{cc}
    \includegraphics[width=0.45\textwidth]{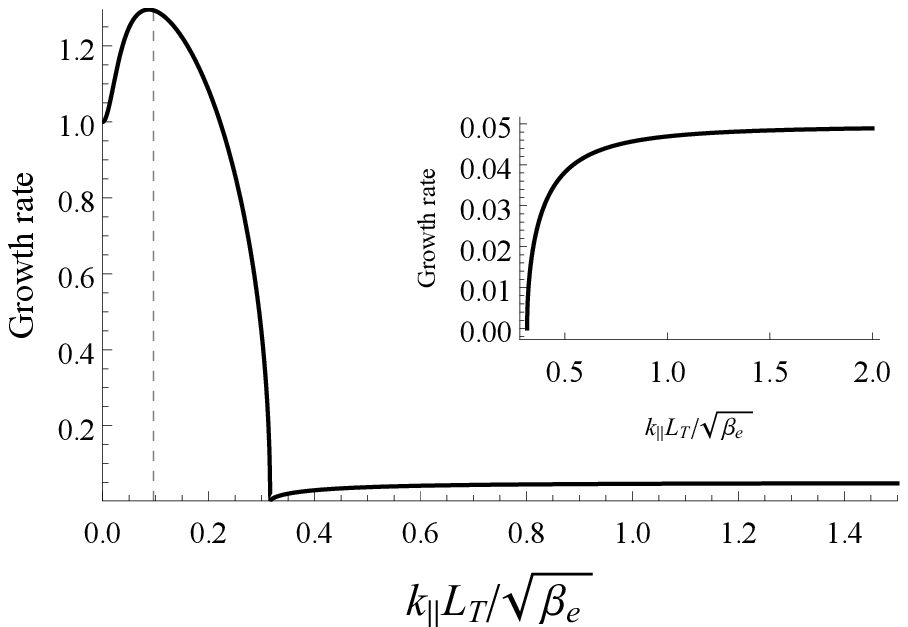} &  
     \includegraphics[width=0.45\textwidth]{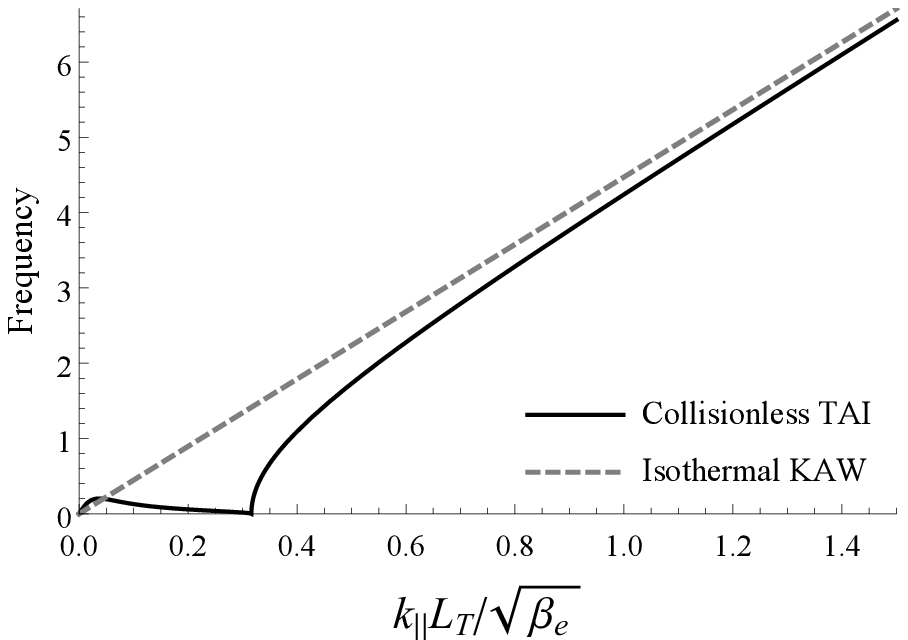} \\\\
     (a) $L_B/L_T =10$, $k_y d_e = 0.1 $ & (b) $L_B/L_T =10$, $k_y d_e = 0.1 $ \\\\\\
     \includegraphics[width=0.45\textwidth]{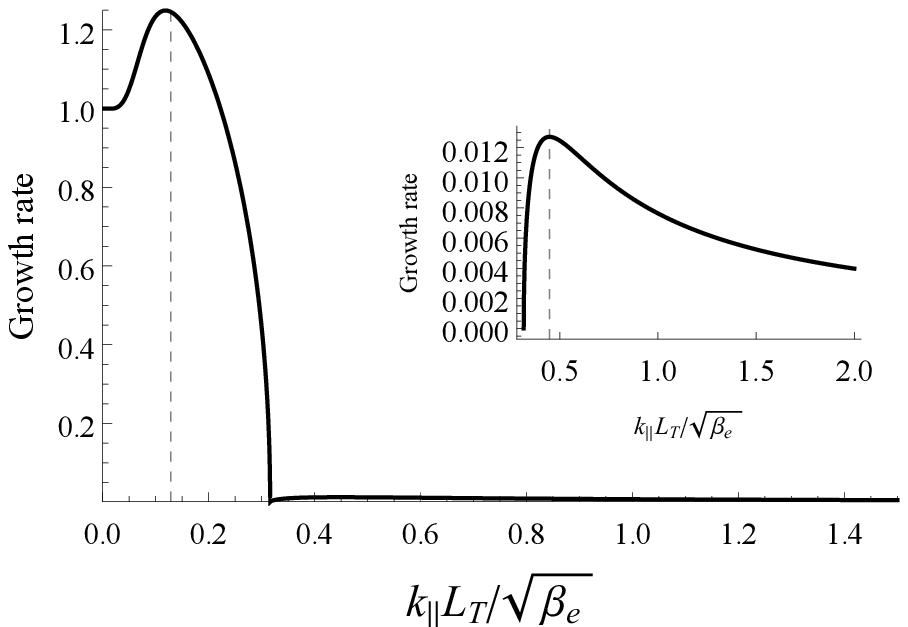} &
    \includegraphics[width=0.45\textwidth]{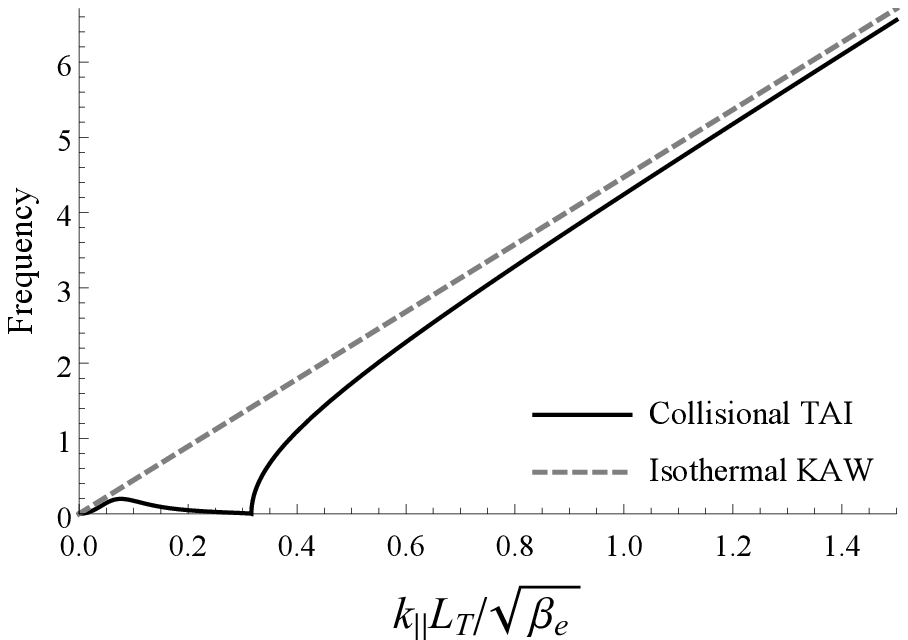} \\\\
     (c) $L_B/L_T =10$, $k_y d_e \chi = 0.002$ & (d) $L_B/L_T =10$, $k_y d_e \chi = 0.002$\\
\end{tabular}
    \caption{(a), (c) The growth-rate and (b), (d) the real frequency of the TAI \eqref{eq:tai_growth_rate_and_real_frequency} in the isothermal limit \eqref{eq:isothermal_limit}, plotted as functions of $k_\parallel L_T/\sqrt{\beta_e}$ and normalised to the cETG growth rate \eqref{eq:cetg_gamma} ($\taubar=1$). (a) and (b) correspond to the collisionless case, while (c) and (d) to the collisional one.
    The vertical dashed lines in panels (a) and (c) are for $k_\parallel = k_{\parallel \text{max}}$ given by \eqref{eq:tai_curvature_maximum_explicit}. The dashed lines in panels (b) and (d) show the isothermal KAW frequency \eqref{eq:kaw_isothermal}.
    Both the growth rate and the real frequency vanish at the critical parallel wavenumber $k_{\parallel c}L_T/\sqrt{\beta_e} = 0.32$, given by \eqref{eq:tai_kpar_critical}. The insets in panels (a) and (c) show details of the behaviour of the growth rate for $k_\parallel > k_{\parallel c}$; the vertical dashed line in the inset of panel (c) is the (secondary) maximum at $k_{\parallel} = \sqrt{2} k_{\parallel c}$ discussed at the end of \secref{sec:isothermal_kaw}. The perpendicular wavenumbers chosen in this figure are all safely below the transition wavenumber \eqref{eq:tai_transition_kperp}, which is $k_{\perp *} d_e = 0.71$ or $k_{\perp *} d_e \chi = 0.03$ in the collisionless or collisional cases, respectively.}
    \label{fig:tai_growth_rate_and_frequency}
    
\end{figure} 

The dispersion relation \eqref{eq:tai_dispersion_relation_general} contains most of the interesting features of the TAI physics (see, however, sections \ref{sec:stabilisation_of_isothermal_slab_TAI} and \ref{sec:stabilisation_of_isobaric_slab_TAI}). The most obvious feature of \eqref{eq:tai_dispersion_relation_general} is that both the growth rate and frequency vanish when 
\begin{align}
    2 \omega_{de} \omega_{*e}  = \omega_\text{KAW}^2 \quad \Rightarrow \quad \frac{k_{\parallel c} L_T}{\sqrt{\beta_e}} = \left( \frac{L_T}{L_B} \right)^{1/2} \frac{k_y}{k_\perp}.
        \label{eq:tai_kpar_critical}
\end{align}
This corresponds to the point of transition from the curvature-dominated regime ($k_{\parallel} < k_{\parallel c}$), on which we will focus in this section, to the KAW-dominated regime ($k_{\parallel} > k_{\parallel c}$), which will be the subject of \secref{sec:isothermal_kaw}.

If we extract the real and imaginary parts of \eqref{eq:tai_dispersion_relation_general}, the (real) frequency $\omega_r = \text{Re}(\omega)$ and the growth rate $\gamma = \text{Im}(\omega)$ of the growing modes can be written as 
\begin{align}
     \omega_r^2 =  \left|2 \omega_{de} \omega_{*e} - \omega_\text{KAW}^2 \right| \taubar \: f_-(\xi_{*}), \quad  \gamma^2 = \left|2 \omega_{de} \omega_{*e} - \omega_\text{KAW}^2 \right| \taubar \: f_+(\xi_{*}),        
    \label{eq:tai_growth_rate_and_real_frequency}
\end{align}
where
\begin{align}
    f_\pm(\xi_{*}) = \frac{1}{2\taubar} \left[ \sqrt{\left( \bar{\tau} + \frac{1}{1+\xi_{*}^2} \right)^2 + \frac{\xi_{*}^2}{(1 + \xi_{*}^2)^2}} \pm \text{sgn}\left(2 \omega_{de} \omega_{*e} - \omega_\text{KAW}^2 \right)\left( \bar{\tau} + \frac{1}{1+\xi_{*}^2} \right)  \right].
    \label{eq:tai_f}
\end{align}
The growth rate and frequency \eqref{eq:tai_growth_rate_and_real_frequency} are plotted as functions of $k_{\parallel} L_T/\sqrt{\beta_e}$ in \figref{fig:tai_growth_rate_and_frequency}. For $k_{\parallel} < k_{\parallel c}$, the growth rate has a maximum for some non-zero $k_{\parallel}$; it is about to turn out that this maximum corresponds to the cTAI growth rate \eqref{eq:isothermal_tai_dispersion_relation}, which was derived in the isothermal limit, $\xi_*\ll 1$. Expanding \eqref{eq:tai_f} in small $\xi_*$ to leading and sub-leading order, and seeking the maximum of the resultant expression with respect to $k_{\parallel}$, we find that this maximum occurs approximately at [see \eqref{eq:isothermal_tai_maximum_solution}]
\begin{align}
    \frac{k_{\parallel \text{max}} L_T}{\sqrt{\beta_e}} = \left\{
    \begin{array}{ll}
         \displaystyle \left[\frac{\pi}{64} \frac{3 + 4 \taubar}{(1+\taubar)^2} \frac{L_T}{L_B} \right]^{1/4} \left( \frac{k_y^2 d_e}{k_\perp} \right)^{1/2}, &  \displaystyle \text{collisionless},   \\[4mm]
          \displaystyle \left[\frac{81}{50}\frac{3 + 4 \taubar}{(1+\taubar)^2}  \frac{L_T}{ L_B} \right]^{1/6}\left( \frac{k_y^2 d_e}{k_\perp} \chi\right)^{1/3}, &  \displaystyle \text{collisional},
    \end{array}
    \right.
    \label{eq:tai_curvature_maximum_explicit}
\end{align}
indicated by the vertical dashed lines in panels (a) and (c) of \figref{fig:tai_growth_rate_and_frequency}; $\chi$ is defined in~\eqref{eq:flux_freezing_scale_col}. Calculating the growth rate \eqref{eq:tai_growth_rate_and_real_frequency} at $k_{\parallel} = k_{\parallel \text{max}}$, one recovers \eqref{eq:isothermal_tai_dispersion_relation} up to small corrections [see \eqref{eq:tai_curvature_maximum_growth_rate_corrections}], as promised. 

This solution, however, is only valid so long as it remains in the isothermal limit \eqref{eq:isothermal_limit}. Evaluating $\xi_*$ at $k_{\parallel} =k_{\parallel \text{max}}$, we find, defining $\alpha =1,2$ in the collisionless and collisional limits, respectively, [see \eqref{eq:isothermal_tai_maximum_xi}]
\begin{align}
    \xi_* \left(k_{\parallel \text{max}}\right) \sim \frac{k_{\parallel \text{max}}}{k_{\parallel c}}  \sim  \left( \frac{k_\perp}{k_{\perp *}} \right)^{1/(1+\alpha)} \ll 1
    \label{eq:xi_at_max}
\end{align}
 provided that $k_\perp \ll k_{\perp *}$, where $k_{\perp *}$ is the perpendicular wavenumber at which $\xi_*\left(k_{\parallel c} \right) \sim 1$, viz., 
\begin{align}
    k_{\perp *} d_e = \left\{
    \begin{array}{ll}
         \displaystyle \frac{4}{\sqrt{\pi}} \left( \frac{L_{T}}{L_B} \right)^{1/2}, &  \displaystyle \text{collisionless},   \\[4mm]
          \displaystyle \frac{5}{9} \frac{L_{T}}{L_B} \chi^{-1} , &  \displaystyle \text{collisional},
    \end{array}
    \right. 
    \quad \Rightarrow \quad \xi_*(k_{\parallel c}) = 
    \left\{
    \begin{array}{ll}
         \displaystyle \frac{k_\perp}{k_{\perp *}}, &  \displaystyle \text{collisionless},   \\[4mm]
          \displaystyle \frac{k_\perp^2}{k_{\perp *} k_y} , &  \displaystyle \text{collisional}.
    \end{array}
    \right.
    \label{eq:tai_transition_kperp}
\end{align}
Thus, the isothermal regime is valid at sufficiently long perpendicular wavelengths. At $k_\perp >k_{\perp *}$, a different, isobaric regime takes over, which will be considered in \secref{sec:isobaric_limit}. 

Lastly, we note that, for $k_{\parallel} < k_{\parallel c}$, the magnitude of the real frequency is vanishingly small when compared to the growth rate: expanding both the growth rate and the real frequency in \eqref{eq:tai_growth_rate_and_real_frequency} for $\xi_* \ll 1$, we find
\begin{align}
    \frac{\omega_r^2}{\gamma^2}  \approx \frac{\xi_*^2}{4(1+\taubar)^2} \ll 1. 
    \label{eq:tai_ratio_of_frequency_to_growth_rate}
\end{align}
Thus, cTAI is, like cETG, an (almost) purely growing mode; this is distinct from the case of the sETG, whose frequency and growth rate are comparable at the latter's maximum (see \secref{sec:electrostatic_regime_etg}).

\subsection{Isothermal KAWs and slab TAI}
\label{sec:isothermal_kaw}

\subsubsection{Isothermal KAWs}
\label{sec:isothermal_kaws}
If $k_\perp \ll k_{\perp}^*$, i.e., $\xi_*\left(k_{\parallel c} \right) \ll 1$, then the isothermal approximation \eqref{eq:isothermal_limit_logt} continues to be satisfied at $k_{\parallel} > k_{\parallel c}$, but the effects of the magnetic drifts become negligible for $k_{\parallel} \gg k_{\parallel c}$. In this region, our system \eqref{eq:density_moment_collisional}-\eqref{eq:t_moment_collisional} becomes, approximately,
\begin{align}
     &\frac{\rmd}{\rmd t} \frac{\dne}{n_{0e}} =  - \vthe \gradd_\parallel d_e^2 \gradd_\perp^2 \mathcal{A} , \quad \frac{\rmd \mathcal{A}}{\rmd t} + \frac{\vthe}{2} \frac{\partial \varphi}{\partial z} =   \frac{\vthe}{2}  \gradd_\parallel  \frac{\dne}{n_{0e}}, \quad \varphi = - \taubar \frac{\dne}{n_{0e}}.
     \label{eq:kaw_equation}
\end{align}
These equations are also valid in the collisionless limit [there is no intrinsically collisional physics in~\eqref{eq:kaw_equation}, as the resistive term in \eqref{eq:velocity_moment_collisional} is negligible under the ordering~\eqref{eq:tai_ordering_collisional}].
We recognise these as the equations of Electron Reduced Magnetohydrodynamics (ERMHD, see \citealt{sch09} or \citealt{boldyrev13}), which describe the dynamics, linear and nonlinear, of kinetic Alfv\'en waves (KAWs). Indeed, linearising and Fourier transforming \eqref{eq:kaw_equation}, we find the dispersion relation
\begin{align}
    \omega^2 = k_\parallel^2 \vthe^2 k_\perp^2 d_e^2 \frac{1+\taubar}{2} =  \omega_\text{KAW}^2 \left(1 + \taubar \right).
    \label{eq:kaw_isothermal}
\end{align}
These are the familiar (isothermal) KAWs that arise in homogeneous systems (\citealt{howes06,sch09,sch19,zocco11,boldyrev13,passot17}). The physics of these waves is as follows. Suppose that a density perturbation $\dne/n_{0e} = -\taubar^{-1} \varphi$ with $k_{\parallel} \neq 0$ is created. This gives rise to a parallel pressure gradient, which manifests itself as a parallel (perturbed) density gradient, as any parallel temperature variation is instantaneously ironed out by rapid parallel streaming or thermal conduction. This parallel pressure gradient must be balanced by the parallel electric field [second equation in \eqref{eq:kaw_equation}], whose inductive part, through Amp\'ere's law \eqref{eq:field_equations}, leads to a parallel current. But a parallel current is a parallel electron flow, which leads to compressional rarefaction along the field that opposes the original density perturbation [first equation in \eqref{eq:kaw_equation}]. This is also the reason for the reduction of the cTAI growth rate at $k_\parallel > k_{\parallel \text{max}}$ and its vanishing at $k_{\parallel} = k_{\parallel c}$ [see panels (a) and (c) in \figref{fig:tai_growth_rate_and_frequency}]: the parallel compression that provides the restoring force for the KAW perturbations increases as $k_\parallel$ increases, weakening the instability mechanism of the cTAI described in \secref{sec:isothermal_ctai}. 

\subsubsection{Isothermal slab TAI}
\label{sec:isothermal_slab_tai}
Remarkably, however, it turns out that isothermal KAW, at $k_{\parallel} > k_{\parallel c}$, are still unstable in the presence of an equilibrium electron temperature gradient: expanding \eqref{eq:tai_dispersion_relation_general} or \eqref{eq:tai_growth_rate_and_real_frequency} for $\xi_* \ll 1$ at $k_{\parallel} \gg k_{\parallel c}$ (the latter in order to drop the $\omega_{de}$ effects), we find
\begin{align}
   \omega_r^2 \approx \omega_\text{KAW}^2 (1+\taubar),\quad \gamma^2 \approx \omega_\text{KAW}^2 \frac{\xi_*^2}{4(1+\taubar)}.
    \label{eq:stai_isothermal_limit}
\end{align}
By analogy with sETG, we shall henceforth refer to this KAW-dominated TAI as the `slab' TAI (sTAI); it was our original motivation for calling the instability `thermo-Alfv\'enic'\footnote{ The sTAI instability appears to be a close relative of the `electron magnetothermal instability' (eMTI) discovered by \cite{xu16} in their treatment of stratified plasma atmospheres, and analysed by them in the high-beta limit appropriate to the astrophysical applications on which they were focused.}.

The precise mechanism by which the isothermal sTAI operates is somewhat subtler than cTAI, relying on the fact that the isothermal condition \eqref{eq:isothermal_limit_logt} that led to \eqref{eq:kaw_isothermal} is, in fact, only approximately satisfied. Indeed, $\gradd_\parallel \log T_e$ is determined, in the collisional limit, at next order in $\xi_*$ by \eqref{eq:isothermal_limit_next_order} which, linearising and Fourier transforming, can be written as 
\begin{align}
    \left(\gradd_\parallel \log T_e \right)_{\vec{k}} = -i \xi_* \left(\gradd_\parallel \frac{\dne}{n_{0e}} \right)_{\vec{k}}.
    \label{eq:stai_logt_correction_isothermal}
\end{align}
This means that a small but finite parallel gradient of temperature effectively introduces a correction to the parallel density gradient in \eqref{eq:kaw_equation} that is $\pi/2$ out of phase with the contribution that enables the isothermal KAWs. This gives rise to the instability \eqref{eq:stai_isothermal_limit} in both the collisional limit, and, it turns out, the collisionless one, where \eqref{eq:stai_logt_correction_isothermal} also holds but with $\xi_*$ given by \eqref{eq:xi_definition_collisionless} [see \eqref{eq:lin_logt_final}]. Restoring finite parallel temperature gradients in \eqref{eq:kaw_equation}, we have 
\begin{align}
     &\frac{\rmd}{\rmd t} \frac{\dne}{n_{0e}} =   -\vthe \gradd_\parallel d_e^2 \gradd_\perp^2 \mathcal{A} , \quad \frac{\rmd \mathcal{A}}{\rmd t} + \frac{\vthe}{2} \frac{\partial \varphi}{\partial z} =  \frac{\vthe}{2}  \gradd_\parallel  \frac{\dne}{n_{0e}} + \frac{\vthe}{2} \gradd_\parallel \log T_e,
     \label{eq:kaw_equation_stai}
\end{align}
with dispersion relation 
\begin{align}
    \omega^2 = \omega_\text{KAW}^2 (1+ \taubar - i \xi_*) \quad \Rightarrow \quad \omega \approx \pm \omega_\text{KAW}\left( \sqrt{1+\taubar} -\frac{i\xi_*}{2 \sqrt{1+\taubar}} + \dots \right),
    \label{eq:kaw_stai_dispersion_relation}
\end{align}
whose real and imaginary parts are exactly the frequency and growth rate \eqref{eq:stai_isothermal_limit}. 

If we restore the magnetic-drift terms in the density equation, we find, in the collisionless limit, that the sTAI growth rate increases from zero at $k_{\parallel} = k_{\parallel c}$ to a finite, $k_{\parallel}$-independent limit \eqref{eq:stai_isothermal_limit} at $k_{\parallel} \gg k_{\parallel c}$, viz., 
\begin{align}
    \gamma \rightarrow \frac{1}{4} \sqrt{\frac{\pi}{2(1+\taubar)}} k_\perp d_e \omega_{*e} = \sqrt{\frac{\omega_{de} \omega_{*e}}{2(1+\taubar)}} \frac{k_\perp}{k_{\perp *}},
    \label{eq:tai_isothermal_collisionless_max}
\end{align}
where $k_{\perp *}$ is given by \eqref{eq:tai_transition_kperp} [see \figref{fig:tai_growth_rate_and_frequency}(a), inset]. As we shall see shortly in \secref{sec:stabilisation_of_isothermal_slab_TAI}, this value only persists up to a certain $k_\parallel$ where sTAI is stabilised by compressional heating, which was neglected in \eqref{eq:tai_dispersion_relation_general}. In the collisional limit, $\gamma \rightarrow 0$ as $k_{\parallel} \rightarrow \infty$ (also, in fact, shown to go to $\gamma <0$ in \secref{sec:stabilisation_of_isothermal_slab_TAI}). The growth rate has a maximum at $k_\parallel = \sqrt{2} k_{\parallel c}$, which is shown by the vertical dashed line in the inset of \figref{fig:tai_growth_rate_and_frequency}(c). The growth rate at this maximum is 
\begin{align}
    \gamma = \frac{k_\perp^2 d_e^2 \vthe^2}{8 \sqrt{2(1+\taubar)} \kappa} \sqrt{\frac{\omega_{*e}}{\omega_{de}}} = \frac{1}{2} \sqrt{\frac{\omega_{de} \omega_{*e}}{2(1+\taubar)}} \frac{k_\perp^2}{k_{\perp *} k_y} .
    \label{eq:tai_isothermal_collisional_max}
\end{align}
These results are derived at the end of \apref{app:isothermal_limit}.
Both the maximum growth rates \eqref{eq:tai_isothermal_collisionless_max} and \eqref{eq:tai_isothermal_collisional_max} are manifestly much smaller than the maximum growth rate of cTAI \eqref{eq:isothermal_tai_dispersion_relation} as long as $k_\perp \ll k_{\perp *}$, i.e., as long as the isothermal approximation, in which all of these results have been derived in the first place, is valid. 

Thus, at long perpendicular wavelengths ($k_\perp \ll k_{\perp *}$), the dominant instability is cTAI, reaching its maximum growth rate \eqref{eq:isothermal_tai_dispersion_relation} at the parallel wavenumber \eqref{eq:tai_curvature_maximum_explicit}.

\subsubsection{Stabilisation of isothermal slab TAI}
\label{sec:stabilisation_of_isothermal_slab_TAI}
The sTAI growth rates do not, in fact, stay positive to infinite parallel wavenumbers. The instability is eventually quenched by the compressional-heating term in the temperature equation [\eqref{eq:tpar_moment} or \eqref{eq:t_moment_collisional} in the collisionless and collisional limits, respectively] that begins to compete with the TAI drive.

To show this, let us consider the collisional limit and, instead of \eqref{eq:tai_ordering_collisional}, the ordering 
\begin{align}
    (k_\perp d_e)^2 \nu_{ei} \ll \omega \sim \omega_{*e} \ll \kappa k_\parallel^2.
    \label{eq:isothermal_stab_tai_ordering}
\end{align}
In this limit, the system is still isothermal to leading order in $\xi_* \ll 1$, but now we must also retain the compressional heating term in \eqref{eq:tai_logt_initial} to determine $\gradd_\parallel \log T_e$ at next order: instead of \eqref{eq:isothermal_limit_next_order}, we have, therefore,
\begin{align}
    \kappa \gradd_\parallel^3 \log T_e = \frac{\rho_e \vthe}{2L_T} \frac{\partial}{\partial y} \gradd_\parallel \frac{\delta n_e}{n_{0e}} + \frac{2}{3} \gradd_\parallel^2 u_{\parallel e}.
    \label{eq:isothermal_stab_logt}
\end{align}
Furthermore, it turns out that we must also retain the resisitve term in \eqref{eq:velocity_moment_collisional} at this order as it will end up making a contribution of the same order as the second term in~\eqref{eq:isothermal_stab_logt}. Thus, the second equation in~\eqref{eq:kaw_equation_stai} is replaced by
\begin{align}
    \frac{\rmd \mathcal{A}}{\rmd t} + \frac{\vthe}{2} \frac{\partial \varphi}{\partial z} =  \frac{\vthe}{2}  \gradd_\parallel  \frac{\dne}{n_{0e}} + \frac{\vthe}{2} \gradd_\parallel \log T_e + \nu_{ei} \frac{u_{\parallel e}}{\vthe},
     \label{eq:isothermal_stab_equations}
\end{align}
Combining \eqref{eq:isothermal_stab_logt} and \eqref{eq:isothermal_stab_equations} with the density equation, still the same as in \eqref{eq:kaw_equation_stai}, we obtain the following dispersion relation
\begin{align}
    \omega^2 - \omega_\text{KAW}^2(1+\taubar - i\xi_*)  = - i \left(\frac{2}{3}+\constant \right) \frac{\omega}{\kappa k_\parallel^2} \omega_\text{KAW}^2,
    \label{eq:isothermal_stab_dispersion_relation}
\end{align}
where $\constant$ is a numerical constant of order unity [see \eqref{eq:lin_col_ignore_terms}]. This is the same as \eqref{eq:kaw_stai_dispersion_relation} apart from the right-hand side, previously neglected. At the stability boundary, the frequency $\omega$ must be purely real, and both the real and imaginary parts of \eqref{eq:isothermal_stab_dispersion_relation} must vanish individually, giving [cf. \eqref{eq:lin_col_isothermal_stai_stabilisation}]
\begin{align}
    \omega^2 = \omega_\text{KAW}^2 (1+\taubar), \quad \omega  = - \frac{\omega_{*e}}{\constant + 2/3}  \quad \Rightarrow \quad  \mp \omega_{\text{KAW}} \sqrt{1+\taubar} = \frac{\omega_{*e}}{\constant + 2/3}.
    \label{eq:isothermal_stab_boundary}
\end{align}
For $k_y \sim k_\perp$, \eqref{eq:isothermal_stab_boundary} are lines of constant $k_\parallel$ in wavenumber space, limiting the isothermal sTAI at large parallel wavenumbers:
\begin{align}
    \frac{k_\parallel L_{T}}{\sqrt{\beta_e}}  = \pm  \frac{1}{(\constant + 2/3)\sqrt{2(1+\taubar)}} \frac{k_y}{k_\perp}.
    \label{eq:isothermal_stai_stabilisation_kpar}
\end{align}
This stabilisation of the isothermal sTAI was not captured in the TAI dispersion relation~\eqref{eq:tai_dispersion_relation_general} because the  ordering~\eqref{eq:tai_ordering_collisional} did not formally allow frequencies comparable to the drift wave frequency, required by \eqref{eq:isothermal_stab_boundary}.

In the collisionless limit, we also find that the sTAI is stabilised above a line of constant~$k_\parallel$ [cf. \eqref{eq:lin_stai_stab_isothermal_boundary}]
\begin{align}
    \omega_\text{KAW} \sim \omega_{*e},
    \label{eq:isothermal_stab_boundary_collisionless}
\end{align}
due again to the competition between the compressional heating in the equation for the parallel temperature \eqref{eq:tpar_moment} and the TAI drive. In \apref{app:stabilisation_of_isothermal_stai}, we detail a collisionless calculation analogous to that performed above in the collisional limit, but the latter is sufficient here for illustrating the physics underlying the stabilisation mechanism. In both cases, the stabilisation of sTAI does not appear in \figref{fig:tai_growth_rate_and_frequency} (and, later, in \figref{fig:tai_growth_rate_and_frequency_isobaric}) because the orderings \eqref{eq:tai_ordering_collisional} or \eqref{eq:tai_ordering_collisionless} that lead to the TAI dispersion relation~\eqref{eq:tai_dispersion_relation_general} do not formally allow this stabilisation; instead, readers will find it in figures~\ref{fig:lin_em_no_drifts}(c) and \ref{fig:lin_col_em_plots_no_drifts}(c) in the collisionless and collisional limits, respectively, where solutions of a more precise dispersion relation are shown.

Though useful for delineating the precise regions of instability in the $(k_\perp, k_\parallel)$ space, this stabilisation of the isothermal sTAI is of secondary importance because it is cTAI that is the dominant instability at long perpendicular wavelengths ($k_\perp \ll k_{\perp *}$), which was the main conclusion of \secref{sec:isothermal_slab_tai}.

\subsection{Isobaric limit}
\label{sec:isobaric_limit}
Let us now consider what happens in the opposite limit of short perpendicular wavenlengths, $k_\perp \gg k_\perp^*$, corresponding to thermal conduction (or its collisionless analogue, parallel streaming) being weak in comparison with the $\omega_{*e}$ driving, viz.,
\begin{align}
    \xi_{*} \gg 1.
    \label{eq:isobaric_limit}
\end{align}
Assuming this in addition to \eqref{eq:tai_ordering_collisional} or \eqref{eq:tai_ordering_collisionless},
we find that the dominant term in \eqref{eq:tai_logt_initial} is the second term on the right-hand side, meaning that, to leading order,
\begin{align}
     \gradd_\parallel \log p_e = \gradd_\parallel \log T_e + \gradd_\parallel \frac{\dne}{n_{0e}} = 0.
    \label{eq:isobaric_limit_logt}
\end{align}
This is the isobaric limit, in which the total \textit{pressure} is constant along the perturbed field lines, rather than just the total temperature. That is, the temperature perturbation has to adjust to cancel not just the variation of the equilibrium temperature along the perturbed field line, as was the case in the isothermal limit, but now also the variation of the perturbed density. At next order in $\xi_*$, from \eqref{eq:tai_logt_initial}, we have
\begin{align}
    \frac{\rho_e \vthe}{2L_T} \frac{\partial}{\partial y}\gradd_\parallel \log p_e =  -\kappa \gradd_\parallel^3 \frac{\dne}{n_{0e}} \quad \Rightarrow \quad \frac{|\gradd_\parallel \log p_e |}{|\gradd_\parallel \delta n_e/n_{0e} |} \sim \frac{1}{\xi_*} \ll 1,
    \label{eq:isobaric_limit_next_order}
\end{align}
so we can now neglect the entire right-hand side of \eqref{eq:velocity_moment_collisional}, reducing the latter equation to $E_{\parallel} = 0$. For $k_{\parallel} \ll k_{\parallel c}$, i.e. neglecting the KAW restoring force, our system \eqref{eq:density_moment_collisional}-\eqref{eq:t_moment_collisional} therefore becomes
\begin{align}
    \frac{\rmd }{\rmd t} \frac{\dne}{n_{0e}} = -\frac{\rho_e \vthe}{L_B} \frac{\partial}{\partial y} \frac{\dTe}{T_{0e}}, \quad \frac{\rmd  \mathcal{A}}{\rmd t} + \frac{\vthe}{2} \frac{\partial \varphi}{\partial z} = 0, 
    \quad \gradd_\parallel \frac{\dTe}{T_{0e}} = \frac{\rho_e}{L_T } \frac{\partial \mathcal{A}}{\partial y} - \gradd_\parallel \frac{\dne}{n_{0e}},
    \label{eq:isobaric_tai_equations}
\end{align}
where $\varphi = - \taubar \dne/n_e$. As with the isothermal cTAI, these equations remain valid in the collisionless limit, because $\delta T_{\parallel e} = \delta T_{\perp e} = \delta T_e$ to leading order in $\omega/\omega_{*e}$.

In \eqref{eq:isobaric_tai_equations}, the temperature perturbation is determined from the third equation, which is simply the isobaric condition \eqref{eq:isobaric_limit_logt}. However, given the ordering \eqref{eq:tai_ordering_collisional}, the correction to $\dTe$ due to the density perturbation is small, viz., $\dne/n_{0e} \sim (\omega_{de}/\omega) \dTe/T_{0e}$, which follows from the first equation in \eqref{eq:isobaric_tai_equations}. That is, to leading order, there is no difference between the isothermal and isobaric conditions when it comes to determining the temperature perturbation. Hence, the associated dispersion relation is
\begin{align}
    \omega^2 = - 2 \omega_{de} \omega_{*e} \taubar \quad \Rightarrow \quad \omega = \pm i \left(2 \omega_{de} \omega_{*e} \taubar\right)^{1/2}.
    \label{eq:isobaric_tai_dispersion_relation}
\end{align}

\begin{figure}

\begin{tabular}{cc}
    \includegraphics[width=0.45\textwidth]{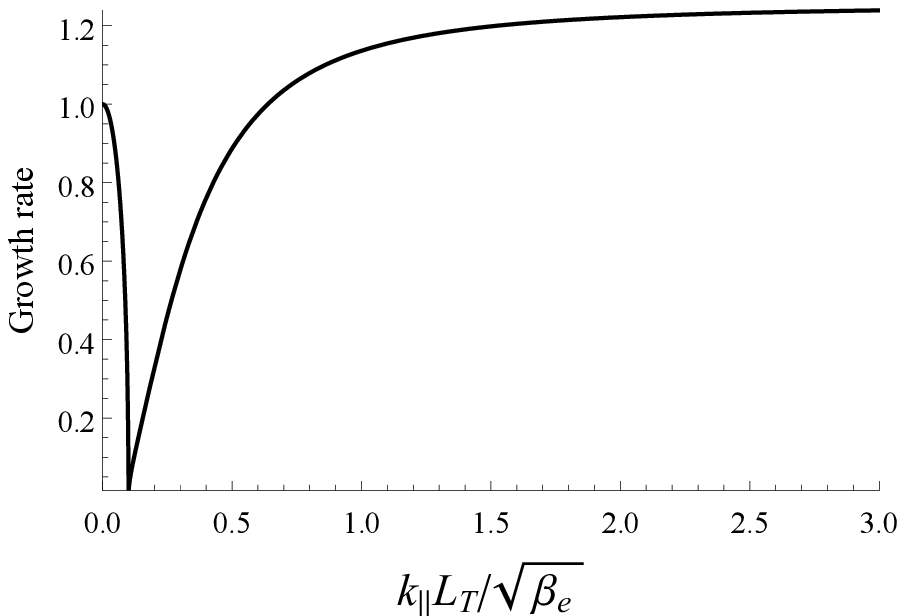} &  
     \includegraphics[width=0.45\textwidth]{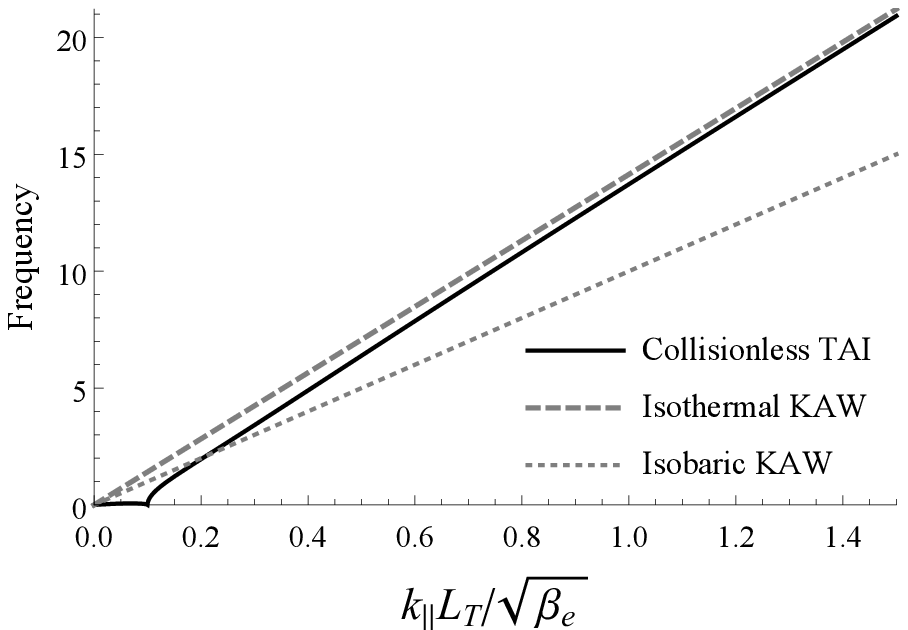} \\\\
     (a) $L_B/L_T =100$, $k_y d_e = 0.8 $ & (b) $L_B/L_T =100$, $k_y d_e = 0.8$ \\\\\\
     \includegraphics[width=0.45\textwidth]{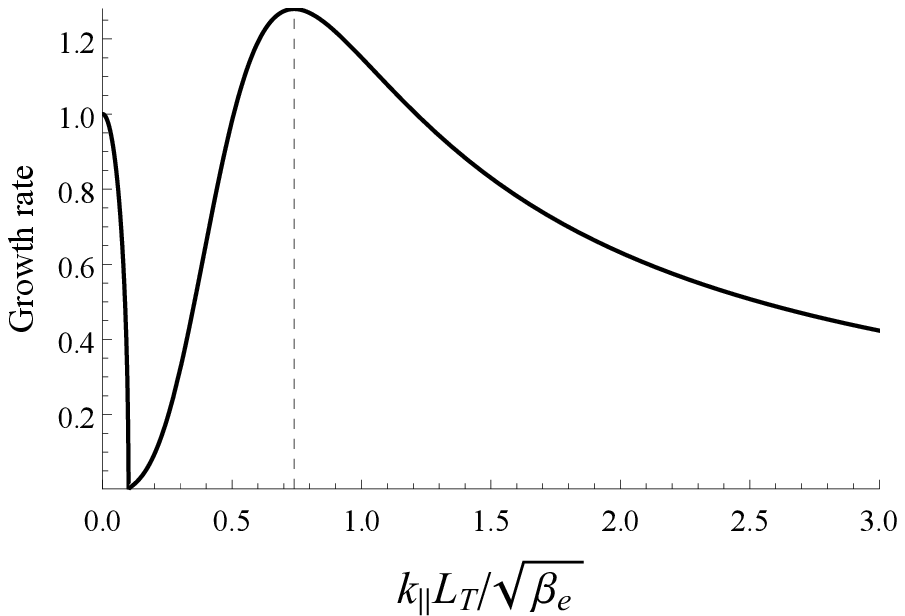} &
    \includegraphics[width=0.45\textwidth]{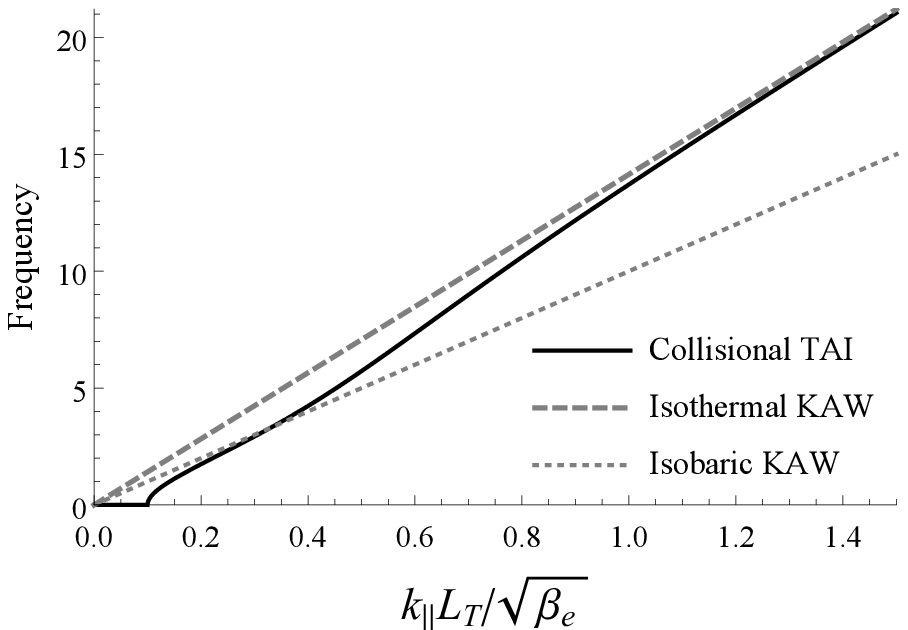} \\\\
     (c) $L_B/L_T =100$, $k_y d_e \chi = 0.1$ & (d) $L_B/L_T =100$, $k_y d_e \chi = 0.1$\\
\end{tabular}
    \caption{ (a), (c) The growth-rate and (b), (d) the real frequency of the TAI \eqref{eq:tai_growth_rate_and_real_frequency} in the isobaric limit \eqref{eq:isobaric_limit}, plotted as functions of $k_\parallel L_T/\sqrt{\beta_e}$ and normalised to the cETG growth rate \eqref{eq:cetg_gamma} ($\taubar=1$). (a) and (b) correspond to the collisionless case, while (c) and (d) the collisional one.
    The vertical dashed line in (c) is for $k_{\parallel \text{max}}$ given by \eqref{eq:stai_max_kpar}. The dashed and dotted lines in panels (b) and (d) are the isothermal \eqref{eq:kaw_isothermal} and isobaric \eqref{eq:kaw_isobaric} KAW frequencies, respectively.
    Both the growth rate and the real frequency vanish at the critical parallel wavenumber $k_{\parallel c}L_T/\sqrt{\beta_e} = 0.1$, given by \eqref{eq:tai_kpar_critical}. The perpendicular wavenumbers chosen in this figure are all safely above the transition wavenumber \eqref{eq:tai_transition_kperp}, which is $k_{\perp *} d_e = 0.23$ or $k_{\perp *} d_e \chi = 0.003$ in the collisionless or collisional cases, respectively. The parallel wavenumber corresponding to the transition between the isobaric and isothermal regimes at a fixed $k_y$ (viz., for $\xi_* \sim 1$) is given by $k_\parallel L_B/\sqrt{\beta_e} = 0.35$ or $0.36$ in the collisionless and collisional cases, respectively. We chose a very large value of $L_B/L_T$ to show the asymptotic regimes clearly. }
    \label{fig:tai_growth_rate_and_frequency_isobaric}
    
\end{figure}

Analysing --- and plotting, in \figref{fig:tai_growth_rate_and_frequency_isobaric} --- the dispersion relation \eqref{eq:tai_growth_rate_and_real_frequency} in the isobaric regime, both collisional and collisionless, we find that the maximum of the growth rate in the region $k_{\parallel} < k_{\parallel c}$ occurs at $k_{\parallel } = 0$, i.e., it is, in fact, the 2D cETG mode that has the fastest growth. At finite $k_{\parallel}$, it is weakened by the presence of the restoring force associated with KAWs, reaching $\gamma = 0$ at $k_{\parallel} = k_{\parallel c}$ --- this is evident in panels (a) and (c) of \figref{fig:tai_growth_rate_and_frequency_isobaric}. 

The dispersion relation \eqref{eq:isobaric_tai_dispersion_relation} is identical to the cETG dispersion relation \eqref{eq:cetg_gamma}. This is because the second equation in \eqref{eq:isobaric_tai_equations} is simply $E_\parallel = 0$, implying that the magnetic field is now frozen into the $\vec{E}\times \vec{B}$ flow, as are the temperature perturbations [see \eqref{eq:flux_freezing_velocity}, wherein the second term vanishes in the isobaric limit]. This is distinct to the case of the isothermal cTAI introduced in \secref{sec:isothermal_ctai}, where the magnetic field was frozen into a \textit{different} velocity field than the temperature perturbations, viz., the mean electron flow \eqref{eq:flux_freezing_velocity_isothermal}. As a result, unlike in the isothermal case, there is no enhancement of the cETG growth rate by the TAI mechanism in the isobaric regime: \eqref{eq:isobaric_tai_dispersion_relation} can simply be regarded as an extension of the familiar cETG into the electromagnetic regime. However, physically, the isobaric cTAI is not an interchange mode, since it involves $k_\parallel \neq 0$. Its mechanism is similar to its isothermal cousin (\figref{fig:ctai}), except the balance along the perturbed field is of pressure rather than temperature. It may therefore be appropriate to regard it as an electron-scale extension of MHD-like modes, such as the kinetic ballooning mode (KBM) --- indeed, the condition $E_\parallel = 0$, which is a direct consequence of pressure balance \eqref{eq:isobaric_limit_logt}, is often invoked as a signature of such modes (\citealt{snyder01,kotschenreuther19}).

\subsubsection{Isobaric slab TAI}
\label{sec:isobaric_slab_TAI}
For $k_\parallel \gg k_{\parallel c}$, and still assuming \eqref{eq:isobaric_limit}, our system \eqref{eq:density_moment_collisional}-\eqref{eq:t_moment_collisional} becomes, approximately,
\begin{align}
     &\frac{\rmd}{\rmd t} \frac{\dne}{n_{0e}} =  - \vthe \gradd_\parallel d_e^2 \gradd_\perp^2 \mathcal{A} , \quad \frac{\rmd \mathcal{A}}{\rmd t} + \frac{\vthe}{2} \frac{\partial \varphi}{\partial z} =  0, \quad \varphi = - \taubar \frac{\dne}{n_{0e}}.
     \label{eq:kaw_equation_isobaric}
\end{align}
As with the isothermal KAW, these equations are also valid in the collisionless limit.
These equations are similar to \eqref{eq:kaw_equation}, except that the parallel electric field is now zero because the parallel gradient of the perturbed pressure vanishes. This new system describes the dynamics of \textit{isobaric} KAWs --- so called because they obey \eqref{eq:isobaric_limit_logt}. Their dispersion relation is
\begin{align}
    \omega^2 = k_{\parallel}^2 \vthe^2 k_\perp^2 d_e^2 \frac{\taubar}{2} =  \omega_\text{KAW}^2 \taubar.
    \label{eq:kaw_isobaric}
\end{align}
These isobaric KAW, which arise in strongly driven systems (large $\omega_{*e}$), work in a similar fashion to their isothermal cousins described at the beginning of \secref{sec:isothermal_kaw}, except the inductive part of the parallel electric field now creates a magnetic perturbation and, therefore, a parallel current, from the electrostatic part of the parallel electric field, rather than from a combination of the latter and the parallel pressure gradient. 

Like the isothermal KAW, the isobaric KAW are unstable to sTAI: expanding \eqref{eq:tai_dispersion_relation_general} or \eqref{eq:tai_growth_rate_and_real_frequency} for $\xi_* \gg 1$ at $k_{\parallel} \gg k_{\parallel c}$, we find
\begin{align}
   \omega_r^2 = \omega_\text{KAW}^2 \taubar, \quad \gamma^2 = \omega_\text{KAW}^2 \frac{1}{4\taubar \xi_*^2},
    \label{eq:stai_isobaric_limit}
\end{align}
which is \eqref{eq:kaw_isobaric} once again but with a small, but finite, growth rate.
In a similar fashion to the isothermal sTAI described in \secref{sec:isothermal_kaw}, the instability arises due to the fact that the isobaric condition \eqref{eq:isobaric_limit_logt} that led to \eqref{eq:kaw_equation_isobaric} is, in fact, only approximately satisfied. In the collisional limit, $\gradd_\parallel \log p_e$ is determined at next order in $\xi_*^{-1}$ by \eqref{eq:isobaric_limit_next_order}, which, linearising and Fourier transforming, can be written as 
\begin{align}
    \left(\gradd_\parallel \log p_e \right)_{\vec{k}} =  \frac{1}{i\xi_*} \left(\gradd_\parallel \frac{\dne}{n_{0e}} \right)_{\vec{k}}.
    \label{eq:stai_log_t_correction_isobaric}
\end{align}
This means that there will be a term in the second equation in \eqref{eq:kaw_equation_isobaric} that is $\pi/2$ out of phase with the electrostatic contribution to the parallel electric field that enables the isobaric KAW. The result is the instability \eqref{eq:stai_isobaric_limit}, which exists also in the collisionless limit, but with $\xi_*$ given by \eqref{eq:xi_definition_collisionless} [see \eqref{eq:lin_logt_final}]. Indeed, restoring finite pressure gradients in \eqref{eq:kaw_equation_isobaric}, we have
\begin{align}
     &\frac{\rmd}{\rmd t} \frac{\dne}{n_{0e}} =  -  \vthe \gradd_\parallel d_e^2 \gradd_\perp^2 \mathcal{A} , \quad \frac{\rmd \mathcal{A}}{\rmd t} + \frac{\vthe}{2} \frac{\partial \varphi}{\partial z} = \frac{\vthe}{2} \gradd_\parallel \log p_e,
     \label{eq:kaw_equation_isobaric_stai}
\end{align}
leading to the dispersion relation
\begin{align}
    \omega^2 = \omega_\text{KAW}^2 \left( \taubar + \frac{1}{i\xi_*} \right) \quad \Rightarrow \quad \omega \approx \pm \omega_\text{KAW} \left( \sqrt{\taubar} - \frac{i}{2\sqrt{\taubar} \xi_*} + \dots \right),
    \label{eq:stai_dispersion_relation_isobaric}
\end{align}
whose real and imaginary parts are exactly the frequency and growth rate \eqref{eq:stai_isobaric_limit}. 

As $k_\parallel$ is increased, the isobaric limit \eqref{eq:isobaric_limit} must eventually break down and be replaced by the isothermal limit \eqref{eq:isothermal_limit}. This means that there will be a transition between isobaric and isothermal KAW, and the associated limits of sTAI, occuring, clearly, at $\xi_* \sim 1$. In the collisionless limit, the growth rate once again asymptotes to a constant value as $k_{\parallel } \rightarrow \infty$ ($\xi_* \rightarrow 0$) --- this is just the isothermal limit \eqref{eq:tai_isothermal_collisionless_max} except now, since $k_\perp \gg k_{\perp *}$, this growth rate is large in comparison with the cETG growth rate achieved at $k_\parallel = 0$ [see \figref{fig:tai_growth_rate_and_frequency_isobaric}(a), noting the normalisation]. Note that $\gamma \sim \omega_\text{KAW}$ near the transition $\xi_* \sim 1$ (i.e., at $k_\parallel \sim \omega_{*e}/\vthe$), but $\gamma \ll \omega_\text{KAW}$ as $k_{\parallel} \rightarrow \infty$. In the collisional limit, there is peak growth at $\xi_* \sim 1$, or
\begin{align}
    k_{\parallel \text{max}} \sim \sqrt{\frac{\omega_{*e}}{\kappa}} \sim \frac{1}{\vthe} \sqrt{\omega_{*e} \nu_e}. 
    \label{eq:stai_max_kpar}
\end{align}
Determining the precise prefactor, which depends only on $\taubar$ and is, thus, order unity, is only possible numerically, but is, at any rate, inessential. The growth rate at this wavenumber is
\begin{align}
    \gamma \sim \omega_\text{KAW} \sim k_\perp d_e \sqrt{\omega_{*e} \nu_e}.
    \label{eq:stai_max_gamma}
\end{align}
Again, this growth rate is large in comparison with the cETG peak growth rate at $k_{\parallel} = 0$:
\begin{align}
   \frac{\gamma}{\sqrt{2\omega_{de} \omega_{*e}}} \sim k_\perp d_e \sqrt{\frac{\nu_e}{\omega_{de}}} \sim \left( \frac{k_\perp}{k_{\perp *}} \right)^{1/2} \gg 1.
    \label{eq:stai_max_comparison}
\end{align}
\Figref{fig:tai_growth_rate_and_frequency_isobaric} illustrates all of this behaviour. We remind the reader that at large $k_\parallel$ (i.e., in the deep isothermal regime), the instability is quenched by compressional heating in both collisional and collisionless limits (see \secref{sec:stabilisation_of_isothermal_slab_TAI}). 

Thus, the isobaric ($k_\perp \gg  k_{\perp *}$) regime of the TAI is quite different from the isothermal one: the dominant instability is again electromagnetic, rather than electrostatic, but it is the slab TAI --- an instability of KAWs reaching peak growth at the parallel wavenumber where the relevant parallel timescale --- either the parallel-streaming or thermal-conduction rate in the collisionless or collisional regimes, respectively --- is comparable to $\omega_{*e}$. It must be appreciated, of course, that this behaviour only occurs in a relatively narrow interval of perpendicular wavelengths satisfying $k_{\perp *} \ll k_\perp \ll d_e^{-1}$ (or $\ll d_e^{-1} \chi^{-1}$ in the collisional regime). For $k_\perp d_e \gtrsim 1$ (or $\chi^{-1}$ in the collisional regime), it is replaced by the electrostatic instabilities described in \secref{sec:electrostatic_regime_etg}. 

\subsubsection{Stabilisation of isobaric slab TAI}
\label{sec:stabilisation_of_isobaric_slab_TAI}
As was the case with the isothermal sTAI, the isobaric sTAI is also stabilised within a certain region of wavenumber space, this time due to the effects of finite resitivity, or finite electron inertia, in the parallel momentum equation --- \eqref{eq:velocity_moment} or \eqref{eq:velocity_moment_collisional} in the collisionless and collisional limits, respectively.

To work out this stabilisation, we once again consider the collisional limit and, instead of \eqref{eq:tai_ordering_collisional}, the ordering
\begin{align}
    (k_\perp d_e)^2 \nu_{ei} \sim  \omega \sim \kappa k_\parallel^2 \ll \omega_{*e}.
    \label{eq:isobaric_stab_tai_ordering}
\end{align}
A direct consequence of this ordering is that one has to retain the resisitive term in the leading-order parallel momentum equation, viz., the second equation in \eqref{eq:kaw_equation_isobaric_stai}, coming from \eqref{eq:velocity_moment_collisional}, is replaced with 
\begin{align}
     \frac{\rmd \mathcal{A}}{\rmd t} + \frac{\vthe}{2} \frac{\partial \varphi}{\partial z} = \frac{\vthe}{2} \gradd_\parallel \log p_e + \nu_{ei} \frac{u_{\parallel e}}{\vthe}.
     \label{eq:isobaric_stab_equations}
\end{align}
This means that, instead of the system being isobaric to leading order in $\xi_* \gg 1$, the parallel pressure gradient now balances the electron-ion frictional force:
\begin{align}
    \gradd_\parallel \log p_e +  \frac{2\nu_{ei} u_{\parallel e}}{\vthe^2} = 0.
    \label{eq:isobaric_stab_pressure_balance}
\end{align}
This is obvious from \eqref{eq:tai_logt_initial} in the limit \eqref{eq:isobaric_stab_tai_ordering}. To next order, we must now retain both the time derivative of $\gradd_\parallel\log T_e$ and the compressional heating term in \eqref{eq:tai_logt_initial}:
\begin{align}
    \frac{\rho_e \vthe}{2L_T}  \frac{\partial}{\partial y} \left( \gradd_\parallel \log p_e +  \frac{2\nu_{ei} u_{\parallel e}}{\vthe^2} \right) = \left( \frac{\rmd}{\rmd t} - \kappa \gradd_\parallel^2 \right) \left( \gradd_\parallel \frac{\delta n_e}{n_{0e}} + \frac{2\nu_{ei} u_{\parallel e}}{\vthe^2} \right) - \frac{2}{3}\gradd_\parallel^2 u_{\parallel e}.
    \label{eq:isobaric_stab_next_order}
\end{align}
Combining \eqref{eq:isobaric_stab_equations}, \eqref{eq:isobaric_stab_next_order} and the density equation from \eqref{eq:kaw_equation_isobaric_stai}, we find the dispersion relation
\begin{align}
    \omega^2 - \omega_\text{KAW}^2 \left(\taubar + \frac{1}{i \xi_*} \right)  = - \frac{1}{i \xi_*} \frac{(k_\perp d_e)^2 \nu_{ei}}{\kappa k_\parallel^2} \omega^2 - \frac{1}{\xi_*} \left( \frac{5}{3} + \constant \right) \frac{\omega}{\kappa k_\parallel^2 } \omega_\text{KAW}^2,
    \label{eq:isobaric_stab_dispersion_relation}
\end{align}
where $\constant$ the same numerical constant as in \eqref{eq:isothermal_stab_dispersion_relation} [see \eqref{eq:lin_col_ignore_terms}]. This is the same as \eqref{eq:stai_dispersion_relation_isobaric}, apart from the right-hand side, previously neglected. The second term on the right-hand side simply leads to a small, in $\xi_*\ll 1$, modification of the (real) frequency, and so can be neglected. 

As usual, at the stability boundary, the frequency $\omega$ must be purely real, and both the real and imaginary parts of \eqref{eq:isobaric_stab_dispersion_relation} must vanish individually, giving [cf. \eqref{eq:lin_col_isobaric_stai_stabilisation}]
\begin{align}
    \omega^2  =\omega_\text{KAW}^2 \taubar, \quad \frac{(k_\perp d_e)^2 \nu_{ei}}{\kappa k_\parallel^2} \omega^2  = \omega_\text{KAW}^2 \quad \Rightarrow \quad  \frac{(k_\perp d_e)^2 \nu_{ei}}{\kappa k_\parallel^2} =  \frac{1}{\taubar}.
    \label{eq:isobaric_stab_boundary}
\end{align}
This is a line $k_\parallel \propto k_\perp$ in wavenumber space; moving from small to large parallel wavenumbers, there is a sliver of stability around this line, above which (viz., towards higher $k_\parallel$) the isobaric sTAI grows again to its peak at $\xi_* \sim 1$: see figures~\ref{fig:lin_col_em_plots_no_drifts}(a) and \ref{fig:lin_col_em_plots}(a) in \apref{app:collisional_linear_theory}, where the stability boundary is worked out exactly. As with the case of the isothermal sTAI, this stabilisation was not captured by the general TAI dispersion relation \eqref{eq:tai_dispersion_relation_general} because the ordering \eqref{eq:tai_ordering_collisional} did not formally allow frequencies comparable to both the heat-conduction and the resistive-dissipation rates, required by \eqref{eq:isobaric_stab_boundary}.

In the collisionless limit, we find that the isobaric sTAI is stabilised at the flux-freezing scale \eqref{eq:flux_freezing_scale} [cf. \eqref{eq:lin_stai_stab_isobaric_boundary}]
\begin{align}
    k_\perp d_e \sim 1.
    \label{eq:isobaric_stab_boundary_collisionless}
\end{align}
This is not via a mechanism analogous to the collisional case, as there are no resistive effects in the collisionless limit, but is instead due to the effect of finite electron inertia appearing in the parallel-momentum equation \eqref{eq:velocity_moment} (see \apref{app:stabilisation_of_isobaric_stai}).

The stabilisation of the isobaric sTAI is somewhat more relevant than the stabilisation of the isothermal sTAI (\secref{sec:stabilisation_of_isothermal_slab_TAI}), owing to the fact that the isobaric sTAI is the dominant instability for $k_{\perp *} \ll k_\perp \ll d_e^{-1}$ (or $\ll d_e^{-1} \chi^{-1}$ in the collisional regime). However, we shall discover in \secref{sec:kaw_dominated_slab_tai_turbulence} that the isobaric sTAI contributes only an order-unity amount to the turbulent energy injection --- rather than introducing significant qualitative differences --- and so the (linear) stabilisation thereof appears to be of little consequence in the nonlinear context.

\section{Summary of linear instabilities}
\label{sec:summary_of_wavenumber_space}
In sections \ref{sec:electrostatic_regime_etg} and \ref{sec:electromagnetic_regime_tai}, we introduced the linear instabilities supported by our low-beta system of equations in the electrostatic and electromagnetic regimes, respectively. In both the collisionless and collisional limits, we found that there were four main instabilities: slab ETG [sETG, \eqref{eq:setg_gamma} or \eqref{eq:col_setg_gamma}], curvature-mediated ETG [cETG, \eqref{eq:cetg_gamma}], slab TAI [sTAI, \eqref{eq:tai_isothermal_collisionless_max} or \eqref{eq:stai_max_gamma}] and curvature-mediated TAI [cTAI, \eqref{eq:isothermal_tai_dispersion_relation}]. Before moving on to our discussions of the turbulence supported by these modes, it will be useful to take stock of what we have learned by surveying the locations of each of these instabilities in wavenumber space. Throughout the discussions that follow, we will assume $k_y \sim k_\perp$, and so consider $(k_\perp, k_\parallel)$ to be the relevant wavenumber-space coordinates. We shall also assume $\taubar\sim 1$, implying that both species have roughly comparable temperatures and, more crucially, that $\taubar$ has no dependence on perpendicular wavenumbers (as it could do, for example, on scales comparable to the ion Larmor radius; see \apref{app:ion_equations}).

\subsection{Collisionless limit}
\label{sec:collisionless_limit}
Let us first focus on the collisionless limit. At electrostatic scales $k_\perp \gtrsim d_e^{-1}$ [i.e., below the flux-freezing scale \eqref{eq:flux_freezing_scale}], we have both the sETG and cETG instabilities. The transition between these two instabilities occurs when their growth rates are comparable,~viz.,
\begin{align}
    (k_\parallel^2 \vthe^2 \omega_{*e} )^{1/3} \sim (\omega_{de} \omega_{*e})^{1/2} \quad \Rightarrow \quad \frac{k_\parallel L_T}{\sqrt{\beta_e}} \sim \left( \frac{L_T}{L_B} \right)^{3/4} k_y d_e.
    \label{eq:setg_vs_cetg}
\end{align}
The sETG instability begins to be quenched by Landau damping when its growth rate becomes comparable to the parallel streaming rate:
\begin{align}
     (k_\parallel^2 \vthe^2 \omega_{*e} )^{1/3} \sim k_\parallel \vthe \quad \Rightarrow \quad \frac{k_\parallel L_T}{\sqrt{\beta_e}} \sim  k_y d_e.
     \label{eq:setg_vs_parallel_streaming}
\end{align}
Note that this is the same line as that corresponding to the maximum of the sETG growth rate, viz., $k_\parallel \vthe \sim \omega_{*e}$. However, it must be stressed that this is only true asymptotically, as is evident from figures \ref{fig:lin_em_no_drifts}(a) and \ref{fig:lin_em}(a).
Furthermore, careful analysis of collisionless dispersion relation reveals that the sETG instability is also effectively stabilised --- with only exponentially small growth rates remaining --- around the flux-freezing scale [see \eqref{eq:lin_es_stability_boundary} and the surrounding discussion]. This `fluid' stabilisation occurs when its growth rate becomes comparable to the KAW frequency: 
\begin{align}
    (k_\parallel^2 \vthe^2 \omega_{*e} )^{1/3} \sim \omega_\text{KAW} \quad \Rightarrow \quad \frac{k_\parallel L_T}{\sqrt{\beta_e}} \sim (k_\perp d_e)^{-2}.
    \label{eq:setg_vs_kaw}
\end{align}

For $k_{\perp *} \lesssim k_\perp \lesssim d_e^{-1}$, the dominant instability is the isobaric sTAI, which is separated from the cETG instability by $k_\parallel = k_{\parallel c}$. The cETG instability in this perpendicular-wavenumber range, and for $ k_\parallel \lesssim k_{\parallel c}$, can also be thought about as either the isobaric version of cTAI or the electron version of KBM (see \secref{sec:isobaric_limit}). The isobaric sTAI instability at $k_\parallel \gtrsim k_{\parallel c}$ is stabilised around the flux-freezing scale $k_\perp d_e \sim 1$ [see \eqref{eq:isobaric_stab_boundary_collisionless}]. The area bounded by the lines $k_\perp d_e \sim 1$, $k_\parallel = k_{\parallel c}$ and \eqref{eq:setg_vs_kaw} thus contains only exponentially small growth rates that would be quenched by the effects of finite dissipation in any real physical system. 

For $k_\perp \lesssim k_{\perp *}$, the cETG (or isobaric cTAI) instability is superseded by the isothermal cTAI, which is now the dominant instability, and is separated from sTAI along the horizontal line $k_\parallel = k_{\parallel c}$. 

The sTAI growth rate is cut off at large parallel wavenumbers due to the effect of parallel compression [see \eqref{eq:isothermal_stab_boundary_collisionless}], viz., when
\begin{align}
    \omega_\text{KAW} \sim \omega_{*e} \quad \Rightarrow \quad \frac{ k_\parallel L_T}{\sqrt{\beta_e}} \sim 1.
    \label{eq:stai_vs_drive}
\end{align}

\begin{figure}

\begin{tikzpicture}[scale=1.6, thick, every node/.style={scale=1.2}]

\def\xaxis{7}
\draw[-latex] (0,0) -- (\xaxis,0);
\draw (\xaxis,0) node[anchor=north,scale=0.9] {$k_\perp  d_e $};

\def\yaxis{4}
\draw[-latex] (0,0) -- (0,\yaxis);
\draw (0,\yaxis) node[anchor=east,scale=0.9] {$k_\parallel L_T/\sqrt{\beta_e}$};

\def\boundarystyle{dashed}
\def\boundarythickness{thin}
\def\ymax{3.8}
\def\xmax{6.5}
\def\fluxfreezing{2.75}
\def\kawstable{1.5}
\def\kparc{0.75}
\def\streamingmax{\xmax*\kawstable/\fluxfreezing}
\def\cetgminx{\fluxfreezing+0.5}
\def\cetgminy{\kparc}
\def\cetgmaxy{2.2}
\def\kperpstar{\kparc*\fluxfreezing/\kawstable}
\def\systemsize{0.5*\kparc}
\def\dotradius{0.05}

\draw[\boundarystyle, \boundarythickness] (\fluxfreezing,0) -- (\fluxfreezing,\cetgminy);
\draw (\fluxfreezing,0) node[anchor=north,scale=0.8] {$1$};
\draw[\boundarystyle,\boundarythickness] (\fluxfreezing,\kawstable) -- (\fluxfreezing,\ymax);

\draw[\boundarystyle,\boundarythickness] (0,\kawstable) -- 
node[midway, anchor = south] {{\scriptsize \eqref{eq:stai_vs_drive}}}
(\fluxfreezing,\kawstable);
\draw (0,\kawstable) node[anchor=east,scale=0.8] {$1$};
\draw (\fluxfreezing/2-0.2,\kawstable/2 + \kparc/2) node[anchor = center] {{\scriptsize sTAI}};
\draw (\fluxfreezing/2,\ymax/2 + \kawstable/2) node[anchor = center] {{\scriptsize Stable KAW}};

\draw[dotted,\boundarythickness] (\kperpstar,\kparc) -- 
node[midway, anchor=south,rotate=atan(\kawstable/\fluxfreezing)] 
{{\scriptsize $\gamma \propto k_y^2$}} 
(\fluxfreezing,\kawstable);

\draw[\boundarystyle,\boundarythickness] (\fluxfreezing,\kawstable) -- 
node[midway, anchor=south,rotate=atan(\kawstable/\fluxfreezing)] 
{{\scriptsize $\gamma \propto k_y$}} 
(\xmax,\streamingmax);

\draw (\fluxfreezing/2+\xmax/2 - 0.5,\kawstable/\fluxfreezing*\fluxfreezing/2+\kawstable/\fluxfreezing*\xmax/2 +0.5 ) node[anchor = center, rotate=atan(\kawstable/\fluxfreezing)] {{\scriptsize Parallel streaming}};

\draw (\fluxfreezing/2+\xmax/2-0.2,\kawstable/\fluxfreezing*\fluxfreezing/2+\kawstable/\fluxfreezing*\xmax/2 ) 
node[anchor = north west, 
] {{\scriptsize \eqref{eq:setg_vs_parallel_streaming}}};

\draw[\boundarystyle,\boundarythickness] (0,\kparc) -- (\kperpstar/4,\kparc);
\draw[\boundarystyle,\boundarythickness] (3*\kperpstar/4,\kparc) -- (\cetgminx,\kparc);
\draw (0,\kparc) node[anchor=east,scale=0.8] {$(L_T/L_B)^{1/2}$};

\draw[\boundarystyle,\boundarythickness] (\fluxfreezing,\kawstable) --
node[midway, anchor = south west]
{{\scriptsize \eqref{eq:setg_vs_kaw}}}
(\cetgminx,\cetgminy);
\def\cetgrise{\cetgmaxy-\kparc}
\def\cetgrun{\xmax-\cetgminx}
\draw[\boundarystyle,\boundarythickness] (\cetgminx,\cetgminy) -- 
node[midway, anchor=north west
] 
{{\scriptsize \eqref{eq:setg_vs_cetg}}}
(\xmax,\cetgmaxy);
\draw (\cetgminx+0.5,3*\kparc/4) node[anchor = north west] {{\scriptsize cETG}};
\draw (\fluxfreezing/2+\xmax/2,\kawstable+0.25) node[anchor = center] {{\scriptsize sETG}};

\draw[\boundarystyle,\boundarythickness] (\fluxfreezing,\kawstable) --  (\fluxfreezing,\cetgminy);

\draw[\boundarystyle,\boundarythickness] (\kperpstar,0) -- (\kperpstar,\kparc);
\draw (\kperpstar,0) node[anchor=north, scale=0.8] {$k_{\perp *} d_e$};
\draw (\kperpstar/2,\kparc/3 ) node[anchor=center] {{\scriptsize cTAI}};
\draw (\kperpstar/2+0.38,\kparc/3 +0.22) node[anchor=center] {{\scriptsize \eqref{eq:tai_curvature_maximum_explicit}}};

\draw[\boundarystyle,\boundarythickness] (\xmax,0) -- (\xmax,\ymax);
\draw (\xaxis-0.25,\yaxis/2) node[anchor = center, rotate=90] {{\scriptsize Ultraviolet cutoff}};

\def\ydiff{\kparc - \systemsize}
\draw (0,\systemsize) node[anchor = east, scale = 0.8] {$ L_T/\sqrt{\beta_e} L_\parallel$};
\draw [-,thick,black] (0,\systemsize) -- 
node[midway, anchor=south,rotate=-atan(\ydiff/\kperpstar)] 
{{\scriptsize $\gamma \propto k_y$}} 
(\kperpstar,\kparc)  -- (\xmax,\streamingmax);
\draw[fill=black] (0,\systemsize) ellipse [x radius=\dotradius, y radius =\dotradius];
\draw[fill=black] (\fluxfreezing,\kawstable) ellipse [x radius=\dotradius, y radius =\dotradius];

\fill [gray, opacity=0.15] (0,\kawstable) -- (\fluxfreezing,\kawstable) -- (\fluxfreezing,\ymax) -- (0,\ymax) -- cycle;
\fill [gray, opacity=0.15] (\fluxfreezing,\kawstable) -- (\xmax,\streamingmax) -- (\xmax,\ymax) -- (\fluxfreezing,\ymax) -- cycle;
\fill [gray, opacity=0.15] (\xmax,0) -- (\xaxis,0) -- (\xaxis,\ymax) -- (\xmax,\ymax) -- cycle;
\fill [gray, opacity=0.15] (\fluxfreezing,\kparc) -- (\fluxfreezing,\kawstable) -- (\cetgminx,\cetgminy)  -- cycle;	
	
\draw[dotted, \boundarythickness] (\kperpstar/2,0) -- (\kperpstar/2,\ymax);
\draw (\kperpstar/2,\ymax) node[anchor=south,scale=0.8] {\Figref{fig:tai_growth_rate_and_frequency}};

\draw[dotted, \boundarythickness] (\kperpstar/2+\fluxfreezing/2,0) -- (\kperpstar/2+\fluxfreezing/2,\ymax);
\draw (\kperpstar/2+\fluxfreezing/2,\ymax) node[anchor=south,scale=0.8] {\Figref{fig:tai_growth_rate_and_frequency_isobaric}};

\end{tikzpicture}

    \caption{ Collisionless modes in the $(k_\perp ,k_\parallel )$ plane, where the axes are plotted on logarithmic scales. The dotted lines are the asymptotic boundaries between the various modes, with the shaded regions indicating stability. The stable region and the stability boundary are derived and plotted in a more quantitative way in \apref{app:exact_stability_boundary} (see \figref{fig:lin_em}a). At electrostatic scales (i.e., those below the flux-freezing scale, $k_\perp d_e >1$), the curvature-mediated ETG [cETG, \eqref{eq:cetg_gamma}] transitions into the slab ETG [sETG,~\eqref{eq:setg_gamma}] along the boundary \eqref{eq:setg_vs_cetg}, while the sETG is damped by parallel streaming above~\eqref{eq:setg_vs_parallel_streaming}. `Fluid' stabilisation of the sETG occurs along \eqref{eq:setg_vs_kaw}. At electromagnetic scales (i.e., those above the flux-freezing scale, $k_\perp d_e <1$), slab TAI [sTAI, \eqref{eq:stai_isobaric_limit}] is stabilised along~$k_\perp d_e \sim 1$, meaning that the region enclosed by the lines $k_\perp d_e \sim 1$, $k_\parallel = k_{\parallel c}$, and \eqref{eq:setg_vs_kaw} contains only exponentially small growth rates, and can thus effectively be considered stable [note that $k_{\parallel c} L_T/\sqrt{\beta_e} =(L_T/L_B)^{1/2}$, see \eqref{eq:tai_kpar_critical}].
    The cETG transitions into the curvature-mediated TAI [cTAI, \eqref{eq:isothermal_tai_dispersion_relation}] along~$k_\perp = k_{\perp *}$, with $k_{\perp *}$ defined in \eqref{eq:tai_transition_kperp}. cTAI is separated from sTAI by the horizontal line $k_\parallel  = k_{\parallel c}$, while sTAI is stabilised by compressional heating at the horizontal line given by~\eqref{eq:stai_vs_drive}, transitioning into purely oscilliatory (isothermal) KAWs~\eqref{eq:kaw_isothermal}. Electron finite-Larmor-radius (FLR) effects eventually provide an ultraviolet cutoff at large perpendicular wavenumbers $k_\perp d_e$, though this is outside the range of validity of our drift-kinetic approximation. Note that the transition to ion-scale physics at small perpendicular wavenumbers $k_\perp \rho_i \lesssim 1$ lies outside our adiabatic-ion approximation. The solid black line indicates the location of the maximum growth rate at each fixed $k_\perp$, while the solid dots are the (possible) locations of the energy-containing scale(s) (see \secref{sec:free_energy_and_turbulence}). The dotted vertical lines indicate the location in $k_\perp$ of figures \ref{fig:tai_growth_rate_and_frequency} and \ref{fig:tai_growth_rate_and_frequency_isobaric}, which show the isothermal and isobaric regimes, respectively.}
    \label{fig:collisionless_phase_space_portrait}
\end{figure} 

This is all illustrated in \figref{fig:collisionless_phase_space_portrait}, where the solid line shows the location of the peak growth rate at each $k_y$ --- following, at $k_\perp \lesssim k_{\perp *}$, the peak growth of the isothermal cTAI \eqref{eq:tai_curvature_maximum_explicit}, and at $k_\perp \geqslant k_{\perp *}$, the boundary $\xi_* \sim 1$ between the isothermal and isobaric regimes. The increase of the growth rate with $k_y$ is unchecked in the drift-kinetic approximation that we have adopted, and requires the damping effects associated with the finite Larmor radius (FLR) of the electrons to be taken into account; this will introduce some ultraviolet cutoff in perpendicular wavenumbers. At the largest scales, we must eventually encounter ion dynamics, but the effects that this may have are outside the scope of this paper. All of these modes are, of course, limited by the finite parallel system size $L_\parallel$, meaning that the smallest accessible parallel wavenumber is $k_\parallel \sim L_{\parallel}^{-1}$.

\subsection{Collisional limit}
\label{sec:collisional_limit}
 \begin{figure}
    \centering
    
        \begin{tikzpicture}[scale=1.6, thick, every node/.style={scale=1.2}]

\def\xaxis{7}
\draw[-latex] (0,0) -- (\xaxis,0);
\draw (\xaxis,0) node[anchor=north,scale=0.9] {$k_\perp  d_e $};

\def\yaxis{4}
\draw[-latex] (0,0) -- (0,\yaxis);
\draw (0,\yaxis) node[anchor=east,scale=0.9] {$k_\parallel L_T/\sqrt{\beta_e}$};

\def\boundarystyle{dashed}
\def\boundarythickness{thin}
\def\ymax{3.8}
\def\xmax{6.5}
\def\fluxfreezing{2.75}
\def\kawstable{1.5}
\def\kparc{0.75}
\def\streamingmax{\xmax*\kawstable/\fluxfreezing}
\def\cetgminx{\fluxfreezing}
\def\cetgminy{\kparc}
\def\cetgmaxy{2.5}
\def\kperpstar{\kparc*\fluxfreezing/\kawstable}
\def\systemsize{0.5*\kparc}
\def\isobaricminx{\fluxfreezing-0.3}
\def\dotradius{0.05}
\def\kawxmax{{\fluxfreezing - (\ymax - \kawstable)*(\fluxfreezing-\isobaricminx)/(\kawstable - \kparc)}}

\draw[\boundarystyle, \boundarythickness] (\fluxfreezing,0) -- (\fluxfreezing,\kawstable);
\draw (\fluxfreezing,0) node[anchor=north,scale=0.8] {$\chi^{-1}$};

\def\deltay{(\ymax - \kawstable)}
\def\deltax{(\kawxmax - \fluxfreezing)}
\draw[\boundarystyle,\boundarythickness] (\fluxfreezing,\kawstable) -- 
node[midway, anchor=east] 
{{\scriptsize \eqref{eq:col_stai_vs_resisitivity}}} 
(\kawxmax,\ymax);
\draw[\boundarystyle,\boundarythickness]  (\isobaricminx,\kparc) --
(\fluxfreezing,\kawstable);

\draw[\boundarystyle,\boundarythickness] (0,\kawstable) -- (17*\fluxfreezing/24,\kawstable);
\draw[\boundarystyle,\boundarythickness]  (5*\fluxfreezing/6,\kawstable) -- (\fluxfreezing,\kawstable);
\draw (\kperpstar,\kawstable)  node[anchor=south] {{\scriptsize \eqref{eq:stai_vs_drive_col}}};
\draw (0,\kawstable) node[anchor=east,scale=0.8] {$1$};
\draw (\fluxfreezing/2-0.2,\kawstable/2 + \kparc/2) node[anchor = center] {{\scriptsize sTAI}};
\draw (\fluxfreezing/2,\ymax/2 + \kawstable/2) node[anchor = center] {{\scriptsize Stable KAW}};

\draw[dotted,\boundarythickness] (\kperpstar,\kparc) -- 
node[midway, anchor=south, 
rotate = atan(\kawstable/\fluxfreezing
] 
{{\scriptsize $\gamma \propto k_y^{3/2}$}} 
(\fluxfreezing,\kawstable);

\draw[\boundarystyle,\boundarythickness] (\fluxfreezing,\kawstable) -- 
node[midway, anchor=south, rotate = atan(\kawstable/\fluxfreezing)] 
{{\scriptsize $\gamma \propto k_y$}} 
(\xmax,\streamingmax);

\draw (\fluxfreezing/2+\xmax/2 -0.1,\kawstable/\fluxfreezing*\fluxfreezing/2+\kawstable/\fluxfreezing*\xmax/2 +0.6 ) node[anchor = center, rotate=atan(\kawstable/\fluxfreezing)] {{\scriptsize Thermal conduction}};

\draw (\fluxfreezing/2+\xmax/2 -0.2,\kawstable/\fluxfreezing*\fluxfreezing/2+\kawstable/\fluxfreezing*\xmax/2) node[anchor = north west
] {\scriptsize \eqref{eq:col_setg_vs_thermal_conduction}};

\draw[\boundarystyle,\boundarythickness] (0,\kparc) -- (\kperpstar/4,\kparc);
\draw[\boundarystyle,\boundarythickness] (3*\kperpstar/4,\kparc) -- (\cetgminx,\kparc);
\draw (0,\kparc) node[anchor=east,scale=0.8] {$(L_T/L_B)^{1/2}$};

\def\rise{(\cetgmaxy - \cetgminy)}  
\def\run{(\xmax - \cetgminx)}  

\draw[\boundarystyle,\boundarythickness] (\cetgminx,\cetgminy) -- 
node[midway, anchor=north west
] 
{{\scriptsize \eqref{eq:col_setg_vs_cetg}}} 
(\xmax,\cetgmaxy);

\draw (\cetgminx+1,\cetgminy) node[anchor = north west] {{\scriptsize cETG}};
\draw (\fluxfreezing/2+\xmax/2-0.5,\kawstable+0.25) node[anchor = center] {{\scriptsize sETG}};

\draw[\boundarystyle,\boundarythickness] (\kperpstar,0) -- (\kperpstar,\kparc);
\draw (\kperpstar,0) node[anchor=north, scale=0.8] {$k_{\perp *} d_e $};
\draw (\kperpstar/2,\kparc/3 ) node[anchor=center] {{\scriptsize cTAI}};
\draw (\kperpstar/2+0.38,\kparc/3 +0.22) node[anchor=center] {{\scriptsize \eqref{eq:tai_curvature_maximum_explicit}}};

\draw[\boundarystyle,\boundarythickness] (\xmax,0) -- (\xmax,\ymax);
\draw (\xaxis-0.25,\yaxis/2) node[anchor = center, rotate=90] {{\scriptsize Perpendicular viscosity}};

\def\ydiff{\kparc - \systemsize}
\draw (0,\systemsize) node[anchor = east, scale = 0.8] {$L_T/ \sqrt{\beta_e} L_\parallel$};
\draw [-,thick,black] (0,\systemsize) -- 
node[midway, anchor=south, rotate = -atan(\ydiff/\kperpstar)] 
{{\scriptsize $\gamma \propto k_y$}} 
(\kperpstar,\kparc)  -- (\xmax,\streamingmax);
\draw[fill=black] (0,\systemsize) ellipse [x radius=\dotradius, y radius =\dotradius];
\draw[fill=black] (\fluxfreezing,\kawstable) ellipse [x radius=\dotradius, y radius =\dotradius];

\fill [gray, opacity=0.15] (0,\kawstable) -- (\fluxfreezing,\kawstable) -- (\kawxmax,\ymax) -- (0,\ymax) -- cycle;
\fill [gray, opacity=0.15] (\fluxfreezing,\kawstable) -- (\xmax,\streamingmax) -- (\xmax,\ymax) -- (\kawxmax,\ymax) -- cycle;
\fill [gray, opacity=0.15] (\xmax,0) -- (\xaxis,0) -- (\xaxis,\ymax) -- (\xmax,\ymax) -- cycle;

\draw[dotted, \boundarythickness] (\kperpstar/2,0) -- (\kperpstar/2,\ymax);
\draw (\kperpstar/2,\ymax) node[anchor=south,scale=0.8] {\Figref{fig:tai_growth_rate_and_frequency}};

\draw[dotted, \boundarythickness] (\kperpstar/2+\fluxfreezing/2,0) -- (\kperpstar/2+\fluxfreezing/2,\ymax);
\draw (\kperpstar/2+\fluxfreezing/2,\ymax) node[anchor=south,scale=0.8] {\Figref{fig:tai_growth_rate_and_frequency_isobaric}};

\end{tikzpicture}

    \caption{Collisional modes in the $(k_\perp ,k_\parallel )$ plane, where the axes are plotted on logarithmic scales. The dotted lines are the asymptotic boundaries between the various modes, with the shaded regions indicating stability. The stable region and the stability boundary are derived and plotted in a more quantitative way in \apref{app:exact_stability_boundary_collisional} (see \figref{fig:lin_col_em_plots}a). At electrostatic scales (i.e., those below the flux-freezing scale, $k_\perp d_e \chi >1$), the curvature-mediated ETG [cETG, \eqref{eq:cetg_gamma}] transitions into the (collisional) slab ETG [sETG, \eqref{eq:col_setg_gamma}] along the boundary~\eqref{eq:col_setg_vs_cetg}. sETG is damped by parallel heat conduction above~\eqref{eq:col_setg_vs_thermal_conduction}. At electromagnetic scales (i.e., those above the flux-freezing scale, $k_\perp d_e \chi <1$), the slab TAI [sTAI, \eqref{eq:stai_isobaric_limit}] is stabilised by the effects of finite resisitivty along~\eqref{eq:col_stai_vs_resisitivity}, while cETG transitions into the curvature-mediated TAI [cTAI, \eqref{eq:isothermal_tai_dispersion_relation}] at $k_\perp = k_{\perp *}$, with $k_{\perp *}$ defined in \eqref{eq:tai_transition_kperp}. cTAI is separated from sTAI by the horizontal line $k_\parallel = k_{\parallel c}$ [note that $k_{\parallel c} L_T/\sqrt{\beta_e} =(L_T/L_B)^{1/2}$, see~\eqref{eq:tai_kpar_critical}], while the sTAI is stabilised by compressional heating at the horizontal line given by \eqref{eq:stai_vs_drive_col}, transitioning into purely oscilliatory (isothermal) KAWs~\eqref{eq:kaw_isothermal}. Perpendicular electron viscosity will eventually provide an ultraviolet cutoff for these modes at large perpendicular wavenumbers $k_\perp d_e$, though this is outside the range of validity of our drift-kinetic approximation. As in \figref{fig:collisionless_phase_space_portrait}, the ion-scale range $k_\perp \rho_i\lesssim 1$ is left outside our considerations. The solid black line indicates the location of maximum growth at each fixed $k_\perp$, while the solid dots are (possible) locations of the energy containing scale(s) (see \secref{sec:free_energy_and_turbulence}). The dotted vertical lines indicate the location in $k_\perp$ of figures \ref{fig:tai_growth_rate_and_frequency} and \ref{fig:tai_growth_rate_and_frequency_isobaric}, which show the isothermal and isobaric regimes, respectively.} 
    \label{fig:collisional_phase_space_portrait}
\end{figure}
The picture is qualitatively similar in the collisional limit, except the transition between the electrostatic and electromagnetic regimes is modified, as discussed in \secref{sec:summary_of_equations}. At electrostatic scales $k_\perp \gtrsim d_e^{-1} \chi^{-1}$ [i.e., those below the flux-freezing scale \eqref{eq:flux_freezing_scale_col}], we once again have both the (collisional) sETG and cETG instabilites, whose growth rates become comparable when
\begin{align}
    \left( \frac{k_\parallel^2 \vthe^2 \omega_{*e}}{\nu_{ei}} \right)^{1/2} \sim \left( \omega_{de} \omega_{*e} \right)^{1/2} \quad \Rightarrow \quad \frac{k_\parallel L_T}{\sqrt{\beta_e}} \sim \left( \frac{L_T}{L_B} \right)^{1/2} (k_y d_e \chi)^{1/2}.
    \label{eq:col_setg_vs_cetg}
\end{align}
The sETG instability is now quenched by thermal conduction at
\begin{align}
    \left( \frac{k_\parallel^2 \vthe^2 \omega_{*e}}{\nu_{ei}} \right)^{1/2} \sim \kappa k_\parallel^2 \quad \Rightarrow \quad \frac{k_\parallel L_T}{\sqrt{\beta_e} } \sim  (k_y d_e \chi)^{1/2}.
    \label{eq:col_setg_vs_thermal_conduction}
\end{align}
Note that this is the same line as that corresponding to the maximum of the collisional-sETG growth rate, viz., $ (k_\parallel \vthe)^2/\nu_{ei}  \sim \omega_{*e}$. As in the collisionless case, this is, of course, only true asymptotically: see figures \ref{fig:lin_col_em_plots_no_drifts}(a) and \ref{fig:lin_col_em_plots}(a).

For $k_{\perp *} \lesssim k_\perp \lesssim d_e^{-1} \chi^{-1}$, the dominant instability is once again the isobaric sTAI, separated from cETG by $k_\parallel = k_{\parallel c}$. As in the collisionless limit, the cETG instability in this perpendicular-wavenumber range, and for $ k_\parallel \lesssim k_{\parallel c}$, can also be thought of as either the isobaric version of cTAI or the electron version of KBM (see \secref{sec:isobaric_limit}). The isobaric sTAI instability is stabilised due to the effects of finite resistivity along the line [see \eqref{eq:isobaric_stab_boundary}]
 \begin{align}
     \kappa k_\parallel^2 \sim (k_\perp d_e)^2 \nu_{ei} \quad \Rightarrow \quad \frac{k_\parallel L_T}{\sqrt{\beta_e}} \sim k_\perp d_e \chi.
     \label{eq:col_stai_vs_resisitivity}
 \end{align}
 
 For $k_\perp \lesssim k_{\perp *}$, the cETG (or isobaric cTAI) instability is superseded by the isothermal cTAI, which is once again the dominant instability, and is separated from the isothermal sTAI by $k_\parallel = k_{\parallel c}$. As in the collisionless case, the isothermal sTAI is cut off at large parallel wavenumbers due to the effects of parallel compression [see~\eqref{eq:isothermal_stab_boundary}], viz., 
\begin{align}
    \omega_\text{KAW} \sim \omega_{*e} \quad \Rightarrow \quad \frac{k_\parallel L_T}{\sqrt{\beta_e}} \sim 1.
    \label{eq:stai_vs_drive_col}
\end{align}

This is all illustrated in \figref{fig:collisional_phase_space_portrait}, where the solid line again shows the location of the fastest growth for each $k_y$. As in the collisionless case, modes are stabilised at large perpendicular numbers, this time by perpendicular electron viscosity, and limited by the parallel system size for small parallel wavenumbers. However, they are now also limited at large parallel wavenumbers by the mean free path $\lambdae$, at which the collisional approximation breaks down. This means that the maximum parallel wavenumber allowed in this collisional limit is $k_\parallel \sim \lambdae^{-1}$.

All of the boundaries between modes derived in this section are, of course, only asymototic illustrations, and do not quantitatively reproduce, for example, the exact stability boundaries in wavenumber space (which are derived in appendices \ref{app:exact_stability_boundary} and \ref{app:exact_stability_boundary_collisional}). However, given that the arguments of the following section rely on scaling estimates, rather than quantitative relationships between parameters, the illustrations of the layout of wavenumber space provided by figures \ref{fig:collisionless_phase_space_portrait} and \ref{fig:collisional_phase_space_portrait} will be sufficient for our purposes.

\section{Electromagnetic turbulence and transport}
\label{sec:free_energy_and_turbulence}

\subsection{Free energy}
\label{sec:free_energy}
Magnetised plasma systems containing small perturbations around a Maxwellian equilibrium nonlinearly conserve free energy, which is a quadratic norm of the magnetic perturbations and the perturbations of the distribution functions of both ions and electrons away from the Maxwellian. In the system that we are considering, the (normalised) free energy takes the form
\begin{equation}
    \frac{W}{n_{0e} T_{0e}} = \int \frac{\rmd^3 \vec{r}}{V} \: \left(\frac{\varphi \bar{\tau}^{-1} \varphi}{2} + \left| d_e \grad_\perp \mathcal{A} \right|^2 + \frac{1}{2}\frac{\dne^2}{n_{0e}^2} + \frac{u_{\parallel e}^2}{\vthe^2} + \frac{1}{4} \frac{\delta T_{\parallel e}^2}{T_{0e}^2} + 
\frac{1}{2} \frac{\delta T_{\perp e}^2}{T_{0e}^2}  + \dots\right).
\label{eq:free_energy_summarised}
\end{equation}
The `$\dots$' stand for the squares of further moments of the perturbed distribution function (such as the parallel and perpendicular heat fluxes $\delta q_{\parallel e}, \delta q_{\perp e}$, etc.). The derivation of~\eqref{eq:free_energy_summarised} can be found in~\apref{app:free_energy}. In the collisional limit, these further moments of the perturbed distribution function are negligible, and \eqref{eq:free_energy_summarised} becomes [see \eqref{eq:energetics_free_energy_0_collisional}]
\begin{align}
    \frac{W}{n_{0e} T_{0e}} = \int \frac{\rmd^3 \vec{r}}{V} \: \left(\frac{\varphi \bar{\tau}^{-1} \varphi}{2} + \left| d_e \grad_\perp \mathcal{A} \right|^2 + \frac{1}{2}\frac{\dne^2}{n_{0e}^2} + \frac{3}{4} \frac{\delta T_{ e}^2}{T_{0e}^2} \right).
\label{eq:free_energy_summarised_collisional}
\end{align}
The free energy is a nonlinear invariant, i.e., it is conserved by nonlinear interactions (\citealt{abel13}), but can be injected into the system by equilibrium gradients, and is dissipated by collisions; even when these are small, they are always eventually accessed via phase-mixing of the distribution function towards small velocity scales and nonlinear interactions towards small spatial scales.

In view of this, the time-evolution of the free energy \eqref{eq:free_energy_summarised} can be written as [see \eqref{eq:energetics_total_derivative}]
\begin{align}
    \frac{1}{n_{0e} T_{0e}} \frac{\rmd W}{\rmd t}= \varepsilon - D,
    \label{eq:free_energy_time_derivative_summarised}
\end{align}
where $D$ stands for the collisional dissipation [see \eqref{eq:energetics_dissipation_electrons} and \eqref{eq:energetics_dissipation_electrons_collisional}], and $\varepsilon$ is the injection rate due, in our system, to the electron-temperature gradient [see \eqref{eq:energetics_energy_injection_electrons} and \eqref{eq:energetics_energy_injection_electrons_collisional}]:
\begin{align}
    \varepsilon = \frac{1}{L_T} \int \frac{\rmd^3 \vec{r}}{V}\left\{
    \begin{array}{ll}
         \displaystyle    \left( \frac{1}{2} \frac{\delta T_{\parallel e}}{T_{0e}} +  \frac{\delta  T_{\perp e}}{T_{0e}} \right) v_{Ex} + \frac{\frac{1}{2} \delta q_{\parallel e}  + \delta q_{\perp e}}{n_{0e} T_{0e}} \frac{\delta \! B_x}{B_0}, &  \displaystyle \text{collisionless},   \\[4mm]
          \displaystyle  \frac{3}{2} \frac{\dTe}{T_{0e}} v_{Ex} + \frac{\delta q_e}{n_{0e} T_{0e}} \frac{\delta \! B_x}{B_0} , &  \displaystyle \text{collisional},
    \end{array}
    \right.
    \label{eq:energy_injection_simplified}
\end{align}
where
\begin{align}
    v_{Ex} = - \frac{\rho_e \vthe}{2} \frac{\partial \varphi}{\partial y}, \quad   \frac{\delta \! B_x}{B_0} = \rho_e \frac{\partial \mathcal{A}}{\partial y}, \quad \frac{\delta q_e}{n_{0e} T_{0e}} = - \frac{3}{2} \kappa \gradd_\parallel \log T_e.
    \label{eq:exb_flow_deltabx}
\end{align}
The expression multiplying $1/L_T$ is the `turbulent' heat flux due to the energy transport by the $\vec{E}\times \vec{B}$ flows and to the heat fluxes along the perturbed field lines. The first term in~\eqref{eq:energy_injection_simplified} is the energy injection by ETG (\secref{sec:electrostatic_regime_etg}), the second by TAI (\secref{sec:electromagnetic_regime_tai}). Evidently, the latter is only present in the electromagnetic regime, when perturbations of the magnetic-field direction are allowed. 

Free energy is normally the quantity whose cascade from large (injection) to small (dissipation) scales determines the properties of a plasma's turbulent state (see \citealt{sch08,sch09}, and references therein). Temperature-gradient-driven turbulence is no exception (\citealt{barnes11}), and so we devote the remainder of this section to working out at what scales and to what saturated amplitudes the ETG-TAI injection \eqref{eq:energy_injection_simplified} will drive turbulent fluctuations.

\subsection{Electrostatic turbulence}
\label{sec:electrostatic_turbulence}

\subsubsection{Collisionless slab ETG turbulence}
\label{sec:collisionless_slab_etg_turbulence}
Following \cite{barnes11}, we shall conjecture that our fully developed electrostatic turbulence always organises itself into a state wherein there is a local cascade of the free energy \eqref{eq:free_energy_summarised} that carries the injected power $\varepsilon$ from the outer scale, through some putative `inerital range', to the dissipation scale. The outer scale is something that we will have to determine, while the dissipation scale will be near $k_\perp \rho_e \sim 1$, and so outside the range of validity of our drift-kinetic approximation. 

The perpendicular nonlinearity in our equations is the advection of fluctuations by the fluctuating $\vec{E}\times \vec{B}$ flows. Therefore, we take the nonlinear turnover time associated with such a cascade to be the nonlinear $\vec{E}\times \vec{B}$ advection rate:
\begin{align}
    t_\text{nl}^{-1} \sim k_\perp v_E \sim  \rho_e \vthe k_\perp^2 \amp{\varphi} \sim \Omega_e (k_\perp \rho_e)^2 \amp{\varphi}.
    \label{eq:turb_nonlinear_time}
\end{align}
Here and in what follows, $\amp{\varphi}$ refers to the characteristic amplitude of the electrostatic potential at the scale $k_\perp^{-1}$, rather than to the Fourier transform of the field. More formally, we shall take $\amp{\varphi}$ to be defined by
\begin{align}
   \amp{\varphi}^2 = \int_{k_\perp}^\infty \rmd k_\perp' \: E_\perp^\varphi(k_\perp'), \quad E_\perp^\varphi(k_\perp) \equiv 2\pi k_\perp \int_{-\infty}^\infty \rmd k_\parallel \left<|\varphi_{\vec{k}}|^2 \right>,
    \label{eq:turb_varphi_definition}
\end{align}
where $ E_\perp^\varphi(k_\perp) $ is the 1D perpendicular energy spectrum, $\varphi_{\vec{k}}$ the spatial Fourier transform of the potential, and the angle brackets denote an ensemble average. Perturbations of other quantities, such as the velocity, parallel temperature, magnetic field, etc., will similarly be taken to refer to their characteristic amplitude at a given perpendicular scale.

Assuming that any possible anisotropy in the perpendicular plane can be neglected\footnote{The existence of such a state is not always guaranteed: e.g., \cite{colyer17} found that the saturated state of electrostatic ETG turbulence existed in a zonally-dominated state, which evidently violates this assumption. In fact, the zonal state is much closer to being 2D isotropic than a streamer-dominated state; \cite{barnes11} explicitly invoked zonal flows to enforce isotropy.}, a Kolmogorov-style constant-flux argument leads to the scaling of the amplitudes in the inertial range:
\begin{equation}
     \frac{\bar{\tau}^{-1}\amp{\varphi}^2}{ t_{\text{nl}}} \sim \varepsilon = \text{const} \quad \Rightarrow \quad \amp{\varphi} \sim \left( \frac{\varepsilon}{\Omega_e} \right)^{1/3} \left(k_\perp \rho_e \right)^{-2/3}.
    \label{eq:turb_setg_constant_flux}
\end{equation}
The scaling \eqref{eq:turb_setg_constant_flux} translates into the following 1D spectrum: 
\begin{equation}
    E_\perp^\varphi(k_\perp) \sim \frac{\amp{\varphi}^2}{k_\perp} \propto k_\perp^{-7/3},
    \label{eq:turb_setg_energy_spectrum}
\end{equation}
the same as was obtained, using a similar argument, and confirmed numerically, by \cite{barnes11} for electrostatic, gyrokinetic ITG turbulence. In making this argument, we have assumed that the free-energy density at a given scale $k_\perp^{-1}$ can be adequately represented by the first term in the integrand of \eqref{eq:free_energy_summarised}, i.e., that all the other fields whose squares contribute to the free energy are either small or comparable to $\varphi$, but never dominant in comparison with it. Whether this is true will depend on the nature of the turbulent fluctuations supported by the system in any given part of the $(k_\perp, \kpar)$ space through which the cascade might be taking free energy on its journey towards dissipation. Let us specialise to the region of the wavenumber space (marked `sETG' in \figref{fig:collisionless_phase_space_portrait}) where the fluctuations are collisionless, electrostatic drift waves described by \eqref{eq:setg_equations}. From the first two equations of \eqref{eq:setg_equations},\footnote{The linear part of the third equation in \eqref{eq:setg_equations} tells us that $\delta \amp{T}_{\parallel e}/T_{0e} \sim (\omega_{*e}/\omega) \amp{\varphi}$ but, as we are about to discover, this is only true at the outer scale, while in the inertial range, the ETG injection term is subdominant.}
\begin{align}
    \taubar^{-1} \amp{\varphi} \sim \frac{k_\parallel \vthe}{\omega} \frac{\amp{u}_{\parallel e}}{\vthe} \sim \left( \frac{k_\parallel \vthe}{\omega} \right)^2 \frac{\delta \amp{T}_{\parallel e}}{T_{0e}}, 
    \label{eq:turb_setg_field_scalings}
\end{align}
where evidently we ought to estimate $\omega \sim t_\text{nl}^{-1}$. Then, all three fluctuating fields do indeed have the same size and the same scaling if we posit 
\begin{align}
    t_\text{nl}^{-1} \sim  k_\parallel \vthe.
    \label{eq:turb_setg_critical_balance}
\end{align}
This is a statement of \textit{critical balance}, whereby the characteristic time associated with propagation along the field lines is assumed comparable to the nonlinear advection rate $t_\text{nl}^{-1}$ at each perpendicular scale $k_\perp^{-1}$ --- \cite{barnes11} justified this by the standard causality argument borrowed from MHD turbulence (\citealt{GS95,GS97}; \citealt{boldyrev05}; \citealt{nazarenko11}): two points along the field line can only remain correlated with one another if information can propagate between them faster than they are decorrelated by the nonlinearity. We have taken the rate of information propagation along the field lines to be $k_\parallel \vthe$; a somewhat involved reasoning is needed to explain why this should work even though $k_\parallel \vthe$ is the rate of phase mixing (which, in the linear theory, is expected to give rise to Landau damping) rather than of wave propagation, and why Landau damping is ineffective in the nonlinear state (see \citealt{sch16}, \citealt{adkins18}). 

Combining \eqref{eq:turb_nonlinear_time}, \eqref{eq:turb_setg_constant_flux} and \eqref{eq:turb_setg_critical_balance}, we find
\begin{align}
   k_\parallel \vthe \sim  t_\text{nl}^{-1} \sim \Omega_e \left( \frac{\varepsilon}{\Omega_e} \right)^{1/3} (k_\perp \rho_e)^{4/3}.
    \label{eq:turb_setg_nonlinear_time}
\end{align}
By comparison, for the most unstable sETG modes, \eqref{eq:setg_max} gives us
\begin{align}
    k_\parallel \vthe \sim \omega_{*e} \sim k_y \rho_e \frac{\vthe}{L_T}.
    \label{eq:turb_setg_most_unstable}
\end{align}
These modes grow at a rate $\omega_{*e} \propto k_y$. This means that the nonlinear interactions must overwhelm the linear instability in the inertial range.\footnote{\label{foot:nonlinear_argument} Here is another way to see this. Imagine that the sETG instability dominates energy injection at each scale and that the energy thus injected is removed to the next smaller scale by the nonlinearity, at a rate $t_\text{nl}^{-1}$. Such a scheme would be consistent if the energy flux injected at each scale by the instability were larger than the flux arriving to this scale from larger scales. Let us see if this is possible. Balancing the nonlinear energy-removal rate \eqref{eq:turb_setg_nonlinear_time} with the injection rate $\omega_{*e}$, we learn that $\amp{\varphi} \sim (k_\perp L_T)^{-1}$ (corresponding to a 1D spectrum $\propto k_\perp^{-3}$). The injected energy flux is then $\varepsilon \sim \omega_{*e} \amp{\varphi}^2 \sim \Omega_e (\rho_e/L_T)^3 (k_\perp \rho_e)^{-1}$. So it declines at smaller scales, and is easily overwhelmed by the nonlinear transfer from larger scales.} The outer scale, i.e., the scale that limits the inertial range on the infrared side and at which the free energy is effectively injected, is then the scale at which the nonlinear cascade rate and the rate of maximum growth of the instability are comparable: balancing \eqref{eq:turb_setg_nonlinear_time} and \eqref{eq:turb_setg_most_unstable}, we get 
\begin{align}
    \Omega_e (k_\perp^o \rho_e)^2 \amp{\varphi}^o \sim k_\parallel^o \vthe \sim \omega_{*e}^o \quad \Rightarrow \quad \amp{\varphi}^o \sim (k_\perp^o L_T)^{-1}, \quad  k_y^o \rho_e \sim k_\parallel^o L_T,
    \label{eq:turb_setg_outerscale_balance}
\end{align}
where the superscript `$o$' refers to quantities at the outer scale. 

Now, in order to determine $k_\perp^o$, we need a further constraint. \cite{barnes11} found it by conjecturing that $k_\parallel^o$ in \eqref{eq:turb_setg_outerscale_balance} would be set by the parallel system size $L_\parallel$ (the connection length $\sim \pi q L_B$, in the case of tokamaks). This was the only reasonable choice because there was no lower cutoff in $k_\perp$ of the (electrostatic) ITG-unstable modes. This is not, however, the case in our model of the sETG instability, which is stabilised at the flux-freezing scale \eqref{eq:flux_freezing_scale}, i.e., at $k_\perp d_e \sim 1$. It appears to be a general rule, confirmed by numerical simulations (\citealt{parra12pc}), that the outer scale is, in fact, determined by the smallest possible $k_y \rho_e$ or the smallest possible $k_\parallel L_T$, whichever is larger. Putting this within the visual context of \figref{fig:collisionless_phase_space_portrait}, the outer scale is set either by $k_\parallel^o \sim L_\parallel^{-1}$ or by $k_\perp^o \sim d_e^{-1}$, whichever is encountered first when moving along the solid black line from the ultraviolet cutoff towards larger scales. The former possibility, $k_\parallel^o \sim L_\parallel^{-1}$, is realised when $L_\parallel \ll L_T/\sqrt{\beta_e}$, and the latter, $k_\perp^o \sim d_e^{-1}$, otherwise. Thus,
\begin{align}
    k_\perp^o d_e \sim \frac{k_\parallel^o L_T}{\sqrt{\beta_e}} \sim \left\{
    \begin{array}{ll}
         \displaystyle \frac{L_T}{L_\parallel \sqrt{\beta_e}}, &  \displaystyle  \frac{L_B}{L_T} \ll \frac{L_B}{L_\parallel \sqrt{\beta_e}},   \\[4mm]
          \displaystyle  1. &  \displaystyle  \frac{L_B}{L_T} \gtrsim \frac{L_B}{L_\parallel \sqrt{\beta_e}}. 
    \end{array}
    \right.
        \label{eq:turb_setg_outer_scale}
\end{align}
We have inserted the normalisation of the temperature gradient to $L_B$ for future convenience. 

Let us now estimate the energy flux that is injected by sETG at the outer scale~\eqref{eq:turb_setg_outer_scale}: considering the first term in the expression for the energy flux \eqref{eq:energy_injection_simplified} (the second, involving finite perturbations to the magnetic field, is negligible in the electrostatic regime) and ignoring any possibility of a non-order-unity contribution from phase factors, we have 
\begin{align}
     \varepsilon \sim \omega_{*e}^o \amp{\varphi}^o \frac{\delta \amp{T}_{\parallel e}^o}{T_{0e}} \sim \frac{\vthe \rho_e^2}{L_T^3 \sqrt{\beta_e}}(k_\perp^o d_e)^{-1},
    \label{eq:turb_setg_energy_flux}
\end{align}
where we have used $ \delta \amp{T}_{\parallel e}^o/T_{0e} \sim \amp{\varphi}^o$ and \eqref{eq:turb_setg_outerscale_balance}. This quantity is directly related to the turbulent heat flux: combining \eqref{eq:turb_setg_energy_flux} with \eqref{eq:turb_setg_outer_scale}, we get 
\begin{align}
    Q^\text{sETG} \sim n_{0e}T_{0e} \varepsilon L_T  \sim  Q_\text{gB}\left\{
    \begin{array}{ll}
         \displaystyle \left( \frac{L_\parallel}{L_B} \right)\left( \frac{L_B}{L_T} \right)^3, &  \displaystyle  \frac{L_B}{L_T} \ll \frac{L_B}{L_\parallel \sqrt{\beta_e}},   \\[4mm]
          \displaystyle  \frac{1}{\sqrt{\beta_e}} \left( \frac{L_B}{L_{T}} \right)^2, &  \displaystyle  \frac{L_B}{L_T} \gtrsim \frac{L_B}{L_\parallel \sqrt{\beta_e}}, 
    \end{array}
    \right.
    \label{eq:turb_setg_heat_flux}
\end{align}
where the `gyro-Bohm' flux is $Q_\text{gB} = n_{0e} T_{0e} \vthe \left(\rho_e/L_B \right)^2$. Note that the $L_B/L_T$ scaling in \eqref{eq:turb_setg_heat_flux} is only valid for sufficiently large $L_B/L_T$ as our analysis ignores any finite critical temperature gradients associated with the sETG instability (see \apref{app:finite_critical_gradients}). The first expression in \eqref{eq:turb_setg_heat_flux} is the same scaling as that obtained by \cite{barnes11}, but this time for electrostatic turbulence driven by an electron temperature gradient\footnote{ \cite{chapman22} found such a scaling of the heat flux with $L_B/L_T$ in their investigations of nonlinear pedestal turbulence driven by ETG modes [see their equation (1), and the following discussion] suggesting, perhaps, that this scaling may hold in more realistic --- and complex --- plasma systems than that considered here.}. In the formal limit of $\beta_e \rightarrow 0$, this is the only possible outcome because the second inequality in \eqref{eq:turb_setg_heat_flux} can never be satisfied. At finite $\beta_e$, however, in the sense in which it is allowed by our ordering and for sufficiently large temperature gradients, we obtain a different, less steep scaling of the turbulent heat flux, given by the second expression in \eqref{eq:turb_setg_heat_flux}.

Whether the scaling \eqref{eq:turb_setg_heat_flux} is relevant in our system depends on the dominant energy injection therein being from the electrostatic sETG drive at $k_\perp d_e \gtrsim 1$. That is, in fact, far from guaranteed if $L_\parallel > L_T/\sqrt{\beta_e}$, i.e., if sufficiently small $k_\parallel$ are allowed for the electromagnetic instabilities to matter --- and so for the outer scale to be located at even larger scales along the thick black line in \figref{fig:collisionless_phase_space_portrait}. Another reason why we must consider the electromagnetic part of the wavenumber space is to do with the cETG instability. At $k_\perp d_e \gtrsim 1$, its growth rate is always small in comparison with the with sETG [for the large $L_B/L_T$ that we are considering here, see~\eqref{eq:setg_vs_cetg_gamma}], but it is a 2D mode, so it is not stabilised at $k_\perp d_e \sim 1$ (it does not bend magnetic fields) and there is no reason to assume that it cannot provide the dominant energy inection at some large scale $k_\perp  d_e  \ll 1$. There, it competes with TAI (\secref{sec:electromagnetic_regime_tai}), so we shall have to examine the TAI turbulence alongside the cETG one. 

These topics are, of course, the raison d'\^etre of this work and we shall tackle them in \secref{sec:electromagnetic_turbulence}, but first we wish, for the sake of completeness, to work out the collisional version of sETG turbulence --- an impatient reader can skip this.

\subsubsection{Collisional slab ETG turbulence}
\label{sec:collisional_slab_etg_turbulence}
For collisional sETG turbulence, the argument proceeds similarly to \secref{sec:collisionless_slab_etg_turbulence}. Instead of \eqref{eq:turb_setg_field_scalings}, we now have, in view of \eqref{eq:col_setg_equations}, 
\begin{align}
    \taubar^{-1} \amp{\varphi} \sim \frac{k_\parallel \vthe}{\omega} \frac{\amp{u}_{\parallel e}}{\vthe} \sim \frac{(k_\parallel \vthe)^2}{\omega \nu_{ei}} \frac{\delta \amp{T}_e}{T_{0e}} \sim \frac{\delta \amp{T}_e}{T_{0e}},
    \label{eq:turb_setg_col_field_orderings}
\end{align}
where we assume that all frequencies, including the nonlinear rate \eqref{eq:turb_nonlinear_time}, are now comparable to the rate of parallel thermal conduction [instead of the parallel streaming rate; see \eqref{eq:col_setg_max}]:
\begin{align}
    t_\text{nl}^{-1}\sim \omega  \sim \frac{(k_\parallel \vthe)^2}{\nu_{ei}}. 
    \label{eq:turb_setg_col_critical_balance}
\end{align}
This condition now replaces \eqref{eq:turb_setg_critical_balance} as the `critical-balance' conjecture, whereby the parallel scale of the perturbations is determined in terms of their perpendicular scale. Note that, since now $\amp{u}_{\parallel e}/\vthe \ll \amp{\varphi}$, it is still reasonable to estimate the free-energy density by $\sim \taubar^{-1} \amp{\varphi}^2$. 

At the outer scale, using \eqref{eq:col_setg_max} and \eqref{eq:turb_setg_col_critical_balance}, we find, analogously to \eqref{eq:turb_setg_outerscale_balance}, 
\begin{align}
    \Omega_e (k_\perp^o \rho_e )^2 \amp{\varphi}^o \sim \frac{( k_\parallel^o \vthe )^2}{\nu_{ei}} \sim \omega_{*e} \quad \Rightarrow \quad \amp{\varphi}^o \sim (k_\perp^o L_T)^{-1}, \quad k_y^o \rho_e \sim (k_\parallel^o)^2 L_T \lambdae.
    \label{eq:turb_col_setg_outerscale_balance}
\end{align}
Note that the relationship between the parallel and perpendicular outer scales can be recast as 
\begin{align}
     \frac{k_\parallel^o L_T}{\sqrt{\beta_e}} \sim \left(k_y^o d_e \chi\right)^{1/2}, \quad \chi \equiv \frac{L_T}{\lambdae \sqrt{\beta_e}}
    \label{eq:turb_col_setg_outerscale_balance_rephrase}
\end{align}
where $\chi$ is defined as in \eqref{eq:flux_freezing_scale_col}.

By analogous logic to the collisionless sETG case, the outer scale can be set either by the parallel system size or by the flux-freezing scale \eqref{eq:flux_freezing_scale_col}, $k_\perp d_e \chi \sim 1$, depending on which is encountered first by the thick black line in \figref{fig:collisional_phase_space_portrait} when descending towards larger scales. The result is 
\begin{align}
    k_\perp^o d_e \chi \sim \left\{
    \begin{array}{ll}
         \displaystyle \left( \frac{L_T}{L_\parallel \sqrt{\beta_e}} \right)^2, &  \displaystyle  \frac{L_B}{L_T} \ll \frac{L_B}{L_\parallel \sqrt{\beta_e}},   \\[4mm]
          \displaystyle  1, &  \displaystyle  \frac{L_B}{L_T} \gtrsim \frac{L_B}{L_\parallel \sqrt{\beta_e}}.
    \end{array}
    \right.
    \label{eq:turb_col_setg_outer_scale}
\end{align}
In view of \eqref{eq:turb_col_setg_outerscale_balance}, the energy flux is again given by \eqref{eq:turb_setg_energy_flux}, which, with the substitution of \eqref{eq:turb_col_setg_outer_scale}, becomes
\begin{align}
    \varepsilon \sim \frac{\vthe \rho_e^2}{L_T^3 \sqrt{\beta_e}} \chi 
    \left\{\begin{array}{ll}
         \displaystyle \left( \frac{L_\parallel \sqrt{\beta_e}}{L_T} \right)^2, &  \displaystyle  \frac{L_B}{L_T} \ll \frac{L_B}{L_\parallel \sqrt{\beta_e}},   \\[4mm]
          \displaystyle  1, &  \displaystyle  \frac{L_B}{L_T} \gtrsim \frac{L_B}{L_\parallel \sqrt{\beta_e}}.
    \end{array}
    \right.
    \label{eq:turb_col_setg_energy_flux}
\end{align}
Therefore, finally, the turbulent heat flux is 
\begin{align}
    Q^\text{sETG}_\nu\sim Q_\text{gB} \left\{
    \begin{array}{ll}
         \displaystyle \left(\frac{L_\parallel}{L_B} \right)^2 \left( \frac{L_B}{\lambdae} \right) \left( \frac{L_B}{L_T} \right)^3 , &  \displaystyle  \frac{L_B}{L_T} \ll \frac{L_B}{L_\parallel \sqrt{\beta_e}},   \\[4mm]
          \displaystyle  \frac{1}{\beta_e} \left( \frac{L_B}{\lambdae} \right) \left( \frac{L_B}{L_T} \right), &  \displaystyle   \frac{L_B}{L_T} \gtrsim \frac{L_B}{L_\parallel \sqrt{\beta_e}}.
    \end{array}
    \right.
    \label{eq:turb_col_setg_heat_flux}
\end{align}
These are the collisional analogues of the scalings \eqref{eq:turb_setg_heat_flux}, and are both proportional to the electron collision frequency ($\propto \lambdae^{-1}$). 
Such a scaling of turbulent heat flux with collisionality was identified by \cite{colyer17} from their simulations of electrostatic ETG turbulence, though their argument relied on consideration of the dynamics of zonal flows within their electron-scale system, and so the comparison is superficial.

\subsection{Electromagnetic turbulence}
\label{sec:electromagnetic_turbulence}

\subsubsection{KAW-dominated, slab TAI turbulence}
\label{sec:kaw_dominated_slab_tai_turbulence}
On the large-scale side of the flux-freezing scales \eqref{eq:flux_freezing_scale} and \eqref{eq:flux_freezing_scale_col}, for $k_{\perp *} \lesssim k_\perp \lesssim d_e^{-1}$ (or $d_e^{-1} \chi^{-1}$ in the collisional limit), the dominant instability is the isobaric sTAI (see \secref{sec:isobaric_limit}), an instability of kinetic Alfv\'en waves. KAW turbulence has been studied quite extensively, both numerically \citep{cho04,cho09,howes11prl,boldyrev12,meyrand13,told15,groselj18prl,groselj19,franci18} and observationally \citep{alexandrova09,sahraoui10,chen13}, in the context of the `kinetic-range' free-energy cascade in the solar wind \citep{sch09,boldyrev13,passot17}. The theory of this cascade proceeds along the same lines as the theory of any critically balanced cascade in a wave-supporting anisotropic medium \citep{nazarenko11} and leads again to a $\kperp^{-7/3}$ energy spectrum \citep{cho04,sch09} or, with some modifications, to a $\kperp^{-8/3}$ one \citep{boldyrev12,meyrand13}, which appears to be closer to what is observed. 

Ignoring the latter nuance, it is easy to see that the re-emergence of the $k_\perp^{-7/3}$ spectrum is unsurprising, as the arguments of \secref{sec:electrostatic_turbulence} that led to \eqref{eq:turb_setg_constant_flux} and \eqref{eq:turb_setg_energy_spectrum} are unchanged for KAWs. What is changed, however, is the linear propagation rate that must be used in the critical-balance conjecture: the parallel scale $k_\parallel^{-1}$ of a perturbation is now the distance that an (isobaric) KAW can travel in one nonlinear time, so, from \eqref{eq:kaw_isobaric}, we have, instead of \eqref{eq:turb_setg_critical_balance} or \eqref{eq:turb_setg_col_critical_balance},
\begin{align}
    \omega_{\text{KAW}} \sim  k_\parallel \vthe k_\perp d_e \sim   t_\text{nl}^{-1} ,
    \label{eq:turb_kaw_critical_balance}
\end{align}
where $t_\text{nl}$ is still given by \eqref{eq:turb_nonlinear_time}\footnote{This $t_\text{nl}$ is the nonlinear time associated with the fluctuating $\vec{E}\times \vec{B}$ flows, coming from the convective time derivative \eqref{eq:convective_derivative}. In the electromagnetic regime, there is, in addition to this, the nonlinearity associated with the parallel gradients being taken along perturbed magnetic field lines, including finite $\delta \! \vec{B}_\perp$, as in \eqref{eq:parallel_derivative}. However, it is straightforward to show [by, e.g., estimating the sizes of the nonlinear terms appearing in \eqref{eq:isothermal_tai_equations}, \eqref{eq:kaw_equation_stai} or \eqref{eq:kaw_equation_isobaric_stai}] that the $\vec{E}\times \vec{B}$ nonlinearity is either comparable to, or larger than, the $\delta \! \vec{B}_\perp$ nonlinearity in all of the regimes of interest, meaning that we may continue to use \eqref{eq:turb_nonlinear_time} as our estimate for the nonlinear time.}. 

This is the standard argument of KAW-turbulence theory (see references above), which, however, was developed for situations in which energy arrived to sub-Lamor scales from larger scales (i.e., from $k_\perp \rho_i < 1$) and cascaded down to smaller scales --- as indeed it typically does in space-physical and astrophysical contexts. In contrast, here we are dealing with an energy source in the form of an ETG-driven instability, the isobaric sTAI, which operates most vigorously at the smallest electromagnetic scales. Indeed, as we saw at the end of \secref{sec:isobaric_slab_TAI}, for a given $k_\perp d_e$, the sTAI growth rate peaks at $\xi_* \sim 1$, and is of the order of the KAW frequency $\omega_\text{KAW}$ at that scale. This gives 
\begin{align}
    \gamma \sim \omega_\text{KAW} \sim  \left\{
    \begin{array}{ll}
         \displaystyle \omega_{*e} k_\perp d_e \sim \frac{\vthe}{L_T \sqrt{\beta_e}} (k_\perp \rho_e)^2 , &  \displaystyle \text{collisionless},   \\[4mm]
          \displaystyle \sqrt{\omega_{*e} \nu_e} k_\perp d_e \sim \frac{\vthe}{\sqrt{L_T \lambdae \beta_e}} \left( k_\perp \rho_e \right)^{3/2} , &  \displaystyle \text{collisional},
    \end{array}
    \right.
    \label{eq:turb_kaw_growth_rates}
\end{align}
where we used $k_{\parallel} \sim \omega_{*e}/\vthe$ and $k_\parallel \sim (\omega_{*e}/\kappa )^{1/2}  \sim (\omega_{*e} \nu_e)^{1/2}/\vthe$ for the collisionless and collisional estimates, respectively. Comparing \eqref{eq:turb_kaw_growth_rates} with \eqref{eq:turb_setg_nonlinear_time}, we see that, in both cases, the instability growth rate increases faster with $k_\perp$ than the nonlinear cascade rate $t_\text{nl}^{-1} \propto k_\perp^{4/3}$. It is intuitively obvious that these two rates reach parity at the flux-freezing scale, $k_\perp d_e \sim 1$ or $k_\perp d_e \chi \sim 1$, in the collisionless and collisional limits, respectively. This can be formally confirmed by a calculation analogous to the one in \secref{sec:electrostatic_turbulence}.
Thus, the dominant injection occurs at the small-scale end of the putative `inertial range'. In the absence of any inverse cascade, there is nothing to push the energy towards larger scales. This means that the balances \eqref{eq:turb_setg_constant_flux}, \eqref{eq:turb_setg_nonlinear_time} and \eqref{eq:turb_kaw_critical_balance} are not, in fact, realised for KAW turbulence driven by the isobaric sTAI. 

In order to predict the power injected by sTAI, and the associated contribution to the turbulent heat flux, we resurrect the argument that, for sETG, we tossed aside in footnote~\ref{foot:nonlinear_argument}. We conjecture that the sTAI instability dominates the energy injection at each scale, and the energy thus injected is removed to the next smaller scale by the nonlinearity, at a rate $t_\text{nl}^{-1}$; we shall confirm a posteriori that this is a consistent scheme. The resulting balance gives us, using \eqref{eq:turb_nonlinear_time} and \eqref{eq:turb_kaw_growth_rates},
\begin{align}
    t_\text{nl}^{-1} \sim \Omega_e (k_\perp \rho_e)^2 \amp{\varphi} \sim \gamma \quad \Rightarrow \quad 
    \amp{\varphi} \sim
    \left\{
    \begin{array}{ll}
         \displaystyle \frac{d_e}{L_T}, &  \displaystyle \text{collisionless},   \\[4mm]
          \displaystyle \frac{d_e}{\sqrt{L_T \lambdae}} (k_\perp \rho_e)^{-1/2} , &  \displaystyle \text{collisional},
    \end{array}
    \right.
    \label{eq:turb_kaw_phi_spectra}
\end{align}
and $\delta \! \amp{B}_\perp/B_0 \sim k_\perp \rho_e \amp{\mathcal{A}} \sim (\rho_e/d_e) \amp{\varphi}$ [where we have used $k_\perp d_e \amp{\mathcal{A}} \sim \amp{\varphi}$, which follows from the first equation in \eqref{eq:kaw_isobaric} with $\omega\sim \omega_{\text{KAW}}$]. The corresponding energy spectra \eqref{eq:turb_varphi_definition} are $\propto k_\perp^{-1}$ and $\propto k_\perp^{-2}$ in the collisionless and collisional regimes, respectively. The injected power is
\begin{align}
    \gamma \amp{\varphi}^2 \sim 
    \frac{\vthe \rho_e^2}{L_T^3 \sqrt{\beta_e}}
    \left\{
    \begin{array}{ll}
         \displaystyle (k_\perp d_e)^2, &  \displaystyle \text{collisionless},   \\[4mm]
          \displaystyle  (k_\perp d_e )^{1/2} \chi^{3/2} , &  \displaystyle \text{collisional},
    \end{array}
    \right.
    \label{eq:turb_kaw_injected_power}
\end{align}
where $\chi$ is defined in \eqref{eq:flux_freezing_scale_col} or \eqref{eq:turb_col_setg_outerscale_balance_rephrase}. This means that, at each scale, the energy that arrives from larger scales can be ignored in comparison with the energy injected locally by sTAI --- unlike for the sETG cascade, this scale-by-scale injection scheme is consistent for `sTAI turbulence'.

It is clear from \eqref{eq:turb_kaw_injected_power} that the injected power is dominated by the flux-freezing scale~\eqref{eq:flux_freezing_scale} or \eqref{eq:flux_freezing_scale_col}, where it reaches parity with the power injected by sETG, \eqref{eq:turb_setg_energy_flux} or \eqref{eq:turb_col_setg_energy_flux}, and where also the sTAI approximation breaks down and sETG takes over. Thus, the turbulent heat flux due to the sTAI turbulence is given by the same expression as that for the sETG turbulence at sufficiently large temperature gradients --- the second expressions in \eqref{eq:turb_setg_heat_flux} and \eqref{eq:turb_col_setg_heat_flux}. The only effect of sTAI is to equip the sETG turbulence spectrum~\eqref{eq:turb_setg_energy_spectrum} with an electromagnetic tail at long wavelengths --- scaling as $k_\perp^{-1}$ and $k_\perp^{-2}$ in the collisionless and collisional cases, respectively --- but without changing by more than an order-unity amount its ability to transport energy\footnote{This conclusion is based on the (asymptotic) assumption that both sTAI and sETG inject energy around the same outer scale $k_\parallel^o L_T/\sqrt{\beta_e} \sim 1$, $k_\perp^o d_e \sim 1$ (or $\sim \chi^{-1}$ in the collisional limit). However, a more quantitative analysis of the stability properties of the collisionless and collisional systems shows that sTAI is stabilised slightly towards the large-scale side of this assumed outer scale, while sETG is stabilised slightly towards the small-scale side of it (see \secref{sec:stabilisation_of_isobaric_slab_TAI} and appendices \ref{app:exact_stability_boundary} and \ref{app:exact_stability_boundary_collisional}). Thus, in principle, it is possible to assess the comparative roles of these two instabilities in a quantitative way (e.g., numerically). Whether such an analysis is interesting qualitatively depends on whether the two modes behave very differently in a nonlinear setting.}.

\subsubsection{Curvature-mediated-TAI turbulence}
\label{sec:curvature_mediated_tai_turbulence}
At $k_\perp \lesssim k_{\perp *}$, the isothermal cTAI replaces the isobaric sTAI as the dominant instability. Since the nonlinear cascading is still done by the $\vec{E}\times \vec{B}$ flows, the nonlinear time is still given by~\eqref{eq:turb_nonlinear_time}. 
However, how to work out the `inertial-range' scalings for this cascade is not obvious: since the real frequency is vanishingly small in comparison to the growth rate at the cTAI maximum [see \eqref{eq:tai_ratio_of_frequency_to_growth_rate}], there is no obvious analogue of the `critical balance' conjectures \eqref{eq:turb_setg_critical_balance} or \eqref{eq:turb_kaw_critical_balance}; indeed, it is not even a given that the cascade will be local in wavenumber space. We shall not be deterred by this uncertainty, as we can, in fact, still calculate the injected free-energy flux \eqref{eq:energy_injection_simplified} by considering solely the fluctuations at the injection scale; we shall then propose a way of determining what that scale is, and hence calculate the turbulent heat flux. 

First, let us assume that the dominant free-energy injection will occur at the wavenumbers \eqref{eq:tai_curvature_maximum_explicit}, where the cTAI growth rate is largest, and given by \eqref{eq:isothermal_tai_dispersion_relation}:
\begin{align}
    \gamma \sim \frac{k_y^o \rho_e \vthe}{\sqrt{L_B L_T} }.
    \label{eq:turb_ctai_maximum}
\end{align}
Unlike for the electrostatic modes, the second, `electromagnetic' term in \eqref{eq:energy_injection_simplified} --- involving energy transport due to heat flux along perturbed field lines --- must contribute to the energy injection by cTAI. Let us estimate its size at the outer scale. The third equation in \eqref{eq:isothermal_tai_equations} gives us 
\begin{align}
     \frac{\delta\! \amp{B}_x^o}{B_0} \sim k_y^o\rho_e \amp{\mathcal{A}}  \sim k_{\parallel}^o L_T \frac{\delta \amp{T}_e^o}{T_{0e}}.
     \label{eq:turb_ctai_outer_scale_magnetic_field}
\end{align}
Recalling \eqref{eq:collisional_heat_flux}, we estimate the size of the perturbed heat flux in the collisional limit from \eqref{eq:stai_logt_correction_isothermal}:
\begin{align}
    \frac{\delta \amp{q}_e^o}{n_{0e} T_{0e} } \sim \kappa\gradd_\parallel \log \amp{T}_e^o \sim \kappa \xi_*^o k_\parallel^o \frac{\delta \! \amp{n}_e^o}{n_{0e}} \sim \frac{\omega_{*e}^o }{ k_\parallel^o} \amp{\varphi}^o.
    \label{eq:turb_ctai_outer_scale_heat_flux_collisional}
\end{align}
Analogously, in the collisionless limit, we find that [see \eqref{eq:lin_qpar} and what follows it]
\begin{align}
     \frac{\delta \amp{q}_{\parallel e}^o}{n_{0e} T_{0e}} \sim  \frac{\delta \amp{q}_{\perp e}^o}{n_{0e} T_{0e}} \sim \xi_*^o \frac{\delta \! \amp{n}_e^o}{n_{0e}} \sim  \frac{\omega_{*e}^o }{ k_\parallel^o}  \amp{\varphi}^o.
     \label{eq:turb_ctai_outer_scale_heat_flux_collisionless}
\end{align}
Thus, in both limits, the electromagnetic contribution to the free-energy injection can be written, at the outer scale, as
\begin{align}
    \varepsilon \sim \frac{1}{L_T} \frac{\delta \amp{q}_e^o}{n_{0e} T_{0e}} \frac{\delta \amp{B}_x^o}{B_0} \sim \omega_{*e}^o \amp{\varphi}^o \frac{\delta \amp{T}_e^o}{T_{0e}} ,
    \label{eq:turb_ctai_outer_injection}
\end{align}
meaning that it is comparable to the first term in \eqref{eq:energy_injection_simplified}, the electrostatic contribution due to energy transport by the $\vec{E}\times \vec{B}$ flow. 

The potential at the outer scale can once again be estimated from the balance of the nonlinear time \eqref{eq:turb_nonlinear_time} with the growth rate \eqref{eq:turb_ctai_maximum}:
\begin{align}
    \rho_e \vthe (k_\perp^o)^2 \amp{\varphi} \sim \gamma \quad \Rightarrow \quad \amp{\varphi}^o \sim \frac{1}{k_\perp^o\sqrt{L_BL_T} },
    \label{eq:turb_ctai_outer_scale_balance}
\end{align}
while the temperature perturbations can be related to $\varphi^o$ via the first equation in \eqref{eq:isothermal_tai_equations}:
\begin{align}
    \frac{\delta \amp{T}_e^o}{T_{0e}} \sim  \frac{\gamma}{\omega_{de}^o} \frac{\delta \amp{n}_{e}^o}{n_{0e}}\sim \left( \frac{L_B}{L_T} \right)^{1/2}  \amp{\varphi}^o \sim  (k_\perp^o L_T)^{-1}.
    \label{eq:turb_ctai_outer_scale_temperature}
\end{align}
Therefore, the injected energy flux \eqref{eq:turb_ctai_outer_injection} is
\begin{align}
    \varepsilon \sim \frac{\vthe \rho_e^2}{L_T^{3} \sqrt{\beta_e}} \left( \frac{L_T}{ L_B} \right)^{1/2} (k_\perp^o d_e)^{-1}.
    \label{eq:turb_ctai_energy_flux}
\end{align}

We must now determine $k_\perp^o$. We conjecture that, like in sETG turbulence, the nonlinear interaction rate in cTAI turbulence will increase faster with $k_\perp$ than the growth rate~\eqref{eq:turb_ctai_maximum}, $\gamma \propto k_y$. This would certainly be the case if the cascade were local, wherein the Kolmogorov-style argument leading to \eqref{eq:turb_setg_constant_flux} applied (in which case $t_\text{nl}^{-1} \propto k_\perp^{4/3}$ again). Then $k_\perp^o$ will be the smallest that it can be. Since it is related to $k_\parallel^o$ via \eqref{eq:tai_curvature_maximum_explicit} (corresponding to the maximum growth rate) viz.,
\begin{align}
    \frac{k_{\parallel}^o L_T}{\sqrt{\beta_e}} \sim \left\{
    \begin{array}{ll}
         \displaystyle \left( \frac{L_T}{ L_B} \right)^{1/4} (k_\perp^o d_e)^{1/2}, &  \displaystyle \text{collisionless},   \\[4mm]
          \displaystyle \left( \frac{L_T}{ L_B} \right)^{1/6} (k_\perp^o d_e \chi)^{1/3}, &  \displaystyle \text{collisional},
    \end{array}
    \right.
    \label{eq:turb_ctai_outer_scale}
\end{align}
we can treat this expression as the analogue of the last expression in~\eqref{eq:turb_setg_outerscale_balance} or \eqref{eq:turb_col_setg_outerscale_balance}. As we did in our treatment of sETG turbulence in sections \ref{sec:collisionless_slab_etg_turbulence} and \ref{sec:collisional_slab_etg_turbulence}, we now posit that the parallel outer scale of cTAI turbulence will be set by the system's parallel size, $k_\parallel^o \sim L_\parallel^{-1}$. Then, from \eqref{eq:turb_ctai_outer_scale},
\begin{align}
    k_\perp^o d_e \sim  \left\{
    \begin{array}{ll}
         \displaystyle \frac{L_T^{3/2} L_B^{1/2} }{\beta_e L_\parallel^2}, &  \displaystyle \text{collisionless},   \\[4mm]
          \displaystyle \frac{L_T^{3/2} L_B^{1/2} \lambdae }{\beta_e L_\parallel^3}, &  \displaystyle \text{collisional}.
    \end{array}
    \right.
    \label{eq:turb_ctai_outer_perpendicular}
\end{align}
This, of course, assumes that there is no dynamics at larger scales that can set the perpendicular outer-scale. We discuss the constraints set by this assumption in \secref{sec:ion_dynamics}.
Using \eqref{eq:turb_ctai_outer_perpendicular} in \eqref{eq:turb_ctai_energy_flux}, we can estimate the heat flux due to cTAI turbulence:
\begin{align}
    Q^\text{cTAI} \sim n_{0e} T_{0e} \varepsilon L_T \sim Q_\text{gB} \left\{
    \begin{array}{ll}
         \displaystyle \sqrt{\beta_e}\left( \frac{L_\parallel}{L_B} \right)^2 \left( \frac{L_B}{L_{T}} \right)^3, &  \displaystyle \text{collisionless},   \\[4mm]
          \displaystyle \sqrt{\beta_e}\left( \frac{L_\parallel}{L_B} \right)^3 \left(\frac{L_B}{\lambdae} \right)\left( \frac{L_B}{L_{T}} \right)^3, &  \displaystyle \text{collisional}.
    \end{array}
    \right.
    \label{eq:turb_ctai_heat_flux}
\end{align}

In order for this construction to be valid, $L_\parallel$ must be large enough for $k_\parallel^o \sim L_\parallel^{-1} \lesssim k_{\parallel c}$, the latter given by \eqref{eq:tai_kpar_critical} --- otherwise the system cannot access the cTAI regime in the first place. The condition for this is 
\begin{align}
    \frac{k_\parallel^o L_T}{\sqrt{\beta_e}} \lesssim \left(\frac{L_T}{L_B} \right)^{1/2} \quad \Leftrightarrow \quad \frac{L_B}{L_T} \gtrsim \left( \frac{L_B}{L_\parallel \sqrt{\beta_e}} \right)^2.
    \label{eq:turb_heat_flux_kparc_condition}
\end{align}
Thus, cTAI turbulence is relevant for temperature gradients that are even larger than those needed to access the sETG and sTAI regimes described by \eqref{eq:turb_setg_heat_flux} and \eqref{eq:turb_col_setg_heat_flux}. By comparing the heat fluxes \eqref{eq:turb_ctai_heat_flux} with the second expressions in \eqref{eq:turb_setg_heat_flux} and \eqref{eq:turb_col_setg_heat_flux}, it is not hard to ascertain that the cTAI fluxes are larger than the sETG-sTAI ones as long as \eqref{eq:turb_heat_flux_kparc_condition} is satisfied. 

\subsection{Summary of turbulent regimes}
\label{sec:summary_of_turbulent_regimes}
In sections \ref{sec:electrostatic_turbulence} and \ref{sec:electromagnetic_turbulence}, we found scaling estimates for the turbulent heat fluxes arising from~sETG, sTAI and cTAI in both the collisionless and collisional limits. Which of these scalings is realised is determined by the size of the electron temperature gradient $L_T$ for given values of $L_\parallel$, $L_B$ and $\beta_e$. There are three distinct regimes. For 
\begin{align}
    k_\perp^o d_e \sim \frac{L_T}{L_\parallel \sqrt{\beta_e}} \gg 1 \quad \Leftrightarrow \quad \frac{L_B}{L_T} \ll \frac{L_B}{L_\parallel \sqrt{\beta_e}},
    \label{eq:turb_regime_1}
\end{align}
the system contains only electrostatic (perpendicular) scales, and the heat flux will simply be that arising from sETG turbulence, given by the first expressions in \eqref{eq:turb_setg_heat_flux} and \eqref{eq:turb_col_setg_heat_flux} in the collisionless and collisional limits, respectively. For 
\begin{align}
    k_{\parallel c} \lesssim \frac{1}{L_\parallel} \lesssim\frac{\sqrt{\beta_e}}{L_T} \quad \Leftrightarrow \quad \frac{L_B}{L_\parallel \sqrt{\beta_e}} \lesssim\frac{L_B}{L_T} \lesssim \left(\frac{L_B}{L_\parallel \sqrt{\beta_e}}\right)^{2},
    \label{eq:turb_regime_2}
\end{align}
the system can access electromagnetic (perpendicular) scales, with the (isobaric) sTAI and stable KAW being added to the collection of possible modes. However, we showed in \secref{sec:kaw_dominated_slab_tai_turbulence} that the only effect of the sTAI was to equip the sETG turbulent spectrum with an electromagnetic tail at longer wavelengths, with at most an order-unity enhancement of the turbulent heat flux. This heat flux is still the same as that arising from the sETG turbulence, but with the outer scaled fixed at the flux-freezing scale --- it is given by the second expressions in~\eqref{eq:turb_setg_heat_flux} and \eqref{eq:turb_col_setg_heat_flux}. Finally, for 
\begin{align}
    \frac{1}{L_\parallel} \lesssim k_{\parallel c} \quad \Leftrightarrow \quad \frac{L_B}{L_T} \gtrsim \left(\frac{L_B}{L_\parallel \sqrt{\beta_e}}\right)^{2},
    \label{eq:turb_regime_3}
\end{align}
the system has a large enough parallel size to activate cTAI. The resultant turbulent heat flux, given by~\eqref{eq:turb_ctai_heat_flux}, dominates over that due to the sETG and sTAI.

\begin{figure}
    \centering
    
    \begin{tabular}{c}
         \scalebox{1}{
    \begin{tikzpicture}[scale=1.6, thick, every node/.style={scale=1.25}]

\def\xaxis{6}
\draw[-latex] (0,0) -- (\xaxis,0);
\draw (\xaxis,0) node[anchor=north,scale=0.9] {$L_B/L_T $};

\def\yaxis{4}
\draw[-latex] (0,0) -- (0,\yaxis);
\draw (0,\yaxis) node[anchor=south east,scale=0.9,rotate=90] {$\text{log}\left(Q/Q^\text{gB} \right)$};

\def\fluxstyle{solid}
\def\fluxthickness{thick}
\def\asymptotestyle{dashed}
\def\temperaturestyle{dotted}

\def\xseparationmiddle{2.5}
\def\yseparationmiddle{0.75}
\def\yjump{0}

\def\xmin{0}
\def\ymin{0}

\def\xa{1.75}
\def\ya{1.25}

\def\xb{\xa + \xseparationmiddle}
\def\yb{\ya + \yseparationmiddle}

\def\xc{\xb}
\def\yc{\yb + \yjump}

\def\scalefactor{1}
\def\xd{\xc +\scalefactor*\xa-\scalefactor*\xmin}
\def\yd{\yc +\scalefactor*\ya-\scalefactor*\ymin}

\def\labsize{0.75}

\draw[-, \fluxstyle, \fluxthickness] (\xmin,\ymin) --
node[midway,anchor=south,scale=\labsize, rotate = atan(\ya/\xa)] 
{$\propto (L_B/L_T)^3$}
(\xa,\ya);

\draw[-, \temperaturestyle, \fluxthickness] (\xa,0) -- (\xa,\ya);
\draw (\xa,0) node[anchor=north,scale=\labsize] {$L_B/\sqrt{\beta_e} L_\parallel$};

\draw[-, \fluxstyle, \fluxthickness] (\xa,\ya) --
node[midway,anchor=south,scale=\labsize, rotate = atan(\yseparationmiddle/\xseparationmiddle)]
{$\propto (L_B/L_T)^2$}
(\xb,\yb);

\draw[-, \temperaturestyle, \fluxthickness] (\xb,0) -- (\xb,\yb);
\draw (\xb,0) node[anchor=north,scale=\labsize] {$(L_B/\sqrt{\beta_e} L_\parallel)^2$};

\draw[-, \fluxstyle, \fluxthickness] (\xc,\yc) -- 
node[anchor=south,scale=\labsize, rotate = atan(\ya/\xa)]
{$\propto (L_B/L_T)^3$}
(\xd,\yd);

 \end{tikzpicture}}   \\\\
            (a) Collisionless limit \eqref{eq:turb_heat_flux_final_collisionless} \\
         \scalebox{1}{
    \begin{tikzpicture}[scale=1.6, thick, every node/.style={scale=1.25}]

\def\xaxis{6}
\draw[-latex] (0,0) -- (\xaxis,0);
\draw (\xaxis,0) node[anchor=north,scale=0.9] {$L_B/L_T $};

\def\yaxis{4}
\draw[-latex] (0,0) -- (0,\yaxis);
\draw (0,\yaxis) node[anchor=south east,scale=0.9,rotate=90] {$\text{log}\left(Q_\nu/Q^\text{gB} \right)$};

\def\fluxstyle{solid}
\def\fluxthickness{thick}
\def\asymptotestyle{dashed}
\def\temperaturestyle{dotted}

\def\xseparationmiddle{2.75}
\def\yseparationmiddle{0.5}
\def\yjump{0.75}

\def\xmin{0}
\def\ymin{0}

\def\xa{1.75}
\def\ya{1.25}

\def\xb{\xa + \xseparationmiddle}
\def\yb{\ya + \yseparationmiddle}

\def\xc{\xb}
\def\yc{\yb + \yjump}

\def\scalefactor{0.8}
\def\xd{\xc +\scalefactor*\xa-\scalefactor*\xmin}
\def\yd{\yc +\scalefactor*\ya-\scalefactor*\ymin}

\def\labsize{0.75}

\draw[-, \fluxstyle, \fluxthickness] (\xmin,\ymin) --
node[midway,anchor=south,scale=\labsize, rotate = atan(\ya/\xa)] 
{$\propto (L_B/L_T)^3$}
(\xa,\ya);

\draw[-, \temperaturestyle, \fluxthickness] (\xa,0) -- (\xa,\ya);
\draw (\xa,0) node[anchor=north,scale=\labsize] {$L_B/\sqrt{\beta_e} L_\parallel$};

\draw[-, \fluxstyle, \fluxthickness] (\xa,\ya) --
node[midway,anchor=north,scale=\labsize, rotate = atan(\yseparationmiddle/\xseparationmiddle)]
{$\propto (L_B/L_T)^1$}
(\xb,\yb);

\draw[-, \temperaturestyle, \fluxthickness] (\xb,0) -- (\xb,\yb);
\draw (\xb,0) node[anchor=north,scale=\labsize] {$(L_B/\sqrt{\beta_e} L_\parallel)^2$};

\draw[-, \fluxstyle, \fluxthickness] (\xb,\yb) -- (\xc,\yc);

\draw[-, \fluxstyle, \fluxthickness] (\xc,\yc) -- 
node[anchor=south,scale=\labsize, rotate = atan(\ya/\xa)]
{$\propto (L_B/L_T)^3$}
(\xd,\yd);

\draw[-, \asymptotestyle, \fluxthickness] (\xc -\scalefactor*\xa + \scalefactor*\xmin,\yc - \scalefactor*\ya + \scalefactor*\ymin) -- (\xc, \yc);

\def\yoffset{0}

\begin{scope}
    \clip (\xc -\scalefactor*\xa + \scalefactor*\xmin-0.05,\yoffset) rectangle (\xc -\scalefactor*\xa + \scalefactor*\xmin +0.05 ,\yc - \scalefactor*\ya + \scalefactor*\ymin - 0.4);
    \draw[-, \temperaturestyle, \fluxthickness] (\xc -\scalefactor*\xa + \scalefactor*\xmin,\yoffset) -- (\xc -\scalefactor*\xa + \scalefactor*\xmin,\yc - \scalefactor*\ya + \scalefactor*\ymin);
\end{scope}

\draw (\xc -\scalefactor*\xa + \scalefactor*\xmin,\yoffset) node[anchor=north,scale=\labsize] {$(L_B/\sqrt{\beta_e} L_\parallel)^{3/2}$};

 \end{tikzpicture}}  \\\\
            (b) Collisional limit \eqref{eq:turb_heat_flux_final_collisional}
    \end{tabular}

    \caption{The scaling of the turbulent heat-flux with $L_B/L_T$ in the (a) collisionless and (b) collisional limits. As $L_B/L_T$ is increased, the electron transport initially becomes less stiff, as flux freezing pins down the ETG injection scale, after which it stiffens again as cTAI takes over.} 
    \label{fig:heat_flux_scalings}
\end{figure}

To summarise, we can write the turbulent heat flux in the collisionless limit as
\begin{align}
    Q \sim Q^\text{gB} 
    \left\{
    \begin{array}{ll}
         \displaystyle \left( \frac{L_\parallel}{L_B} \right)\left( \frac{L_B}{L_T} \right)^3, &  \displaystyle  \frac{L_B}{L_T} \ll \frac{L_B}{L_\parallel \sqrt{\beta_e}},   \\[4mm]
          \displaystyle  \frac{1}{\sqrt{\beta_e}} \left( \frac{L_B}{L_{T}} \right)^2, &  \displaystyle   \frac{L_B}{L_\parallel \sqrt{\beta_e}} \lesssim \frac{L_B}{L_T}  \lesssim \left( \frac{L_B}{L_\parallel \sqrt{\beta_e}} \right)^2, \\[4mm]
          \displaystyle \sqrt{\beta_e}  \left( \frac{L_\parallel}{L_B} \right)^2 \left( \frac{L_B}{L_{T}} \right)^3, &  \displaystyle    \frac{L_B}{L_T}  \gtrsim \left(\frac{L_B}{L_\parallel \sqrt{\beta_e}} \right)^2,
    \end{array}
    \right.
    \label{eq:turb_heat_flux_final_collisionless}
\end{align}
or, in the collisional limit, as
\begin{align}
    Q_\nu \sim Q^\text{gB} 
    \left\{
    \begin{array}{ll}
         \displaystyle \left(\frac{L_\parallel}{L_B} \right)^2 \left( \frac{L_B}{\lambdae} \right) \left( \frac{L_B}{L_T} \right)^3, &  \displaystyle   \frac{L_B}{L_T} \ll \frac{L_B}{L_\parallel \sqrt{\beta_e}},   \\[4mm]
          \displaystyle  \frac{1}{\beta_e} \left( \frac{L_B}{\lambdae} \right) \left( \frac{L_B}{L_T} \right), &  \displaystyle  \frac{L_B}{L_\parallel \sqrt{\beta_e}} \lesssim \frac{L_B}{L_T}  \lesssim \left( \frac{L_B}{L_\parallel \sqrt{\beta_e}} \right)^2, \\[4mm]
          \displaystyle  \sqrt{\beta_e}  \left( \frac{L_\parallel}{L_B} \right)^3 \left(\frac{L_B}{\lambdae} \right)\left( \frac{L_B}{L_{T}} \right)^3, &  \displaystyle    \displaystyle    \frac{L_B}{L_T}  \gtrsim \left(\frac{L_B}{L_\parallel \sqrt{\beta_e}} \right)^2.
    \end{array}
    \right.
    \label{eq:turb_heat_flux_final_collisional}
\end{align}
Notably, this implies that the effect of increasing $\beta_e$ (or increasing $L_\parallel/L_B \sim \pi q$, as in a tokamak edge), is first to make the electron heat transport less stiff, as flux freezing pins down the ETG injection scale, and then to stiffen it back again, as cTAI takes over. This is sketched in \figref{fig:heat_flux_scalings}. A striking (and perhaps disturbing) feature of these results is the discontinuity in the collisional turbulent heat flux around the transition between the sTAI- and cTAI-dominated regimes, described by the last two expressions in~\eqref{eq:turb_heat_flux_final_collisional}. Comparing these, it is easy to see that the latter is larger than the former for 
\begin{align}
    \frac{L_B}{L_T} \gtrsim \left(\frac{L_B}{L_\parallel \sqrt{\beta_e}} \right)^{3/2}.
    \label{eq:turb_heat_flux_discontinuity}
\end{align}
This condition is obviously met before the parallel system size is large enough in order to activate the cTAI, meaning that the sTAI regime must persist --- despite it supporting a notionally lower flux than that predicted by the cTAI scaling --- until the inequality in~\eqref{eq:turb_regime_3} is satisfied, at which point the cTAI takes over, leading to the discontinuity. Whether this and the other simple `twiddle-algebra' considerations that led to \eqref{eq:turb_heat_flux_final_collisionless} and \eqref{eq:turb_heat_flux_final_collisional} survive the encounter with quantitative reality is left for future numerical investigations to determine.

\section{Discussion} 
\label{sec:discussion}

\subsection{Summary}
\label{sec:summary}
We have considered electromagnetic instabilities and turbulence driven by the electron-temperature gradient in a local slab model of a tokamak-like plasma with constant equilibrium gradients (including magnetic drifts but not magnetic shear, see \secref{sec:magnetic_equilibrium_and_geometry}), with the governing equations (\secref{sec:summary_of_equations}) derived in a low-beta asymptotic limit of gyrokinetics. Central to these considerations was the electron inertial scale $d_e$, which divided our system into two distinct physical regimes: electrostatic (perpendicular scales below $d_e$, $k_\perp \gg d_e^{-1}$, or $d_e^{-1} \chi^{-1}$ in the collisional limit, where $\chi = L_T/\lambdae \sqrt{\beta_e}$) and electromagnetic (perpendicular scales above $d_e$, but still smaller than the ion gyroradius, $\rho_i^{-1} \ll k_\perp \ll d_e^{-1}$, or $d_e^{-1} \chi^{-1}$ in the collisional limit), distinguished by whether or not the magnetic field lines were frozen into the electron flow \eqref{eq:flux_freezing_velocity}. 

In the electrostatic regime, magnetic field lines are decoupled from the electron flow, and so electrons are free to flow across field lines without perturbing them. In this regime, we recovered both the familiar electrostatic electron-temperature-gradient (sETG, sections \ref{sec:setg} and \ref{sec:col_setg}) and curvature-mediated ETG (cETG, \secref{sec:curvature_ETG}) instabilities, noting in particular that the mechanism responsible for the extraction of free energy from the (radial) equilibrium temperature gradient was the fluctuating $\vec{E}\times \vec{B}$ flow --- the usual electrostatic linear drive --- in that it converted the equilibrium temperature variation into perturbations of the electron temperature [see, e.g., the third equation in \eqref{eq:setg_equations}]. 

In the electromagnetic regime, the magnetic field lines are frozen into the electron flow \eqref{eq:flux_freezing_velocity}, meaning that perpendicular magnetic-field perturbations $\delta \! \vec{B}_\perp$ are created as electrons move across field lines and drag the latter along. Crucially, this means that the equilibrium temperature gradient has a component along the perturbed field line, viz., its projection onto the radial component of the perturbed magnetic field [see, e.g., the second term in~\eqref{eq:logt_definition_intro}], which proved to be responsible for the electromagnetic destabilisation associated with the novel thermo-Alfv\'enic instability (TAI)~(\secref{sec:electromagnetic_regime_tai}). We showed that the TAI exists in both a slab version (sTAI, destabilising kinetic Alfv\'en waves, sections \ref{sec:isothermal_kaws} and \ref{sec:isobaric_slab_TAI}) and a curvature-mediated version (cTAI, sections \ref{sec:isothermal_ctai} and \ref{sec:isobaric_limit}). The transition between these two occurs at the critical parallel wavenumber $k_{\parallel c}$ \eqref{eq:tai_kpar_critical}: from sTAI at $k_{\parallel} \gg k_{\parallel c}$ to cTAI at $k_{\parallel} \lesssim k_{\parallel c}$. Another important scale for the TAI is the perpendicular wavenumber $k_{\perp *}$~\eqref{eq:tai_transition_kperp}, which controls the transition between the isobaric ($k_{\perp *} \lesssim k_{\perp} \lesssim d_e^{-1}$, or $d_e^{-1} \chi^{-1}$ in the collisional limit) and isothermal ($\rho_i^{-1} \ll k_\perp \lesssim k_{\perp *}$) limits. In the isobaric limit (\secref{sec:isobaric_limit}), we demonstrated that cTAI is subdominant to sTAI, and can be regarded as an electron-scale extension of MHD-like modes, such as kinetic-ballooning modes (KBMs). In contrast, in the isothermal limit (\secref{sec:isothermal_ctai}), we found, most importantly for transport, that the cTAI is the dominant instability, with a peak growth rate \eqref{eq:isothermal_tai_dispersion_relation} greater than that of the cETG \eqref{eq:cetg_gamma}, exciting electromagnetic perturbations with a specific parallel wavenumber \eqref{eq:tai_curvature_maximum_explicit} (unlike the cETG, which is two-dimensional). This isothermal cTAI's physical mechanism hinges on the fact that --- in the presence of either dominant parallel streaming $k_\parallel \vthe$ (in the collisionless limit) or thermal conduction $\kappa k_\parallel^2 \propto k_\parallel^2\vthe^2/\nu_e$ (in the collisional one) --- perturbations of the magnetic field are coupled to those of the electron temperature as the latter must always adjust to cancel the variation of the equilibrium temperature along the perturbed field line [see, e.g., the isothermal condition \eqref{eq:isothermal_limit_logt}]. Such an instability mechanism can only be present in the electromagnetic regime, when perturbations of the magnetic field's direction are significant.

Given that the dominant source of turbulent energy injection is often associated with the largest scales of a given system, the presence of such a large-scale, electromagnetic instability suggested that the picture of electromagnetic turbulence would depart significantly from the electrostatic one. This is indeed what we found: using a critical-balance phenomenology analogous to \cite{barnes11} to construct a turbulent-cascade theory for the free energy injected by both the electrostatic and electromagnetic instabilities (\secref{sec:free_energy_and_turbulence}), we demonstrated that the cTAI dominated the turbulent transport for temperature gradients $L_B/L_T$ larger than $\beta_e^{-1}(L_B/L_\parallel)^2$ (\secref{sec:summary_of_turbulent_regimes}). Moreover, the turbulent electron heat flux carried by the fluctuations at the cTAI injection scale \eqref{eq:turb_ctai_outer_injection} turned out to scale more steeply with the temperature gradient than the heat flux due to the electrostatic sETG turbulence in this regime, thus giving rise to stiffer transport [see \eqref{eq:turb_heat_flux_final_collisionless} in the collisionless limit and \eqref{eq:turb_heat_flux_final_collisional} in the collisional one]. These results would appear to be particularly relevant in the context of the edge regions of a tokamak, where both the safety factor and the temperature gradients are large (see, e.g., \citealt{ham21} and references therein). 

These results demonstrate that if finite perturbations of the magnetic-field direction are allowed in the presence of a radial equilibrium electron temperature gradient, then the system is able to extract free energy from the equilibrium temperature gradient via a route that is distinct from the usual $\vec{E}\times \vec{B}$ feedback, and that this extraction channel can be dominant. Given that all realistic plasmas are at least somewhat electromagnetic, no matter how small the plasma beta, this physics should be of some concern, or at least interest, to those attempting to model the effect of electromagnetic turbulence in tokamak-relevant configurations.

\subsection{Open issues}
\label{sec:open_issues}
The results and conclusions of this paper were derived within the context of a reduced model, as doing so allowed us to focus directly on the fundamental physical processes behind electromagnetic destabilisation on electron scales in the presence of an electron temperature gradient. Such simplifications, however, always come at a cost to general practical applicability, and so we will here devote some space to a discussion of the most pressing questions and lines of investigation left open, or opened up, by this work. 

\subsubsection{Ion dynamics}
\label{sec:ion_dynamics}
All of the results of this paper have been derived in the limit where the ion density response is Boltzmann, as in \eqref{eq:quasineutrality_initial}. In terms of perpendicular scales, this is equivalent to the assumption that $k_\perp \gg \rho_i^{-1}$. Simultaneously, the electromagnetic physics --- our main subject --- occurs on the scales at which magnetic-field perturbations can be created by electron motions, viz., below the flux-freezing scale, $k_\perp \lesssim d_e^{-1}$. Therefore, in order for the adiabatic-ion assumption to remain valid, we need a sufficient separation between $\rho_i$ and the largest perpendicular scale within our system. For the outer scale \eqref{eq:turb_ctai_outer_perpendicular} of our cTAI turbulence, this implies a restriction on the electron beta of
\begin{align}
     \frac{Z^2 m_e}{\tau m_i} \ll \beta_e \ll  
     \frac{\tau m_i}{Z^2 m_e} \left( \frac{L_T}{L_B} \right)^3 \left\{
    \begin{array}{ll}
         \displaystyle \left( \frac{L_B}{L_\parallel} \right)^4, &  \displaystyle \text{collisionless},   \\[4mm]
          \displaystyle  \left( \frac{L_B}{L_\parallel} \right)^6 \left( \frac{\lambdae}{L_B} \right)^2, &  \displaystyle   \text{collisional},
    \end{array}
    \right.
    \label{eq:beta_restriction}
\end{align}
with the lower bound following from demanding that $\rho_i \gg d_e$.
This scale separation is never going to be very large in a realistic plasma, and thus an important question is whether the TAI mechanism --- that provides an electromagnetic source of free energy on the largest electron scales --- survives at, or indeed across, the ion-Larmor transition, for $k_\perp \rho_i \lesssim 1$. Answering this will require both a careful handling of finite-ion-Larmor-radius (FLR) effects and the introduction of an ion-temperature gradient, in addition to the electron one. These extensions have been left for future investigation. We note that, in what is perhaps a preview of the result of such an investigation, \cite{maeyama21} found that there was very little difference in the electrostatic potential $\varphi$ between the cases of adiabatic and kinetic ions when electromagnetic effects were taken into account (see their figure 4), which suggests that at least qualitatively, the nonlinear results of \secref{sec:free_energy_and_turbulence} may not be significantly modified by the presence of non-adiabatic ions for the plasma parameters considered here.

\subsubsection{Microtearing modes and magnetic shear}
\label{sec:microtearing_modes_and_magnetic_shear}
As mentioned in \secref{sec:introduction}, much of the research into electromagnetic microinstabilities and turbulence in fusion contexts has focused on two microinstability classes: micro-tearing modes (MTMs) and KBMs. While we have already discussed the latter within the context of this work (\secref{sec:isobaric_limit}), we have little to say about MTMs. This is because we did not include in our model any shear of the equilibrium magnetic field --- often thought to be a crucial ingredient in MTM dynamics, which encourages the associated tearing of magnetic field lines (see, e.g., \citealt{zocco15} and references therein). Note that the effect sometimes viewed as responsible for driving slab MTMs in the absence of magnetic shear, the so-called `time-dependent thermal force' (\citealt{hassam80a}), is negligible within our analysis (see \apref{app:time_dependent_thermal_force}). As a result, we conclude that the TAI cannot be classed as a particular branch of the MTM zoo. It is, naturally, an interesting question how the results of this paper would be modified in the presence of magnetic shear; given that the TAI mechanism leads to a growth of perturbations of the magnetic field's direction, it is possible that the TAI could drive tearing in a sheared setting. In any case, introducing magnetic shear into our reduced system should provide an appropriately simple model for an investigation of MTM dynamics. An analogue of such a system in full tokamak geometry is the electromagnetic extension of \cite{hardman22}, currently in preparation. 

\subsubsection{Nonlinear saturation of electromagnetic simulations}
\label{sec:nonlinear_saturation_of_electromagnetic_simulations}
An aspect of turbulent transport that has baffled tokamak modellers in recent years is the failure to find a saturated state in local nonlinear electromagnetic simulations (see, e.g., \citealt{pueschel13b,pueschel13a,pueschel14} and references therein). It is believed that this failure is due to the presence of MTMs or KBMs, and to their interactions with zonal flows, though relatively little is understood about whether this issue is a truly physical one --- related to the mechanisms of saturation of electromagnetic turbulence --- or is due to numerical subtleties and difficulties. Given that the model equations considered in this paper are clearly electromagnetic, their nonlinear numerical investigation should be able to provide some insight into this issue. Should these equations experience a blow-up similar to gyrokinetics, then they are sufficiently simple --- in comparison to the full gyrokinetic system employed by the simulations cited above --- that making theoretical sense of this saturation failure should be more amenable.

The issue of the blow-up aside, there is of course the broader question of the structure of the saturated state of electromagnetic turbulence in tokamak plasmas --- or even the much simpler tokamak-inspired ones, like ours. The \textit{a priori} analysis provided in \secref{sec:free_energy_and_turbulence} is but a preliminary step towards a more thorough numerical investigation, based on the model derived here, of cTAI turbulence, its saturation, its transport properties, its ability to support reduced transport states (cf. \citealt{ivanov20,ivanov22}), etc. These questions will be addressed in a future publication, for which the present article provides the nessecary theoretical background. 
\\

We are indebted to 
G.~Acton,
M.~Barnes,
S.~Cowley,
I.~Dodin,
W.~Dorland,
M.~Hardman, 
D.~Hosking,
M.~Kunz,
N.~Loureiro,
L.~Milanese,
J.~Parisi,
and
F.~Parra 
for helpful discussions and suggestions at various stages of this project.
This work has been carried out within the framework of the EUROfusion Consortium and has received funding from the Euratom research and training programme 2014–2018 and 2019–2020 under Grant Agreement No. 633053, and from the UKRI Energy Programme (EP/T012250/1). The views and opinions expressed herein do not necessarily reflect those of the European Commission. TA was supported by a UK EPSRC studentship. The work of AAS was supported in part by UK EPSRC (EP/R034737/1). Declaration of interests: the authors report no conflict of interest.


\begin{appendix}

\section{Derivation of low-beta equations}
\label{app:derivation_of_low_beta_equations}
We would like to work with a set of equations that, while representing a correct approximation to plasma dynamics in some physically realisable limit and containing all the physics that is of interest to us, have a minimum of features that increase technical complexity without being qualitatively essential. This attitude was taken in \cite{zocco11}, who were interested in electron kinetics in the context of magnetic reconnection; the optimal regime to consider turned out to be the low-beta limit of ion gyrokinetics and electron drift kinetics. A similar regime will serve our purposes here, but, as we now wish to include also energy injection due to an equilibrium temperature gradient and the magnetic drifts associated with a magnetic geometry of locally constant curvature, we will present a self-contained derivation of the relevant equations. 

In what follows, \apref{app:equilibrium_and_fluctuations} introduces the nature of the equilibrium and fluctuations that we consider in our system, including the constraints on the equilibrium lengthscales due to the magnetic geometry defined in \secref{sec:magnetic_equilibrium_and_geometry}. \Apref{app:low_beta_ordering} describes and physically motivates our asymptotic ordering. \Apref{app:the_gyrokinetic_equation} introduces the gyrokinetic system of equations. Our low-beta ordering is then implemented to derive equations describing both ion and electron dynamics in appendices \ref{app:ion_equations} and \ref{app:electron_equations}, respectively. The collisional limit of the resultant equations is then derived in \apref{app:collisional_limit}. Finally, \apref{app:strongly_driven_limit} details the reduction of our equations --- both collisionless and collisional --- to those considered in the main text. Readers merely interested in the latter equations can skip ahead to \apref{app:strongly_driven_limit}, working backwards where further clarification is required. 

\subsection{Equilibrium and fluctuations}
\label{app:equilibrium_and_fluctuations}
We will describe both species ($\s =e$ for electrons and $\s=i$ for ions) kinetically, with their distribution functions sought in the form
\begin{align}
    f_\s = f_{0\s} + \df_\s.
    \label{eq:distribution_function}
\end{align}
Although we neglected both density and ion-temperature gradients in the main text, here we shall, for the sake of generality, allow our local equilibria $f_{0\s}$ to support radial gradients, which are assumed to be constant across our domain, in both density and temperature for both species, viz.,
\begin{equation}
    \gradd f_{0\s} =  - \left[ \frac{1}{L_{n_\s}} + \frac{1}{L_{T_\s}} \left( \frac{v^2}{\vths^2} - \frac{3}{2} \right) \right] \hat{\vec{x}} f_{0\s}, \quad L_{n_\s}^{-1} = - \frac{1}{n_{0\s}} \frac{\rmd n_{0\s}}{\rmd x}, \quad L_{T_\s}^{-1} = - \frac{1}{T_{0\s}} \frac{\rmd T_{0\s}}{\rmd x},
    \label{eq:equilibrium_distributions}
\end{equation}
where $n_{0\s}$ and $T_{0\s}$ are the equilibrium density and temperature of species $\s$, respectively, $\vths = \sqrt{2 T_{0\s}/m_\s}$ is their thermal speed and $m_s$ their mass. It is assumed that all equilibrium quantities, of typical lengthscale $L$, evolve on the (long) transport timescale $\tau_E^{-1} \sim (\rho_\s/L)^3 \Omega_\s$, and so can be considered static. Here, $\rho_\s = \vths/|\Omega_s|$ is the thermal Larmor radius and $\Omega_s = q_\s B_0/m_\s c$ the cyclotron frequency of species $\s$ with charge $q_\s$ ($q_i = Ze$, $q_e = -e$), with $B_0$ the equilibrium magnetic field strength.  Note that quasineutrality ($n_{0e} = Zn_{0i}$) implies that $L_{n_{e}} = L_{n_i} = L_n$. 

The perturbations $\df_\s$ around these equilibria have characteristic frequency $\omega$ and wavenumbers $k_\parallel$ and $k_\perp$ parallel and perpendicular, respectively, to the magnetic field $\vec{B}$. The magnetic field consists of an equilibrium part that is oriented in the $\vec{b}_0$ direction and varies radially, plus a time- and space-dependent fluctuating part:
\begin{equation}
    \vec{B}(\vec{r},t) = B_0(x) \vec{b}_0 + \dB(\vec{r},t).
    \label{eq:magnetic_field_appendix}
\end{equation}
The equilibrium (mean) magnetic field has the scale length and radius of curvature
\begin{equation}
    L_B^{-1} =- \frac{1}{B_0}\frac{\rmd B_0}{\rmd x}, \quad R^{-1}  = \left| \vec{b}_0 \cdot \gradd \vec{b}_0\right|,
    \label{eq:magnetic_field_scale_length_appendix}
\end{equation}
respectively, both of which are assumed to be constant across our domain, while the fluctuating part $\dB$ has the same characteristic frequency and wavenumbers as $\df_s$. The electric field $\vec{E}$ is assumed to have no mean part.

For a non-relativistic plasma, the equilibrium magnetic field is described by Amp\`ere's law and force balance:
\begin{equation}
    \vec{j}_0 = \frac{c}{4\pi} \curll \vec{B}_0, \spc \frac{1}{c}\vec{j}_0 \times \vec{B}_0 = \gradd_\perp \sum_\s n_{0\s} T_{0\s} . 
    \label{eq:pressure_balance}
\end{equation}
Combining these two equations, we arrive at the usual expression of force balance between the pressures of all plasma species, the equilibrium magnetic pressure, and magnetic curvature force due to field-line bending:
\begin{equation}
    \gradd_\perp \left( \sum_\s n_{0\s} T_{0\s} + \frac{B_0^2}{8\pi} \right) = \frac{B_0^2}{4\pi} ( \vec{b}_0 \cdot \gradd) \vec{b}_0.
    \label{eq:mhd_equilibrium}
\end{equation}
Adopting the geometry described in section \ref{sec:magnetic_equilibrium_and_geometry}, with the generalisation \eqref{eq:equilibrium_distributions}, this gives us a constraint by which the equilibrium lengthscales of our system are related:
\begin{equation}
    \frac{\beta_e}{2}\left(\frac{1}{L_n} +  \frac{1}{L_{T_{e}}}  \right) +  \frac{\tau\beta_e}{2Z} \left(\frac{1}{L_n} +  \frac{1}{L_{T_i}}  \right) + \frac{1}{L_B} = \frac{1}{R},
    \label{eq:equilibrium_constraint}
\end{equation}
where $\tau = T_{0i}/T_{0e}$ is the temperature ratio and $\beta_e = 8\pi n_{0e} T_{0e} /B_0^2$ the electron beta. Consideration of such constraints is important at finite beta: e.g., the so-called `Gradient Drift Coupling' (GDC) instability found by \cite{pueschel15} was demonstrated as spurious by \cite{rogers18}, with the growth rate of the instability disappearing once the equilibrium constraint had been taken into account. At vanishingly small beta, which will be assumed in \eqref{eq:ordering_of_beta}, however, \eqref{eq:equilibrium_constraint} simply becomes $R = L_B$, and the remaining equilibrium lengthscales may be chosen arbitrarily. In what follows, we shall no longer distinguish between $R$ and $L_B$. 

\subsection{Low-beta gyrokinetic ordering}
\label{app:low_beta_ordering}
We want our equations to be simple as possible, but sufficiently complete in order to retain the parallel streaming of electrons (and their associated kinetic effects, such as Landau damping; \citealt{landau46}), kinetic Alfv\'en waves (KAW), drift waves, perpendicular advection by both magnetic drifts and $\vec{E} \times \vec{B}$ flows, and electron collisions. Therefore, we postulate an asymptotic ordering in which the characteristic frequency of the perturbations $\omega$ and the characteristic frequencies of all of the above phenomena are formally comparable:
\begin{align}
    \omega & \sim k_\parallel \vthe \sim \omega_\text{KAW}  \sim \omega_{*\s} \sim \omega_{d\s}   \sim k_\perp v_E  \sim \nu_{ee} \sim \nu_{ei},
    \label{eq:ordering_timescales}
\end{align}
where 
\begin{align}
    \omega_\text{KAW} = \frac{1}{\sqrt{2}} k_\parallel \vthe k_\perp d_e, \quad \omega_{*\s} = \frac{k_y \rho_\s \vths}{2L_{T_\s}}, \quad \omega_{d\s} = \frac{k_y \rho_\s \vths}{2L_B}, 
    \label{eq:definition_timescales}
\end{align}
are the kinetic Alfv\'en wave frequency, the drift frequency, and magnetic-drift frequency respectively, $\vec{v}_E = c \vec{E} \times \vec{B}/B^2$ is the $\vec{E} \times \vec{B}$ drift velocity ($c$ is the speed of light), and
\begin{align}
     \nu_{ei} = \frac{4\sqrt{2\pi}}{3} \frac{e^4 n_{0e} \log  \Lambda}{m_e^{1/2} T_{0e}^{3/2}} ,\quad  \nu_{ee}= \frac{\nu_{ei}}{Z},
     \label{eq:definition_collision_frequencies}
 \end{align}
are the electron-ion and electron-electron collision frequencies, respectively, with $\log\Lambda$ the usual Coulomb logarithm (\citealt{braginskii65}, \citealt{helander05}). 

In \eqref{eq:definition_timescales}, we also used the electron skin depth (inertial length) $d_e = \rho_e/\sqrt{\beta_e}$. This lengthscale will be of key significance for us because it regulates the transition between the electrostatic and electromagnetic regimes. Indeed, the ordering of parallel streaming with respect to KAW implies that
\begin{align}
    k_{\parallel} \vthe \sim \omega_\text{KAW} \quad \Rightarrow \quad k_\perp d_e \sim 1,
    \label{eq:ordering_streaming_vs_kaw}
\end{align}
meaning that we will retain the effects of electron inertia. Our ordering of $\beta_e$ with respect to other physical (dimensionless) parameters is determined by our choice of ordering of perpendicular wavenumbers $k_\perp$ with respect to the electron and ion Larmor radii. We choose to work in the drift-kinetic approximation for electrons, ordering our perpendicular wavenumbers so that
\begin{align}
    k_\perp \rho_i \sim 1 \quad \Rightarrow \quad k_\perp \rho_e \sim \sqrt{\frac{m_e}{m_i}} \sim k_\perp d_e \sqrt{\beta_e} \quad \Rightarrow \quad \beta_e \sim \frac{m_e}{m_i},
    \label{eq:ordering_of_beta}
\end{align}
the last relation following from \eqref{eq:ordering_streaming_vs_kaw}.
We stress that this choice, while an analytically convenient one, it is by no means the unique possible route to the minimalist equations that we are going to derive here. 

The ordering of the drift and collision frequencies with respect to the parallel streaming rate gives us the ordering of parallel wavenumbers:
\begin{align}
   & k_\parallel \vthe \sim \omega_{*\s} \sim \omega_{d\s} \sim k_\perp\rho_e \frac{\vthe}{L}\quad \Rightarrow \quad k_{\parallel} L \sim \sqrt{\beta_e}, \label{eq:ordering_streaming_vs_drifts} \\
  &  k_\parallel \vthe \sim \nu_{ee} \sim \nu_{ei} \sim \frac{\vthe}{\lambdae} \quad \Rightarrow \quad k_\parallel \lambdae \sim 1, \label{eq:ordering_streaming_vs_collisions}
\end{align}
where $\lambdae = \vthe/\nu_{e}$ is the electron mean-free path, and $\nu_e = \nu_{ee} + \nu_{ei}$.

The ordering of the $\vec{E}\times \vec{B}$ drifts with respect to parallel streaming determines the size of perpendicular flows within our system:
\begin{align}
    k_\parallel \vthe \sim k_\perp v_E \quad \Rightarrow \quad \frac{v_E}{\vthe} \sim \frac{k_\parallel}{k_\perp} \sim \frac{d_e}{L} \sqrt{\beta_e} \equiv \epsilon \sqrt{\beta_e}, \label{eq:ordering_streaming_vs_exb} 
\end{align}
where $\epsilon = d_e/L \sim \rho_i/L$ is the gyrokinetic small parameter (see, e.g., \citealt{abel13}), which need not be ordered with respect to $\beta_e$. It mandates small-amplitude, anisotropic perturbations. The frequency of these perturbations is small compared to the Larmor frequencies of both electrons and ions:
\begin{align}
    \frac{\omega}{\Omega_e} \sim \frac{k_\perp v_E}{\Omega_e} \sim k_\perp \rho_e \epsilon \sqrt{\beta_e} \sim \epsilon\beta_e, \quad \frac{\omega}{\Omega_i} = \frac{m_i}{Z m_e} \frac{\omega}{\Omega_e} \sim \epsilon.
    \label{eq:ordering_frequencies_vs_larmor}
\end{align}
The ordering of $v_E$ allows us to order the amplitude of the perturbed scalar potential $\phi$:
\begin{align}
    \frac{v_E}{\vthe} \sim \frac{c}{B_0} \frac{k_\perp \phi}{\vthe} \sim k_\perp\rho_e \frac{e\phi}{T_{0e}} \quad \Rightarrow \quad \frac{e\phi}{T_{0e}} \sim \epsilon.
    \label{eq:ordering_phi}
\end{align}
The density perturbations $\dn_\s$ are ordered anticipating a Boltzmann density response and the temperature perturbations $\delta T_s$ are assumed comparable to them:
    \begin{align}
     \frac{\dTe}{T_{0e}} \sim \frac{\delta T_i}{T_{0i}} \sim \frac{\delta n_{i}}{n_{0i}} =\frac{\dne}{n_{0e}}  \sim \frac{e\phi}{T_{0e}} \sim  \epsilon.
    \label{eq:ordering_field_amplitudes_electrostatic}
\end{align}

Finally, the perpendicular magnetic-field perturbations are ordered so as to allow field variation along the exact (perturbed) field lines to be order-unity different from the variation along the direction of the equilibrium magnetic field, viz.,
\begin{align}
   \frac{\partial}{\partial z} \sim  \frac{\dBperp}{B_0} \cdot \gradd_\perp \quad \Rightarrow \quad \frac{\dBperp}{B_0} \sim \frac{k_\parallel}{k_\perp} \sim  \frac{k_\parallel L}{k_\perp d_e} \frac{d_e}{L} \sim \epsilon \sqrt{\beta_e},
    \label{eq:ordering_dBperp}
\end{align}
whereas the (compressive) parallel magnetic-field perturbations are ordered anticipating pressure balance:
\begin{align}
    \frac{\dBpar}{B_0}  = \frac{4\pi}{B_0^2} \delta \left( \frac{B^2}{8\pi} \right) \sim \frac{4\pi}{B_0^2} \delta (n_\s T_\s) \sim \beta_e \frac{\dTe}{T_{0e}} \sim \epsilon \beta_e.
    \label{eq:ordering_dBpar}
\end{align}
This will allow us to ignore $\dBpar$ everywhere. 

By ordering the characteristic frequencies of the perturbations $\omega$ to timescales relevant to the physics that we are interested in [see \eqref{eq:ordering_timescales}] and adopting a particular ordering of perpendicular wavenumbers [see \eqref{eq:ordering_of_beta}], we have found that all relevant quantities are naturally ordered with respect to either $\beta_e$ or the gyrokinetic small parameter $\epsilon = d_e/L$, where $L \sim L_{n_\s} \sim L_{T_\s} \sim L_B \sim R$. To summarise, we postulate the following ordering of frequencies:
\begin{align}
    \frac{\omega}{\Omega_e} \sim \epsilon \beta_e, \quad \frac{\omega}{\Omega_i} \sim \epsilon,
    \label{eq:ordering_frequencies}
\end{align}
lengthscales:
\begin{align}
    k_\perp \rho_i \sim k_\perp d_e \sim 1, \quad k_\perp \rho_e \sim \sqrt{\beta_e}, \quad k_\parallel L \sim \sqrt{\beta_e}, \quad k_\parallel \lambdae \sim 1, \quad \frac{k_\parallel}{k_\perp} \sim  \epsilon \sqrt{\beta_e},
    \label{eq:ordering_lengthscales}
\end{align}
and fluctuation amplitudes:
\begin{align}
     \frac{e\phi}{T_{0e}} \sim \frac{\dne}{n_{0e}} \sim \frac{\delta n_{i}}{n_{0i}} \sim \frac{\dTe}{T_{0e}} \sim \frac{\delta T_i}{T_{0i}} \sim  \epsilon, \quad \frac{\dBperp}{B_0} \sim \epsilon \sqrt{\beta_e}, \quad  \frac{\dBpar}{B_0} \sim \epsilon \beta_e.
     \label{eq:ordering_amplitudes}
\end{align}

The above ordering of frequencies, lengthscales and amplitudes with respect to the small parameter $\epsilon$ is the standard gyrokinetic ordering (see, e.g., \citealt{abel13}).  
We choose to treat the ordering in $\beta_e$, and thus in the electron-ion mass ratio, as subsidiary to this, viz., 
\begin{align}
    \epsilon \ll \sqrt{\beta_e} \sim \sqrt{\frac{m_e}{m_i}} \ll 1,
    \label{eq:expansion_parameter_beta}
\end{align}
with all other dimensionless parameters, such as the ratios between different equilibrium scales, being treated as finite (i.e., independent of $\beta_e$), although we will introduce further subsidiary expansions in these parameters later on. In \secref{app:the_gyrokinetic_equation}, we introduce the gyrokinetic approximation, which will serve as the starting point for further reduction of our equations by means of the low-beta ordering.

\subsection{Gyrokinetics}
\label{app:the_gyrokinetic_equation}
Under the gyrokinetic ordering, the perturbed distribution function for species $\s$ consists of a Boltzmann and gyrokinetic parts:
\begin{equation}
    \df_\s(\vec{r},\vec{v},t) = - \frac{q_\s \phi(\vec{r},t)}{T_{0\s}} f_{0\s}(x,\vec{v}) + h_\s(\vec{R}_\s, \vpar,  v_\perp, t),
    \label{eq:perturbed_distribution_function}
\end{equation}
where $\vec{R}_\s = \vec{r} - \vec{b}_0 \times \vec{v}_\perp/\Omega_\s$ is the guiding-centre position, and $h_\s$ evolves according to the gyrokinetic equation
\begin{equation}
    \frac{\partial}{\partial t} \left( h_\s - \frac{q_\s \left< \chi \right>_{\vec{R}_\s}}{T_{0\s}} f_{0\s} \right) + \left(\vpar \vec{b}_0 + \vec{v}_{d \s} \right) \cdot \gradd h_\s + \vec{v}_\chi \cdot \gradd_\perp \left( h_\s + f_{0\s} \right) = \left( \frac{\partial h_\s}{\partial t} \right)_c.
    \label{eq:gyrokinetic_equation}
\end{equation}
Here, $\chi = \phi - \vec{v}\cdot \vec{A}/c$ is the gyrokinetic potential ($\phi$ and $\vec{A}$ are the scalar and vector potential, respectively). It gives rise to the nonlinear drift
\begin{equation}
    \vec{v}_\chi \cdot \gradd_\perp  h_\s =  \frac{c}{B_0}\vec{b}_0 \cdot \left( \frac{\partial \left< \chi \right>_{\vec{R}_\s}}{\partial \vec{R}_\s} \times \frac{\partial h_\s}{\partial \vec{R}_\s} \right),
    \label{eq:nonlinearity}
\end{equation}
which includes the $\vec{E} \times \vec{B}$ drift, the parallel streaming along perturbed field lines, and the $\gradd B$ drift associated with the perturbed magnetic field (see \citealt{howes06}). There are also important linear terms: energy injection due to gradients of the equilibrium distribution [see \eqref{eq:equilibrium_distributions}]
\begin{align}
   \vec{v}_\chi \cdot \gradd_\perp  f_{0\s}  = -\frac{c}{B_0} \frac{\partial \left< \chi \right>_{\vec{R}_\s}}{\partial Y_\s} \frac{\partial f_{0\s}}{\partial x} = \frac{c}{B_0}\frac{\partial \left< \chi \right>_{\vec{R}_\s}}{\partial Y_\s} \left[ \frac{1}{L_{n_\s}} + \frac{1}{L_{T_\s}} \left( \frac{v^2}{\vths^2} - \frac{3}{2} \right)\right]f_{0s},
    \label{eq:linear_drive}
\end{align}
and the magnetic drifts associated with the equilibrium field
\begin{equation}
    \vec{v}_{d\s} = \frac{\vec{b}_0}{\Omega_\s} \times \left[ \vpar^2 \vec{b}_0\cdot \grad\vec{b}_0 + \frac{1}{2}\vperp^2 \grad\log B_0 \right] = - \sgn(q_\s) \frac{\rho_\s \vths}{2} \left[\frac{2}{R} \frac{v_\parallel^2}{\vths^2} + \frac{1}{L_B}\frac{v_\perp^2}{\vths^2} \right] \hat{\vec{y}},
    \label{eq:magnetic_drifts}
\end{equation}
where $R$ and $L_B$ are defined in \eqref{eq:magnetic_field_scale_length_appendix}. The last term on the right-hand side of \eqref{eq:gyrokinetic_equation} is the (linearised) collision operator; we shall specify its explicit form in \apref{app:electron_equations}. 

The gyrokinetic equation \eqref{eq:gyrokinetic_equation} is closed by the quasineutrality condition
\begin{equation}
 0 = \sum_{\s} q_\s \delta n_\s = \sum_\s q_\s \left[ -\frac{q_\s \phi}{T_{0\s}}  n_{0\s}    + \int \rmd^3 \vec{v}  \left< h_\s \right>_{\vec{r}}\right],
 \label{eq:quasineutrality}
\end{equation}
and by the parallel and perpendicular parts of Amp\`ere's law, which are, respectively,
\begin{align}
    \gradd_\perp^2 A_\parallel &= - \frac{4\pi}{c} \sum_\s q_\s \int \rmd^3 \vec{v} \: \vpar \left< h_\s \right>_{\vec{r}},
    \label{eq:parallel_amperes_law} \\
     \gradd_\perp^2 \dBpar& = - \frac{4\pi}{c} \vec{b}_0 \cdot \left[ \gradd_\perp \times \sum_\s q_\s \int \rmd^3 \vec{v} \left< \vec{v}_\perp h_\s \right>_{\vec{r}} \right].
    \label{eq:perpendicular_amperes_law}
\end{align}
However, given the ordering \eqref{eq:ordering_dBpar}, we are able to neglect parallel magnetic field perturbations everywhere, meaning that the gyrokinetic potential reduces to
\begin{align}
    \chi = \phi - \frac{v_\parallel A_\parallel}{c}.
    \label{eq:gyrokinetic_potential_simplified}
\end{align}
We thus only need $\phi$ and $A_\parallel$ to determine the other fields to lowest order, and so \eqref{eq:perpendicular_amperes_law} can be dropped from our system of equations. 

In the above and throughout this appendix, $\left< ... \right>$ denotes averages with respect to the gyroangle $\vartheta$: for any function $g$,
\begin{align}
    \left< g (\vec{R}_\s) \right>_{\vec{r}} & = \left< g(\vec{r} - \vec{\rho}_\s(\vartheta)) \right> = \int_0^{2\pi} \frac{d\vartheta}{2\pi} \: g(\vec{r} - \vec{\rho}_\s(\vartheta)), \label{eq:gyroaverage_at_constant_r} \\
    \left< g (\vec{r}) \right>_{\vec{R}_\s} & = \left< g(\vec{R}_s + \vec{\rho}_\s(\vartheta)) \right> = \int_0^{2\pi} \frac{d\vartheta}{2\pi} \: g(\vec{R}_s + \vec{\rho}_\s(\vartheta)) \label{eq:gyroaverage_at_constant_R},
\end{align}
where $\vec{\rho}_s(\vartheta) = \vec{b}_0 \times \vec{v}_\perp/\Omega_\s$ is the velocity-dependent gyroradius, $\vec{v} = \vpar \vec{b}_0 + v_\perp(\cos\vartheta \hat{\vec{y}} - \sin \vartheta \hat{\vec{x}})$, and the unit vectors $\{ \hat{\vec{x}},\hat{\vec{y}},\vec{b}_0 \}$ form a right-handed orthonormal basis. 

In appendices \ref{app:ion_equations} and \ref{app:electron_equations}, we systematically expand the gyrokinetic system of equations \eqref{eq:gyrokinetic_equation}, \eqref{eq:quasineutrality} and~\eqref{eq:parallel_amperes_law} to obtain a closed system to leading order in the low-beta expansion \eqref{eq:expansion_parameter_beta}.

\subsection{Ion kinetics and field equations}
\label{app:ion_equations}
We can neglect the parallel-streaming term in \eqref{eq:gyrokinetic_equation} for the ions, because
\begin{equation}
     \kpar \vthi  \sim \sqrt{\frac{m_e}{m_i}} \kpar \vthe .
    \label{eq:ion_streaming}
\end{equation}
The gyrokinetic potential reduces to the electrostatic potential in the case of the ions, $\chi \approx \phi$, because
\begin{equation}
    \frac{\vthi A_\parallel}{c\phi} \sim \sqrt{\beta_e} \sim \sqrt{\frac{m_e}{m_i}}. 
    \label{eq:small_a_term_ions}
\end{equation}
Finally, we can neglect any contributions arising from the collision operator, because ion collision rates are small within our expansion:
\begin{align}
    \nu_{ii} \sim \sqrt{\frac{m_e}{m_i}} \nu_{ei}, \quad \nu_{ie} \sim \frac{m_e}{m_i}\nu_{ei}.
    \label{eq:ion_collisions}
\end{align}
Introducing the decomposition
\begin{equation}
    h_i = g_i + \frac{Z}{\tau} \left< \varphi \right>_{\vec{R}_i} f_{0i}, \quad \varphi = \frac{e\phi}{T_{0e}},
    \label{eq:non_adiabatic_response}
\end{equation}
we can, therefore, write our ion gyrokinetic equation as follows
\begin{align}
    \left(\frac{\partial}{\partial t} + \vec{v}_{di} \cdot \dperp \right)g_i +  \vec{v}_{di} \cdot \gradd_\perp  \left( \frac{Z}{\tau}\left< \varphi \right>_{\vec{R}_i}  f_{0i} \right)+ \frac{\rho_e \vthe}{2}\left\{ \left< \varphi \right>_{\vec{R}_i}, g_i + f_{0i} \right\} =0.
    \label{eq:ion_gyrokinetic_equation}
\end{align}

In general, we must solve \eqref{eq:ion_gyrokinetic_equation} for $g_i$ in order to determine $h_i$, and thus the ion contribution to the field equations \eqref{eq:quasineutrality} and~\eqref{eq:parallel_amperes_law}. However, since all the parallel dynamics have been neglected in \eqref{eq:ion_gyrokinetic_equation}, its solution $g_i$, and hence $h_i$, will be an even function of ~$\vpar$. Therefore, the ion contribution in \eqref{eq:parallel_amperes_law} vanishes, and we obtain a field equation for $A_\parallel$ in terms of electron dynamics (the electron parallel current) only:
\begin{align}
     \frac{u_{\parallel e}}{\vthe} = d_e^2 \gradd_\perp^2 \mathcal{A}, \quad \mathcal{A} = \frac{A_\parallel}{\rho_e B_0},
     \label{eq:parallel_amperes_law_final}
\end{align}
where we have defined $\mathcal{A}$ as the dimensionless counterpart to $A_\parallel$, as in \eqref{eq:parallel_derivative}. Thus, the only place where ion dynamics enter into our equations is through the quasineutrality condition \eqref{eq:quasineutrality}, which, with the decomposition \eqref{eq:non_adiabatic_response}, becomes
\begin{align}
    \frac{\dne}{n_{0e}} = - \taubar^{-1} \varphi + \frac{1}{n_{0i}} \int \rmd^3 \vec{v} \: \left< g_i \right>_{\vec{r}},
    \label{eq:quasineutrality_final}
\end{align}
where $\taubar^{-1}$ is an operator defined as follows:
\begin{align}
    - \taubar^{-1} \varphi = - \frac{Z}{\tau} (1 - \hat{\Gamma}_0) \varphi \approx \left\{
    \begin{array}{cc}
    \displaystyle\frac{Z}{2\tau} \rho_i^2 \gradd_\perp^2 \varphi, & \displaystyle k_\perp \rho_i \ll 1,\\[4mm]
    \displaystyle - \frac{Z}{\tau} \varphi, & \displaystyle k_\perp \rho_i \gg 1,
    \end{array}
    \right.
    \label{eq:taubar_definition}
\end{align}
and the operator $\hat{\Gamma}_0$ can be expressed, in Fourier space, in terms of the modified Bessel function of the first kind: $\Gamma_0 = I_0(\alpha_i)e^{-\alpha_i}$, where $\alpha_i = (k_\perp \rho_i)^2/2$.

Throughout this paper, we will be concerned with two physical limits in which \eqref{eq:ion_gyrokinetic_equation} is rendered solvable and the quasineutrality constraint \eqref{eq:quasineutrality_final} simplified. The first of these is the limit $k_\perp \rho_i \gg 1$. Under this assumption, and with the ordering $\omega \sim \omega_{di} \sim \omega_{*i}$ [see \eqref{eq:ordering_timescales}], the solution of \eqref{eq:ion_gyrokinetic_equation} has the size
\begin{align}
    g_i \sim \frac{1}{\sqrt{k_\perp\rho_i}} \varphi f_{0i},
    \label{eq:ion_gi_size}
\end{align}
because all the drive (inhomogeneous) terms in \eqref{eq:ion_gyrokinetic_equation} involve the gyroaveraged potential $\left< \varphi \right>_{\vec{R}_i} \sim \varphi/\sqrt{k_\perp \rho_i}$. There is another gyroaveraging in \eqref{eq:quasineutrality_final}, so the contribution 
\begin{align}
    \frac{1}{n_{0i}} \int \rmd^3 \vec{v} \: \left< g_i \right>_{\vec{r}} \sim  \frac{\varphi }{k_\perp \rho_i}
    \label{eq:ion_contribution_size}
\end{align}
can be safely neglected in this limit. The remaining equation relating $\dne$ to $\varphi$ is, therefore,
\begin{align}
    \frac{\dne}{n_{0e} } = - \taubar^{-1} \varphi = - \frac{Z}{\tau} \varphi,
    \label{eq:adiabatic_ions_appendix}
\end{align}
which is \eqref{eq:quasineutrality_initial}, an approximation of `adiabatic ions'. 

The second useful limit is one of strong ETG drive. Let us introduce a subsidiary ordering of equilibrium gradients 
\begin{align}
    L_n \sim L_{T_i} \sim  L_B \ll L_{T_e}
    \label{eq:strongly_driven_lengthscales}
\end{align}
and frequencies
\begin{align}
    \omega_{di} \sim \omega_{*i} \sim \omega_{de} \ll \omega \sim \omega_{*e}.
    \label{eq:strongly_driven_limit}
\end{align}
If this is satisfied, then, still allowing $k_\perp \rho_i \sim 1$,
\begin{align}
     g_i \sim  \frac{\omega_{*i}}{\omega} \left< \varphi \right>_{\vec{R}_i} f_{0i}
     \label{eq:ion_gi_size_strongly_driven}
\end{align}
and, consequently, in \eqref{eq:quasineutrality_final},
\begin{align}
    \frac{1}{n_{0i}} \int \rmd^3 \vec{v} \: \left< g_i \right>_{\vec{r}} \sim \frac{\omega_{*i}}{\omega} \varphi \ll \varphi.
    \label{eq:ion_contribution_size_strongly_driven}
\end{align}
Neglecting this term leaves us again with a simple linear relationship between $\dne$ and $\varphi$, but $\taubar^{-1}$ is still the Bessel operator defined in \eqref{eq:taubar_definition}, keeping the effects of finite ion Larmor radius (FLR) without the need to solve the ion gyrokinetic equation. Ion-FLR modifications do not play a crucial physical role in the majority of this paper, but, for completeness, we have retained $\taubar$ dependencies where they may be relevant for future investigations.

\subsection{Electron equations}
\label{app:electron_equations}

\subsubsection{Electron kinetic equation}
\label{app:electron_kinetic_equation}
Our ordering of perpendicular lengthscales \eqref{eq:ordering_lengthscales} means that the electrons are drift-kinetic to leading order in our expansion \eqref{eq:expansion_parameter_beta}. It is convenient to revert to working with the total perturbed distribution function $\df_e$ [\eqref{eq:perturbed_distribution_function} for $s=e$], instead of $h_e$. In the limit $k_\perp \rho_e \ll 1$, all gyroaverages in \eqref{eq:gyrokinetic_equation} turn into unity operators, and, making use of the simplification
\begin{align}
    \frac{\partial}{\partial t} + \vpar \frac{\partial}{\partial z} + \vec{v}_\chi \cdot \gradd_\perp = \frac{\rmd}{\rmd t} + \vpar \gradd_\parallel,
    \label{eq:gk_operators}
\end{align}
where the operators $\rmd/\rmd t$ (convective derivative with respect to the $\vec{E}\times \vec{B}$ flow) and $\gradd_\parallel$ (parallel derivative along the exact field line) are defined in \eqref{eq:convective_derivative} and \eqref{eq:parallel_derivative}, respectively, we find
\begin{equation}
    \left( \frac{\rmd}{\rmd t} + v_\parallel \gradd_\parallel + \vec{v}_{de} \cdot \dperp \right)\df_e = (\vec{v}_{de} \cdot \gradd_\perp \varphi) f_{0e}-  \vec{v}_\chi \cdot \gradd_\perp f_{0e} - \frac{v_\parallel eE_\parallel}{T_{0e}}f_{0e} + \left( \frac{\partial \df_e}{\partial t} \right)_c.
    \label{eq:electron_drift_kinetic_equation}
\end{equation}
In terms of our dimensionless field variables, the parallel electric field is 
\begin{equation}
    - \frac{e E_\parallel}{T_{0e}}  =  \frac{2}{\vthe} \frac{\rmd \mathcal{A}}{\rmd t} + \frac{\partial \varphi}{\partial z}.
    \label{eq:parallel_electric_field_appendix}
\end{equation}
Following \eqref{eq:linear_drive}, the linear drive term is 
\begin{align}
   \vec{v}_\chi \cdot \gradd_\perp f_{0e} = \frac{\rho_e \vthe}{2} \frac{\partial}{\partial y} \left( \varphi - 2 \frac{\vpar}{\vthe} \mathcal{A} \right) \left[ \frac{1}{L_n} + \frac{1}{L_{T_e}} \left( \frac{v^2}{\vthe^2} - \frac{3}{2} \right) \right] f_{0e}.
    \label{eq:linear_drive_electrons}
\end{align}

\subsubsection{Electron collision operator}
\label{app:electron_collision_operator}
We now wish to specify the form of the collision operator in \eqref{eq:electron_drift_kinetic_equation}. Given that our primary concern is not precise quantitative capture of collisional transport, we shall eschew the most general Landau collision operator in favour of something more analytically convenient, while still retaining the correct conservation properties, as well as capturing the effects of friction between electrons and ions. Namely, we adopt a modified version of the \cite{dougherty64} operator:
\begin{align}
   \left( \frac{\partial \df_e}{\partial t} \right)_c = \nu_e  \left[\frac{1}{2} \frac{\partial}{\partial \vpar} \left( \vthe^2 \frac{\partial}{\partial \vpar} + 2\vpar \right)+ 2 \frac{\partial}{\partial v_\perp^2} v_\perp^2\left(\vthe^2 \frac{\partial}{\partial v_\perp^2} + 1 \right)\right]\df_e \nonumber  \\
     + \nu_e \left[ \left( \frac{ 2\vpar^2}{\vthe^2} - 1\right) + 2\left( \frac{ v_\perp^2}{\vthe^2}-1 \right)\right] \frac{\delta T_{\parallel e} + 2 \delta T_{\perp e}}{3T_e} f_{0e} + 2\nu_{ee} \frac{\vpar u_{\parallel e}}{\vthe^2 }f_{0e},
     \label{eq:collision_operator}
\end{align}
where
\begin{equation}
    \frac{\delta T_{\parallel e}}{T_{0e}} = \frac{1}{n_{0e}} \int \rmd^3 \vec{v} \: \left( \frac{2v_\parallel^2}{\vthe^2} -1 \right)\df_e, \quad \frac{\delta T_{\perp e}}{T_{0e}} = \frac{1}{n_{0e}} \int \rmd^3 \vec{v} \: \left( \frac{v_\perp^2}{\vthe^2}-1 \right)\df_e
    \label{eq:temperature_perturbations}
\end{equation}
are the parallel and perpendicular electron temperature perturbations, respectively. The terms in \eqref{eq:collision_operator} involving $\delta T_{\parallel e}$ and $\delta T_{\perp e}$ are there to ensure that the operator conserves particle number and energy. It does not conserve momentum:
\begin{align}
    \frac{1}{n_{0e}} \int \rmd^3 \vec{v} \: \vpar \left( \frac{\partial \df_e}{\partial t} \right)_c = - \nu_{ei} u_{\parallel e},
    \label{eq:collision_operator_momentum}
\end{align}
reflecting the effect of electrons experiencing friction against the motionless ion background. 
This collision operator is identical to that adopted in \cite{mandell18} (up to velocity normalisations; see their appendix A), except we have neglected all electron FLR contributions, consistent with the ordering \eqref{eq:ordering_lengthscales} and the resultant drift-kinetic approximation.

\subsubsection{Hermite-Laguerre expansion}
\label{app:hermite_laguerre_expansion}
It will be useful to consider a `fluid' description of the plasma, by expanding $\df_e$ in an appropriate polynomial basis. It will prove convenient to use the Hermite-Laguerre moments of $\df_e$, defined by
\begin{align}
    g_{\ell,m}(\vec{r},t) & = \frac{1}{n_{0e}} \int \rmd^3\vec{v} \: (-1)^\ell \frac{ H_m(v_\parallel/\vthe) L_\ell(v_\perp^2/\vthe^2)  }{\sqrt{2^m m!}} \: \df_e(\vec{r},v_\parallel,v_\perp^2,t), \label{eq:transformation} \\
    \df_e(\vec{r},v_\parallel,v_\perp^2,t) & = \sum_{\ell=0}^\infty \sum_{m=0}^\infty (-1)^\ell \frac{H_m(v_\parallel/\vthe) L_\ell(v_\perp^2/\vthe^2) f_{0e}}{\sqrt{2^m m!}} \: g_{\ell,m}(\vec{r},t), \label{eq:inverse_transformation}
\end{align}
where $H_m$ are the Hermite polynomials
\begin{equation}
    H_m(\hat{v}) = (-1)^m e^{\hat{v}^2}  \frac{\rmd^m}{\rmd \hat{v}^m} e^{-\hat{v}^2}, \quad \frac{1}{\sqrt{\pi}}\int \rmd \hat{v} \: H_m(\hat{v})H_{m'}(\hat{v}) e^{-\hat{v}^2} = 2^m m! \: \delta_{mm'},
    \label{eq:hermite_polynomials}
\end{equation}
and $L_\ell$ are the Laguerre polynomials 
\begin{equation}
    L_\ell(\mu) = \frac{e^\mu  }{\ell ! }\frac{\rmd^\ell}{d\mu^\ell}(e^{-\mu}\mu^\ell), \quad \int \rmd \mu \: L_\ell(\mu) L_{\ell'}(\mu) e^{-\mu} = \delta_{\ell \ell'}.
    \label{eq:laguerre_polynomials}
\end{equation}
The use of Hermite polynomials as a (parallel) velocity basis for gyrokinetics has seen much application in the slab-type geometry that we are considering in this paper (\citealt{smith97,watanabe04,zocco11,zocco15,hatch13,loureiro16viriato}), as they are orthogonal with respect to a (parallel) Maxwellian weight function, as in \eqref{eq:hermite_polynomials}. The Laguerre polynomials are a convenient extension of this basis to perpendicular velocities, given that they are also orthogonal with respect to a (perpendicular) Maxwellian weight function, as in \eqref{eq:laguerre_polynomials}. Our choice of collision operator \eqref{eq:collision_operator} was motivated by the fact that the Hermite-Laguerre basis functions are its eigenfunctions. 

Applying the transformation \eqref{eq:transformation} to \eqref{eq:electron_drift_kinetic_equation} and making use of the recurrence relations 
\begin{align}
    &\hat{v}H_m = \frac{1}{2}H_{m+1} + m H_{m-1}, \quad \frac{\rmd H_m}{\rmd \hat{v}} = 2m H_{m-1},
    \label{eq:hermite_recurrence_relations} \\
    &\mu L_\ell = (2\ell +1) L_\ell - (\ell+1) L_{\ell+1} - \ell L_{\ell-1}, \quad \frac{\rmd L_\ell}{\rmd \mu} = \frac{\rmd L_{\ell-1}}{\rmd \mu} - L_{\ell-1},
    \label{eq:laguerre_recurrence_relations}
\end{align}
we arrive at the following equation for the Hermite-Laguerre moments of $\df_e$:
\begin{equation}
    \frac{\rmd g_{\ell,m}}{\rmd t} + \frac{\vthe}{\sqrt{2}} \dpar (\sqrt{m+1}\: g_{\ell,m+1} + \sqrt{m}\:  g_{\ell,m-1})+ \omega_{de}[g_{\ell,m}]- C[g_{\ell,m} ]  = I_{\ell,m},
    \label{eq:laguerre_hermite_moments}
\end{equation}
where, introducing the short-hand $\delta_{\ell',m'}= \delta_{\ell\ell'}\delta_{mm'}$, we define
\begin{align}
    C[g_{\ell,m}] & = - \nu_e(m+2\ell)g_{\ell,m} + \nu_{ee} g_{0,1} \delta_{0,1}  + \frac{\nu_e}{3}(\sqrt{2}g_{0,2} + 2 g_{1,0})(\sqrt{2}\delta_{0,2} +2 \delta_{1,0}), \label{eq:moment_collision_operator} \\
    \omega_{de}[g_{\ell,m}] & = \frac{\rho_e \vthe}{2 L_B} \frac{\partial}{\partial y} \left[\sqrt{(m+1)(m+2)} g_{\ell,m+2} +  (\ell+1) g_{\ell+1,m} + 2(m+\ell+1) g_{\ell,m}  \right. \nonumber \\
                                              & \quad\quad\quad\quad  + \sqrt{m(m-1)} g_{\ell,m-2} + \ell g_{\ell-1,m}  \Bigr], \label{eq:moment_magnetic_drifts}\\
    I_{\ell,m} & = -\frac{\rho_e \vthe}{2 } \frac{\partial \varphi  }{\partial y }\left[ \frac{\delta_{0,0}}{L_n} + \frac{1}{L_{T_e}} \left( \delta_{1,0} + \frac{1}{\sqrt{2}} \delta_{0,2} \right) - \frac{1}{L_B} \left(\sqrt{2} \delta_{0,2} + \delta_{1,0} +2 \delta_{0,0} \right) \right]  \nonumber \\
    & \quad  + \frac{\rho_e \vthe}{\sqrt{2}} \frac{\partial \mathcal{A}  }{\partial y }\left[\frac{\delta_{0,1}}{L_n} + \frac{1}{L_{T_e}}\left( \delta_{0,1} +\delta_{1,1} + \sqrt{\frac{3}{2}}\delta_{0,3}  \right)\right] \nonumber\\
    & \quad + \frac{\vthe}{\sqrt{2}} \left( \frac{2}{\vthe} \frac{\rmd \mathcal{A}}{\rmd t} + \frac{\partial \varphi}{\partial z} \right) \delta_{0,1},\label{eq:moment_energy_injection}
\end{align}
The second term in \eqref{eq:laguerre_hermite_moments} is responsible for linear (parallel) phase-mixing in the Hermite moments $m$ at a rate $\kpar \vthe$ (see \citealt{parker16, sch16, adkins18}), while the magnetic-drift term $\omega_{de}[g_{\ell,m}]$ is responsible for coupling between both Hermite and Laguerre moments, adding another mechanism of parallel phase mixing as well as introducing perpendicular phase-mixing in $\ell$. Note that the coupling to the perpendicular moment hierarchy only occurs in the presence of the magnetic drifts. The collision operator $C[g_{\ell,m}]$ is responsible for regulating fine structure in phase space by introducing a collisional cutoff for high $m$'s and $\ell$'s. Lastly, $I_{\ell,m}$ represents the energy injection from equilibrium gradients and momentum injection from the parallel electric field.

\subsubsection{`Fluid' equations}
\label{app:moment_equations}
In general, \eqref{eq:laguerre_hermite_moments} represents an infinite hierarchy of coupled moments through which the injected energy flows. However, it will be useful for our main discussion to separate a particular set of `fluid' moments: the perturbations of density $\dne/n_{0e} = g_{0,0}$,
\begin{align}
    \frac{\rmd}{\rmd t} \frac{\dne}{n_{0e}}  + \gradd_\parallel u_{\parallel e} + \frac{\rho_e \vthe}{2L_B} \frac{\partial}{\partial y}\left(2 \frac{\dne}{n_{0e}} - 2 \varphi + \frac{\delta T_{\parallel e}}{T_{0e}} + \frac{\delta T_{\perp e}}{T_{0e}} \right) = - \frac{\rho_e \vthe}{2 L_n} \frac{\partial \varphi}{\partial y}, \label{eq:g00}
\end{align}
parallel velocity $u_{\parallel e}/\vthe =g_{0,1}/\sqrt{2}$,
\begin{align}
    \frac{\rmd}{\rmd t} \frac{u_{\parallel e}}{\vthe} & + \frac{\vthe}{2} \gradd_\parallel  \left( \frac{\dne}{n_{0e}} +\frac{\delta T_{\parallel e}}{T_{0e}}   \right) + \frac{\rho_e \vthe}{2 L_B} \frac{\partial}{\partial y} \left( 4 \frac{u_{\parallel e}}{\vthe} + \frac{\delta q_{\parallel e} + \delta q_{\perp e}}{n_{0e} T_{0e} \vthe} \right) \nonumber \\
   &  + \nu_{ei} \frac{u_{\parallel e}}{\vthe} = \frac{\rho_e \vthe}{2} \left( \frac{1}{L_n} + \frac{1}{L_{T_{e}}} \right) \frac{\partial \mathcal{A}}{\partial y} + \frac{\rmd \mathcal{A}}{\rmd t} + \frac{\vthe}{2} \frac{\partial \varphi}{\partial z}, \label{eq:g01}
\end{align}
parallel temperature $\delta T_{\parallel e}/T_{0e} =\sqrt{2}g_{0,2}$ [cf. the first equation in \eqref{eq:temperature_perturbations}],
\begin{align}
     \frac{\rmd}{\rmd t} \frac{\delta T_{\parallel e}}{T_{0e}} & + \vthe \gradd_\parallel \left( \frac{\delta q_{\parallel e}}{n_{0e} T_{0e} \vthe} + 2 \frac{u_{\parallel e}}{\vthe} \right) + \frac{4}{3} \nu_{e} \frac{\delta T_{\parallel e} - \delta T_{\perp e}}{T_{0e}} \nonumber \\
   &  + \frac{\rho_e \vthe}{2L_B} \frac{\partial}{\partial y} \left(2 \frac{\dne}{n_{0e}} - 2 \varphi  + 6 \frac{\delta T_{\parallel e}}{T_{0e}}  + 2\sqrt{6} g_{04} + \sqrt{2} g_{12}\right) = - \frac{\rho_e \vthe}{2 L_{T_e}} \frac{\partial \varphi}{\partial y} \label{eq:g02},
\end{align}
perpendicular temperature $\delta T_{\perp e}/T_{0e} =  g_{1,0}$ [cf. the second equation in \eqref{eq:temperature_perturbations}],
\begin{align}
    \frac{\rmd}{\rmd t} \frac{\delta T_{\perp e}}{T_{0e}} & + \vthe \gradd_\parallel \frac{\delta q_{\perp e} }{n_{0e} T_{0e} \vthe} +\frac{2}{3} \nu_e \frac{\delta T_{\perp e} - \delta T_{\parallel e}}{ T_{0e}} \nonumber \\
   &  + \frac{\rho_e \vthe}{2 L_B} \frac{\partial}{\partial y} \left(\frac{\dne}{n_{0e}} - \varphi  + 4 \frac{\delta T_{\perp e}}{T_{0e}}  + \sqrt{2}g_{12} + 2 g_{20}\right) = -  \frac{\rho_e \vthe}{2 L_{T_e}} \frac{\partial \varphi}{\partial y}, \label{eq:g10}
\end{align}
parallel heat flux $\delta q_{\parallel e}/n_{0e}T_{0e}\vthe = \sqrt{3}g_{0,3}$,
\begin{align}
    \frac{\rmd }{\rmd t } \frac{\delta q_{\parallel e}}{n_{0e} T_{0e} \vthe} & + \vthe \gradd_\parallel \left( \sqrt{2}g_{04} + \frac{3}{2} \frac{\delta T_{\parallel e}}{T_{0e}} \right) + 3 \nu_e \frac{\delta q_{\parallel e}}{n_{0e} T_{0e} \vthe}  \nonumber \\
    & + \frac{\rho_e \vthe}{2 L_B} \frac{\partial}{\partial y} \left(2 \sqrt{15} g_{05} + 8 \frac{\delta q_{\parallel e}}{n_{0e} T_{0e} \vthe} + 6 \frac{u_{\parallel e}}{\vthe} + \sqrt{3}g_{13} \right) = \frac{3\rho_e \vthe}{2 L_{T_e}} \frac{\partial \mathcal{A}}{\partial y}, \label{eq:g03}
 \end{align}
and perpendicular heat flux $\delta q_{\perp e}/n_{0e}T_{0e}\vthe = g_{1,1}/\sqrt{2}$,
\begin{align}
    \frac{\rmd}{\rmd t} \frac{\delta q_{\perp e}}{n_{0e} T_{0e} \vthe} & + \vthe \gradd_\parallel \left( \frac{1}{\sqrt{2}} g_{12} + \frac{1}{2 }\frac{\delta T_{\perp e}}{T_{0e}} \right) + 3 \nu_e\frac{\delta q_{\perp e}}{n_{0e} T_{0e} \vthe}  \nonumber \\
    & +\frac{\rho_e \vthe}{2 L_B} \frac{\partial}{\partial y} \left( \sqrt{3} g_{13} + 6 \frac{\delta q_{\perp e}}{n_{0e} T_{0e} \vthe} + \sqrt{2}g_{21} + \frac{u_{\parallel e}}{\vthe} \right) = \frac{\rho_e \vthe}{2 L_{T_e} } \frac{\partial \mathcal{A}}{\partial y}. \label{eq:g11}
\end{align}
Equations \eqref{eq:g00} and \eqref{eq:g01} are the standard density and parallel-momentum equations for electrons, including the effects of electron inertia, equilibrium gradients of density, temperature and magnetic field, and the non-isothermality of electrons. But for this last feature, they would have been closed equations, as without it, there is no coupling to the perturbations of temperature and heat flux.

A hybrid fluid-kinetic system consisting of \eqref{eq:g00}, \eqref{eq:g01} and \eqref{eq:electron_drift_kinetic_equation}, with the kinetic equation \eqref{eq:electron_drift_kinetic_equation} used to close the fluid ones by calculating the temperature and heat-flux moments, would be ideologically similar to the `Kinetic MHD' description of plasma dynamics (\citealt{kulsrud83}). 

\subsection{Collisional limit}
\label{app:collisional_limit}

\subsubsection{Subsidiary collisional ordering}
\label{app:subsidiary_collisional_ordering}
We now consider the collisional limit of our system of equations \eqref{eq:g00}-\eqref{eq:g11}, in which $\nu_{ee}$ and $\nu_{ei}$ are the dominant frequencies, viz., $ \nu_{ei} \sim \nu_{ee} \gg \omega$. Given that we wish to retain kinetic Alfv\'en waves, drift waves, perpendicular advection by both magnetic drifts and $\vec{E} \times \vec{B}$ flows, as well as finite heat conduction and resistivity, we postulate, analogously to \eqref{eq:ordering_timescales}:
\begin{align}
    \nu_{ee} \sim \nu_{ei} \gg \omega \sim \omega_\text{KAW} \sim \omega_{*\s} \sim \omega_{d\s} \sim k_\perp v_E \sim (k_\perp d_e)^2 \nu_{ei} \sim \kappa k_{\parallel}^2,
    \label{eq:ordering_collisional_frequencies}
\end{align}
where $\kappa \sim \vthe^2/\nu_{e}$ is the electron thermal diffusivity. The parallel stremaing rate $k_\parallel \vthe$ is no longer the relevant parallel frequency; the new ordering can be worked out by following the same logic as in \apref{app:low_beta_ordering} but replacing $k_\parallel \vthe$ with the parallel conduction rate.

Namely, instead of \eqref{eq:ordering_streaming_vs_kaw}, we have
\begin{align}
    \kappa k_\parallel^2 \sim \omega_\text{KAW} \quad \Rightarrow \quad k_\perp d_e \sim k_\parallel \lambdae,
    \label{eq:ordering_thermal_conduction_vs_kaw}
\end{align}
where $\lambdae$ is once again the electron mean free path. The same relation guarantees $\kappa k_\parallel^2 \sim (k_\perp d_e)^2 \nu_{ei}$. Ordering $\kappa k_\parallel^2$ with respect to the drift frequencies gives us, with the aid of \eqref{eq:ordering_thermal_conduction_vs_kaw},
\begin{align}
    \kappa k_\parallel^2 \sim \omega_{*\s} \sim \omega_{d\s} \sim k_\perp \rho_e \frac{\vthe}{L} \sim \sqrt{\beta_e} k_\parallel \lambdae \frac{\vthe}{L} \quad \Rightarrow \quad k_\parallel L \sim \sqrt{\beta_e},
    \label{eq:ordering_thermal_conduction_vs_drifts}
\end{align}
so \eqref{eq:ordering_streaming_vs_drifts} survives unscathed. Combining \eqref{eq:ordering_thermal_conduction_vs_drifts} with \eqref{eq:ordering_thermal_conduction_vs_kaw} gives us 
\begin{align}
    k_\perp d_e \sim k_\parallel \lambdae \sim \sqrt{\beta_e} \frac{\lambdae}{L} \equiv \chi^{-1},
    \label{eq:ordering_lengthscales_collisional}
\end{align}
i.e., the perpendicular wavelengths must be ordered comparable to the flux-freezing scale anticipated in \eqref{eq:flux_freezing_scale_col} --- the collisional analogue of what was $k_\perp d_e \sim 1$ in the collisionless case [see \eqref{eq:ordering_streaming_vs_kaw}].

To obtain the ordering of the fluctuation amplitudes, we let, analogously to \eqref{eq:ordering_streaming_vs_exb}, and using \eqref{eq:ordering_thermal_conduction_vs_kaw} and \eqref{eq:ordering_thermal_conduction_vs_drifts}, 
\begin{align}
    \kappa k_\parallel^2 \sim k_\perp v_E \quad \Rightarrow \quad \frac{v_E}{\vthe} \sim \frac{k_\parallel}{k_\perp} k_\parallel\lambdae \sim \epsilon \sqrt{\beta_e}.
    \label{eq:ordering_thermal_conduction_vs_exb}
\end{align}
Knowing this and noting that \eqref{eq:ordering_lengthscales_collisional} implies
\begin{align}
    k_\perp \rho_e \sim \chi^{-1} \sqrt{\beta_e},
    \label{eq:ordering_collisional_rhoe}
\end{align}
we find, by the same logic as \eqref{eq:ordering_frequencies_vs_larmor}-\eqref{eq:ordering_dBperp}, the ordering of the frequencies 
\begin{align}
    \frac{\omega}{\Omega_e} \sim k_\perp \rho_e \: \epsilon \sqrt{\beta_e} \sim \chi^{-1} \epsilon \beta_e, \spc \frac{\omega}{\Omega_i} \sim \chi^{-1} \epsilon,
    \label{eq:ordering_frequencies_collisional}
\end{align}
and of the fluctuation amplitudes
\begin{align}
    \frac{\delta T_e}{T_{0e}} \sim \frac{\delta T_i}{T_{0i}} \sim \frac{\delta n_i}{n_{0i}} = \frac{\dne}{n_{0e}} \sim \frac{e \varphi}{T_{0e}} \sim \chi \epsilon, \quad \frac{\dBperp}{B_0} \sim \chi \epsilon \sqrt{\beta_e}.
    \label{eq:ordering_amplitudes_collisional}
\end{align}
To summarise, \eqref{eq:ordering_lengthscales_collisional}, \eqref{eq:ordering_frequencies_collisional} and \eqref{eq:ordering_amplitudes_collisional} represent once again an ordering of lengthscales, frequencies, and amplitudes with respect to $\epsilon$ and $\beta_e$, but now also to the subsidiary expansion parameter $\chi^{-1}$ --- that this parameter should be small follows straightforwardly from, e.g., $\nu_{ei} \gg (k_\perp d_e) \nu_{ei}$. Thus, the formal heirarchy of our expansions is now 
\begin{align}
    \epsilon \ll \sqrt{\beta_e} \ll \chi^{-1} \ll 1,
    \label{eq:ordering_of_chi}
\end{align}
with all other dimensionless parameters being treated as finite.

\subsubsection{Collisional limit of low-beta equations}
\label{app:low_beta_collisional_equations}
We begin by considering the equations for the temperature perturbations \eqref{eq:g02} and \eqref{eq:g10}, in which the terms responsible for collisional temperature isotropisation are now dominant: to leading order in $\chi^{-1}$,
\begin{align}
  \nu_e \frac{\delta T_{\parallel e} - \delta T_{\perp e}}{T_{0e}} = 0 \quad \Rightarrow \quad \frac{\delta T_{\parallel e}}{T_{0e}} = \frac{\delta T_{\perp e}}{T_{0e}} \equiv \frac{\dTe}{T_{0e}},
    \label{eq:col_temperature_linear_combination}
\end{align}
so we no longer need to distinguish between the parallel and perpendicular temperature perturbations. We then obtain the equation for $\dTe$ by adding $(1/2)$\eqref{eq:g02}$+$\eqref{eq:g10}:
\begin{align}
    \frac{3}{2} \frac{\rmd}{\rmd t} \frac{\dTe}{T_{0e}}  +  \gradd_\parallel \left( \frac{\frac{1}{2}\delta q_{\parallel e} + \delta q_{\perp e}}{n_{0e} T_{0e} }+ u_{\parallel e} \right) + \frac{\rho_e \vthe}{L_B} \frac{\partial}{\partial y} \left(  \frac{\dne}{n_{0e}} - \varphi + \frac{7}{2} \frac{\dTe}{T_{0e}} \right)  = - \frac{3}{2} \frac{\rho_e \vthe}{2 L_{T_e}} \frac{\partial \varphi}{\partial y}. \label{eq:col_temperature_equation_initial}
\end{align}
We have neglected the higher-order moments ($g_{0,4}$, $g_{1,2}$, and $g_{2,0}$) in \eqref{eq:g02} and \eqref{eq:g10} because, from the balance of the collision and parallel streaming (or magnetic drift) terms in \eqref{eq:laguerre_hermite_moments}, they are small in $\chi^{-1}$:
\begin{align}
    g_{\ell,m+1} \sim \chi^{-1} g_{\ell,m}, \quad g_{\ell+1,m} \sim \chi^{-2} g_{\ell,m}. 
    \label{eq:col_higher_order_moments}
\end{align}
The parallel and perpendicular heat fluxes can be calculated from \eqref{eq:g03} and \eqref{eq:g11}, where the collisional terms are again dominant and the higher-order moments ($g_{05}$, $g_{13}$, and $g_{21}$) are negligible by \eqref{eq:col_higher_order_moments}, viz.,
\begin{align}
    \frac{\delta q_{\parallel e}}{n_{0e} T_{0e}} = 3 \frac{\delta q_{\perp e}}{n_{0e} T_{0e}} = - \frac{\vthe^2}{2\nu_e}\left(\gradd_\parallel \frac{\dTe}{T_{0e}} - \frac{\rho_e}{ L_{T_e}} \frac{\partial \mathcal{A}}{\partial y} \right). \label{eq:col_heat_fluxes} 
\end{align}
The combined heat flux that appears in \eqref{eq:col_temperature_equation_initial} is, therefore,
\begin{align}
    \frac{\delta q_{e}}{n_{0e} T_{0e} } =  \frac{\frac{1}{2}\delta q_{\parallel e} + \delta q_{\perp e}}{n_{0e} T_{0e} } = - \frac{3}{2}\kappa \gradd_\parallel \log T_e,
    \label{eq:col_combined_heat_flux}
\end{align}
where we have introduced the parallel derivative of the total temperature $\gradd_\parallel \log T_e$, as in \eqref{eq:logt_definition_intro} and $\kappa = 5\vthe^2/18\nu_e$.

The density equation \eqref{eq:g00} keeps all of its terms under the collisional ordering, whereas in the parallel-velocity equation \eqref{eq:g01}, the electron-inertia and magnetic-drift terms are all small by a factor of $\chi^{-2}$, and so can be neglected. Assembling all this together, we obtain the following system of equations
\begin{align}
    &\frac{\rmd}{\rmd t} \frac{\dne}{n_{0e}}  + \gradd_\parallel u_{\parallel e} + \frac{\rho_e \vthe}{L_B} \frac{\partial}{\partial y}\left( \frac{\dne}{n_{0e}} -  \varphi + \frac{\delta T_{ e}}{T_{0e}} \right) = - \frac{\rho_e \vthe}{2 L_n} \frac{\partial \varphi}{\partial y}, \label{eq:col_density_equation} \\
    &\frac{\rmd \mathcal{A}}{\rmd t} + \frac{\vthe}{2} \frac{\partial \varphi}{\partial z} =  \frac{\vthe}{2} \gradd_\parallel \left( \frac{\dne}{n_{0e}}+ \frac{\delta T_{ e}}{T_{0e}} \right) - \frac{\rho_e \vthe}{2} \left( \frac{1}{L_n} + \frac{1}{L_{T_e}} \right) \frac{\partial \mathcal{A}}{\partial y} + \nu_{ei} d_e^2 \gradd_\perp^2 \mathcal{A}, \label{eq:col_velocity_equation} \\
    & \frac{\rmd}{\rmd t} \frac{\dTe}{T_{0e}} -\kappa \gradd_\parallel^2 \log T_e + \frac{2}{3} \gradd_\parallel u_{\parallel e} + \frac{2}{3}\frac{\rho_e \vthe}{L_B} \frac{\partial}{\partial y} \left(  \frac{\dne}{n_{0e}} -  \varphi + \frac{7}{2} \frac{\dTe}{T_{0e}} \right) =-\frac{\rho_e \vthe}{2 L_{T_e}} \frac{\partial \varphi}{\partial y} , \label{eq:col_temperature_equation}
\end{align}
with $\varphi$ and $\mathcal{A}$ still satisfying \eqref{eq:quasineutrality_final} and \eqref{eq:parallel_amperes_law_final}, respectively. The ion gyrokinetic equation \eqref{eq:ion_gyrokinetic_equation} is unchanged by this ordering because ion-ion and ion-electron collisions were already neglected in the low-beta ordering [see \eqref{eq:ion_collisions}].

A consequence of the collisional ordering  --- evident from \eqref{eq:col_velocity_equation} --- is that electron inertia has been neglected, as we are considering perpendicular scales smaller than the electron inertial scale $d_e$ [cf. \eqref{eq:ordering_lengthscales_collisional}]. However, as we demonstrate in sections \ref{sec:electrostatic_regime_etg} and~\ref{sec:electromagnetic_regime_tai}, these equations support `collisional' analogues of the instabilities found in the full kinetic system, making them a useful (more analytically tractable) model for illustrating the underlying physical mechanisms of these instabilities without the (kinetic) technical detail. Some readers may be concerned about the fact that we have used a model collision operator \eqref{eq:collision_operator} in the derivation of \eqref{eq:col_density_equation}-\eqref{eq:col_temperature_equation}, as the velocity dependence of the collision frequency when using the Landau collision operator could lead to additional terms that have not been captured in our analysis. In \apref{app:choice_of_collision_operator}, we show that these terms do not change our results, as none of the (collisional) physics that we discuss in this paper relies on the exact details of the collision operator.

\subsection{Strongly driven limit}
\label{app:strongly_driven_limit}
Finally, we would like to make a further step to simplify the equations derived in \apref{app:moment_equations} and their collisional counterparts \eqref{eq:col_density_equation}-\eqref{eq:col_temperature_equation}. This consists of adopting the strongly driven limit introduced in \eqref{eq:strongly_driven_lengthscales} and \eqref{eq:strongly_driven_limit}. As already explained at the end of \apref{app:ion_equations}, all remaining ion physics in this limit is contained in the closure~\eqref{eq:quasineutrality_final} without the $g_i$ contribution. In the equations for the fluid moments \eqref{eq:g00}-\eqref{eq:g10} and \eqref{eq:col_density_equation}-\eqref{eq:col_temperature_equation}, this limit allows one to drop some magnetic-drift terms that never contribute in a qualitatively important way. 

Namely, consider \eqref{eq:g00}. Since $L_n  \sim L_B$, $\omega_{de} \ll \omega$, and $\varphi \sim \dne/n_{0e}$, we can always reduce it to 
\begin{align}
    \frac{\rmd}{\rmd t} \frac{\dne}{n_{0e}} + \gradd_\parallel u_{\parallel e} + \frac{\rho_e \vthe}{2 L_B} \frac{\partial}{\partial y} \left( \frac{\delta T_{\parallel e}}{T_{0e} } + \frac{\delta T_{\perp e}}{T_{0e}} \right) = 0.
    \label{eq:strongly_driven_density}
\end{align}
This is \eqref{eq:density_moment}, the first equation of our minimalist collisionless system, or \eqref{eq:density_moment_collisional} in the collisional case. The surviving magnetic-drift term provides the feedback for the curvature-mediated instabilities that are the focus of \secref{sec:curvature_ETG} and much of \secref{sec:electromagnetic_regime_tai}. Clearly, it can only be non-negligible if the temperature perturbations 
\begin{align}
    \frac{\delta T_{\parallel e}}{T_{0e}} \sim \frac{\delta T_{\perp e}}{T_{0e}} \sim \frac{\delta T_e}{T_{0e}} \sim \frac{\omega}{\omega_{de}} \frac{\delta n_e}{n_{0e}}
    \label{eq:strongly_driven_temperature}
\end{align}
are large compared to the density ones, which they will be, in some subsidiary limits (when $\omega \ll \omega_{*e}$). If they are not, the magnetic-drift term in \eqref{eq:strongly_driven_density} is as small as the terms that we have already neglected and so must also be dropped, but the important point is that the neglected terms are never large enough to need retaining. 

By a similar argument, if $L_n  \sim L_B$ and $\omega \gg \omega_{de}$, the terms containing $L_B$ and $L_n$ can all be dropped from \eqref{eq:col_velocity_equation} and \eqref{eq:col_temperature_equation}, giving us \eqref{eq:velocity_moment_collisional} and \eqref{eq:t_moment_collisional}, the two remaining equations in our minimalist collisional system. In the collisionless case, \eqref{eq:velocity_moment}-\eqref{eq:tperp_moment} are obtained from \eqref{eq:g01}-\eqref{eq:g10} in a similar way, but one must stipulate also $\omega_{de} \ll k_\parallel \vthe$ and assume that none of the Hermite-Laguerre moments involved can be much larger than $\delta T_{\parallel e}/T_{0e}$ or $\delta T_{\perp e} /T_{0e}$.

We reiterate that the strongly driven limit is not formally an ordering --- in the sense that some of the terms that are retained can, in certain meaningful limits, turn out to be as small as those terms that have been neglected --- but the latter are negligible always, and so the remaining equations are always no worse off for not having them. Cautious readers may be reassured by the fact that all of the instabilities considered in sections \ref{sec:electrostatic_regime_etg} and \ref{sec:electromagnetic_regime_tai} are derived in a limit in which this is a valid approximation.

Given that throughout the majority of this paper we will be concerned with the dynamics arising from the electron temperature gradient $L_{T_e}$, we shall henceforth adopt the notation $L_T = L_{T_e}$, apart from where there is possible ambiguity about which temperature gradient is being referred to, such as in \apref{app:conservation_laws}.

\section{Conservation laws}
\label{app:conservation_laws}
In this appendix, we derive the free energy associated with our low-beta equations~\eqref{eq:ion_gyrokinetic_equation} and \eqref{eq:electron_drift_kinetic_equation} [or, equivalently, the hierarchy of moments \eqref{eq:laguerre_hermite_moments}].

\subsection{Free energy}
\label{app:free_energy}
Plasma systems containing small perturbations around a Maxwellian equilibrium nonlinearly conserve free energy, defined as 
\begin{equation}
	W = U - \sum_\s T_{0\s} \delta S_\s, \quad U = \int \frac{\rmd^3 \vec{r}}{V} \: \frac{\left| \delta \! \vec{B} \right|^2}{8\pi}, \quad - \delta S_\s = \int \frac{\rmd^3 \vec{r}}{V} \int \rmd^3 \vec{v} \: \frac{\df_\s^2}{2 f_{0\s}},
	\label{eq:free_energy_definition}
\end{equation}
where $\delta S_\s$ is the entropy of the perturbed distribution function of species $\s$ (see \citealt{sch08,sch09} and references therein), and $V = L_xL_yL_\parallel$ is the volume of the system. Given the ordering of the parallel magnetic field perturbations \eqref{eq:ordering_dBpar}, the internal energy consists only of the perpendicular magnetic field perturbations
\begin{align}
    U = \int\frac{\rmd^3 \vec{r}}{V} \: \frac{|\dB_\perp|^2}{8\pi } = n_{0e} T_{0e} \int\frac{\rmd^3 \vec{r}}{V} \: \left| d_e \grad_\perp \mathcal{A} \right|^2.
    \label{eq:internal_energy}
\end{align}
We now consider the contributions of each of the kinetic species to the free energy. 
Noting that the Hermite-Laguerre basis \eqref{eq:transformation}-\eqref{eq:inverse_transformation} has a Parseval theorem, we may write
\begin{align}
    - T_{0e} \delta S_e = n_{0e} T_{0e} \int \frac{\rmd^3 \vec{r}}{V} \: \frac{1}{2}\sum_{\ell=0}^\infty \sum_{m=0}^\infty  {g_{\ell,m}^2}.
    \label{eq:energetics_electron_entropy}
\end{align}
Recalling, from \eqref{eq:perturbed_distribution_function} and \eqref{eq:non_adiabatic_response}, that
\begin{equation}
    	\df_i =  \frac{Z}{\tau} \left( \left< \varphi \right>_{\vec{R}_i } - \varphi \right) f_{0i} + g_i, 
    	\label{eq:ions_perturbed_distribution_function}
\end{equation}
we can express the ion contribution to the entropy as
\begin{align}
	-T_{0i} \delta S_i = T_{0i} \int \frac{\rmd^3 \vec{r}}{V} \int \rmd^3 \vec{v} \: \frac{\left< \df_i^2 \right>_{\vec{r}}}{2 f_{0i}} = n_{0e} T_{0e} \int \frac{\rmd^3 \vec{r}}{V} \: \frac{\varphi \taubar^{-1} \varphi}{2} + T_{0i} \int \frac{\rmd^3 \vec{r}}{V} \int \rmd^3 \vec{v} \: \frac{\left<g_i^2\right>_{\vec{r}}}{2f_{0i}}.
	\label{eq:energetics_ion_entropy}
\end{align}
Here the operator $\taubar$, which contains only even powers of $\dperp$ [see \eqref{eq:taubar_definition}], is understood to act on both sides of itself, with the powers of $\dperp$ distributed evenly. 

Putting \eqref{eq:energetics_electron_entropy} and \eqref{eq:energetics_ion_entropy} together, we can write the overall free energy of the system as
\begin{equation}
	\frac{W}{n_{0e} T_{0e}} = \frac{W_0}{n_{0e} T_{0e}} + \frac{\tau}{Z n_{0i}} \int \frac{\rmd^3 \vec{r}}{V} \int \rmd^3 \vec{v} \: \frac{\left< g_i^2 \right>_{\vec{r}}}{2f_{0i}},
	 \label{eq:energetics_free_energy}
\end{equation}
 where $W_0$ is the free energy of the system for $g_i = 0$:
\begin{align}
	 \frac{W_0}{n_{0e} T_{0e}} =  \int \frac{\rmd^3 \vec{r}}{V} &\left( \frac{\varphi { \taubar}^{-1} \varphi}{2} + \left| d_e \grad_\perp \mathcal{A} \right|^2 + \frac{1}{2}\sum_{\ell=0}^\infty \sum_{m=0}^\infty  g_{\ell,m}^2 \right).  	 
	 \label{eq:energetics_free_energy_0}
\end{align}
As expected, this free energy is a sum of the quadratic norms of the electromagnetic fields and the Hermite-Laguerre moments $g_{\ell,m}$ of the electron distrubution function $\df_e$. 

In the collisional limit, we do not need to retain all of the latter contributions to~\eqref{eq:energetics_free_energy_0}, because, according to \eqref{eq:col_higher_order_moments}, higher-order moments are small in the collisional expansion. To leading order in $\chi^{-1}$, we find
\begin{align}
	 \frac{W_0}{n_{0e} T_{0e}} & =  \int \frac{\rmd^3 \vec{r}}{V} \left( \frac{\varphi { \taubar}^{-1} \varphi}{2} + \left| d_e \grad_\perp \mathcal{A} \right|^2 + \frac{1}{2}\frac{\dne^2}{n_{0e}^2} + \frac{3}{4} \frac{\dTe^2}{T_{0e}^2}\right).
	 \label{eq:energetics_free_energy_0_collisional}
\end{align}
This is \eqref{eq:free_energy_summarised_collisional}.

\subsection{Free-energy budget}
\label{app:free_energy_budget}
Let us now work out the time derivative of the free energy. To calculate the time derivative of the last term in \eqref{eq:energetics_free_energy_0}, we multiply \eqref{eq:laguerre_hermite_moments} by $g_{\ell,m}$ and sum over $\ell$~and~$m$. Neither the parallel-streaming nor the magnetic-drift terms in \eqref{eq:laguerre_hermite_moments} make any contribution, viz.,
\begin{align}
       \sum_{\ell = 0}^\infty \sum_{m=0}^\infty \int \frac{\rmd^3 \vec{r}}{V} \: g_{\ell,m}\left[ \frac{\vthe}{\sqrt{2}} \dpar \left(\sqrt{m+1}\: g_{\ell,m+1} + \sqrt{m}\:  g_{\ell,m-1} \right) + \omega_{de}[g_{\ell,m}]\right] = 0,
       \label{eq:energetics_parallel_streaming_and_drifts}
\end{align}
because pairwise terms of the form
\begin{align}
    g_{\ell,m} \gradd_\parallel g_{\ell,m'} + g_{\ell,m'}\gradd_\parallel g_{\ell,m}, \quad  g_{\ell,m} \frac{\partial}{\partial y} g_{\ell,m'} + g_{\ell,m'} \frac{\partial}{\partial y} g_{\ell,m},
    \label{eq:integration_by_parts_identity}
\end{align}
vanish identically when integrated over all space. The contribution from the collision term in \eqref{eq:laguerre_hermite_moments} is
\begin{align}
   D_e & =  -\sum_{\ell = 0}^\infty \sum_{m=0}^\infty \int \frac{\rmd^3 \vec{r}}{V} \: g_{\ell,m} C[g_{\ell,m} ]  \nonumber \\
   & = 2\nu_{ei} \int \frac{\rmd^3 \vec{r}}{V} \: \left|d_e^2 \gradd_\perp^2 \mathcal{A} \right|^2 + \nu_e \int \frac{\rmd^3 \vec{r}}{V}\left[\frac{2}{3} \left( \frac{\delta T_{\parallel e} - \delta T_{\perp e}}{T_{0e}} \right)^2   \right. 	\nonumber \\
	  &\quad + \left. \sum_{m=3}^\infty m{g_{0,m}^2}  +  \sum_{m=1}^\infty (m+2) {g_{1,m}^2} + \sum_{\ell=2}^\infty \sum_{m=0}^\infty (m+ 2\ell ) {g_{\ell,m}^2} \right] \geqslant 0 ,\label{eq:energetics_dissipation_electrons}
\end{align}
where we have used the fact that 
\begin{align}
    2 g_{0,2}^2 + 2g_{1,0}^2 - \frac{1}{3} \left( \sqrt{2} g_{0,2} + 2g_{1,0} \right)^2 = \frac{2}{3}\left( \frac{\delta T_{\parallel e} - \delta T_{\perp e}}{T_{0e}} \right)^2
    \label{eq:energetics_temperature_equalisation}
\end{align}
to simplify the temperature terms in \eqref{eq:moment_collision_operator}. The contribution from the injection term on the right-hand side of \eqref{eq:laguerre_hermite_moments} can be written as follows, after integrating by parts and using \eqref{eq:g00},
\begin{align}
     \sum_{\ell = 0}^\infty \sum_{m=0}^\infty \int \frac{\rmd^3 \vec{r}}{V} \: g_{\ell,m} I_{\ell,m} = \varepsilon_e + \int \frac{\rmd^3 \vec{r}}{V} \left(\varphi \frac{\rmd}{\rmd t} \frac{\dne}{n_{0e}} - \frac{\rmd}{\rmd t}\left| d_e \grad_\perp \mathcal{A}\right|^2 \right),
     \label{eq:energetics_injection}
\end{align}
where the energy injection due to the electron density and temperature gradients is:
\begin{equation}
	\varepsilon_e = \frac{1}{L_n} \int \frac{\rmd^3 \vec{r}}{V} \: \frac{\dne}{n_{0e}} v_{Ex} +\frac{1}{L_{T_e}} \int \frac{\rmd^3 \vec{r}}{V} \: \left[ \left( \frac{1}{2} \frac{\delta T_{\parallel e}}{T_{0e}} +  \frac{\delta  T_{\perp e}}{T_{0e}} \right) v_{Ex} + \frac{\frac{1}{2} \delta q_{\parallel e}  + \delta q_{\perp e}}{n_{0e} T_{0e}} \frac{\delta \! B_x}{B_0} \right].
	\label{eq:energetics_energy_injection_electrons}
\end{equation}
This is the first expression in \eqref{eq:energy_injection_simplified},
with $v_{Ex}$ and $\delta \! B_x/B_0$ defined in \eqref{eq:exb_flow_deltabx}. We recognise the terms proportional to $v_{Ex}$ as the transport of density and temperature perturbations by the $\vec{E}\times \vec{B}$ flow, while the terms proportional to $\delta \! B_x/B_0$ are the fluxes of temperature along the perturbed field lines. Note that \eqref{eq:energetics_injection} contains a contribution equal to the time derivative of the second term in \eqref{eq:energetics_free_energy_0} (the magnetic energy), which we transfer to the left-hand side of our emerging free-energy budget.  

We now turn to the first term in \eqref{eq:energetics_free_energy_0}.
Using \eqref{eq:ion_gyrokinetic_equation} and \eqref{eq:quasineutrality_final}, its time derivative is
\begin{align}
    \frac{\rmd}{\rmd t} \int \frac{\rmd^3 \vec{r}}{V} \: \frac{\varphi \taubar^{-1} \varphi}{2} 
    =  - \int \frac{\rmd^3 \vec{r}}{V} \: \varphi \frac{\rmd}{\rmd t} \frac{\dne}{n_{0e}} + D_\text{x},
    \label{eq:energetics_phi_squared_derivative}
\end{align}
where a term has arisen that represents energy exchange between electrons and ions due to equilibrium magnetic field gradients:
\begin{equation}
	 D_\text{x} =  -\frac{1}{n_{0i}} \int \frac{\rmd^3 \vec{r}}{V}   \int \rmd^3 \vec{v} \: \varphi\:\vec{v}_{di} \cdot \gradd_\perp \left<g_i + \frac{Z}{\tau} \left< \varphi \right>_{\vec{R}_i} f_{0i} \right>_{\vec{r}}.
	\label{eq:energetics_energy_exchange}
\end{equation}
If we had retained ion collisions in \eqref{eq:ion_gyrokinetic_equation}, a collisional energy-exchange term would also have had to be included in \eqref{eq:energetics_phi_squared_derivative}. 

Assembling \eqref{eq:energetics_dissipation_electrons}, \eqref{eq:energetics_injection} and \eqref{eq:energetics_phi_squared_derivative}, we find 
\begin{align}
    \frac{1}{n_{0e} T_{0e}}\frac{\rmd W_0}{\rmd t} = \varepsilon_e - D_e + D_\text{x}.
    \label{eq:energetics_w0_derivative}
\end{align}

In the collisional limit, the expressions for $D_e$ and $\varepsilon_e$ are significantly simplified. Since, in this limit, $\delta T_{\parallel e} = \delta T_{\perp e} = \dTe$, the first term in the square brackets in \eqref{eq:energetics_dissipation_electrons} vanishes. Then, recalling the definition of the collisional heat flux \eqref{eq:collisional_heat_flux} and neglecting terms of order $\chi^{-2} (\delta q_e/n_{0e} T_{0e} \vthe)^2$ and higher, which are small by \eqref{eq:col_higher_order_moments}, we find that \eqref{eq:energetics_dissipation_electrons} becomes
\begin{align}
   D_e 
   & = 2\nu_{ei} \int \frac{\rmd^3 \vec{r}}{V} \: \left|d_e^2 \gradd_\perp^2 \mathcal{A} \right|^2 + \frac{12}{5} \nu_e \int \frac{\rmd^3 \vec{r}}{V}\: \left(\frac{\delta q_e}{n_{0e} T_{0e} \vthe} \right)^2. 	\label{eq:energetics_dissipation_electrons_collisional}
\end{align}
The expression for the collisional energy injection follows similarly from \eqref{eq:energetics_energy_injection_electrons}:
\begin{equation}
	\varepsilon_e = \frac{1}{L_n} \int \frac{\rmd^3 \vec{r}}{V} \: \frac{\dne}{n_{0e}} v_{Ex} +\frac{1}{L_{T_e}} \int \frac{\rmd^3 \vec{r}}{V} \: \left( \frac{3}{2} \frac{\dTe}{T_{0e}} v_{Ex} + \frac{\delta q_e}{n_{0e} T_{0e}} \frac{\delta \! B_x}{B_0} \right).
	\label{eq:energetics_energy_injection_electrons_collisional}
\end{equation}
This is the second expression in \eqref{eq:energy_injection_simplified}.
Naturally, both \eqref{eq:energetics_dissipation_electrons_collisional} and \eqref{eq:energetics_energy_injection_electrons_collisional} can also be obtained by direct calculation from~\eqref{eq:energetics_free_energy_0_collisional} using the collisional equations \eqref{eq:col_density_equation}-\eqref{eq:col_temperature_equation}. 

Finally, to calculate the $g_i$ contribution to \eqref{eq:energetics_free_energy}, we multiply the ion gyrokinetic equation \eqref{eq:ion_gyrokinetic_equation} by $g_i/f_{0i}$ and integrate over the entire phase space. After some manipulations, we obtain
\begin{equation}
     \frac{\tau}{Z n_{0i}}\int \frac{\rmd^3 \vec{r}}{V} \int \rmd^3 \vec{v} \: \frac{\rmd}{\rmd t} \frac{\left< g_i^2 \right>_{\vec{r}}}{2f_{0i}} = \varepsilon_i - D_\text{x} , 
	\label{eq:energetics_variance_of_g_i_evolution}
\end{equation}
where $D_\text{x}$ is as defined in \eqref{eq:energetics_energy_exchange}, and the energy injection due to ion equilibrium gradients~is 
\begin{equation}
	\varepsilon_i = -\frac{\rho_i \vthi}{2n_{0i}} \int \frac{\rmd^3 \vec{r}}{V} \int \rmd^3 \vec{v} \left[ \frac{1}{L_{n}} + \frac{1}{L_{T_i}} \left( \frac{v^2}{\vthi^2} - \frac{3}{2} \right)\right] \left< \frac{\partial \left< \varphi \right>_{\vec{R}_i}}{\partial Y_i} g_i \right>_{\vec{r}}.
	\label{eq:energetics_energy_injection_ions}
\end{equation}

Combining this result with \eqref{eq:energetics_w0_derivative}, we arrive at
\begin{align}
    \frac{1}{n_{0e} T_{0e} } \frac{\rmd W}{\rmd t} = \varepsilon_i + \varepsilon_e - D_e.
    \label{eq:energetics_total_derivative}
\end{align}
In the absence of any ion equilibrium gradients, all of the energy injection is due to the equilibrium electron gradients. Then \eqref{eq:energetics_total_derivative} becomes \eqref{eq:free_energy_time_derivative_summarised}. 

\section{Magnetic-flux conservation}
\label{app:magnetic_flux_conservation}
The conservation of magnetic flux is guaranteed if there exists some effective velocity field $\vec{u}_\text{eff}$ such that material loops moving with this velocity always enclose the same amount of magnetic flux. Should such a $\vec{u}_\text{eff}$ exist, then it also preserves magnetic field lines and their topology (\citealt{newcomb58}). Following \cite{cowley85}, we consider
\begin{align}
    \vec{u}_\text{eff} = \frac{c}{B} (\vec{E} - \gradd \Phi ) \times \vec{b},
    \label{eq:flux_u_eff}
\end{align}
where $\vec{B}$ is the total magnetic field, $B = |\vec{B}|$ and $\vec{b} = \vec{B}/B$ are its magnitude and direction, respectively, and $\Phi$ is some single-valued scalar function. Physically, \eqref{eq:flux_u_eff} can be interpreted as the $\vec{E}\times \vec{B}$ flow resulting from an effective electric field~$\vec{E} - \gradd \Phi$. It can then be shown by direct substitution that the electric field satisfies
\begin{align}
    \vec{E} + \frac{\vec{u}_\text{eff} \times \vec{B}}{c} = \gradd \Phi + (E_\parallel - \gradd_\parallel \Phi ) \vec{b},
    \label{eq:flux_electric_field}
\end{align}
where $E_\parallel = \vec{b} \cdot \vec{E}$, $\gradd_\parallel  = \vec{b}\cdot \gradd$. Faraday's law can then be written as 
\begin{align}
    \frac{\partial \vec{B}}{\partial t} = - c \curll \vec{E} = \curll(\vec{u}_\text{eff} \times \vec{B}) - c \curll \left[ (E_\parallel - \gradd_\parallel \Phi ) \vec{b} \right].
    \label{eq:flux_faradays_law}
\end{align}
If we recognise $\vec{u}_\text{eff}$ as the flux- and field-line-preserving velocity, then, following the standard proof of flux conservation (reproduced in numerous MHD textbooks), we find (see, e.g., \citealt{eyink06})
\begin{align}
    \frac{\rmd}{\rmd t} \int_{S(t)}  \vec{B} \cdot \rmd\vec{S}  = - c \oint_{\partial S(t)}   (E_\parallel - \gradd_\parallel \Phi)  \vec{b} \cdot \rmd\vec{\ell}, 
    \label{eq:flux_conservation_broken}
\end{align}
where $S(t)$ is the surface advected by the velocity $\vec{u}_\text{eff}$, and $\partial S(t)$ its boundary. This implies that the conservation of magnetic flux is broken only by the non-zero parallel projection of the effective electric field that gives rise to $\vec{u}_\text{eff}$, i.e., $E_\parallel - \gradd_\parallel \Phi$, meaning that we must look to parallel force balance to determine whether or not the magnetic flux is conserved. 

In our system of equations, this is given by~\eqref{eq:g01}, which, recalling the definition of~$E_\parallel$~\eqref{eq:parallel_electric_field_appendix} and parallel Amp\`ere's law~\eqref{eq:parallel_amperes_law_final}, can be written as
\begin{align}
    E_\parallel - \gradd_\parallel \Phi = - \frac{\rho_e B_0 }{c} \left[ \frac{\rmd}{\rmd t}\frac{u_{\parallel e}}{\vthe} + \eta \gradd_\perp^2 \mathcal{A} + \frac{\rho_e \vthe}{2L_B} \frac{\partial}{\partial y} \left( 4 \frac{u_{\parallel e}}{\vthe} + \frac{\delta q_{\parallel e} + \delta q_{\perp e}}{n_{0e} T_{0e} \vthe} \right)\right],
    \label{eq:flux_parallel_velocity}
\end{align}
where we have defined the Ohmic resitivity $\eta = \nu_{ei} d_e^2$, and identified $\Phi$ to be the `potential' associated with the total parallel pressure, viz.,
\begin{align}
   - \frac{e}{T_{0e}}\gradd_\parallel \Phi = \frac{1}{ n_{0e} T_{0e}} \gradd_\parallel p_{\parallel e} = \gradd_\parallel \left( \frac{\delta n_e}{n_{0e}} + \frac{\delta T_{\parallel e}}{T_{0e}} \right) - \left( \frac{ \rho_e}{L_n} + \frac{ \rho_e}{L_{T_e}} \right) \frac{\partial \mathcal{A}}{\partial y},
    \label{eq:flux_psi}
\end{align}
where $p_{\parallel e} = n_e T_{\parallel e}$, $n_e = n_{0e} + \delta n_e$, and $T_{\parallel e} = T_{0e} + \delta T_{\parallel e}$ are the total parallel pressure, density, and parallel temperature, respectively.
It is clear that the conservation of magnetic flux is broken by the terms on the right-hand side of \eqref{eq:flux_parallel_velocity}; namely, from left to right, finite electron inertia, finite resistivity and magnetic drifts. 

In the collisionless limit, $\eta \rightarrow 0$, allowing us to neglect the resistive term. Below the flux-freezing scale \eqref{eq:flux_freezing_scale}, $k_\perp d_e \ll 1$, the remaining terms on the right-hand side are negligible in comparison to those on the left [the magnetic drifts also vanishing in the strongly driven limit, cf. \eqref{eq:velocity_moment}], meaning that, on electromagnetic scales, the magnetic field becomes frozen into the effective velocity 
\begin{align}
    \vec{u}_\text{eff}  = \frac{\rho_e \vthe }{2} \vec{b}_0 \times  \gradd \left( \varphi + \frac{e\Phi}{T_{0e}} \right) = \vec{v}_E - \frac{\rho_e \vthe}{2} \frac{\vec{b}_0 \times \gradd p_{\parallel e}}{n_{0e} T_{0e}},
    \label{eq:flux_u_collisionless}
\end{align}
where $\vec{v}_E$ is defined in \eqref{eq:convective_derivative}, and we have evaluated \eqref{eq:flux_u_eff} to leading order in the gyrokinetic expansion and used \eqref{eq:flux_psi}. This is \eqref{eq:flux_freezing_velocity}.

In the collisional limit, we retain only finite resitivity on the right-hand side of \eqref{eq:flux_parallel_velocity} [cf. \eqref{eq:col_velocity_equation}], while \eqref{eq:flux_psi} remains valid under the replacement $\delta T_{\parallel e} \rightarrow \delta T_e$. Below the flux-freezing scale \eqref{eq:flux_freezing_scale_col}, the resistive term can also be ignored, and the magnetic field once again becomes frozen into \eqref{eq:flux_u_collisionless}.

\section{Collisionless linear theory}
\label{app:collisionless_linear_theory}
We begin with our field equations, namely, quasineutrality~\eqref{eq:quasineutrality_final} [with $g_i = 0$; see~\eqref{eq:strongly_driven_lengthscales} and what follows it] and parallel Amp\`ere's law~\eqref{eq:parallel_amperes_law_final}:
\begin{align}
    \frac{\delta \tilde{n}_e}{n_{0e}} & = \frac{1}{n_{0e}} \int \rmd^3 \vec{v} \: \delta \! \tilde{f}_e = \tilde{g}_{0,0} = - \taubar^{-1} \tilde{\varphi} , \label{eq:lin_qn_fourier} \\
    \frac{\tilde{u}_{\parallel e}}{\vthe} & = \frac{1}{n_{0e}} \int \rmd^3 \vec{v} \: \frac{\vpar}{\vthe} \delta \! \tilde{f}_e = \frac{1}{\sqrt{2}} \tilde{g}_{0,1} = -(k_\perp d_e)^2 \tilde{\mathcal{A}} ,
    \label{eq:lin_amperes_fourier}
\end{align}
where tildes indicate the Fourier components of the fields, and we have expressed the perturbations of the electron density and parallel velocity in terms of the Hermite-Laguerre moments $\tilde{g}_{\ell,m}$ of $\delta \! \tilde{f}_e$, defined in \eqref{eq:transformation}. To calculate these moments, we linearise and Fourier-transform the electron kinetic equation \eqref{eq:electron_drift_kinetic_equation}. Neglecting collisions and normalising all frequencies to the parallel streaming rate, viz.,
\begin{align}
    \zeta = \frac{\omega}{|k_\parallel| \vthe}, \quad \zeta_* = \frac{\omega_{*e}}{|k_\parallel| \vthe}, \quad \zeta_d = \frac{\omega_{de}}{|k_\parallel| \vthe},
    \label{eq:lin_frequency_normalisations}
\end{align}
with $\omega_{*e}$ and $\omega_{de}$ defined in \eqref{eq:definition_timescales}, \eqref{eq:electron_drift_kinetic_equation} can be written as 
\begin{align}
    & \left[-\zeta + \frac{k_\parallel}{|k_\parallel |} \frac{v_\parallel}{\vthe} + \zeta_d \left(\frac{2v_\parallel^2}{\vthe^2}+ \frac{v_\perp^2}{\vthe^2} \right) \right] \frac{\delta \! \tilde{f}_e}{f_{0e}} \nonumber \\
    & = \left[ -\zeta + \frac{k_\parallel}{|k_\parallel |} \frac{v_\parallel}{\vthe} + \zeta_d \left(\frac{2v_\parallel^2}{\vthe^2}+ \frac{v_\perp^2}{\vthe^2} \right) \right] \tilde{\varphi} + \left[\zeta - \zeta_* \left( \frac{1}{\eta_e} +  \frac{v^2}{\vthe^2} - \frac{3}{2}\right) \right]\left( \tilde{\varphi} -  \frac{2v_\parallel}{\vthe} \tilde{\mathcal{A}}\right),
    \label{eq:lin_kinetic_equation_fourier_transform}
\end{align}
where $\eta_e = L_n/L_T$.
Introducing the dimensionless velocity variables
\begin{align}
    \hat{v} = \frac{k_\parallel}{| k_\parallel |} \frac{v_\parallel}{\vthe}, \quad \mu = \frac{v_\perp^2}{\vthe^2},
    \label{eq:velocity_variables_dimensionless}
\end{align}
we can write the Hermite-Laguerre moments as follows:
\begin{align}
    \tilde{g}_{\ell,m} = \left( \frac{k_\parallel}{|k_\parallel |}\right)^m \left( M_{\ell,m} \tilde{\varphi} - N_{\ell,m} \frac{k_\parallel}{|k_\parallel |} \tilde{\mathcal{A}}\right) + \tilde{\varphi} \: \delta_{0,0},
    \label{eq:lin_glm}
\end{align}
where the coefficient-matrix elements are
\begin{align}
    M_{\ell,m} & = \frac{1}{\sqrt{\pi}} \int_{-\infty}^\infty \rmd \hat{v} \: e^{-\hat{v}^2}\int_0^\infty \rmd \mu \: e^{-\mu}(-1)^\ell \frac{H_m(\hat{v}) L_\ell(\mu)}{\sqrt{2^m m!}} \frac{\zeta - \zeta_*\left(\eta_e^{-1} + \hat{v}^2 + \mu - \frac{3}{2} \right)}{\hat{v} -\zeta + \zeta_d(2\hat{v}^2 + \mu)}, \label{eq:lin_matrices_phi} \\
      N_{\ell,m} & = \frac{1}{\sqrt{\pi}} \int_{-\infty}^\infty \rmd \hat{v} \: e^{-\hat{v}^2}\int_0^\infty \rmd \mu \: e^{-\mu}(-1)^\ell \frac{H_m(\hat{v}) L_\ell(\mu)}{\sqrt{2^m m!}} 2\hat{v}\frac{\zeta - \zeta_*\left(\eta_e^{-1} + \hat{v}^2 + \mu - \frac{3}{2} \right)}{\hat{v} -\zeta + \zeta_d(2\hat{v}^2 + \mu)}. \label{eq:lin_matrices_a}
\end{align}
Using \eqref{eq:lin_glm} in \eqref{eq:lin_qn_fourier} and \eqref{eq:lin_amperes_fourier}, and combining the resultant expressions, we find, after some algebra, the dispersion relation:
\begin{align}
     \left( 1+ \frac{1}{\taubar} + M_{0,0} \right) \left( k_\perp^2 d_e^2 - \frac{1}{\sqrt{2}} N_{0,1} \right) + \frac{1}{\sqrt{2}} M_{0,1} N_{0,0} = 0.
    \label{eq:lin_dispersion_relation_full}
\end{align}
As we shall shortly demonstrate by recovering some familiar limits, this is the ETG dispersion relation \citep{liu71,lee87} coupled to the KAW one, and including the effects of magnetic drifts.

\subsection{Evaluation of coefficient-matrix elements}
\label{sec:evalutation_of_coefficient_matrix_elements}
In its form \eqref{eq:lin_dispersion_relation_full}, our dispersion relation is not particularly amenable to analytical solution, owing to the complexity of the coefficient-matrix elements \eqref{eq:lin_matrices_phi} and \eqref{eq:lin_matrices_a}. We devote this section to an approximate evaluation of these coefficients in order to express \eqref{eq:lin_dispersion_relation_full} in terms of known functions; readers interested in only the outcome of this procedure can skip ahead to \eqref{eq:lin_dispersion_relation}

Following \cite{biglari89}, we write the coefficient-matrix elements appearing in~\eqref{eq:lin_dispersion_relation_full} as follows:
\begin{align}
    M_{0,0} & = \lim_{a,b \rightarrow 1} \left[ \zeta - \zeta_* \left( \frac{1}{\eta_e} - \partial_a - \partial_b - \frac{3}{2} \right) \right] I_{a,b} \label{eq:m00_derivative} \\
    N_{0,0} = \sqrt{2} M_{0,1} & =  \lim_{a,b \rightarrow 1} 2\left[ \zeta - \zeta_* \left( \frac{1}{\eta_e} - \partial_a - \partial_b - \frac{3}{2} \right) \right] J_{a,b} \label{eq:n00_derivative}\\
    N_{0,1} & = \lim_{a,b \rightarrow 1} -2\sqrt{2}\left[ \zeta - \zeta_* \left( \frac{1}{\eta_e} - \partial_a - \partial_b - \frac{3}{2} \right) \right]  \partial_a I_{a,b}, \label{eq:n01_derivative}
\end{align}
where we have defined
\begin{align}
    I_{a,b}(\zeta,\zeta_d) & = \frac{1}{\sqrt{\pi}} \int_{-\infty}^\infty \rmd \hat{v} \int_0^\infty \rmd \mu \:\frac{ e^{-a\hat{v}^2} e^{-b\mu}}{\hat{v} -\zeta + \zeta_d(2\hat{v}^2 + \mu)}, \label{eq:lin_iab} \\
    J_{a,b}(\zeta,\zeta_d) & = \frac{1}{\sqrt{\pi}} \int_{-\infty}^\infty \rmd \hat{v} \int_0^\infty \rmd \mu \:\frac{\hat{v} e^{-a\hat{v}^2} e^{-b\mu}}{\hat{v} -\zeta + \zeta_d(2\hat{v}^2 + \mu)}, \label{eq:lin_jab}
\end{align}
with positive, real constants $a$ and $b$ (ensuring integral convergence). By using a partial-fraction expansion of its integrand, the latter of these can be written in terms of derivatives of the former with respect to $a$ and $b$:
\begin{align}
    J_{a,b} = \frac{1}{a^{1/2} b} +\zeta I_{a,b} + \zeta_d(2\partial_a + \partial_b) I_{a,b}.
    \label{eq:lin_jab_identity}
\end{align}
In writing the coefficient-matrix elements in this way, we have reduced our problem to determining \eqref{eq:lin_iab} in terms of functions that can be either computed numerically or expanded analytically in sensible limits. In the absence of magnetic drifts, $I_{a,b}$ reduces trivially to the well-studied plasma dispersion function \citep{faddeeva54,fried61}:
\begin{align}
  I_{1,1} (\zeta,0) = \mathcal{Z}(\zeta) = \frac{1}{\sqrt{\pi}}\int\rmd \hat{v} \: \frac{e^{-{\hat{v}}^2}}{\hat{v} - \zeta}
    \label{eq:plasma_z_function}
\end{align}
with the integral understood to be along the Landau contour --- while for two-dimensional modes with $k_\parallel \rightarrow 0$, $I_{a,b}$ can be also written in terms of products of plasma dispersion functions (see, e.g., \citealt{similon84,biglari89,ricci06,zocco18,mishchenko18}). How to calculate $I_{a,b}$ analytically in the presence of both parallel streaming and magnetic drifts without approximation remains an open research question, despite some progress being made numerically (\citealt{gurcan14,gultekin18, gultekin20,parisi20}).

Given that we are most interested in the strongly driven limit (see \apref{app:strongly_driven_limit}), we choose to expand the resonant denominator\footnote{A careful reader may be concerned about the potential breakdown of this expansion in the region of the resonance, viz., for $|\hat{v}- \zeta| \sim \zeta_d \ll \zeta $. However, if one removes this potential resonance by changing variables to $u =  \hat{v} + \zeta_d(2\hat{v}^2 + \mu)$, with $\rmd \hat{v} \rmd \mu = \rmd u \rmd \mu/\sqrt{1+8u\zeta_d - 8\mu \zeta_d^2}$, and performs a similar expansion for $\zeta_d\ll 1$, one finds the same result as the `naive' expansion \eqref{eq:lin_resonant_denominator_expansion}, to linear order.} in \eqref{eq:lin_iab} as a series in $\zeta_d \ll 1 \sim \zeta$:
\begin{align}
    \frac{1}{\hat{v} -\zeta + \zeta_d(2\hat{v}^2 + \mu)} =  \frac{1}{\hat{v} - \zeta} \sum_{n=0}^\infty \left(\frac{2 \hat{v}^2 + \mu}{\hat{v} - \zeta} \right)^n (-\zeta_d)^n \approx \frac{1}{\hat{v} -\zeta} - \frac{2\hat{v}^2 + \mu}{(\hat{v} - \zeta)^2} \zeta_d + \dots 
    \label{eq:lin_resonant_denominator_expansion}
\end{align}
We will discuss the validity and consequences of this approximation in \apref{app:exact_stability_boundary}.
Substituting \eqref{eq:lin_resonant_denominator_expansion} into \eqref{eq:lin_iab} and retaining only terms linear in $\zeta_d$, we find, after integrating by parts in the second term and evaluating the integral over $\mu$, that $I_{a,b}$ can be expressed entirely in terms of the plasma dispersion function \eqref{eq:plasma_z_function}:
\begin{align}
    I_{a,b} (\zeta,\zeta_d) = \frac{1}{b} \mathcal{Z}(\sqrt{a}\zeta) + \frac{4}{b}\left[\frac{1}{2a^{1/2}} + \left( a\zeta^2 + \frac{a}{2b} -1 \right) \left(\frac{1}{a^{1/2}} + \zeta \mathcal{Z}(\sqrt{a}\zeta) \right) \right] \zeta_d.
    \label{eq:lin_iab_final}
\end{align}
Finally, substituting \eqref{eq:lin_iab_final} into \eqref{eq:m00_derivative}-\eqref{eq:n01_derivative}, via \eqref{eq:lin_jab_identity} where necessary, and making use of the identities
\begin{align}
        \mathcal{Z}' = - 2(1+\zeta \mathcal{Z}),\quad \mathcal{Z}'' = \frac{2}{\zeta } - \frac{2}{\zeta} \left( \zeta^2 - \frac{1}{2} \right) \mathcal{Z}',
    \label{eq:plasma_z_function_identities}
\end{align}
we find, neglecting density gradients ($\eta_e \rightarrow \infty$):
\begin{align}
    M_{0,0} & = -\zeta \zeta_* + \left[ \zeta - \zeta_* \left( \zeta^2 - \frac{1}{2} \right) \right] \mathcal{Z} \nonumber \\
    & \quad \:\: + \left\{ 4\zeta^3 + (4\zeta^4 - 2\zeta^2) \mathcal{Z}  \right. \left. - \zeta_*\left[4\zeta^4-6\zeta^2 + (4\zeta^5 - 8\zeta^3 +\zeta)\mathcal{Z} \right] \right\}\zeta_d,   \label{eq:m00_final}  \\
     N_{0,0} & = - \zeta_* + 2 \left[\zeta - \zeta_* \left(\zeta^2 - \frac{1}{2} \right) \right](1+ \zeta\mathcal{Z})  \nonumber \\
     & \quad \:\: + \left\{8\zeta^4-4\zeta^2 + (8\zeta^5 -8\zeta^3 -2\zeta) \mathcal{Z}  \right. \nonumber \\
     & \quad \: \: \: \left. - \zeta_* \left[8\zeta^5 - 16\zeta^3 -2\zeta +(8\zeta^6 - 20\zeta^4 + 2\zeta^2 -1) \mathcal{Z} \right] \right\} \zeta_d, \label{eq:n00_final}\\
     N_{0,1} & = - \sqrt{2} \zeta \zeta_* + 2 \sqrt{2} \left[ \zeta  - \zeta_* \left( \zeta^2 - \frac{1}{2} \right)\right] \zeta (1 + \zeta \mathcal{Z}) \nonumber \\
     & \quad \: \: + 2\sqrt{2}\left\{ 4\zeta^5 - 4\zeta^3 -2\zeta  + (2\zeta^5 - 3\zeta^3-\zeta)  \mathcal{Z} \right. \nonumber \\
     & \quad \: \: \: \left. - \zeta_* \left[4\zeta^6 - 10\zeta^4 -2\zeta^2-2+ (4\zeta^7-12\zeta^5+\zeta^3-\zeta)\mathcal{Z} \right] \right\} \zeta_d. \label{eq:n01_final}
\end{align}
Together with \eqref{eq:lin_dispersion_relation_full}, re-written here using the first equality in \eqref{eq:n00_derivative} as
\begin{align}
     \left( 1+ \frac{1}{\taubar} + M_{0,0} \right) \left( k_\perp^2 d_e^2 - \frac{1}{\sqrt{2}} N_{0,1} \right) + \frac{1}{2} N_{0,0}^2 = 0,
    \label{eq:lin_dispersion_relation}
\end{align}
these expressions for the coefficient-matrix elements give us the dispersion relation for our kinetic system in the limit $\zeta_d \ll 1$, written in terms of $\zeta$, $\zeta_*$, $\zeta_d$ and the plasma dispersion function $\mathcal{Z}$. 
As ever in linear plasma (kinetic) theory, physically transparent cases arise when the plasma dispersion function is expanded in large or small argument --- as we shall see, these are the natural limits for recovering characteristic electrostatic and electromagnetic phenomena, respectively.

\subsection{Two-dimensional perturbations}
\label{app:lin_two_dimensional_perturbations}
Let us first consider purely two-dimensional perturbations --- which amounts to setting $k_\parallel = 0$ everywhere --- without ordering $k_\perp d_e$ with respect to unity.
In this limit, $\zeta \propto k_\parallel^{-1} \rightarrow \infty$, so the plasma dispersion function can be expanded as: 
\begin{align}
    \mathcal{Z}(\zeta) \approx i\sqrt{\pi} e^{-\zeta^2} - \frac{1}{\zeta} \left(1 + \frac{1}{2\zeta^2} + \frac{3}{4\zeta^4} + \dots \right).
    \label{eq:z_function_zeta_large}
\end{align}
Ignoring the exponentially small term --- and thus working within the `fluid' approximation --- \eqref{eq:m00_final}-\eqref{eq:n01_final} can be expanded as
\begin{align}
    M_{0,0} & \approx -1 + \frac{1}{2\zeta^2}\left( \frac{\zeta_*}{\zeta} - 1 \right) + \frac{2\zeta_d \zeta_*}{\zeta^2 } + \dots,  \label{eq:m00_expansion_large} \\
    N_{0,0} & \approx \frac{1}{\zeta} \left( \frac{\zeta_*}{\zeta} -1 \right) + \frac{16 \zeta_d \zeta_*}{\zeta^3} + \dots, \label{eq:n00_expansion_large} \\
  \frac{1}{\sqrt{2}}  N_{0,1} & \approx -1 + \frac{\zeta_*}{\zeta } + \frac{8 \zeta_d \zeta_*}{\zeta^2} + \dots, \label{eq:n01_expansion_large}
\end{align}
where we have kept $\zeta_d$ only where it multiplies $\zeta_*$, consistent with the strongly driven limit. Then \eqref{eq:lin_dispersion_relation} becomes
\begin{align}
    \left( \frac{1}{\taubar} + \frac{2\zeta_d \zeta_*}{\zeta^2} \right) \left( 1 + k_\perp^2 d_e^2 - \frac{\zeta_*}{\zeta} \right) = 0,
     \label{eq:lin_2d_dispersion_relation}
\end{align}
where we have ignored all higher-order terms in $\zeta^{-1} \propto k_\parallel \rightarrow 0$. This dispersion relation, of course, could have been obtained without resorting to the kinetic formalism that we have adopted in this appendix; setting $k_\parallel= 0$ in \eqref{eq:density_moment}-\eqref{eq:tperp_moment}, and solving the resultant fluid equations, one obtains exactly~\eqref{eq:lin_2d_dispersion_relation}. The dispersion relation~\eqref{eq:lin_2d_dispersion_relation} admits two solutions.

\subsubsection{Magnetic drift wave}
\label{app:lin_magnetic_drift_wave}
From the second bracket in \eqref{eq:lin_2d_dispersion_relation}, we find a `magnetic drift wave'
\begin{align}
    \zeta = \frac{\zeta_*}{1+k_\perp^2 d_e^2} \quad \Rightarrow \quad \omega = \frac{\omega_{*e}}{1+k_\perp^2 d_e^2}.
    \label{eq:lin_mdw}
\end{align}
This is a purely linear magnetic oscillation involving the balance between the inductive part of the parallel electric field, the electron inertia, and the gradient of the equilibrium pressure along the perturbed field line:
\begin{align}
    \frac{\partial }{\partial t} \left( \mathcal{A} - \frac{u_{\parallel e}}{\vthe} \right) = \frac{\partial}{\partial t} \left(\mathcal{A} - d_e^2 \gradd_\perp^2 \mathcal{A} \right)  = - \frac{\rho_e \vthe}{2L_T} \frac{\partial \mathcal{A}}{\partial y}.
    \label{eq:lin_mdw_equations}
\end{align}
In setting $k_\parallel = 0$, we have decoupled perturbations of the magnetic field --- or, in the electrostatic regime, of the parallel velocity --- from those of density and temperature.

\subsubsection{Curvature-mediated ETG instability}
\label{app:lin_curvature_mediated_etg}
From the first bracket in \eqref{eq:lin_2d_dispersion_relation}, we find 
\begin{align}
    \zeta^2 =  - 2 \zeta_d \zeta_{*e} \taubar \quad \Rightarrow \quad \omega = \pm  i \left(2 \omega_{de} \omega_{*e} \taubar \right)^{1/2}.
    \label{eq:lin_cetg}
\end{align}
This is the cETG growth rate \eqref{eq:cetg_gamma}. We note that there is no critical gradient for the cETG instability, i.e., formally, the $k_\parallel = 0$ mode is unstable at all values of the equilibrium temperature gradient. This is because, in adopting the strongly driven limit (\apref{app:strongly_driven_limit}), we dropped the density gradient, leaving the critical gradient for any instability, including the cETG, to be formally $L_B/L_T=0$ (in other words, there are no finite critical temperature gradients because there is nothing to compare $L_B/L_T$ to). Finite critical temperature gradients, and how they related to the main body of this work, are discussed in \apref{app:finite_critical_gradients}. 

Both modes \eqref{eq:lin_mdw} and \eqref{eq:lin_cetg} persist at all perpendicular wavenumbers because there is no distinction between the electrostatic and electromagnetic regimes for purely two-dimensional phenomena. Indeed, the dispersion relation \eqref{eq:lin_2d_dispersion_relation} is formally valid for $k_\perp d_e \sim 1$, and thus in both the electrostatic ($k_\perp d_e \gg 1$) and electromagnetic ($k_\perp d_e \ll 1$) limits. Restoring finite $k_\parallel$, however, significantly alters this behaviour, as it allows coupling between perturbations of the magnetic field and those of density and temperature, which introduces new instabilities in both the electrostatic and electromagnetic regimes. 

\subsection{Electrostatic 3D perturbations: collisionless sETG}
\label{app:electrostatic_3d_pertrubations_setg}
Let us consider perturbations below the flux-freezing scale \eqref{eq:flux_freezing_scale}, viz., with 
\begin{align}
    k_\perp d_e \rightarrow \infty,
    \label{eq:lin_electrostatic_limit}
\end{align}
for which \eqref{eq:lin_dispersion_relation} reduces to the electrostatic ETG dispersion relation (cf. \citealt{liu71,lee87}):
\begin{align}
      1+  \frac{1}{\taubar}  + M_{0,0} = 0. 
      \label{eq:lin_dispersion_relation_es_etg}
\end{align}
If, in addition to \eqref{eq:lin_electrostatic_limit}, we adopt the limit of long parallel wavelengths and small magnetic drifts, viz.,
\begin{align}
    \omega_{de} \ll k_\parallel \vthe \ll \omega \ll \omega_{*e} \quad \Leftrightarrow \quad  \zeta_d \sim \zeta_*^{-1/3} \ll 1 \ll \zeta \sim \zeta_*^{1/3} \ll \zeta_*,
    \label{eq:lin_electrostatic_limit_ordering}
\end{align}
and once again make use of the expansion \eqref{eq:m00_expansion_large}, we find, retaining only the leading-order terms,
\begin{align}
    \zeta^3 + 2\zeta_d \zeta_*\taubar \zeta + \frac{\taubar \zeta_*}{2} = 0.
    \label{eq:lin_es_dispersion_relation}
\end{align}
This has three roots, whose behaviour is easy to deduce by balancing terms in various limits. The balance of the first two terms in \eqref{eq:lin_es_dispersion_relation} recovers the cETG instability \eqref{eq:lin_cetg}; the balance of the first and third terms yields
\begin{align}
    \zeta = \left( - \frac{\taubar \zeta_*}{2} \right)^{1/3} \quad \Rightarrow \quad \omega = \text{sgn}(k_y) \left( -1, \frac{1}{2} \pm i \frac{\sqrt{3}}{2} \right) \left( \frac{k_\parallel^2 \vthe^2 |\omega_{*e}| \taubar}{2}\right)^{1/3}.
    \label{eq:lin_setg}
\end{align}
This is the collisionless sETG growth rate \eqref{eq:setg_gamma} --- the one unstable root of the three (of the other two, one is damped, and another is a pure drift wave) --- which we would expect to recover in the electrostatic regime (magnetic field lines and electron flows are liberated from one another as flux is unfrozen by finite electron inertia). 

Being a cubic equation with real coefficients and a negative-definite discriminant, \eqref{eq:lin_es_dispersion_relation} has at least one unstable solution at all parallel and perpendicular wavenumbers: there is no region of stability between the cETG and sETG modes --- with the former transitioning into the latter as $k_\parallel$ is increased --- and the sETG is formally unstable for $k_\parallel \rightarrow \infty$. This is because we have thus far neglected the exponentially small resonant term in \eqref{eq:z_function_zeta_large} that is responsible for the Landau damping of sETG at larger parallel wavenumbers. It is relatively obvious that this will occur for $\zeta \sim \zeta_* \sim 1$, where the rate of parallel streaming and energy injection are comparable; this is confirmed in \apref{app:exact_stability_boundary}.

\subsection{Electromagnetic stabilisation of sETG}
\label{app:electromagnetic_stabilisation_of_setg}
\begin{figure}
    
\centering
\begin{tabular}{cc}
     \includegraphics[width=0.48\textwidth]{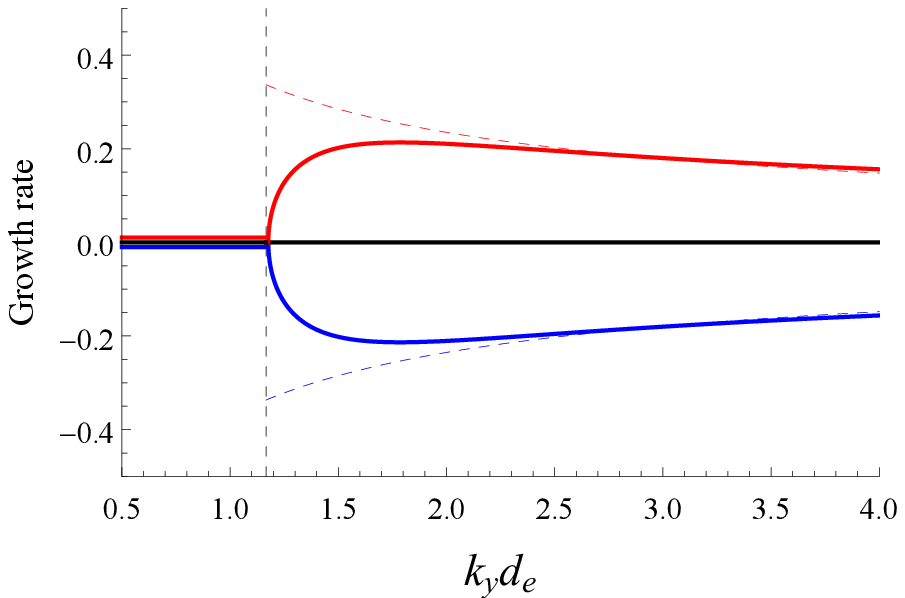}&  
     \includegraphics[width=0.48\textwidth]{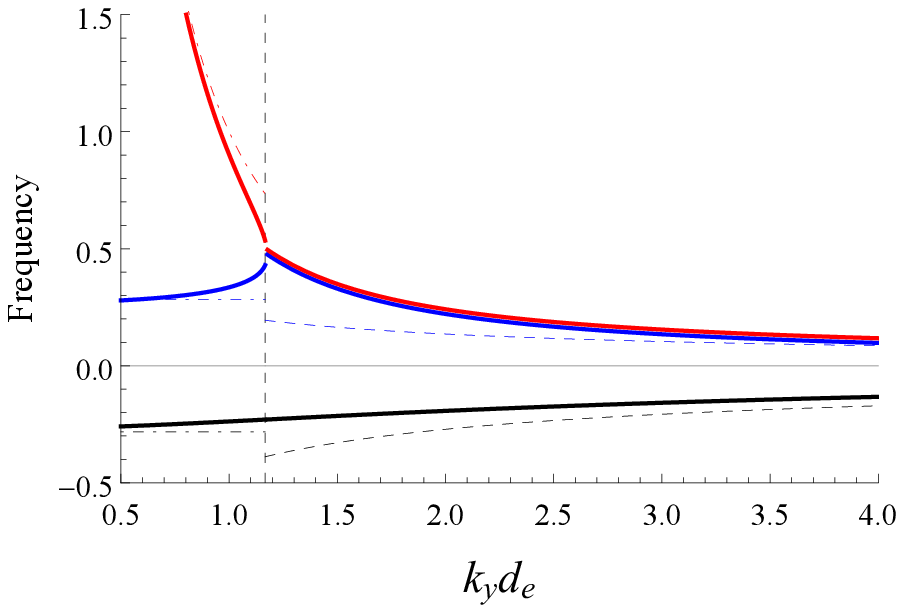} \\\\
     (a) $L_B/L_{T}=250$, $k_x d_e = 0$ & (b) $L_B/L_{T}=250$, $k_x d_e = 0$ 
\end{tabular}
 \caption{The (a) growth rates and (b) frequencies of the collisionless electrostatic instabilities, normalised to $\omega_{*e}$, for $k_\parallel L_T/\sqrt{\beta_e}=0.2$ and $\taubar = 1$. In both plots, the red, blue and black solid lines are the three solutions to the cubic dispersion relation~\eqref{eq:lin_es_dispersion_finite_de}, while the vertical grey dashed line is the `fluid' stability boundary~\eqref{eq:lin_es_stability_boundary}. At perpendicular wavenumbers smaller than~\eqref{eq:lin_es_stability_boundary}, there are only stable modes, as expected, corresponding to the electrostatic limit of the magnetic drift wave [\eqref{eq:lin_es_mdw}, red dot-dashed line] and two isobaric KAW modes [\eqref{eq:lin_es_isobaric_kaw}, blue and black dot-dashed lines]. At perpendicular wavenumbers greater than~\eqref{eq:lin_es_stability_boundary}, the positive-frequency KAW and the magnetic drift wave transition into the two positive-frequency ETG modes [\eqref{eq:lin_setg}, red and blue dashed lines] --- one growing, one damped --- while the negative-frequency KAW transitions into the negative-frequency ETG drift wave [\eqref{eq:lin_setg}, black dashed line].  We chose a very large value of $L_B/L_T$ in order to show the asymptotic regimes clearly. 
 }
    \label{fig:lin_es_plots}
\end{figure}

Formally, the dispersion relation \eqref{eq:lin_es_dispersion_relation} is derived in the limit $k_\perp d_e \rightarrow \infty$, the electrostatic limit. Restoring finite but large $k_\perp d_e$, viz.,
\begin{align}
    \zeta_d \sim \zeta_*^{-1/3} \ll 1 \ll \zeta \sim k_\perp d_e \sim \zeta_*^{1/3} \ll \zeta_*,
    \label{eq:lin_electrostatic_limit_ordering_finite_de}
\end{align}
and using \eqref{eq:m00_expansion_large}-\eqref{eq:n01_expansion_large} in \eqref{eq:lin_dispersion_relation}, we have, instead of \eqref{eq:lin_es_dispersion_relation},
\begin{align}
  \left( \zeta^2 + 2\zeta_d \zeta_* \taubar \right) \left( \zeta - \frac{\zeta_*}{k_\perp^2 d_e^2} \right) + \frac{\taubar}{2} \zeta_* = 0.
    \label{eq:lin_es_dispersion_finite_de}
\end{align}
In addition to the cETG \eqref{eq:lin_cetg} and sETG \eqref{eq:lin_setg} instabilities, \eqref{eq:lin_es_dispersion_finite_de} admits two further solutions: from the second bracket, we obtain the electrostatic (i.e., $k_\perp d_e \gg 1$) limit of the magnetic drift wave \eqref{eq:lin_mdw},
\begin{align}
    \zeta = \frac{\zeta_*}{k_\perp^2 d_e^2} \quad \Rightarrow \quad \omega = \frac{\omega_{*e}}{k_\perp^2 d_e^2},
    \label{eq:lin_es_mdw}
\end{align}
while the balance of the first term in the first bracket and second term in the second bracket with the last term gives rise to two isobaric KAW modes:
\begin{align}
    \zeta^2 = \frac{1}{2} k_\perp^2 d_e^2 \taubar \quad \Rightarrow \quad \omega = \pm \omega_\text{KAW} \sqrt{\taubar}.
    \label{eq:lin_es_isobaric_kaw}
\end{align}
These are a $\zeta \gg 1$ continuation of the isobaric KAWs that arise at lower frequencies, in the electromagnetic regime (see \secref{sec:isobaric_slab_TAI}). These solutions of \eqref{eq:lin_es_dispersion_finite_de} are plotted in \figref{fig:lin_es_plots}.

Together, \eqref{eq:lin_es_mdw} and \eqref{eq:lin_es_isobaric_kaw} conspire to stabilise the sETG mode \eqref{eq:lin_setg} at longer perpendicular wavelengths, around the flux-freezing scale \eqref{eq:flux_freezing_scale}. To show this, we consider the stability boundary associated with \eqref{eq:lin_es_dispersion_finite_de}: assuming $\zeta$ to be purely real, with $\text{Im}(\zeta) \rightarrow + 0$, and demanding that the real and imaginary parts of the resultant expression must vanish individually, we find that the real part is given by \eqref{eq:lin_es_dispersion_finite_de}, while the imaginary part is
\begin{align}
      3\zeta^2 - \frac{2\zeta_*}{k_\perp^2 d_e^2} \zeta + 2 \zeta_d \zeta_* \taubar = 0.
      \label{eq:lin_es_stab_1}
\end{align}
At the stability boundary, $\zeta$ is purely real, meaning that the discriminant of \eqref{eq:lin_es_stab_1} must be positive. This places a restriction on the perpendicular wavenumbers at which sETG is stabilised:
\begin{align}
    \left( \frac{2\zeta_*}{k_\perp^2 d_e^2} \right)^2 - 24 \zeta_d \zeta_* \taubar \geqslant 0 \quad \Rightarrow \quad k_\perp d_e \leqslant \left(\frac{1}{6\taubar} \frac{L_B}{L_T} \right)^{1/4}.
    \label{eq:lin_es_stab_2}
\end{align}

Now considering perpendicular wavenumbers much smaller than \eqref{eq:lin_es_stab_2} --- which amounts to ignoring the effects of the magnetic drifts --- we can solve \eqref{eq:lin_es_dispersion_finite_de} and \eqref{eq:lin_es_stab_1} simultaneously for the stability boundary: 
\begin{align}
  (k_\perp d_e)^2 = \frac{2}{3} \left( \frac{\zeta_*}{\sqrt{\taubar}} \right)^{2/3} \quad \Rightarrow \quad  \frac{k_\parallel L_T}{\sqrt{\beta_e}} = \frac{1}{2\sqrt{\taubar}} \left( \frac{2}{3} \right)^{3/2} \frac{1}{(k_\perp d_e)^2} \frac{k_y}{k_\perp}. 
    \label{eq:lin_es_stability_boundary}
\end{align}
This is the slanted black dashed line in figures \ref{fig:lin_em_no_drifts} and \ref{fig:lin_em}. It is worth noting that this `fluid' stability boundary is, in fact, only approximate: in our treatment of the $\zeta \gg 1$ limit, we have neglected the exponentially small resonant term in \eqref{eq:z_function_zeta_large} that can lead to exponentially small growth rates below the line \eqref{eq:lin_es_stability_boundary}\footnote{These growth rate are only exponentially small as long as the limit $\zeta \gg 1$ is satisfied. At $\zeta \lesssim 1$, however, this is no longer true, and the resonant term in \eqref{eq:z_function_zeta_large} can have a significantly destabilising effect, as in the electromagnetic regime.}. However, these exponentially small growth rates would easily be erased by the effects of finite dissipation in any realistic physical system (or, indeed, simulation), meaning that \eqref{eq:lin_es_stability_boundary} can be interpreted as a criterion for the electromagnetic stabilisation of the sETG instability due to the effects of finite $\beta_e$. This was the conclusion of \cite{maeyama21}, who also derived~\eqref{eq:lin_es_stability_boundary} [their equation (23)] via similar methods to those used here.

\subsection{Electromagnetic 3D perturbations: collisionless TAI}
\label{app:lin_electromagnetic_3d_perturbations_tai}
Moving towards larger scales, we now consider perturbations above the flux-freezing scale \eqref{eq:flux_freezing_scale}, viz., 
\begin{align}
    k_\perp d_e \ll 1.
    \label{eq:lin_electromagnetic_limit}
\end{align}
As we shall see shortly, long perpendicular wavelengths correspond to low frequencies. Let us consider the ordering
\begin{align}
    \omega_{de}   \ll \omega \ll \omega_{*e} \sim k_\parallel \vthe \quad \Leftrightarrow \quad \zeta_d \sim \zeta^2 \ll \zeta \sim k_\perp d_e \ll \zeta_* \sim 1,
    \label{eq:lin_electromagnetic_ordering}
\end{align}
under which the plasma dispersion function can again be expanded, this time in small argument:
\begin{align}
    \mathcal{Z}(\zeta) \approx  i\sqrt{\pi} e^{-\zeta^2} - 2\zeta \left(1 - \frac{2\zeta^2}{3} + \frac{4\zeta^4}{15} + \dots \right).
    \label{eq:plasma_z_function_zeta_small}
\end{align}
Then, \eqref{eq:m00_final}-\eqref{eq:n01_final} can be expanded as 
\begin{align}
    M_{0,0} & \approx i \frac{\sqrt{\pi}}{2} \zeta_* +\left( i \sqrt{\pi} - 2 \zeta_* \right) \zeta + \dots , \label{eq:m00_expansion_small} \\
    N_{0,0} & \approx 2 \left(1 + i \frac{\sqrt{\pi}}{2} \zeta_* \right) \zeta 
    +\dots,\label{eq:n00_expansion_small} \\
  \frac{1}{\sqrt{2}}  N_{0,1} &  \approx 2 \left(1 + i \frac{\sqrt{\pi}}{2} \zeta_* \right) \zeta^2  + 4 \zeta_d \zeta_* + \dots, \label{eq:n01_expansion_small}
\end{align}
where we have once again only kept $\zeta_d$ where it multiplies $\zeta_*$. Retaining only leading-order terms, \eqref{eq:lin_dispersion_relation} becomes
\begin{align}
    \left( 1 + \frac{1}{\taubar} + i\frac{\sqrt{\pi}}{2} \zeta_* \right) \left( k_\perp^2 d_e^2 - 4 \zeta_d \zeta_* \right) - \frac{2}{\taubar}\left(  1 + i\frac{\sqrt{\pi}}{2} \zeta_* \right) \zeta^2 = 0,
    \label{eq:lin_tai_dispersion_initial}
\end{align}
or, after straightforward manipulations, 
\begin{align}
    \zeta^2 + \left( 2\zeta_d \zeta_* - \frac{1}{2} k_\perp^2 d_e^2 \right)\left( \taubar + \frac{1}{1+i\xi_*} \right)=0, \quad \xi_* = \frac{\sqrt{\pi }}{2} \zeta_*.
    \label{eq:lin_tai}
\end{align}
This is the dispersion relation of the collisionless thermo-Alfv\'enic instability (TAI), which we treat in detail in \secref{sec:electromagnetic_regime_tai} and \apref{app:tai_dispersion_relation}. The TAI dispersion relation \eqref{eq:lin_tai} captures all of the properties of the more general dispersion relation \eqref{eq:lin_dispersion_relation} in the electromagnetic regime, with the important exception of the stabilisation of isothermal and isobaric sTAI --- see \eqref{eq:stai_isothermal_limit} and \eqref{eq:stai_isobaric_limit}, respectively --- that we shall work out in the next section.

\subsection{General stability boundary}
\label{app:exact_stability_boundary}
Let us now consider the stability boundary associated with the dispersion relation~\eqref{eq:lin_dispersion_relation}. At the stability boundary, $\zeta$ is purely real, so the real and imaginary parts of \eqref{eq:lin_dispersion_relation} must vanish individually. For a purely real $\zeta$, imaginary terms can only enter through the plasma dispersion functions $\mathcal{Z}(\zeta)$, implying that the coefficient in front of it must vanish, as must, separately, the remainder of the dispersion relation. This yields two equations for the frequency $\zeta$ and wavenumber at the stability boundary, which can then be solved simultaneously to find the corresponding curve in the wavenumber (and parameter) space.

\subsubsection{Stability boundary without magnetic drifts}
\label{app:without_magnetic_drifts}
It will prove instructive to consider first the simplified case of no magnetic drifts ($\zeta_d = 0$), in which, making use of \eqref{eq:m00_final}-\eqref{eq:n01_final}, the dispersion relation \eqref{eq:lin_dispersion_relation} can be simplified to 
\begin{align}
    \left( \frac{2\zeta^2}{k_\perp^2 d_e^2} - \taubar \right) \left[ 1 + \zeta \mathcal{Z} -\zeta \zeta_* - \zeta_* \left( \zeta^2 - \frac{1}{2} \right) \mathcal{Z} \right] = 1.
    \label{eq:lin_dispersion_relation_no_drifts}
\end{align}
Following the steps laid out above, we find, at the stability boundary,
\begin{align}
    \left( \zeta^2 - \frac{1}{2} \right)\zeta_* = \zeta, \quad \left( \frac{2\zeta^2}{k_\perp^2 d_e^2} - \taubar \right) (1 - \zeta \zeta_*) = 1.
    \label{eq:lin_stab_equations}
\end{align}
Substituting $\zeta_*$ from the first equation into the second, we find the real frequency at the stability boundary:
\begin{align}
    \zeta^2 = \frac{1+\taubar}{2} \frac{k_\perp^2 d_e^2}{1+k_\perp^2 d_e^2}.
    \label{eq:lin_stab_frequency}
\end{align}
In view of \eqref{eq:lin_stab_frequency}, $\zeta$ at the stability boundary can be either small or order unity, but never large, for any perpendicular wavenumber. This means that no mode with $\zeta \gg 1$ is, in fact, stable --- as we discussed at the end of \apref{app:electromagnetic_stabilisation_of_setg}, the curve \eqref{eq:lin_es_stability_boundary} was where the `fluid' stability was achieved, but exponentially small growth rates feeding off Landau resonances were still allowed. This is also why we were unable to capture the Landau damping of the sETG in our previous analysis. 
\begin{figure}
    
\centering
\hspace{-0.5cm}
\begin{tabular}{cc}
     \includegraphics[width=0.55\textwidth]{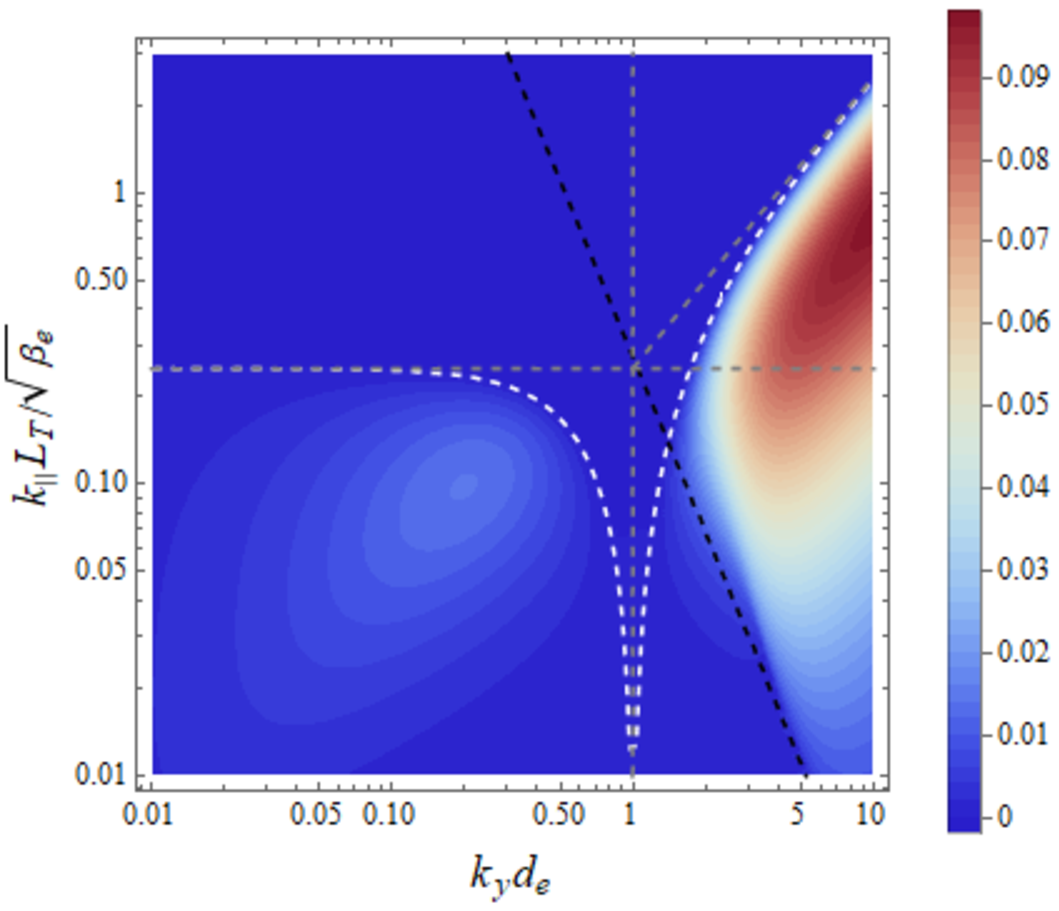}&  
        \includegraphics[width=0.45\textwidth]{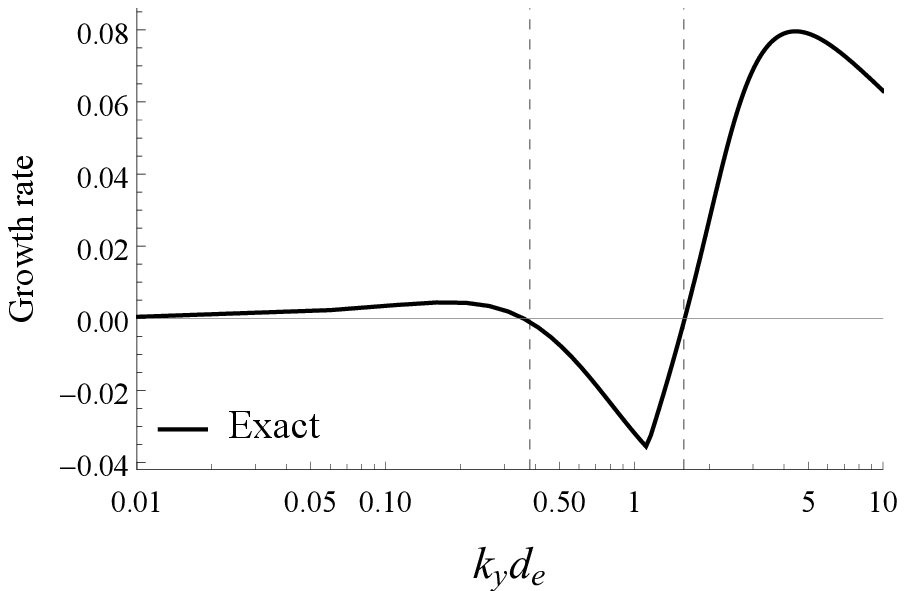} \\\\
     (a) $\omega_{de} = 0$, $k_x d_e  = 0$  & (b) $k_\parallel L_T/\sqrt{\beta_e} =0.2 $  \\\\\\
    \includegraphics[width=0.45\textwidth]{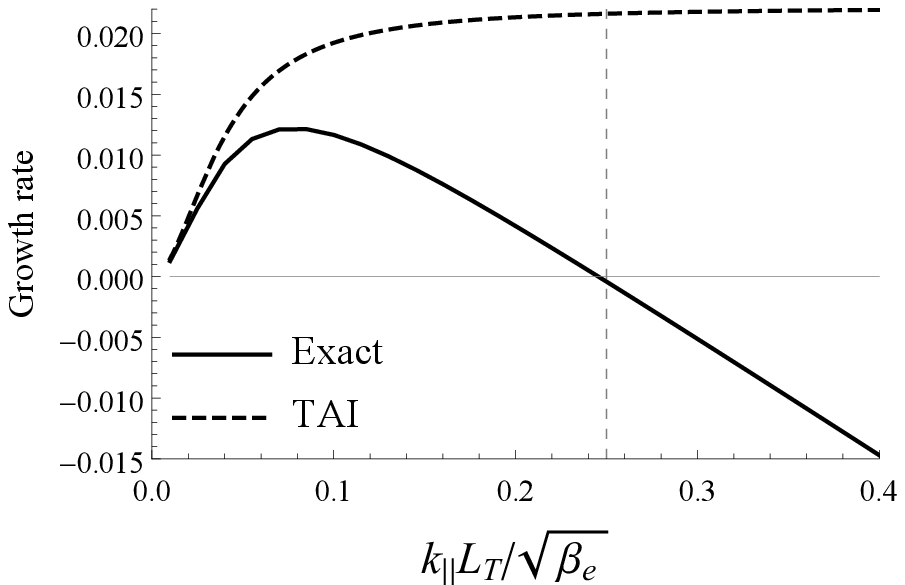}&  
     \includegraphics[width=0.45\textwidth]{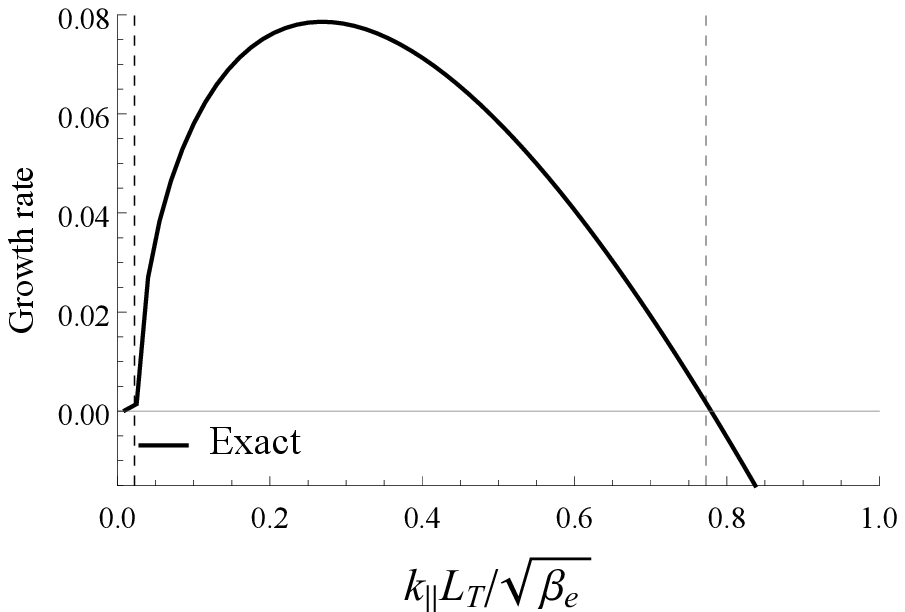} \\\\
          (c)  $k_y d_e  = 0.1 $  & (d) $k_y d_e  = 3.5$ 
\end{tabular}
 \caption{The growth rates of the collisionless instabilities normalised to $\omega_{*e}$, in the absence of magnetic drifts and with $\taubar =1$. Panel (a) is a contour plot of the positive growth rates ($\gamma >0$) in the $(k_y,k_\parallel )$ plane. The white dashed line is the exact stability boundary~\eqref{eq:lin_stab_boundary}, while the horizontal grey dashed line is~\eqref{eq:lin_stab_isothermal_stai}, corresponding to the stabilisation of the isothermal sTAI at large parallel wavenumbers. The vertical grey dashed line is~\eqref{eq:lin_stab_isobaric_stai}, around which the isobaric sTAI is stabilised; the slanted grey dashed line on the right is the sETG stability boundary \eqref{eq:lin_stab_setg}; the slanted black dashed line is the `fluid' sETG stability boundary~\eqref{eq:lin_es_stability_boundary}. In the remaining plots, the solid lines represent the exact growth rate obtained by solving the (collisionless) linear dispersion relation~\eqref{eq:lin_dispersion_relation}, while the dashed line in panel (c) is the growth rate predicted by the approximate TAI dispersion relation \eqref{eq:lin_tai}. Panel (b) is a cut of the growth rate along $k_\parallel L_T/\sqrt{\beta_e} =0.2 $ (plotted against a logarithmic scale), in which the vertical gray dashed lines correspond to the two branches of the exact stability boundary~\eqref{eq:lin_stab_boundary}, between which the growth rate is negative. Panels (c) and (d) are cuts of the growth rate for $k_y d_e  = 0.1$ and $k_y d_e  = 3.5$, respectively. In panel (c), the vertical grey dashed line is \eqref{eq:lin_stab_isothermal_stai}, while the same line in panel (d) is \eqref{eq:lin_stab_setg}. Lastly, the vertical black dashed line on the left of panel (d) is \eqref{eq:lin_es_stability_boundary}.}
    \label{fig:lin_em_no_drifts}
\end{figure}

Substituting \eqref{eq:lin_stab_frequency} into the first equation in \eqref{eq:lin_stab_equations}, we find the expression for the stability boundary in the wavenumber space:
\begin{align}
  \pm \omega_\text{KAW} \sqrt{1+\taubar} =  \frac{1-k_\perp^2 d_e^2 \taubar}{2\sqrt{1+ k_\perp^2 d_e^2}} \omega_{*e},
  \label{eq:lin_stab_boundary}
\end{align}
where $\omega_\text{KAW}$ is defined in \eqref{eq:definition_timescales}. This is the white dashed curve in figures \ref{fig:lin_em_no_drifts}(a) and~\ref{fig:lin_em}(a). In the absence of magnetic drifts, \eqref{eq:lin_stab_boundary} is an exact result. 

In the limit of large perpendicular wavenumbers, \eqref{eq:lin_stab_boundary} gives us the stabilisation of the electrostatic sETG mode due to Landau damping (at large parallel wavenumbers), viz., for $k_\perp d_e \gg 1$, it becomes
\begin{align}
  \mp k_\parallel \vthe \sqrt{\frac{1+\taubar}{2}} = \frac{\taubar \omega_{*e}}{2} \quad \Rightarrow \quad \frac{k_\parallel L_T}{\sqrt{\beta_e}} = \pm \frac{\taubar}{2\sqrt{2(1+\taubar)}} k_y d_e.
  \label{eq:lin_stab_setg}
\end{align}
This is the slanted grey dashed line in the top right-hand corner of figures \ref{fig:lin_em_no_drifts}(a) and \ref{fig:lin_em}(a).
\eqref{eq:lin_stab_setg} also confirms the assertion made in \apref{app:electrostatic_3d_pertrubations_setg} that this stabilisation occurs when the rates of parallel streaming and energy injection are comparable, $k_\parallel \vthe \sim \omega_{*e}$.

In the opposite limit of small perpendicular wavenumbers, \eqref{eq:lin_stab_boundary} asymptotes to a line of constant $k_\parallel$, viz., for $k_\perp d_e \ll 1$, it becomes
\begin{align}
    \pm \omega_{\text{KAW}} \sqrt{1+\taubar} = \frac{\omega_{*e}}{2} \quad \Rightarrow \quad \frac{k_\parallel L_T}{\sqrt{\beta_e}} = \pm \frac{1}{2 \sqrt{2(1+\taubar)}} \frac{k_y}{k_\perp}. 
    \label{eq:lin_stab_isothermal_stai}
\end{align}
This is the upper horizontal grey dashed line in figures \ref{fig:lin_em_no_drifts}(a) and \ref{fig:lin_em}(a). It corresponds to the stabilisation of the isothermal sTAI mode \eqref{eq:stai_isothermal_limit} at large parallel wavenumbers due to compressional heating, as explained in \secref{sec:stabilisation_of_isothermal_slab_TAI} and \apref{app:stabilisation_of_isothermal_stai}.  

Similarly, the stabilisation of the isobaric sTAI mode \eqref{eq:stai_isobaric_limit} can be deduced from \eqref{eq:lin_stab_boundary} in the limit $k_\perp d_e \sim 1$. To see this, we note that left-hand side of \eqref{eq:lin_stab_boundary} is proportional to $\omega_\text{KAW} \propto k_\parallel k_\perp$, whereas the right-hand side is proportional to $\omega_{*e} \propto k_y$. This means that as $k_\parallel \rightarrow 0$, the left-hand side approaches zero faster than the right-hand side, unless the numerator of the right-hand side similarly approaches zero. This means that both branches of the stability boundary will asymptotically approach:
\begin{align}
    1 - k_\perp^2 d_e^2 \taubar = 0 \quad \Rightarrow \quad  k_\perp d_e = \frac{1}{\sqrt{\taubar}}.
    \label{eq:lin_stab_isobaric_stai}
\end{align}
Thus, there is a thin sliver of stability around the flux-freezing scale \eqref{eq:flux_freezing_scale}, where the isobaric sTAI is quenched as $k_\perp$ is increased. This is due to the effects of finite electron inertia coming into play, and competing with parallel streaming, as explained in \secref{sec:stabilisation_of_isobaric_slab_TAI} and \apref{app:stabilisation_of_isobaric_stai}. \eqref{eq:lin_stab_isobaric_stai} also describes the stabilisation of the exponentially small sETG growth rates that occur below the line \eqref{eq:lin_es_stability_boundary} on the small-scale side of the flux-freezing scale, as shown in figures \ref{fig:lin_em_no_drifts}(a) and \ref{fig:lin_em}(a). In \apref{app:lin_fluid_derivation}, we reproduce the boundaries \eqref{eq:lin_stab_isothermal_stai} and \eqref{eq:lin_stab_isobaric_stai} via `fluid' arguments similar to those used in \secref{sec:stabilisation_of_isothermal_slab_TAI} and \secref{sec:stabilisation_of_isobaric_slab_TAI} in the collisional limit.

\subsubsection{Stability boundary with finite magnetic drifts}
\label{app:with_magnetic_drifts}
\begin{figure}
    
\centering
\hspace{-0.5cm}
\begin{tabular}{cc}
     \includegraphics[width=0.55\textwidth]{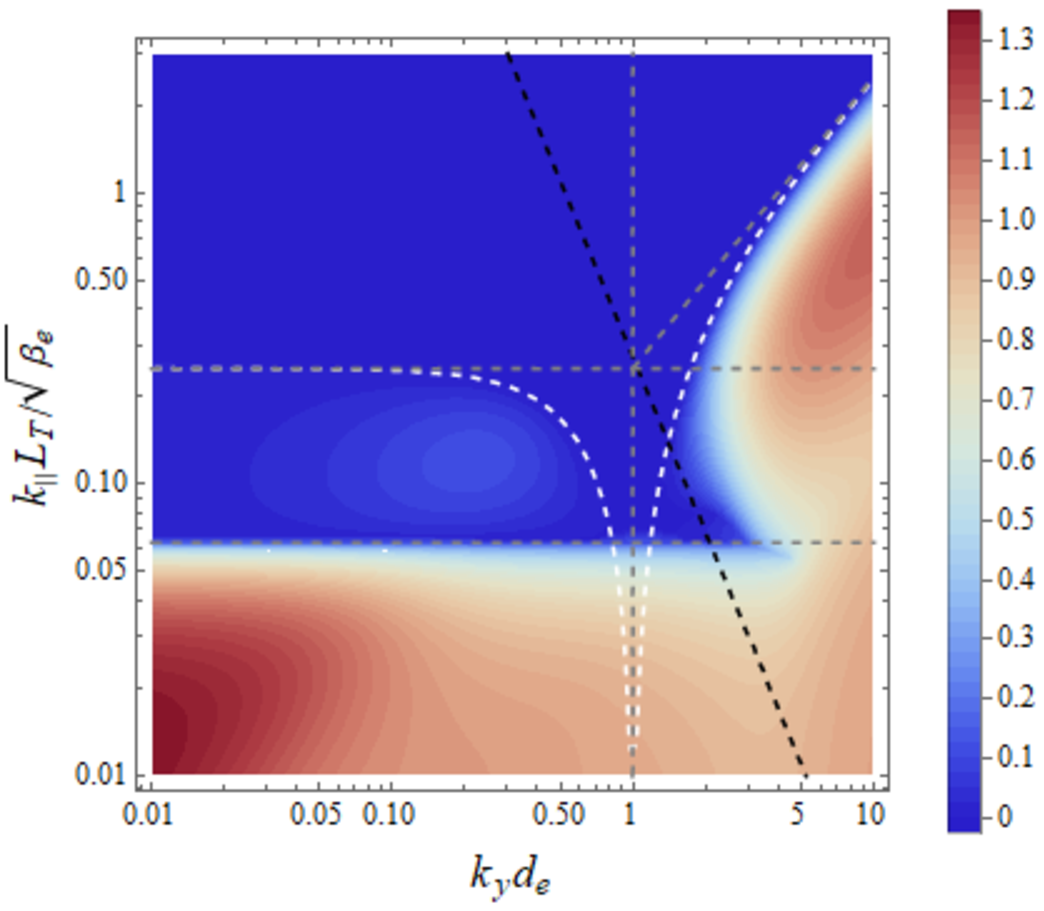}&  
        \includegraphics[width=0.45\textwidth]{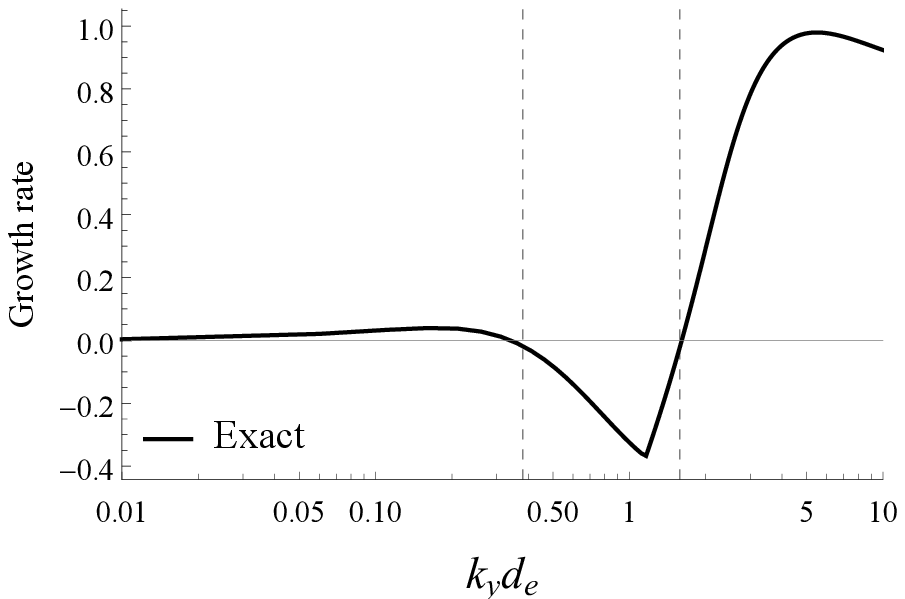} \\\\
     (a) $L_B/L_T = 250$, $k_x d_e = 0$  & (b) $k_\parallel L_T/\sqrt{\beta_e} =0.2 $  \\\\\\
    \includegraphics[width=0.45\textwidth]{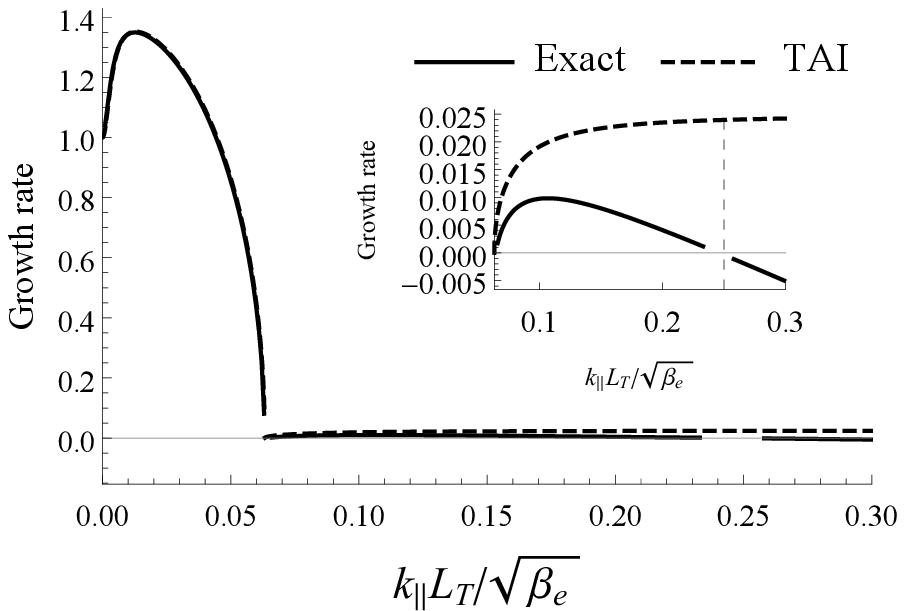}&  
     \includegraphics[width=0.45\textwidth]{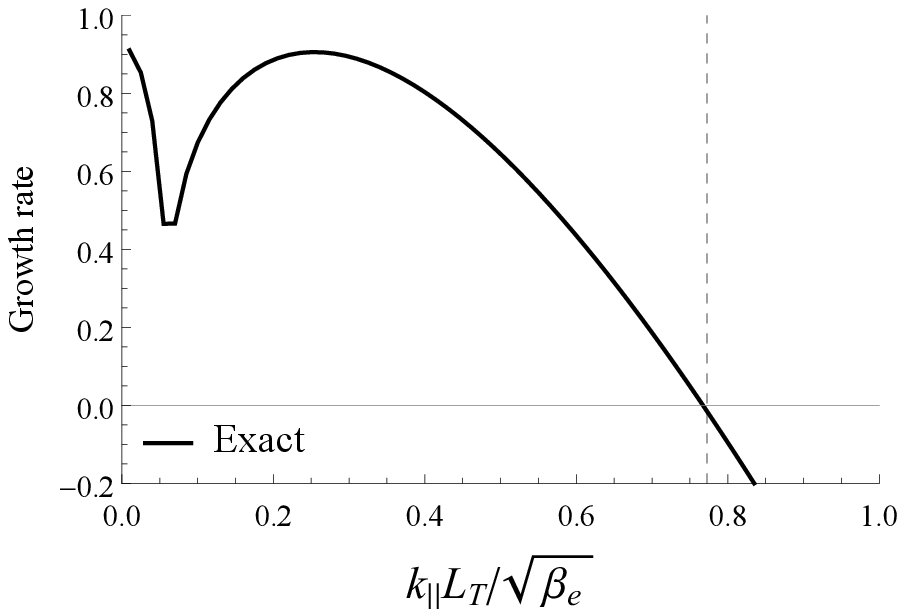} \\\\
          (c)  $k_y d_e = 0.01 $  & (d) $k_y d_e = 3.5$ 
\end{tabular}
 \caption{The same as \figref{fig:lin_em_no_drifts} but with magnetic drifts restored, and now normalised to the cETG growth rate~\eqref{eq:lin_cetg}. We chose a large value of $L_B/L_T$ in order to show the asymptotic regimes clearly. The lower horizontal grey dashed line in panel (a) is $k_\parallel = k_{\parallel c}$, as defined in~\eqref{eq:tai_kpar_critical}. The inset in panel (c) shows the growth rate for $k_\parallel > k_{\parallel c}$; the vertical grey dashed line is~\eqref{eq:lin_stab_isothermal_stai}. The small discontinuity in the growth rate to the left of $k_\parallel = k_{\parallel c}$ in panel (c) is due to the difficulty of resolving such a rapid change in the growth rate over a small range of $k_\parallel$ on a finite grid. From this figure, it is clear that the stability properties of the system at higher $k_\parallel$ are not modified in the presence of finite magnetic drifts. We draw the reader's attention to the enhancement of the cETG growth rate by the cTAI mechanism that can be seen from the red contours in the bottom left-hand corner of panel (a).}
    \label{fig:lin_em}
\end{figure}

Let us now consider how this picture of stability is modified in the presence of magnetic drifts. Though we could, in principle, apply the procedure that resulted in~\eqref{eq:lin_stab_equations} to the full dispersion relation~\eqref{eq:lin_dispersion_relation}, this will not actually yield the correct stability boundary for our system: \eqref{eq:lin_dispersion_relation} is only approximate, owing to the fact that we have expanded the resonant denominators in \eqref{eq:lin_matrices_phi} and \eqref{eq:lin_matrices_a} for $\zeta_d \ll 1$, as in \eqref{eq:lin_resonant_denominator_expansion}, in order to obtain~\eqref{eq:m00_final}-\eqref{eq:n01_final}. While this does not have any significant consequences for the instabilities derived in appendices \ref{app:lin_two_dimensional_perturbations}-\ref{app:lin_electromagnetic_3d_perturbations_tai} --- since they all sit in regimes where this approximation holds --- it does mean that the stability properties of \eqref{eq:lin_dispersion_relation} are not the exact stability properties of the kinetic system.

In particular, \eqref{eq:lin_dispersion_relation} does not retain the (nonlinear) property of gyrokinetics --- inherited by the system of equations derived in \apref{app:derivation_of_low_beta_equations} --- that local gradients of the equilibrium magnetic field cannot inject free energy (see, e.g., \citealt{abel13}). This is because the argument that led to the expression of free-energy conservation \eqref{eq:energetics_total_derivative} relied on the magnetic-drift terms vanishing at every order in the Hermite-Laguerre moment hierarchy \eqref{eq:laguerre_hermite_moments} [see \eqref{eq:energetics_parallel_streaming_and_drifts} and \eqref{eq:integration_by_parts_identity}]; in order to preserve this property in our dispersion relation, we would have to retain the magnetic drifts, without approximation, everywhere, including in the resonant denominators of \eqref{eq:lin_matrices_phi} and \eqref{eq:lin_matrices_a}. Instead, solving \eqref{eq:lin_dispersion_relation} directly leads to spurious growth rates at small parallel wavenumbers, whose magnitudes are inversely proportional to $L_B/L_T$ and vanish only at $\omega_{de} = 0$. However, given that $L_B/L_T$ is large in the strongly driven limit, we find contributions of these growth rates to be everywhere negligible. 

In \figref{fig:lin_em}, we plot the growth rates from the solutions of \eqref{eq:lin_dispersion_relation}. From panel (a), it is clear that the stability properties at large parallel wavenumbers are not significantly modified by the presence of magnetic drifts and both sETG and sTAI are still stabilised along \eqref{eq:lin_stab_setg} (slanted grey dashed line on the right) and \eqref{eq:lin_stab_isothermal_stai} (horizontal grey dashed line), respectively. At lower parallel wavenumbers, sTAI and sETG are replaced by cTAI and cETG, respectively, with their growth rate becoming equal to the cETG growth rate \eqref{eq:lin_cetg} at $k_\parallel = 0$, as evident from panels (c) and (d).  
We draw the reader's attention to the similarity between \figref{fig:lin_em}(a) and the wavenumber-space portrait associated with our collisionless equations (\figref{fig:collisionless_phase_space_portrait}), in that \figref{fig:lin_em}(a) reproduces all the key features that were predicted using the na\"ive estimates of \secref{sec:summary_of_wavenumber_space}.

\subsection{Fluid derivation of collisionless TAI results}
\label{app:lin_fluid_derivation}
In \secref{sec:electromagnetic_regime_tai}, we illustrated the physical mechanisms that led to the instabilities associated with the general TAI dispersion relation~\eqref{eq:tai_dispersion_relation_general}, namely cTAI and sTAI, by considering a series of fluid equations in the collisional limit. While the physical mechanisms in the collisionless limit are almost identical --- with collisional conduction being replaced by parallel particle streaming, $\kappa k_\parallel^2 \rightarrow (2/\sqrt{\pi}) |k_\parallel| \vthe$ --- we seek here to demonstrate these mechanisms explicitly by reproducing many of the key results of \secref{sec:electromagnetic_regime_tai} from a set of equivalent fluid equations in the collisionless limit. 

In particular, we will recover the stabilisation of isothermal and isobaric sTAI via methods similar to those used in sections \ref{sec:stabilisation_of_isothermal_slab_TAI} and \ref{sec:stabilisation_of_isobaric_slab_TAI}. Given that we are interested in sTAI physics --- that occurs at $k_{\parallel} \gg k_{\parallel c}$ [see \eqref{eq:tai_kpar_critical}] --- we shall, in what follows, neglect any incidence of the magnetic drifts. With this simplification, \eqref{eq:lin_matrices_phi} and \eqref{eq:lin_matrices_phi} can be expressed exactly in terms of derivatives of the plasma dispersion function~\eqref{eq:plasma_z_function}: neglecting the density gradient ($\eta_e \rightarrow \infty$), as in \eqref{eq:m00_final}-\eqref{eq:n01_final},
\begin{align}
    M_{\ell,m} & = -\zeta_* \left( \zeta \delta_{0,0} + \frac{1}{\sqrt{2}} \delta_{0,1} \right) + \left\{ \left[\zeta - \zeta_* \left( \zeta^2 - \frac{1}{2} \right) \right] \delta_{0,m} - \zeta_* \delta_{1,m} \right\} \frac{(-1)^m }{\sqrt{2^m m!}} \mathcal{Z}^{(m)}(\zeta), \label{eq:lin_matrices_phi_simplified} \\
      N_{\ell,m} & = 2\zeta \delta_{0,0} -\zeta_* (2 \delta_{1,0} +\sqrt{2} \delta_{0,2}) + 2\zeta M_{\ell,m}, \label{eq:lin_matrices_a_simplified}
\end{align}
where we have used the orthogonality properties \eqref{eq:hermite_polynomials} and \eqref{eq:laguerre_polynomials} of the Hermite-Laguerre basis and the associated recurrence relations \eqref{eq:hermite_recurrence_relations} and \eqref{eq:laguerre_recurrence_relations}, as well as the identity
\begin{align}
    \mathcal{Z}^{(m)} (\zeta) = \frac{\rmd ^m \mathcal{Z}}{\rmd \zeta^m} = \frac{(-1)^m}{\sqrt{\pi}} \int \rmd \hat{v} \: \frac{ e^{-\hat{v}^2}}{\hat{v} - \zeta} H_m(\hat{v}),
    \label{eq:z_function_derivative}
\end{align}
where the integral is once again taken along the Landau contour. 

\subsubsection{Parallel gradient of total parallel temperature}
\label{app:parallel_gradient_of_total_parallel_temperature}
In \secref{sec:electromagnetic_regime_tai}, the parallel gradient of the total temperature along the perturbed field line~\eqref{eq:logt_definition_intro} was a key quantity in understanding the physics associated with the TAI in the collisional limit, and satisfied \eqref{eq:tai_logt_initial}. The equivalent quantity in the collisionless limit is, unsurprisingly, the parallel gradient of the total parallel temperature along the perturbed field line:
\begin{align}
    \gradd_\parallel \log T_{\parallel e} = \gradd_\parallel \frac{\delta T_{\parallel e}}{T_{0e}} - \frac{\rho_e}{L_{T}} \frac{\partial \mathcal{A}}{\partial y}.
    \label{eq:logt_parallel_definition}
\end{align}
Subtracting $\gradd_\parallel$\eqref{eq:tpar_moment}$-(\rho_e/L_{T})\cdot$\eqref{eq:velocity_moment} and using \eqref{eq:communting_identity}, we find the evolution equation for~\eqref{eq:logt_parallel_definition}
\begin{align}
    \frac{\rmd}{\rmd t} \gradd_\parallel \log T_{\parallel e}  + \rho_e \left\{ \frac{\rmd \mathcal{A}}{\rmd t} + \frac{\vthe}{2} \frac{\partial \varphi}{\partial z}, \frac{\delta T_{\parallel e}}{T_{0e}}  \right\} + 2\gradd_\parallel^2 u_{\parallel e} + \frac{\rho_e}{L_T} \frac{\partial}{\partial y} \frac{\rmd}{\rmd t} \frac{u_{\parallel e}}{\vthe} \nonumber \\
    =  - \gradd_{\parallel}^2 \frac{\delta q_{\parallel e}}{n_{0e} T_{0e}} - \frac{\rho_e \vthe}{2 L_{T}} \frac{\partial}{\partial y} \gradd_\parallel \log p_{\parallel e}, \label{eq:tai_logt_parallel_initial}
\end{align}
where we have recognised the parallel derivative of the total parallel pressure
\begin{align}
    \gradd_{\parallel} \log p_{\parallel e} = \gradd_\parallel \frac{\dne}{n_{0e}} + \gradd_\parallel \log T_{\parallel e}.
    \label{eq:logp_parallel_definition}
\end{align}
This is the same as \eqref{eq:tai_logt_initial}, except for the replacements $\dTe \rightarrow \delta T_{\parallel e}, \: -\kappa \gradd_\parallel \log T_e \rightarrow \delta q_{\parallel e}/n_{0e} T_{0e}, \: (2/3)\gradd_\parallel ^2 u_{\parallel e} \rightarrow 2 \gradd_\parallel^2 u_{\parallel e}, \: \nu_{ei} u_{\parallel e} \rightarrow \rmd u_{\parallel e}/\rmd t $, as promised in \secref{sec:general_tai_dispersion_relation}.

The parallel heat flux in \eqref{eq:tai_logt_parallel_initial} must be determined kinetically. In the spirit of 'Landau-fluid' closures (\citealt{hammett90,hammett92,hammett93,dorland93,beer96,snyder97,passot04,goswami05,passot17}), let us seek an expression for $\delta q_{\parallel e}$ in terms of of $\gradd_\parallel \log T_{\parallel e}$. Recalling that $\delta T_{\parallel e}/T_{0e} = \sqrt{2} g_{0,2}$,  $\delta q_{\parallel e}/n_{0e} T_{0e} \vthe = \sqrt{3} g_{0,3}$ and that, using \eqref{eq:plasma_z_function_identities} in \eqref{eq:lin_matrices_phi_simplified} and \eqref{eq:lin_matrices_a_simplified}, 
\begin{align}
     \sqrt{2} M_{02} & = 1 + 2 \left(\zeta^2 - \frac{1}{2} \right) \left[ 1+\zeta \mathcal{Z} - \zeta \zeta_* - \zeta_* \left(\zeta^2 - \frac{1}{2} \right)\mathcal{Z} \right], \label{eq:m02} \\
     N_{02}  & = -\sqrt{2}\zeta_* + 2 \zeta  M_{02}, \label{eq:n02} \\
  \sqrt{3} M_{0,3} & = \zeta \sqrt{2} M_{0,2} - 2 \left[ \zeta - \zeta_* \left( \zeta^2 - \frac{1}{2} \right) \right] (1+ \zeta \mathcal{Z} ),   
  \label{eq:m03} \\
  N_{0,3} & = 2\zeta  M_{0,3}, \label{eq:n03}
\end{align}
we can use \eqref{eq:lin_glm} to write
\begin{align}
    \frac{\delta \tilde{q}_{\parallel e}}{n_{0e} T_{0e} \vthe} = \sqrt{3}\left(\frac{k_\parallel}{|k_\parallel |} M_{0,3} \tilde{\varphi} - N_{0,3} \tilde{\mathcal{A}} \right) = \frac{k_\parallel}{|k_\parallel |} \left( \tilde{\varphi} - \frac{k_\parallel}{|k_\parallel |} 2\zeta \tilde{\mathcal{A}} \right) \sqrt{3} M_{0,3}.
    \label{eq:lin_qpar_moments}
\end{align}
Similarly, $\gradd_\parallel \log T_{\parallel e}$ can, in Fourier space, be written as 
\begin{align}
    \left( \gradd_\parallel \log T_{\parallel e} \right)_{\vec{k}} = i k_\parallel \frac{\delta \tilde{T}_{\parallel e}}{T_{0e}} - i \frac{k_y \rho_e}{L_T} \tilde{\mathcal{A}} = ik_\parallel \left(\tilde{\varphi} - \frac{k_\parallel}{|k_\parallel |} 2 \zeta \tilde{\mathcal{A}} \right)\sqrt{2} M_{0,2}.
    \label{eq:lin_gpar_logt_moments}
\end{align}
Combining \eqref{eq:lin_qpar_moments} and \eqref{eq:lin_gpar_logt_moments}, we obtain the desired expression for the heat-flux in terms of the parallel gradient of the total parallel temperature:
\begin{align}
    \frac{\delta \tilde{q}_{\parallel e}}{n_{0e} T_{0e} \vthe}  = - \frac{1}{|k_\parallel|} \mu(\zeta) \left( \gradd_\parallel \log T_{\parallel e} \right)_{\vec{k}},
    \label{eq:lin_qpar}
 \end{align}
where the collisionless heat-conduction coefficient is
\begin{align}
    \mu(\zeta) = i\frac{\sqrt{3} M_{0,3}}{\sqrt{2}M_{0,2}} = i \left[\zeta - \frac{1+\zeta \mathcal{Z}}{\zeta + \left( \zeta^2 - \frac{1}{2} \right) \mathcal{Z}} \right] \approx \left\{
    \begin{array}{cc}
    \displaystyle \frac{2}{\sqrt{\pi}} , & \displaystyle \zeta \ll 1,\\[4mm]
    \displaystyle - \frac{3i}{2\zeta}, & \displaystyle \zeta \gg 1.
    \end{array}
    \right.
    \label{eq:lin_diffusion_coefficient}
\end{align}
This is identical to the Landau-fluid closure derived in \cite{wang19}, and is formally valid over the entire range of frequencies $\zeta$. 

Let us now return to \eqref{eq:tai_logt_parallel_initial}. Under the ordering \eqref{eq:lin_electromagnetic_ordering}, its left-hand side is negligible in its entirety, meaning that
\begin{align}
    -\gradd_\parallel^2 \frac{\delta q_{\parallel e}}{n_{0e} T_{0e}} = \frac{\rho_e \vthe}{2 L_{T}} \frac{\partial}{\partial y} \gradd_\parallel \log p_{\parallel e}.
    \label{eq:lin_tai_qpar_balance}
\end{align}
As in the collisional case, the competition between these two terms is controlled by $\xi_*$, now defined by \eqref{eq:lin_tai}.
Using \eqref{eq:logp_parallel_definition}, \eqref{eq:lin_qpar} and \eqref{eq:lin_diffusion_coefficient} (the latter for $\zeta \ll 1$), \eqref{eq:lin_tai_qpar_balance} can be recast as an expression for $\gradd_\parallel \log T_{\parallel e}$ in terms of the parallel gradient of the density perturbation:
\begin{align}
    \left( \gradd_\parallel \log T_{\parallel e} \right)_{\vec{k}} = -\frac{i\xi_*}{1+ i\xi_*} \left( \gradd_\parallel \frac{\delta n_e}{n_{0e}} \right)_{\vec{k}} ,
    \label{eq:lin_logt_final}
\end{align}
where $\xi_*$ is defined in \eqref{eq:lin_tai}. This is the collisionless equivalent of \eqref{eq:tai_logt_final_explicit}, which reduces to~\eqref{eq:stai_logt_correction_isothermal} and~\eqref{eq:stai_log_t_correction_isobaric} in the isothermal ($\xi_* \ll 1$) and isobaric ($\xi_* \gg 1$) limits, respectively. We have thus demonstrated how both isothermal and isobaric sTAI, given by \eqref{eq:stai_isothermal_limit} and \eqref{eq:stai_isobaric_limit}, respectively, arise as corrections to isothermality and isobaricity not only in the collisional limit, but in the collisionless one as well. 

\subsubsection{Stabilisation of isothermal sTAI}
\label{app:stabilisation_of_isothermal_stai}
As discussed in \secref{sec:stabilisation_of_isothermal_slab_TAI}, the isothermal sTAI~\eqref{eq:stai_isothermal_limit} is eventually quenched by the compressional heating term in the temperature equation \eqref{eq:tpar_moment} that begins to compete with the TAI drive.

To show this, let us adopt, instead of \eqref{eq:lin_electromagnetic_ordering}, the ordering
\begin{align}
    \omega \sim \omega_{*e} \ll k_\parallel \vthe \quad \Leftrightarrow \quad \zeta \sim \zeta_* \ll 1,
    \label{eq:lin_stai_stab_isothermal_ordering}
\end{align}
but still consider perturbations above the flux-freezing scale, $k_\perp d_e  \ll 1$.
In this limit, the system is still isothermal to leading order in $\xi_* \ll 1$, but now we must also retain the compressional heating term in \eqref{eq:tai_logt_parallel_initial} to determine $\gradd_\parallel \log T_{\parallel e}$ at next order: instead of \eqref{eq:lin_tai_qpar_balance}, we have, therefore,
\begin{align}
    - \left(\gradd_\parallel^2 \frac{\delta q_{\parallel e}}{n_{0e} T_{0e}} \right)_{\vec{k}} = - \frac{2}{\sqrt{\pi}}|k_\parallel | \vthe \left( \gradd_\parallel \log T_{\parallel e} \right)_{\vec{k}}  = \left( \frac{\rho_e \vthe}{2L_T} \frac{\partial}{\partial y} \gradd_\parallel \frac{\delta n_e}{n_{0e}} + 2 \gradd_\parallel^2 u_{\parallel e} \right)_{\vec{k}},
    \label{eq:lin_stai_stab_isothermal_logt}
\end{align}
where we have used \eqref{eq:lin_diffusion_coefficient} for $\zeta \ll 1$. Combining \eqref{eq:lin_stai_stab_isothermal_logt} with the equations for density and parallel momentum, still the same as \eqref{eq:kaw_equation_stai}, we obtain the following dispersion relation
\begin{align}
    \omega^2 - \omega_\text{KAW}^2(1+\taubar - i\xi_*) =  -i \sqrt{\pi} \frac{\omega}{|k_\parallel| \vthe} \omega_\text{KAW}^2.
    \label{eq:lin_stai_stab_isothermal_dispersion}
\end{align}
This is the same as \eqref{eq:kaw_stai_dispersion_relation} apart from the right-hand side, previously neglected. At the stability boundary, the frequency $\omega$ must be purely real, and both the real and imaginary parts of \eqref{eq:lin_stai_stab_isothermal_dispersion} must vanish individually, giving
\begin{align}
    \omega^2 = \omega_\text{KAW}^2 (1+\taubar), \quad \omega = - \frac{\omega_{*e}}{2} \quad \Rightarrow \quad \mp \omega_\text{KAW} \sqrt{1+\taubar} = \frac{\omega_{*e}}{2}.
    \label{eq:lin_stai_stab_isothermal_boundary}
\end{align}
This is \eqref{eq:lin_stab_isothermal_stai}. This stabilisation of isothermal sTAI was not captured by the TAI dispersion relation \eqref{eq:lin_tai} because the ordering \eqref{eq:lin_electromagnetic_ordering} did not formally allow frequencies comparable to $\omega_{*e}$, required by \eqref{eq:lin_stai_stab_isothermal_boundary}.

\subsubsection{Stabilisation of isobaric sTAI}
\label{app:stabilisation_of_isobaric_stai}
As discussed in \secref{sec:stabilisation_of_isobaric_slab_TAI}, the isobaric sTAI~\eqref{eq:stai_isobaric_limit} is stabilised within a certain region of wavenumber space, due to the effects of finite electron inertia in the parallel momentum equation \eqref{eq:velocity_moment}. 

To work out this stabilisation, let us consider, instead of \eqref{eq:lin_electromagnetic_ordering}, the ordering
\begin{align}
    \omega\sim k_\parallel \vthe \ll \omega_{*e} \quad \Leftrightarrow \quad  \zeta \sim 1 \ll \zeta_*,
    \label{eq:lin_stai_stab_isobaric_ordering}
\end{align}
while allowing perpendicular wavenumbers to sample the flux-freezing scale, $k_\perp d_e \sim 1$. A direct consequence of this ordering is that one has to retain the electron inertia in the leading-order parallel momentum equation, viz., the second equation in~\eqref{eq:kaw_equation_isobaric_stai} is replaced with
\begin{align}
    \frac{\rmd \mathcal{A}}{\rmd t} + \frac{\vthe}{2} \frac{\partial \varphi}{\partial z} = \frac{\vthe}{2} \gradd_\parallel \log p_{\parallel e} + \frac{\rmd}{\rmd t} \frac{u_{\parallel e}}{\vthe}.
    \label{eq:lin_stai_stab_isobaric_momentum}
\end{align}
This means that, instead of the system being isobaric to leading order in $\xi_* \gg 1$, the parallel pressure gradient now balances the electron-inertial force:
\begin{align}
    \gradd_\parallel \log p_{\parallel e} + \frac{2}{\vthe} \frac{\rmd}{\rmd t} \frac{u_{\parallel e}}{\vthe} = 0.
    \label{eq:lin_stai_stab_isobaric_leading_order}
\end{align}
This is obvious from \eqref{eq:tai_logt_parallel_initial} in the limit \eqref{eq:lin_stai_stab_isobaric_ordering}. To the next order in this limit, we must retain both the time derivative of $\gradd_\parallel \log T_{\parallel e}$ and the compressional-heating term in \eqref{eq:tai_logt_parallel_initial}:
\begin{align}
    & \left[\frac{\rho_e \vthe}{2L_T} \frac{\partial}{\partial y} \left(\gradd_\parallel \log p_{\parallel e} + \frac{2}{\vthe} \frac{\rmd}{\rmd t} \frac{u_{\parallel e}}{\vthe} \right) \right]_{\vec{k}} \nonumber \\
    & = \left( \frac{\rmd}{\rmd t} + \mu(\zeta) |k_\parallel| \vthe \right) \left( \gradd_\parallel \frac{\delta n_e}{n_{0e}} + \frac{2}{\vthe} \frac{\rmd}{\rmd t} \frac{u_{\parallel e}}{\vthe} \right)_{\vec{k}} - \left(2 \gradd_\parallel^2 u_{\parallel e} \right)_{\vec{k}}.
    \label{eq:lin_stai_stab_isobaric_next_order}
\end{align}
Combining \eqref{eq:lin_stai_stab_isobaric_momentum}, \eqref{eq:lin_stai_stab_isobaric_next_order} and the density equation from \eqref{eq:kaw_equation_isobaric_stai}, we find the dispersion relation
\begin{align}
    \omega^2 - \omega_\text{KAW}^2 \left( \taubar + \frac{1}{i\xi_*} \right) = - \frac{1}{i\xi_*} k_\perp^2 d_e^2 \omega^2  - \left( 3 \omega_\text{KAW}^2 - k_\perp^2 d_e^2 \omega \right) \frac{\omega}{\omega_{*e}},
    \label{eq:lin_stai_stab_isobaric_dispersion}
\end{align}
where, since $\zeta \sim 1$, we have here defined $\xi_* = \omega_{*e}/(\mu |k_\parallel| \vthe )$. This is the same as \eqref{eq:stai_dispersion_relation_isobaric}, apart from the right-hand side, previously neglected, and up to the definition of $\xi_*$. The second term on the right-hand side simply leads to a small, in $\xi_* \ll 1$, modification of the (real) frequency, and so can be neglected. 

As usual, at the stability boundary, the frequency $\omega$ must be purely real, and both the real and imaginary parts of \eqref{eq:lin_stai_stab_isobaric_dispersion} must vanish individually, giving 
\begin{align}
    \omega^2 = \omega_\text{KAW}^2 \taubar, \quad k_\perp^2 d_e^2 \omega^2  = \omega_\text{KAW}^2 \quad \Rightarrow \quad k_\perp^2 d_e^2 = \frac{1}{\taubar}.
    \label{eq:lin_stai_stab_isobaric_boundary}
\end{align}
This is \eqref{eq:lin_stab_isobaric_stai}. As with the case of the isothermal sTAI, this stabilisation was not captured by the general TAI dispersion relation \eqref{eq:lin_tai} because the ordering \eqref{eq:lin_electromagnetic_ordering} did not formally allow frequencies comparable to the parallel streaming rate, required by \eqref{eq:lin_stai_stab_isobaric_boundary}.

\section{Collisional linear theory}
\label{app:collisional_linear_theory}
We begin by linearising and Fourier-transforming our equations for the density \eqref{eq:col_density_equation}, velocity \eqref{eq:col_velocity_equation} and temperature \eqref{eq:col_temperature_equation} in the collisional limit:
\begin{align}
   & \left[\frac{\omega}{\taubar} -2 \left(1+ \frac{1}{\taubar} \right)\omega_{de} +\frac{1}{\eta_e} \omega_{*e} \right]\tilde{\varphi} -  k_\parallel \vthe (k_\perp d_e)^2 \tilde{\mathcal{A}} + 2 \omega_{de} \frac{\delta \tilde{T}_e}{T_{0e}} = 0, \label{eq:density_moment_collisional_fourier_full} \\
   & \left[\omega - \left(1+ \frac{1}{\eta_e} \right)\omega_{*e} + i (k_\perp d_e)^2 \nu_{ei} \right] \tilde{\mathcal{A}} + \frac{k_\parallel \vthe}{2} \left[ - \left( 1+ \frac{1}{\taubar} \right) \tilde{\varphi}+ \frac{\delta \tilde{T}_e}{T_{0e}} \right] =0, 
   \label{eq:velocity_moment_collisonal_fourier_full} \\
   & \left(\omega  - \frac{14}{3} \omega_{de} +  i \kappa k_\parallel^2  \right) \frac{\delta \tilde{T}_e}{T_{0e}} + \frac{2}{k_\parallel \vthe}\left[\frac{1}{3} (k_\parallel  \vthe k_\perp d_e )^2 - \kappa k_\parallel^2 i \omega_{*e} \right] \tilde{\mathcal{A}}  \nonumber\\
      & \quad\quad\quad\quad\quad\quad\quad\quad\quad\quad\quad  - \left[\omega_{*e}  - \frac{4}{3} \left(1+ \frac{1}{\taubar} \right) \omega_{de}\right]\tilde{\varphi} = 0, 
   \label{eq:t_moment_collisional_fourier_full}
\end{align}
where tildes indicate the Fourier components of the fields, $\eta_e = L_n/L_T$, and we have used \eqref{eq:logt_definition_intro} in order to express $\gradd_\parallel \log T_e$ in terms of $\delta T_e/T_{0e}$ and $\mathcal{A}$, as well as \eqref{eq:quasineutrality_final} with $g_i = 0$ in order to express $\delta n_e/n_{0e}$ in terms of $\varphi$. Note that \eqref{eq:density_moment_collisional_fourier_full}-\eqref{eq:t_moment_collisional_fourier_full} are formally only valid in the adiabatic-ion limit \eqref{eq:adiabatic_ions_appendix}, i.e., at $k_\perp \rho_i \gg 1$. The dispersion relation is
\begin{align}
    & \left[ \omega -   \left(1+ \frac{1}{\eta_e} \right)\omega_{*e} + i(k_\perp d_e)^2 \nu_{ei} \right]\left( M_{\varphi \varphi} M_{TT} -M_{\varphi T}M_{T \varphi}  +  i \kappa k_\parallel^2 M_{\varphi \varphi} \right) \nonumber \\
    & \spc - \left(\frac{2}{3}\omega_\text{KAW}^2 - \kappa k_\parallel^2 i \omega_{*e} \right)\left[ M_{\varphi \varphi} + \left(1 + \frac{1}{\taubar} \right) M_{\varphi T} \right] \nonumber \\
    & \spc - \left(1 + \frac{1}{\taubar} \right) \omega_\text{KAW}^2 \left( M_{TT} +  i \kappa k_\parallel^2 \right) + \omega_\text{KAW}^2 M_{T \varphi} = 0 ,
    \label{eq:finite_lb_dispersion_relation_cubic}\end{align}
where we have defined the coefficients independent of $k_\parallel$ by
\begin{align}
    M_{\varphi \varphi} & = \frac{\omega}{\taubar} -2 \left(1+ \frac{1}{\taubar} \right)\omega_{de} +\frac{1}{\eta_e} \omega_{*e} , \label{eq:finite_lb_matrix_elements}\\
    M_{TT}  &= \omega  - \frac{14}{3} \omega_{de},  \nonumber \\
    M_{T \varphi} & = -\omega_{*e}  + \frac{4}{3} \left(1+ \frac{1}{\taubar} \right) \omega_{de}, \nonumber\\
      M_{\varphi T}  & =2  \omega_{de}, \nonumber
\end{align}
and $\omega_{*e}$, $\omega_{de}$ and $\omega_\text{KAW}$ are as defined in \eqref{eq:definition_timescales}. Though an exact solution of the cubic \eqref{eq:finite_lb_dispersion_relation_cubic} is, in principle, possible to write explicitly, it is not particularly useful or enlightening in its full generality. Therefore, we shall consider various asymptotic limits of \eqref{eq:finite_lb_dispersion_relation_cubic} in order to highlight the important aspects of the linear physics supported by our reduced system of equations, as we did in \apref{app:collisionless_linear_theory}.

\subsection{Two-dimensional perturbations}
\label{app:lin_col_two_dimesional_perturbations}
Let us first consider purely two-dimensional perturbations --- which amounts to setting $k_\parallel = 0$ everywhere --- without ordering $k_\perp d_e \chi$ with respect to unity.
In this case, \eqref{eq:finite_lb_dispersion_relation_cubic} reduces instantly to 
\begin{align}
    \left[ \omega -   \left(1+ \frac{1}{\eta_e} \right)\omega_{*e} + i(k_\perp d_e)^2 \nu_{ei} \right] \left(M_{\varphi \varphi} M_{TT} -M_{\varphi T}M_{T \varphi} \right) = 0.
    \label{eq:finite_lb_dispersion_relation_2d}
\end{align}
The dispersion relation \eqref{eq:finite_lb_dispersion_relation_2d} admits two solutions.

\subsubsection{Magnetic drift wave}
\label{app:magnetic_drift_wave}
From the first bracket in \eqref{eq:finite_lb_dispersion_relation_2d}, we have 
\begin{align}
    \omega = \left(1+ \frac{1}{\eta_e} \right)\omega_{*e} - i (k_\perp d_e)^2\nu_{ei},
    \label{eq:lin_col_mdw_damped}
\end{align}
which is a (damped) version of the 'magnetic drift wave' described in \apref{app:lin_magnetic_drift_wave}, a purely magnetic oscillation involving the balance between the inductive part of the parallel electric field, the gradient of the equilibrium pressure along the perturbed field line, and the resistive force in \eqref{eq:col_velocity_equation} [or indeed \eqref{eq:velocity_moment_collisional}]:
\begin{align}
    \frac{\partial \mathcal{A}}{\partial t} = - \frac{\rho_e \vthe}{2}\left( \frac{1}{L_n} + \frac{1}{L_T} \right) \frac{\partial \mathcal{A}}{\partial y} + \nu_{ei} d_e^2 \gradd_\perp^2 \mathcal{A}.
    \label{eq:lin_col_em_mdw_equations}
\end{align}
By setting $k_\parallel = 0$, we have decoupled perturbations of the magnetic field --- or, in the electrostatic regime, of the parallel velocity --- from those of the density and temperature, as in \apref{app:lin_magnetic_drift_wave}. Note that this mode can potentially go unstable at lower collisionality, where it is sometimes referred to as a slab micro-tearing mode (see, e.g., \citealt{hassam80b,drake80,larakers20}) --- this is discussed in \apref{app:choice_of_collision_operator}.

\subsubsection{Finite critical gradients}
\label{app:finite_critical_gradients}
Focusing on the second bracket in \eqref{eq:finite_lb_dispersion_relation_2d}, and solving for the growth rate $\gamma = \text{Im}(\omega)$, we find 
\begin{align}
    \gamma = \pm \sqrt{2 \omega_{de} \omega_{*e} \taubar} \sqrt{1+ \frac{1}{2\eta_e} \left(\taubar - \frac{4}{3} \right) - \frac{\taubar}{8\eta_e^2}  \frac{L_B}{L_T}  - \frac{1}{2} \left( \taubar+ \frac{40}{9} \frac{1}{\taubar} \right) \frac{L_T}{L_B} }.
    \label{eq:finite_lb_growth_rate}
\end{align}
Clearly, in order to have an instability, we need the expression under the square root to be positive-definite, which gives us a condition on $L_B/L_T$:
\begin{align}
    \left( \frac{L_B}{L_T} \right)_- < \frac{L_B}{L_T} < \left( \frac{L_B}{L_T} \right)_+,
    \label{eq:finite_lb_critical_gradient}
\end{align}
where the critical gradients are
\begin{align}
    \left( \frac{L_B}{ L_T} \right)_\pm = 2\eta_e^2 \left[ \frac{2}{\taubar} + \frac{1}{\eta_e}\left(1 - \frac{4}{3\taubar} \right) \pm \frac{2}{\taubar} \sqrt{1 + \frac{1}{\eta_e} \left(\taubar - \frac{4}{3} \right) - \frac{2}{3\eta_e^2} (1+\taubar)}\right].
    \label{eq:finite_lb_upper_and_lower}
\end{align}
These solutions exist only if the temperature gradient is sufficiently steep compared to the density gradient: demanding that the expression under the square root in \eqref{eq:finite_lb_upper_and_lower} be positive semi-definite, we find that $\eta_e$ must satisfy
\begin{align}
    \eta_e  \geqslant\frac{4(1+\taubar)}{3\taubar -4 + \sqrt{40+9 \taubar^2}}.
    \label{eq:finite_lb_eta_condition}
\end{align}
This result is consistent with the long-established understanding that the critical temperature gradient for the ETG instability is proportional to the density gradient (\citealt{jenko01}). 

In the limit of $\eta_e \rightarrow \infty$, we find that \eqref{eq:finite_lb_critical_gradient} becomes
\begin{align}
  \frac{L_B}{L_T} > \frac{1}{2} \left( \taubar + \frac{40}{9} \frac{1}{\taubar^2} \right).
  \label{eq:finite_lb_critical_gradient_infinite_eta}
\end{align}
For $\taubar \sim 1$, this lower bound is a quantity of order unity. Therefore, the condition~\eqref{eq:finite_lb_critical_gradient_infinite_eta} can perhaps be readily satisfied in steep-temperature-gradient regions, such as the tokamak edge/pedestal. It is not a forgone conclusion, however, that the limit of $\eta_e \rightarrow \infty$ can be achieved in experimentally relevant conditions. Though an emerging paradigm for JET-ILW (ITER-like wall) pedestal transport appears to be that the ILW conditions modify the pedestal density in ways that preferentially decrease its gradient (\citealt{hatch19,ham21}) --- thereby increasing both $\eta_e$ and $\eta_i = L_n/L_{T_i}$ --- other recent studies have found that the average value of $\eta_e$ in the pedestal appears to saturate at $\eta_e \sim 1-2$ during the inter-ELM (edge-localised mode) period (\citealt{field20,guttenfelder21}). Within the latter context, the stiff heat-flux scalings derived in \secref{sec:free_energy_and_turbulence} can be viewed as an argument against electron-temperature gradients much above the critical linear threshold being achievable. Whether these considerations are relevant in the nonlinear context will be addressed in future work.

\subsubsection{Curvature-mediated ETG instability}
\label{app:curvature_mediated_ETG}
For the remainder of \apref{app:collisional_linear_theory}, we specialise to the strongly driven limit (see \apref{app:strongly_driven_limit}), wherein $L_B/L_T$ is assumed to be sufficiently far above the lower bound \eqref{eq:finite_lb_critical_gradient_infinite_eta}. Then, \eqref{eq:finite_lb_growth_rate} reduces to the cETG growth rate \eqref{eq:cetg_gamma}, viz.,
\begin{align}
    \gamma = \pm \sqrt{2 \omega_{de} \omega_{*e} \taubar} \quad \Leftrightarrow \quad \omega =\pm i (2 \omega_{de} \omega_{*e} \taubar )^{1/2},
    \label{eq:lin_col_es_cetg}
\end{align}
which is now a purely growing mode that is formally unstable at all values of the equilibrium temperature gradient.

Furthermore, the dispersion relation \eqref{eq:finite_lb_dispersion_relation_cubic} becomes significantly simplified in this limit: neglecting $\omega_{de}$ and $\omega_{*e}/\eta_e$ where they are directly compared with larger terms (viz., with $\omega$ or $\omega_{*e}$) in \eqref{eq:finite_lb_matrix_elements}, one finds
\begin{align}
  & \left[ \omega - \omega_{*e} + i (k_\perp d_e)^2 \nu_{ei} \right]\left( \omega^2  + i \kappa k_\parallel^2 \omega + 2 \omega_{de} \omega_{*e} \taubar \right) \label{eq:lin_col_dispersion_relation}\\
    & -\left[ \frac{2}{3} \omega_\text{KAW}^2 - \kappa k_\parallel^2 i \omega_{*e} \right] \left[ \omega + 2 \omega_{de} (1+\taubar)\right]  - \omega_\text{KAW}^2 \left[ \left( \omega +i\kappa k_\parallel^2 \right)(1+\taubar) - \omega_{*e} \taubar \right] = 0. \nonumber 
 \end{align}
This dispersion relation could also have been obtained directly from the linearisation of the collisional, strongly-driven system of equations \eqref{eq:density_moment_collisional}-\eqref{eq:t_moment_collisional}. Note that it is important to retain the magnetic-drift term $\omega_{de}$ in the fourth bracket: despite it being formally smaller than the frequency $\omega$ with which it shares that bracket, it is required for some leading-order cancellations in certain limits. In what follows, we shall neglect magnetic-drift terms where they are not multiplied by $\omega_{*e}$, consistently with the strongly driven limit.

Both two-dimensional modes \eqref{eq:lin_col_mdw_damped} and \eqref{eq:lin_col_es_cetg} persist at all perpendicular wavenumbers because there is no distinction between the electrostatic and electromagnetic regimes for purely two-dimensional phenomena. Restoring finite $k_\parallel$, however, significantly alters this behaviour, as it allows coupling between perturbations of the magnetic field and those of density and temperature, which introduces new instabilities in both the electrostatic and electromagnetic regimes. 

\subsection{Electrostatic 3D perturbations: collisional sETG}
\label{app:lin_col_electrostatic_regime}
Let us consider perturbations below the flux-freezing scale \eqref{eq:flux_freezing_scale_col}, viz., with
\begin{align}
    k_\perp d_e \chi \gg 1 \quad \Leftrightarrow \quad (k_\perp d_e)^2 \nu_{ei} \gg \omega_{*e}. 
    \label{eq:lin_col_electrostatic_limit}
\end{align}
Then, given that we always have $\omega \lesssim \omega_{*e}$, the first two terms in the first bracket of \eqref{eq:lin_col_dispersion_relation} can be neglected in comparison to the third. Similarly, noting that, from \eqref{eq:lin_col_electrostatic_limit}, 
\begin{align}
    \omega_\text{KAW}^2  = \frac{(k_\parallel \vthe)^2}{2\nu_{ei}} (k_\perp d_e)^2 \nu_{ei} = \frac{1}{a}\kappa k_\parallel^2 (k_\perp d_e)^2\nu_{ei} \gg \kappa k_\parallel^2 \omega_{*e},
    \label{eq:lin_col_ignore_terms}
\end{align}
where $\constant = 2 \nu_{ei} \kappa /\vthe^2 = 5\nu_{ei}/9\nu_e = 5/[9(1+1/Z)]$,
the second term in the third bracket can also be neglected. Dividing throughout by $(k_\perp d_e)^2 \nu_{ei}$, we can therefore write \eqref{eq:lin_col_dispersion_relation} as a quadratic in $\omega$:
\begin{align}
    \omega^2 & + \left( \taubar + \constant + \frac{5}{3} \right)i \frac{(k_\parallel \vthe)^2}{2 \nu_{ei}} \omega + 2 \omega_{de}\omega_{*e} \taubar     \nonumber\\
    &- \left[  (1+\taubar) \constant\frac{(k_\parallel \vthe)^2}{2 \nu_{ei}}  + i \omega_{*e} \taubar\right]\frac{(k_\parallel \vthe)^2}{2 \nu_{ei}} = 0. \label{eq:lin_col_es_dispersion_relation}
\end{align}
The originally cubic dispersion relation has reduced to a quadratic because our collisional system of equations becomes a two-field system ($\varphi$ and $\dTe$) in the electrostatic limit, with $\mathcal{A}$ no longer a dynamic field: neglecting $\omega$ in the first bracket of \eqref{eq:lin_col_dispersion_relation} is equivalent to neglecting the inductive part of the parallel electric field in \eqref{eq:velocity_moment_collisional}, meaning that $\mathcal{A}$ is determined instantaneously from the parallel pressure balance. 

\begin{figure}
    
\centering
\begin{tabular}{cc}
     \includegraphics[width=0.5\textwidth]{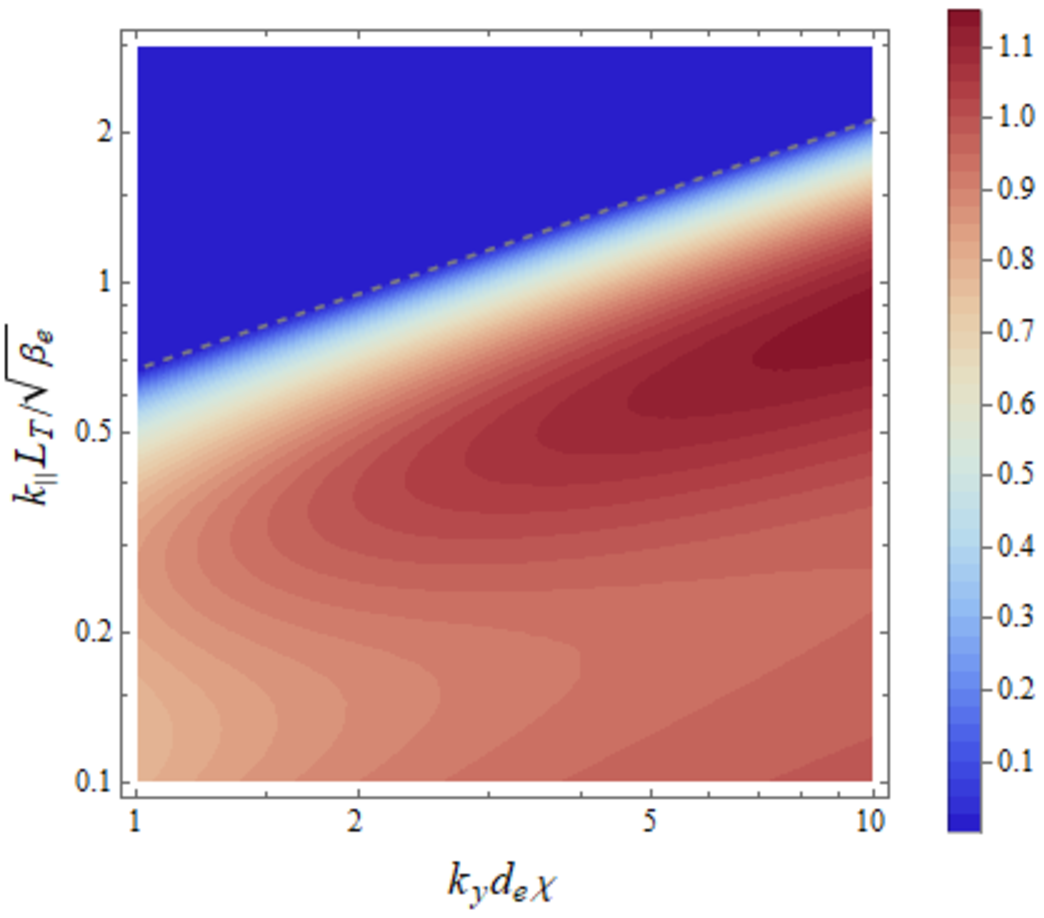}&  
     \includegraphics[width=0.45\textwidth]{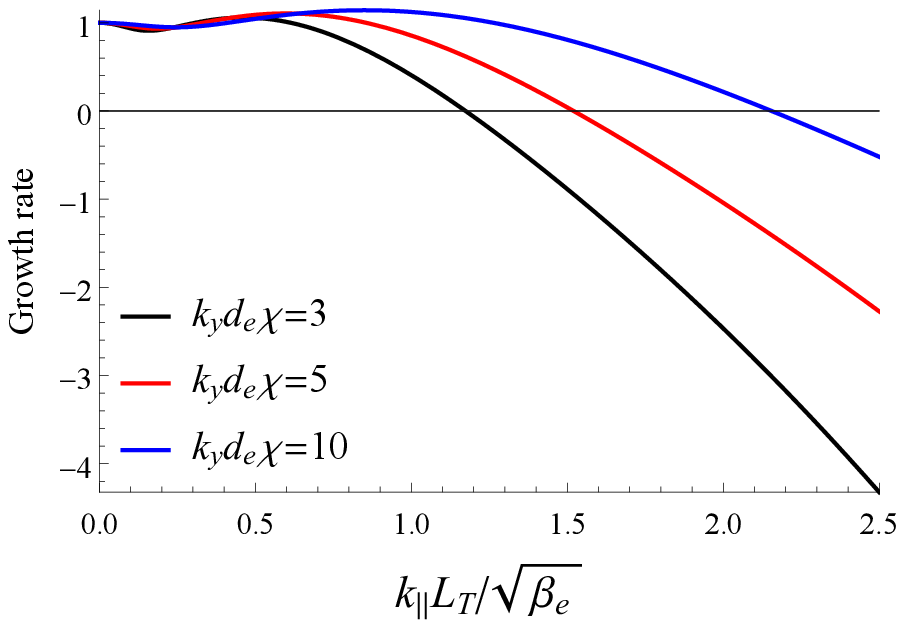} \\\\
     (a) $L_B/L_{T}=250$, $k_x d_e = 0$  & (b) $L_B/L_{T}=250$, $k_x d_e = 0$ 
\end{tabular}
 \caption{Growth rate of the ETG instability in the collisional, electrostatic regime: these are solutions of \eqref{eq:lin_col_es_dispersion_relation} with $\taubar = 1$. Panel (a) is a contour plot of the positive growth rates ($\gamma >0$) in the $(k_y,k_\parallel )$ plane; panel (b) shows the growth rate plotted as a function of $k_\parallel L_{T}/\sqrt{\beta_e}$. We have normalised to the cETG growth rate \eqref{eq:lin_col_es_cetg} in both cases. The stability boundary \eqref{eq:lin_col_es_stability_boundary} is indicated by the grey dashed line in panel (a). We chose a very large value of $L_B/L_T$ in order to show the asymptotic regimes clearly.}
    \label{fig:lin_col_es_plots}
\end{figure}

If, in addition to \eqref{eq:lin_col_electrostatic_limit}, we consider the limit of short parallel wavelengths, which amounts to ignoring the magnetic-drift terms everywhere, viz., 
\begin{align}
  \omega_{de}  \ll \frac{(k_\parallel \vthe)^2}{\nu_{ei}}  \ll \omega \ll \omega_{*e},
    \label{eq:lin_col_es_short_wavelength_ordering}
\end{align}
then the balance of the first and last terms in \eqref{eq:lin_col_es_dispersion_relation} gives us
\begin{align}
    \omega^2 = i \omega_{*e} \frac{(k_\parallel \vthe)^2}{2\nu_{ei}} \taubar \quad \Rightarrow \quad \omega = \pm \frac{1-i\sgn(k_y)}{\sqrt{2}}\left(\frac{k_\parallel^2 \vthe^2 |\omega_{*e}| \taubar}{2\nu_{ei}} \right)^{1/2}.
     \label{eq:lin_col_es_setg}
\end{align}
We recognise this as the collisional sETG growth rate \eqref{eq:col_setg_gamma}, which we would expect to recover in the electrostatic regime (magnetic field lines and electron flows are liberated from one another as flux is unfrozen by, in this case, resistivity). 

At short enough parallel wavelengths, however, the sETG instability is quenched by rapid thermal conduction that leads to the damping of the associated temperature perturbation. To see this, we relax the assumption \eqref{eq:lin_col_es_short_wavelength_ordering} and consider the exact stability boundary of~\eqref{eq:lin_col_es_dispersion_relation}: assuming that $\omega$ is purely real, the real and imaginary parts of \eqref{eq:lin_col_es_dispersion_relation} are, respectively, 
\begin{align}
    \omega^2 +2 \omega_{de} \omega_{*e} \taubar - (1+\taubar) \constant \frac{\left( k_\parallel \vthe \right)^4}{(2\nu_{ei})^2}  = 0 , \quad  \left(  \taubar + \constant + \frac{5}{3} \right) \omega - \omega_{*e} \taubar = 0.
    \label{eq:lin_col_es_stability_boundary_real_and_imaginary}
\end{align}
Given that the second equation in \eqref{eq:lin_col_es_stability_boundary_real_and_imaginary} implies that the frequency at the stability boundary is of order $\sim \omega_{*e}$, the second term in the first equation will always be negligible in comparison to the first, and so can be dropped. The resultant equations can be straightforwardly combined to yield:
\begin{align}
    \left( \frac{k_\parallel L_{T}}{\sqrt{\beta_e}} \right)^4 = \frac{\taubar^2}{ (1+\taubar)\constant\left( \taubar + \constant + 5/3 \right)^2\left(1+1/Z \right)^2}(k_y d_e \chi)^2.
    \label{eq:lin_col_es_stability_boundary}
\end{align}
This is the stability boundary in the electrostatic limit, plotted as the grey dashed line in \figref{fig:lin_col_es_plots}(a). Above this line, corresponding to the limit $ (k_\parallel \vthe)^2/\nu_{ei} \gg \omega_{*e}$, all modes are purely damped due to rapid thermal conduction, as in \figref{fig:lin_col_es_plots}(b). 

The maximum growth rate of the collisional sETG instability is, therefore, reached at $\omega_{*e} \sim (k_\parallel \vthe)^2/\nu_{ei}$, as claimed in~\eqref{eq:col_setg_max}. Apart from factors of order unity, this is the same scaling as \eqref{eq:lin_col_es_stability_boundary}. Indeed, ignoring the magnetic-drift term in~\eqref{eq:lin_col_es_dispersion_relation} and maximising the resultant growth rate with respect to $(k_\parallel \vthe)^2/\nu_{ei}$, one finds
\begin{align}
    \gamma_\text{max} =  C(\taubar) \omega_{*e},
    \label{eq:lin_col_es_setg_max_gamma}
\end{align}
where $C(\taubar)$ is a constant formally of order unity, e.g., $C(1) \approx 0.096$. 

Given that \eqref{eq:lin_col_es_stability_boundary} is the only stability boundary in the electrostatic limit, there is no intermediate region of stability between the cETG and sETG instabilities: as $k_\parallel$ is increased, the cETG mode gradually transitions into the sETG mode (see \figref{fig:lin_col_es_plots}b). Furthermore, \eqref{eq:lin_col_es_setg_max_gamma} implies that, for large temperature gradients, the sETG growth rate will always be asymptotically larger than the cETG one:
\begin{align}
    \frac{\gamma_\text{max}}{\sqrt{2\omega_{de} \omega_{*e}}} \sim \left( \frac{L_B}{L_T} \right)^{1/2}.
    \label{eq:lin_col_es_setg_dominant}
\end{align}
This is \eqref{eq:setg_vs_cetg}. Thus, maximum growth in the electrostatic limit occurs at a finite $k_\parallel$, which scales the same as the stability boundary \eqref{eq:lin_col_es_stability_boundary}.

\subsection{Electromagnetic 3D perturbations: collisional TAI}
\label{app:lin_col_electromagnetic_regime}
Moving towards larger scales, we now consider perturbations above the flux-freezing scale \eqref{eq:flux_freezing_scale_col}, viz.,
\begin{align}
   k_\perp d_e \chi \ll 1 \quad \Leftrightarrow \quad (k_\perp d_e)^2 \nu_{ei} \ll \omega_{*e},
    \label{eq:lin_col_electromagnetic_limit}
\end{align}
meaning that the resistive term in the first bracket in \eqref{eq:lin_col_dispersion_relation} can be neglected, with all other terms retained. 
In order to demonstrate how the two-dimensional perturbations of \apref{app:lin_col_two_dimesional_perturbations} are modified in the presence of finite $k_\parallel$, we consider perturbations satisfying
\begin{align}
    (k_\perp d_e)^2 \nu_{ei} \sim \omega_{de} \ll \omega \ll \omega_{*e} \sim \kappa k_\parallel^2.
    \label{eq:lin_col_em_tai_ordering}
\end{align}
Under this ordering, we ignore the frequency in the first bracket in \eqref{eq:lin_col_dispersion_relation}, except for where is multiples the (large) term proportional to $\kappa k_\parallel^2$ in the second bracket, and, as usual, drop all incidences of $\omega_{de}$ where it is not multiplied by $\omega_{*e}$. The result is
\begin{align}
    (-\omega_{*e} + i \kappa k_\parallel^2 ) \omega^2 + (2 \omega_{de} \omega_{*e} - \omega_\text{KAW}^2) \left[- \omega_{*e} \taubar + i \kappa k_\parallel^2 (1+ \taubar) \right] = 0. 
    \label{eq:lin_col_em_tai_dispersion_relation_initial}
\end{align}
With some straightforward manipulations, this can be rearranged to give 
\begin{align}
    \omega^2 + (2\omega_{de} \omega_{*e} - \omega_\text{KAW}^2 ) \left( \taubar + \frac{1}{1+i\xi_*} \right) = 0, \quad \xi_* = \frac{\omega_{*e}}{\kappa k_\parallel^2}.
    \label{eq:lin_col_em_tai_dispersion_relation}
\end{align}
This is the dispersion relation of the collisional TAI, which we treat in detail in \secref{sec:electromagnetic_regime_tai} and \apref{app:tai_dispersion_relation}. The TAI dispersion relation \eqref{eq:lin_col_em_tai_dispersion_relation} captures all of the properties of the more general dispersion relation \eqref{eq:lin_col_dispersion_relation} in the electromagnetic regime\footnote{Quantitatively well at low $k_\parallel$ (especially in the case of $\omega_{de} \neq 0$), but only qualitatively at higher $k_\parallel$ --- as is evident from figures \ref{fig:lin_col_em_plots_no_drifts} and \ref{fig:lin_col_em_plots}.}, with the important exception of the stabilisation of isothermal and isobaric sTAI --- see \eqref{eq:stai_isothermal_limit} and \eqref{eq:stai_isobaric_limit}, respectively --- that we shall discover in the next section.  

\subsection{Exact stability boundary}
\label{app:exact_stability_boundary_collisional}
Let us now consider the exact stability boundary associated with our collisional dispersion relation \eqref{eq:lin_col_dispersion_relation}. As in \secref{app:lin_col_electrostatic_regime}, we assume that $\omega $ is purely real, and demand that the real and imaginary parts of \eqref{eq:lin_col_dispersion_relation} must vanish individually. The real part gives 
\begin{align}
    \omega^2 = \frac{\left( \omega_\text{KAW}^2 - 2\omega_{de} \omega_{*e} \right) (1+\taubar) - 2\omega_{de} \omega_{*e} \taubar \xi_\eta}{1+\xi_\eta}.
    \label{eq:lin_col_exact_stability_real}
\end{align}
where we have defined
\begin{align}
    \xi_\eta = \frac{(k_\perp d_e)^2 \nu_{ei}}{\kappa k_\parallel^2} = \frac{2}{\constant}\left( \frac{k_\perp d_e \chi }{1+1/Z} \right)^2 \left( \frac{k_\parallel L_{T}}{\sqrt{\beta_e}} \right)^{-2},
    \label{eq:lin_col_xi_eta}
\end{align}
which is the resistive dissipation rate normalised to the thermal conduction rate. Given that, according to \eqref{eq:lin_col_es_stability_boundary}, we expect unstable modes in the electrostatic regime to exist only for $(k_\parallel L_T/\sqrt{\beta_e})^2 \lesssim k_\perp d_e \chi$, it follows that $\xi_\eta \gg 1$ corresponds to the electrostatic limit \eqref{eq:lin_col_electrostatic_limit}, while $\xi_\eta \lesssim 1$, in general, corresponds to the electromagnetic limit \eqref{eq:lin_col_electromagnetic_limit}. 

\begin{figure}
    
\centering
\hspace{-0.5cm}
\begin{tabular}{cc}
     \includegraphics[width=0.55\textwidth]{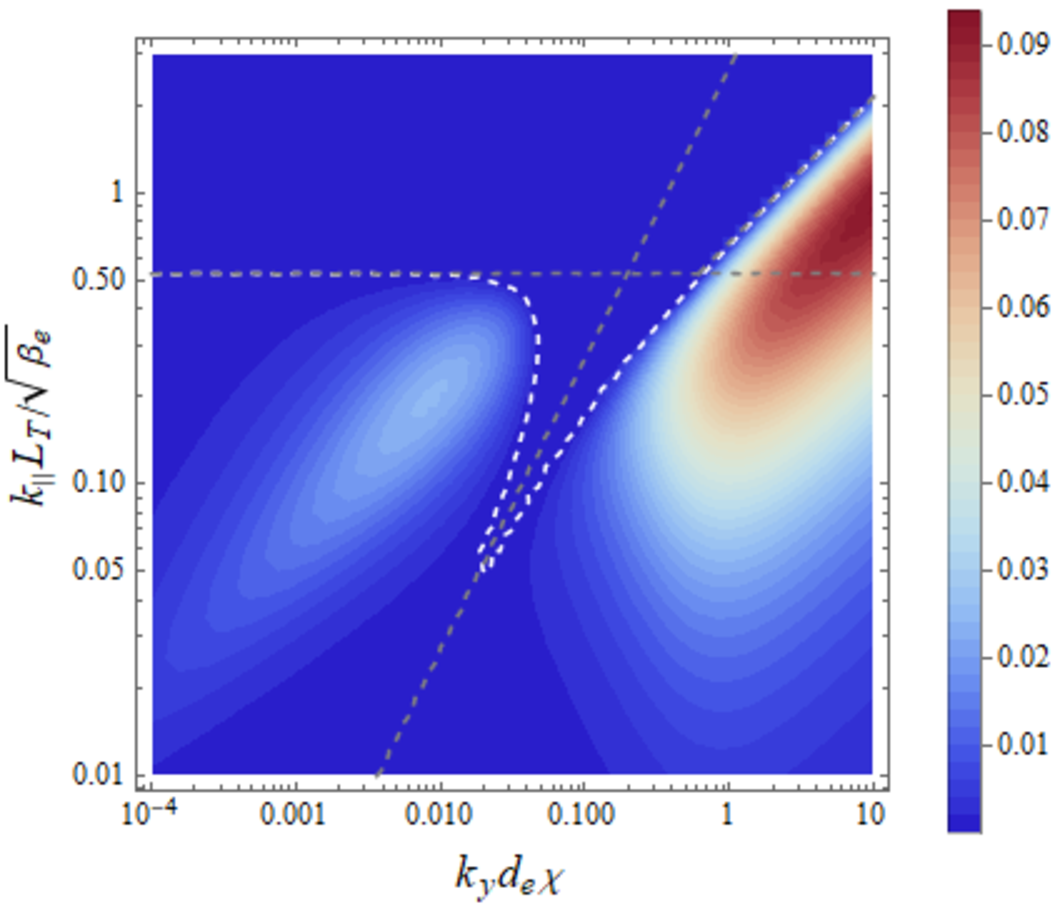}&  
        \includegraphics[width=0.45\textwidth]{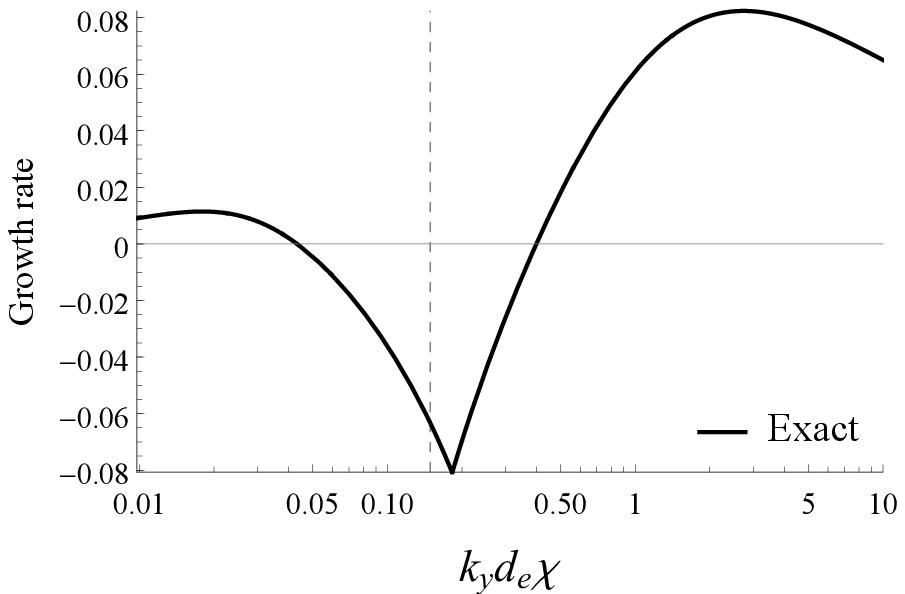} \\\\
     (a) $\omega_{de} = 0$, $k_x d_e \chi = 0$  & (b) $k_\parallel L_T/\sqrt{\beta_e} =0.4 $  \\\\\\
    \includegraphics[width=0.45\textwidth]{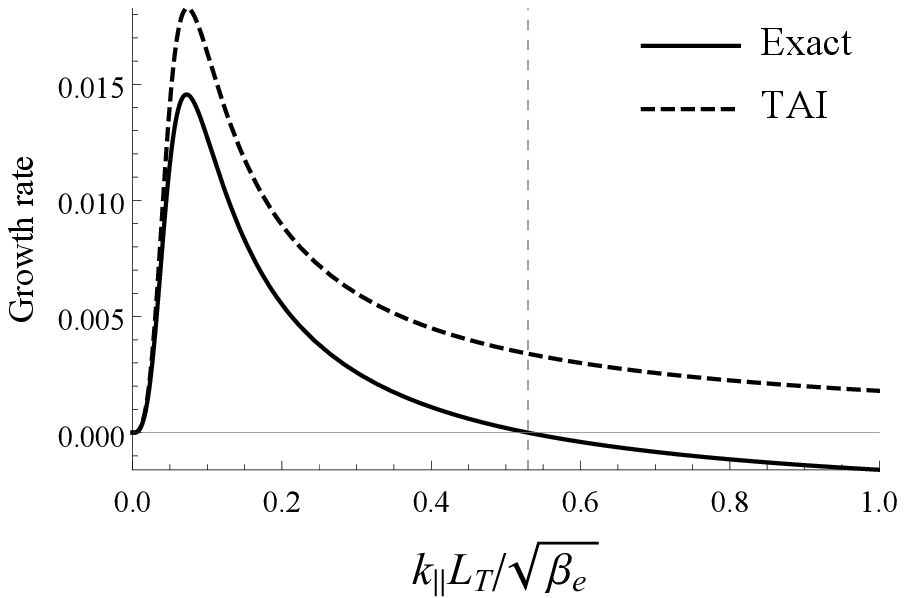}&  
     \includegraphics[width=0.45\textwidth]{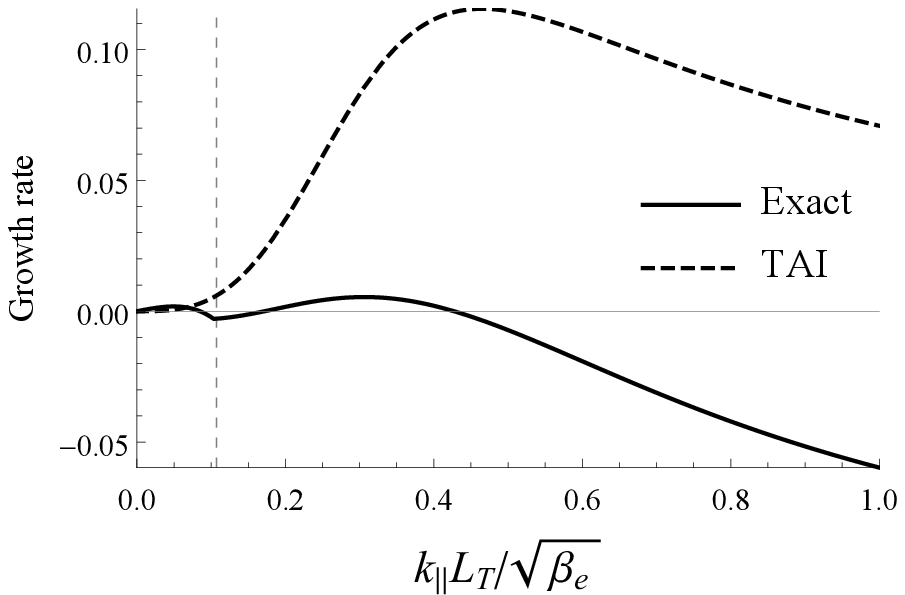} \\\\
          (c)  $k_y d_e \chi = 0.001 $  & (d) $k_y d_e \chi = 0.04$ 
\end{tabular}
 \caption{The growth rates of the collisional instabilities in the absence of magnetic drifts and with $\taubar =1$, normalised to $\omega_{*e}$. Panel (a) is a contour plot of the  positive growth rates ($\gamma > 0$) in the $(k_y,k_\parallel )$ plane. The white dashed line is the exact stability boundary \eqref{eq:lin_col_exact_stability_boundary}, while the horizontal grey dashed line is \eqref{eq:lin_col_isothermal_stai_stabilisation}, corresponding to the stabilisation of the isothermal sTAI at large parallel wavenumbers. The slanted grey dashed line on the left is \eqref{eq:lin_col_isobaric_stai_stabilisation}, around which the isobaric sTAI is briefly stabilised; the slanted grey dashed line on the right is the electrostatic stability boundary \eqref{eq:lin_col_es_stability_boundary}. Panel (b) is a cut of the growth rate along $k_\parallel L_T/\sqrt{\beta_e} =0.4 $ (plotted against a logarithmic scale); panels (c) and (d) are cuts of the growth rate for $k_y d_e \chi = 0.001$ and $k_y d_e \chi = 0.04$, respectively. The growth rates are normalised to $\omega_{*e}$ in all three plots. The solid lines represent the exact growth rate obtained by solving the (collisional) linear dispersion relation \eqref{eq:lin_col_dispersion_relation}, while the dashed lines are the growth rates predicted by the approximate TAI dispersion relation \eqref{eq:lin_col_em_tai_dispersion_relation}. In panels (b) and (d), the vertical grey dashed line is \eqref{eq:lin_col_isobaric_stai_stabilisation}, while the same line in panel (c) is \eqref{eq:lin_col_isothermal_stai_stabilisation}.}
    \label{fig:lin_col_em_plots_no_drifts}
\end{figure}

\begin{figure}
    
\centering
\hspace{-0.5cm}
\begin{tabular}{cc}
   \includegraphics[width=0.55\textwidth]{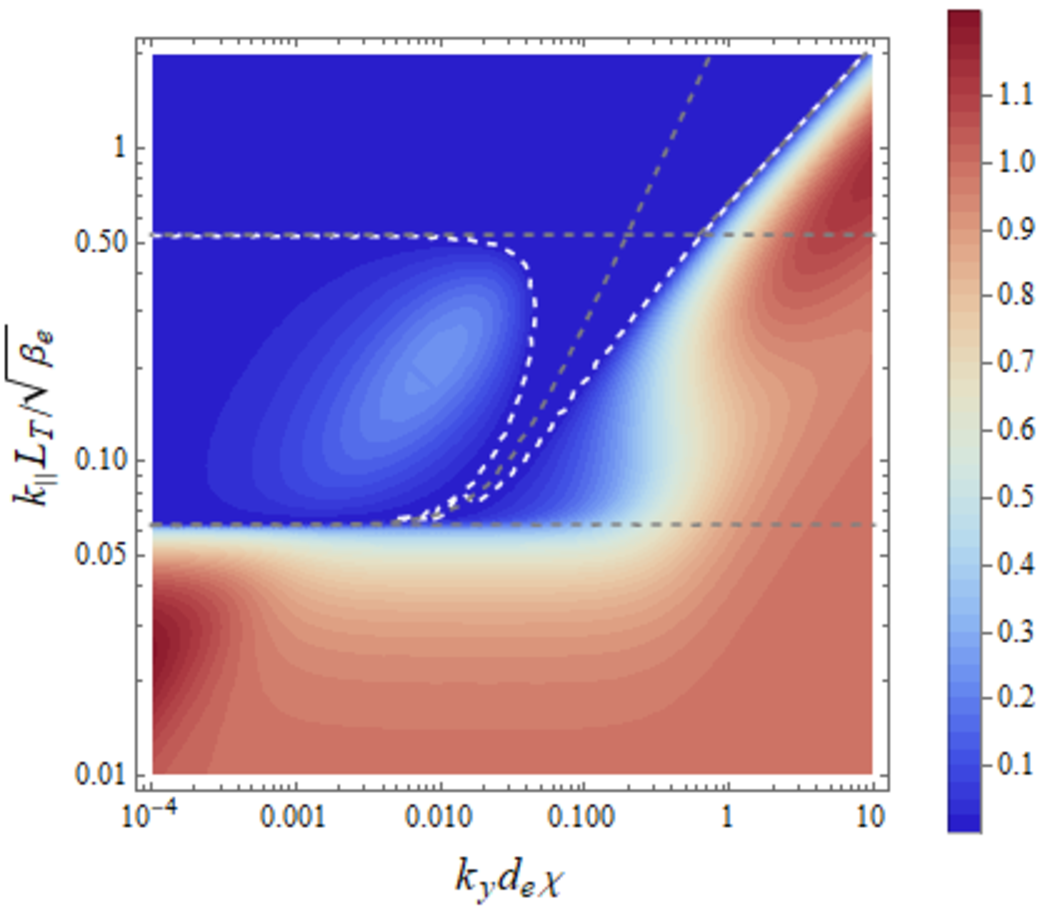}&  
        \includegraphics[width=0.45\textwidth]{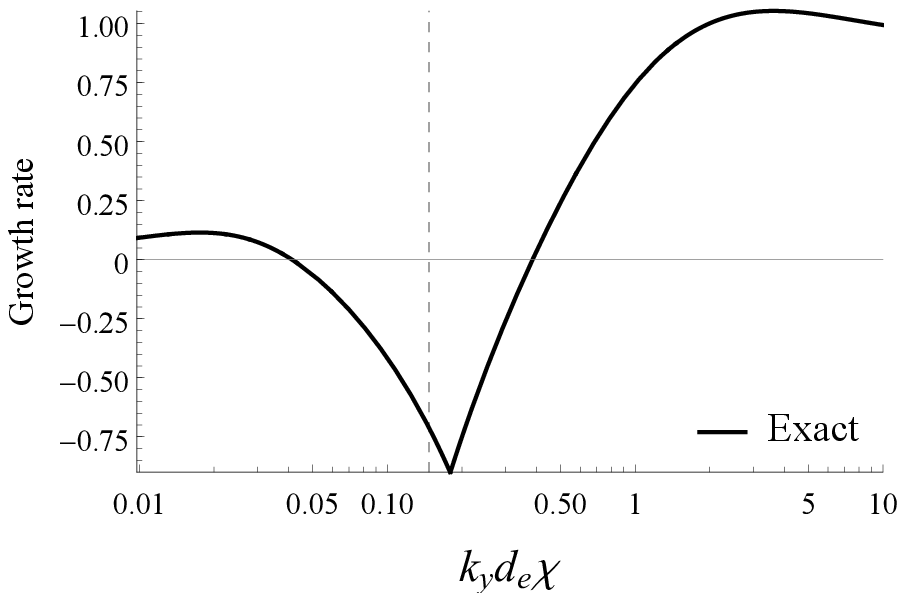} \\\\
     (a)  $L_B/L_T = 250$, $k_x d_e \chi = 0$  & (b) $k_\parallel L_T/\sqrt{\beta_e} =0.4 $  \\\\\\
    \includegraphics[width=0.45\textwidth]{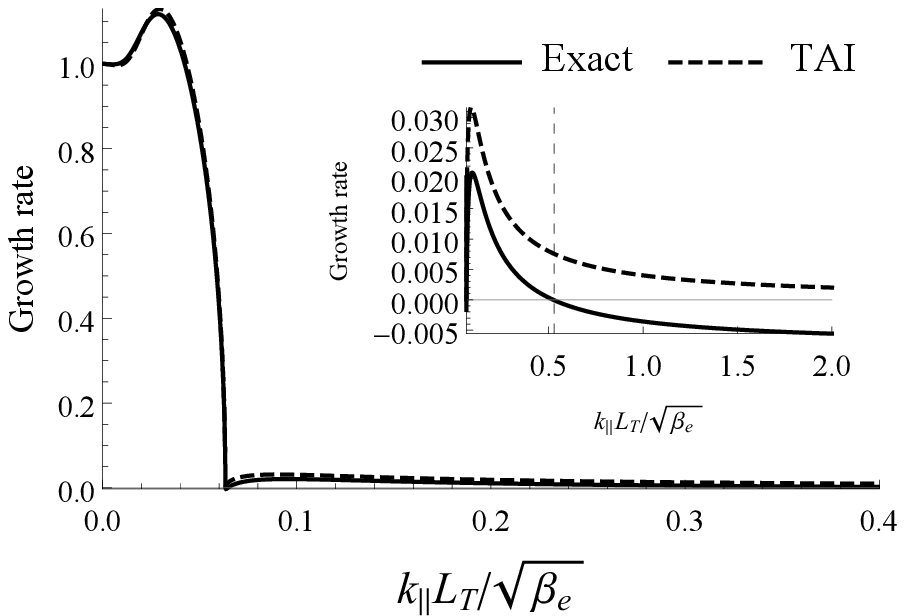}&  
     \includegraphics[width=0.45\textwidth]{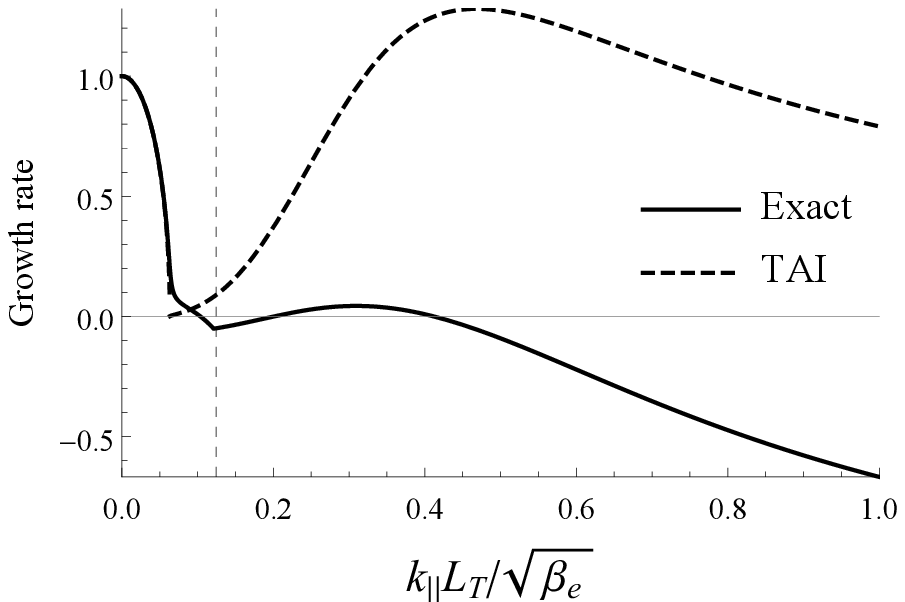} \\\\
          (c)  $k_y d_e \chi = 0.0002 $  & (d) $k_y d_e \chi = 0.04$ 
\end{tabular}
 \caption{The same as \figref{fig:lin_col_em_plots_no_drifts} but with magnetic drifts restored, and now normalised to the cETG growth rate \eqref{eq:lin_col_es_cetg}. We chose a large value of $L_B/L_T$ in order to show the asymptotic regimes clearly. The grey dashed curved line in panel (a) is now \eqref{eq:lin_col_isobaric_stai_general_stabilisation}, while the lower horizontal grey dashed line is $k_{\parallel } = k_{\parallel c}$, as defined in \eqref{eq:tai_kpar_critical}. The inset in panel (c) shows the growth rate for $k_{\parallel} \geqslant k_{\parallel c}$, within which the vertical grey dashed line is \eqref{eq:lin_col_isothermal_stai_stabilisation}. We draw the reader's attention to the enhancement of the cETG growth rate by the cTAI mechanism that can be seen from the red contours in the bottom left-hand corner of panel (a).}
    \label{fig:lin_col_em_plots}
\end{figure}

Extracting the imaginary part of \eqref{eq:lin_col_dispersion_relation}, and using the solution \eqref{eq:lin_col_exact_stability_real} for the frequency $\omega$ at the stability boundary, we find:
\begin{align}
    \pm &\sqrt{\frac{\left( \omega_\text{KAW}^2 - 2\omega_{de} \omega_{*e} \right) (1+\taubar) - 2\omega_{de} \omega_{*e} \taubar \xi_\eta}{1+\xi_\eta}} \nonumber\\
   &\quad \quad  = \omega_{*e} \frac{\omega_\text{KAW}^2(1-\taubar \xi_\eta) - 2\omega_{de} \omega_{*e}}{\omega_\text{KAW}^2 (1 +\taubar) - \omega_\text{KAW}^2 \left( \taubar + \constant+5/3\right)(1+\xi_\eta) -2 \omega_{de} \omega_{*e}}.
   \label{eq:lin_col_exact_stability_boundary}
\end{align}
This is the white dashed curve in figures \ref{fig:lin_col_em_plots_no_drifts}(a) and \ref{fig:lin_col_em_plots}(a). 

In the limit of large perpendicular wavenumbers, \eqref{eq:lin_col_exact_stability_boundary} asymptotes to the electrostatic stability boundary \eqref{eq:lin_col_es_stability_boundary}, which is the slanted grey dashed line in the top right-hand corner of figures \ref{fig:lin_col_em_plots_no_drifts}(a) and \ref{fig:lin_col_em_plots}(a). To show this, anticipating the balance $\omega_{*e} \sim (k_\parallel \vthe)^2/\nu_{ei} \gg 2 \omega_{de} \omega_{*e}$, we neglect all incidences of the magnetic-drift frequency $\omega_{de}$. Then, assuming the resistive rate to be the dominant frequency ($\xi_\eta \gg 1$), and making use of \eqref{eq:lin_col_ignore_terms}, we find that \eqref{eq:lin_col_exact_stability_boundary} reduces to
\begin{align}
    \frac{(k_\parallel \vthe)^4}{(2\nu_{ei})^2} = \frac{\taubar^2 \omega_{*e}^2}{(1+\taubar)\constant\left( \taubar + \constant + 5/3\right)^2}.
    \label{eq:lin_col_exact_stability_boundary_es_limit}
\end{align}
This is \eqref{eq:lin_col_es_stability_boundary} up to normalisations.

In the limit of small perpendicular wavenumbers, \eqref{eq:lin_col_exact_stability_boundary} asymptotes to lines of constant~$k_\parallel$. To show this, we consider the limit of vanishing resistivity ($\xi_\eta \ll 1$), in which \eqref{eq:lin_col_exact_stability_boundary} becomes
\begin{align}
    \mp \sqrt{\left( \omega_\text{KAW}^2 - 2\omega_{de} \omega_{*e} \right) (1+\taubar)} = \omega_{*e} \frac{\omega_\text{KAW}^2 - 2\omega_{de} \omega_{*e}}{ \omega_\text{KAW}^2 \left( \constant  + 2/3\right) + 2 \omega_{de} \omega_{*e}}.
    \label{eq:lin_col_exact_stability_boundary_em_limit}
\end{align}
Clearly, the line 
\begin{align}
    \omega_\text{KAW}^2 = 2 \omega_{de} \omega_{*e} \quad \Rightarrow \quad k_{\parallel } = k_{\parallel c},
    \label{eq:lin_col_kparc}
\end{align}
with $k_{\parallel c}$ defined in \eqref{eq:tai_kpar_critical}, is a solution to this equation, corresponding to the lower horizontal grey dashed line in \figref{fig:lin_col_em_plots}(a). 

In fact, in order for the stability boundary to exist at all, we require that $k_{\parallel} \geqslant k_{\parallel c}$; this follows from the fact that the expression under the square-root on the left-hand side of \eqref{eq:lin_col_exact_stability_boundary_em_limit} must be positive semi-definite. Going back to \eqref{eq:lin_col_exact_stability_real} (i.e., assuming nothing about $k_\perp$) and demanding that the numerator is positive semi-definite, we find a more general condition for the stability boundary to exist:
\begin{align}
  k_\parallel^2 \geqslant k_{\parallel c}^2\left(1 + \frac{\taubar}{1+\taubar } \xi_\eta \right).
    \label{eq:lin_col_kparc_condition}
\end{align}
Since $\xi_\eta \geqslant 0$, this implies that our system can never be stable for $k_{\parallel} \leqslant k_{\parallel c}$. Returning again to \eqref{eq:lin_col_exact_stability_boundary_em_limit} and considering the limit of $k_\parallel \gg k_{\parallel c}$, we find that the stability boundary asymptotically approaches
\begin{align}
    \mp \omega_{\text{KAW}} \sqrt{1+\taubar} = \frac{\omega_{*e}}{\constant + 2/3}  \quad \Rightarrow \quad \frac{k_\parallel L_{T}}{\sqrt{\beta_e}}  = \pm  \frac{1}{(\constant + 2/3)\sqrt{2(1+\taubar)}} \frac{k_y}{k_\perp}.
    \label{eq:lin_col_isothermal_stai_stabilisation}
\end{align}
This is the upper horizontal grey dashed line in figures \ref{fig:lin_col_em_plots_no_drifts}(a) and \ref{fig:lin_col_em_plots}(a).
It corresponds to the stabilisation of the isothermal sTAI \eqref{eq:stai_isothermal_limit} at large parallel wavenumbers due to compressional heating, as explained in \secref{sec:stabilisation_of_isothermal_slab_TAI}.

Similarly, the stabilisation of the isobaric sTAI \eqref{eq:stai_isobaric_limit} can be extracted from \eqref{eq:lin_col_exact_stability_boundary} in the limit $\xi_\eta \sim 1$. Let us initially consider the limit of $k_{\parallel} \gg k_{\parallel c}$, in which we can neglect magnetic drifts, so~\eqref{eq:lin_col_exact_stability_boundary} becomes
\begin{align}
    \pm \omega_\text{KAW} \sqrt{\frac{1+\taubar}{1+\xi_\eta}} = \omega_{*e} \frac{1-\taubar \xi_\eta}{1+\taubar - \left( \taubar + \constant +5/3 \right)(1+\xi_\eta)}.
    \label{eq:lin_col_isobaric_stai_initial}
\end{align}
The left-hand side of \eqref{eq:lin_col_isobaric_stai_initial} is proportional to $\omega_\text{KAW} \propto k_\parallel k_\perp$, whereas the right-hand side is proportional to $\omega_{*e} \propto k_y$. This means that as $k_\parallel, k_y \rightarrow 0$, while maintaining $\xi_\eta \sim 1$, the left-hand side approaches zero faster that the right-hand side, unless the numerator of the right-hand side similarly approaches zero. This means that both branches of the stability boundary will asymptotically approach
\begin{align}
    1 - \taubar \xi_\eta = 0 \quad \Rightarrow \quad \left( \frac{k_\parallel L_T}{\sqrt{\beta_e}} \right)^{2} = \frac{2\taubar}{\constant}\left( \frac{k_\perp d_e \chi }{1+1/Z} \right)^2.
        \label{eq:lin_col_isobaric_stai_stabilisation}
\end{align}
This means that there is a thin sliver of stability within the otherwise unstable isobaric sTAI region: the growth rate briefly dips below zero, before picking back up again and reaching a maximum around $\xi_* \sim 1$ [cf. \eqref{eq:tai_growth_rate_expanded_isobaric_secondary} and the following discussion], as shown in figures \ref{fig:lin_col_em_plots_no_drifts}(d) and \ref{fig:lin_col_em_plots}(d). This is due to finite-resistivity effects coming into play, and competing with thermal conduction, as explained in \secref{sec:stabilisation_of_isobaric_slab_TAI}. 

In the more general case including magnetic drifts, viz., for $k_{\parallel} \gtrsim k_{\parallel c}$, the curve that both branches of the exact stability boundary asymptotically approach is well described by the vanishing of the numerator of the right-hand side of \eqref{eq:lin_col_exact_stability_boundary}:
\begin{align}
    \omega_\text{KAW}^2(1 - \taubar \xi_\eta) - 2 \omega_{de} \omega_{*e} = 0 \quad \rightarrow \quad \left( \frac{k_\parallel L_T}{\sqrt{\beta_e}} \right)^{2} = \frac{L_T}{L_B} \left( \frac{k_y}{k_\perp} \right)^{2} + \frac{2\taubar}{\constant}\left( \frac{k_\perp d_e \chi }{1+1/Z} \right)^2.
    \label{eq:lin_col_isobaric_stai_general_stabilisation}
\end{align}
This is indicated by the grey dashed curved line in \figref{fig:lin_col_em_plots}(a). 
It reproduces \eqref{eq:lin_col_kparc} and \eqref{eq:lin_col_isobaric_stai_stabilisation} in the appropriate limits.

We have thus used the expression for the exact stability boundary \eqref{eq:lin_col_exact_stability_boundary} to derive the boundaries that limit the unstable regions of wavenumber space. From figures \ref{fig:lin_col_em_plots_no_drifts} and \ref{fig:lin_col_em_plots}, it is clear that \eqref{eq:lin_col_es_stability_boundary} bounds the electrostatic instabilities at large parallel wavenumbers, while the electromagnetic region of instability at $k_\parallel > k_{\parallel c}$, corresponding to the sTAI, is bounded by~\eqref{eq:lin_col_isothermal_stai_stabilisation} and \eqref{eq:lin_col_isobaric_stai_stabilisation}. 
We also draw the reader's attention to the similarity between \figref{fig:lin_col_em_plots}(a) and the wavenumber-space portrait associated with our collisional equations (\figref{fig:collisional_phase_space_portrait}), in that it reproduces all the key features that were predicted using the naive estimates of \secref{sec:summary_of_wavenumber_space}.

\section{Analysis of TAI dispersion relation}
\label{app:tai_dispersion_relation}
In this appendix, we consider the mathematical details of the TAI dispersion relation~\eqref{eq:tai_dispersion_relation_general}
\begin{align}
    \omega^2 = -  \left(2 \omega_{de} \omega_{*e} - \omega_\text{KAW}^2 \right) \left( \taubar + \frac{1}{1+i\xi_*} \right), 
    \label{eq:tai_dispersion_relation_general_appendix}
\end{align}
where $\xi_*$ is given by 
\begin{align}
    \xi_{*} = \left\{ 
    \begin{array}{ll}
         \displaystyle \frac{\sqrt{\pi}}{2} \frac{\omega_{*e}}{\left|k_\parallel \right| \vthe}, & \text{collisionless},   \\[4mm]
         \displaystyle \frac{\omega_{*e}}{\kappa k_\parallel^2} = \frac{18}{5} \frac{\omega_{*e}}{(k_\parallel \vthe)^2/ \nu_e} , & \text{collisional}.
    \end{array}
    \right.
    \label{eq:tai_xi_expicit}
\end{align}
Defining 
\begin{align}
    \sigma = \text{sgn}\left(2 \omega_{de} \omega_{*e} - \omega_\text{KAW}^2 \right) = \text{sgn}(k_{\parallel c} - k_\parallel) \equiv e^{i n \pi},
    \label{eq:sigma}
\end{align}
with $k_{\parallel c}$ defined in \eqref{eq:tai_kpar_critical}, 
we can write \eqref{eq:tai_dispersion_relation_general_appendix} as
\begin{align}
    -i \omega = \left| 2 \omega_{de} \omega_{*e} - \omega_\text{KAW}^2 \right|^{1/2} A^{1/2} e^{i(\theta + n \pi)/2},
    \label{eq:tai_frequency}
\end{align}
where
\begin{align}
    A = \sqrt{\left( \taubar + \frac{1}{1+ \xi_*^2}\right)^2 + \frac{\xi_*^2}{(1+ \xi_*)^2}}, \quad \theta = \tan^{-1} \left[ \frac{\xi_*/(1+\xi_*^2)}{\taubar + 1/(1+\xi_*^2)} \right].
    \label{eq:tai_complex_number}
\end{align}
Taking the real and imaginary parts of \eqref{eq:tai_frequency}, and using the fact that 
\begin{align}
    \cos^2 \left( \frac{\theta + n\pi}{2} \right) = \frac{1}{2} \left( 1+ \sigma \cos \theta \right), \quad \sin^2 \left( \frac{\theta + n\pi}{2} \right) = \frac{1}{2 }\left(1 -\sigma  \cos \theta\right),
    \label{eq:tai_trig_identities}
\end{align}
we find the real frequency $\omega_r = \text{Re}(\omega)$ and the growth rate $\gamma = \text{Im}(\omega)$ that satisfy \eqref{eq:tai_dispersion_relation_general_appendix}:
\begin{align}
    \omega_r^2 = | 2\omega_{de} \omega_{*e} - \omega_\text{KAW}^2 | \taubar f_{-}(\xi_*), \quad \gamma^2 = | 2\omega_{de} \omega_{*e} - \omega_\text{KAW}^2 | \taubar f_{+}(\xi_*),
    \label{eq:tai_growth_rate_and_frequency}
\end{align}
where we have defined the functions
\begin{align}
    f_{\pm}(\xi_*) = \frac{1}{2\taubar} \left[ \sqrt{\left( \bar{\tau} + \frac{1}{1+\xi_{*}^2} \right)^2 + \frac{\xi_{*}^2}{(1 + \xi_{*}^2)^2}} \pm \sigma\left( \bar{\tau} + \frac{1}{1+\xi_{*}^2} \right)  \right].
    \label{eq:tai_functions}
\end{align}
These are exactly the formulae \eqref{eq:tai_growth_rate_and_real_frequency} and \eqref{eq:tai_f}. Note that the correspondence between the signs of $\gamma$ and $\omega_r$ was lost in \eqref{eq:tai_growth_rate_and_frequency}. Most of the time, this information will not be important but, if needed, it can be recovered by going back to \eqref{eq:tai_dispersion_relation_general_appendix} and extracting its imaginary part:
\begin{align}
    2\gamma \omega_r =  \left| 2 \omega_{de} \omega_{*e} - \omega_\text{KAW}^2 \right| \frac{\sigma \xi_*}{1+\xi_*^2}.
    \label{eq:tai_dispersion_imaginary_part}
\end{align}
Therefore, for an unstable mode ($\gamma >0$), 
\begin{align}
    \text{sgn} (\omega_r) = \text{sgn} (\sigma \xi_*). 
    \label{eq:tai_real_frequency_sign}
\end{align}

In what follows, it will be useful to consider the asymptotic expansions of \eqref{eq:tai_functions} for small and large argument:
\begin{align}
    f_\pm (\xi_*) = \left\{ 
    \begin{array}{ll}
         \displaystyle \frac{1}{2} \left(1 + \frac{1}{\taubar} \right) \left(1 \pm \sigma \right) - \frac{2\taubar +1 \pm 2\sigma(1+\taubar)}{4\taubar(1+\taubar)}\xi_*^2 + \dots, & \xi_* \ll 1,   \\[4mm]
         \displaystyle \frac{1}{2}\left(1 \pm \sigma \right) + \frac{1+ 2\taubar (1\pm \sigma)}{4\taubar^2} \frac{1}{\xi_*^2 } + \dots , & \xi_* \gg 1.
    \end{array}
    \right.
    \label{eq:tai_functions_expansion}
\end{align} 
Given that we are interested in the behaviour of \eqref{eq:tai_functions} as functions of $k_\parallel$ --- at constant perpendicular wavenumber --- we define a parallel wavenumber $k_{\parallel *}$ such that 
\begin{align}
    \xi_* = \left( \frac{k_{\parallel *}}{\left| k_\parallel \right| } \right)^\alpha,
    \label{eq:xi_kpar_transition_definition}
\end{align}
where $\alpha = 1,2$ in the collisionless and collisional cases, respectively. The wavenumber $k_{\parallel *}$ can be read off~\eqref{eq:tai_xi_expicit} and depends on $k_y$. 
Then, $k_\parallel \sim k_{\parallel *}$ corresponds to the transition between the isothermal range of wavenumbers \eqref{eq:isothermal_limit} ($k_{\parallel} \gg k_{\parallel *}$) and the isobaric one~\eqref{eq:isobaric_limit} ($k_{\parallel} \ll k_{\parallel *}$).

\subsection{Isothermal limit}
\label{app:isothermal_limit}
\begin{figure}
    \centering

\begin{tabular}{cc}
     \includegraphics[width=0.48\textwidth]{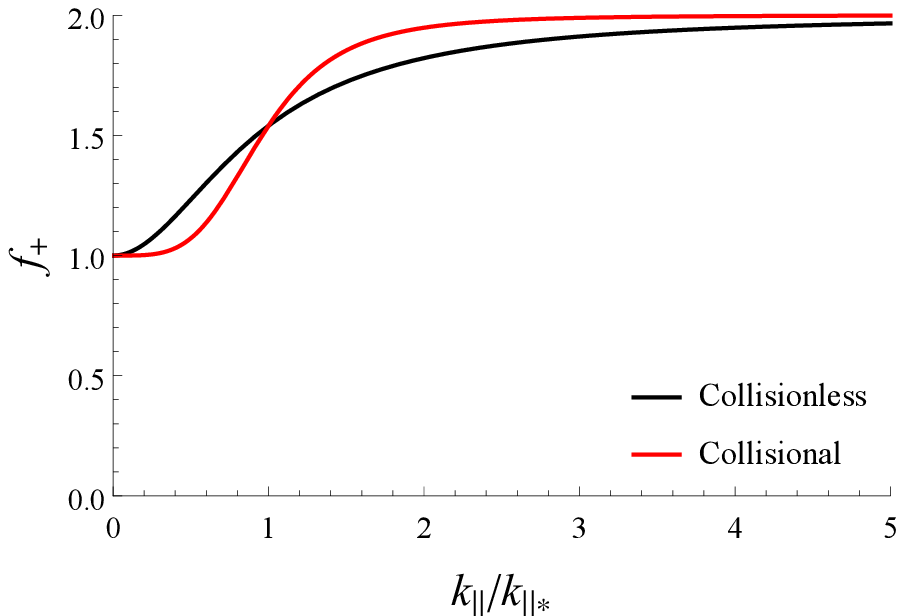}&  
     \includegraphics[width=0.48\textwidth]{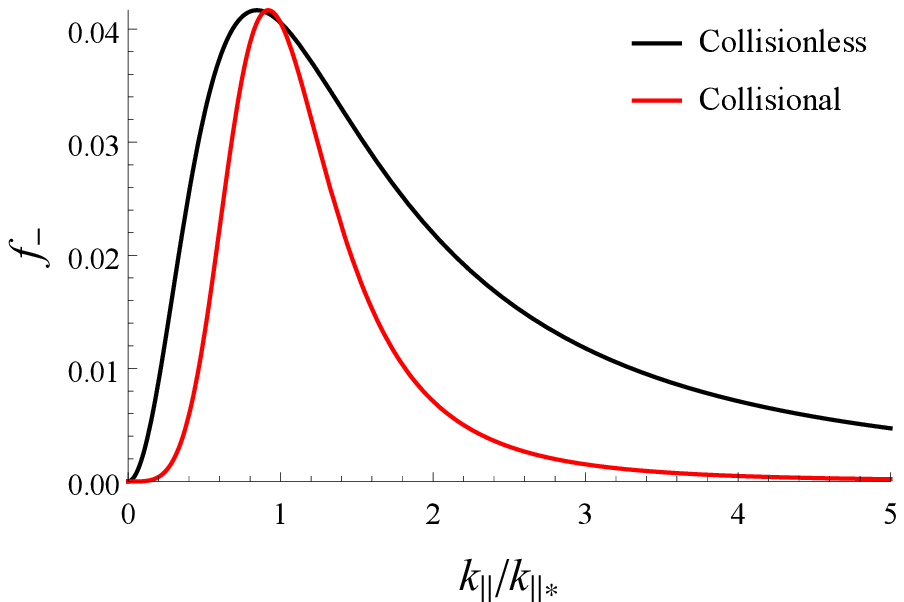}
\end{tabular}
    \caption{The functions \eqref{eq:tai_functions}: (a) $f_+$ and (b) $f_-$ plotted as functions of $k_{\parallel}/k_{\parallel *}$, for $\taubar =1 $ and $\sigma =1$ (for $\sigma = -1$, $f_+ \leftrightarrow f_-$).  It is clear that the region of maximum variation of $f_\pm$ occurs around  $k_{\parallel} \sim k_{\parallel *} \Leftrightarrow \xi_*\sim 1$. }
    \label{fig:tai_function_dependence}
\end{figure}

We first consider the isothermal limit $k_{\parallel } \gg k_{\parallel *}$.
Examining the second expression in \eqref{eq:tai_growth_rate_and_frequency} in the region $k_\parallel < k_{\parallel c}$, we notice that $f_+(\xi_*)$ is a monotonically increasing function of $k_\parallel$ (\figref{fig:tai_function_dependence}), while the prefactor $| 2\omega_{de} \omega_{*e} - \omega_\text{KAW}^2|$ is obviously a monotonically decreasing function of it.
Thus, the growth rate may have a maximum in the isothermal range, the condition for which we will check \textit{a posteriori}. Using~\eqref{eq:tai_functions_expansion} with $\sigma=1$, we expand the growth rate in the isothermal limit $\xi_* \ll 1$:
\begin{align}
    \gamma^2 & = 2 \omega_{de} \omega_{*e} \taubar \left[1 -  \left(\frac{k_\parallel}{k_{\parallel c}} \right)^2 \right] \left[\left(1 + \frac{1}{\taubar} \right) - \frac{3+4\taubar}{4\taubar(1+\taubar)} \xi_*^2 + \dots  \right] \nonumber \\
    & = 2 \omega_{de} \omega_{*e} (1+ \taubar) \left[1 -  \left(\frac{k_\parallel}{k_{\parallel c}} \right)^2 - \frac{3+4\taubar}{4(1+\taubar)^2} \left( \frac{k_{\parallel *}}{k_{\parallel c}}\right)^{2\alpha} \left( \frac{k_{\parallel c}}{k_\parallel} \right)^{2\alpha} + \dots \right],
    \label{eq:tai_growth_rate_expanded}
\end{align}
where we have used \eqref{eq:xi_kpar_transition_definition} and assumed that $k_{\parallel} \ll k_{\parallel c}$. Maximising \eqref{eq:tai_growth_rate_expanded} with respect to $(k_{\parallel}/k_{\parallel c})^2$, we find 
\begin{align}
     \frac{k_{\parallel \text{max}}}{k_{\parallel c}}  = \left[\frac{3+4\taubar}{4(1+\taubar)^2} \alpha \left( \frac{k_{\parallel *}}{k_{\parallel c}} \right)^{2\alpha} \right]^{1/2(1+\alpha)}.
    \label{eq:isothermal_tai_maximum_solution}
\end{align}
Using the definition of $k_{\parallel *}$ \eqref{eq:xi_kpar_transition_definition}, we obtain~\eqref{eq:tai_curvature_maximum_explicit}. This solution is valid provided $k_{\parallel *} \ll k_{\parallel \text{max}} \ll k_{\parallel c}$ (i.e., provided it lies in the isothermal regime and $\sigma = 1$). This translates into \eqref{eq:xi_at_max} by noticing that \eqref{eq:isothermal_tai_maximum_solution} implies
\begin{align}
    \xi_*(k_{\parallel \text{max}} ) = \left( \frac{k_{\parallel *}}{k_{\parallel \text{max}}} \right)^\alpha = \frac{2(1+\taubar)}{\sqrt{\alpha(3+4\taubar)}} \frac{k_{\parallel \text{max}}}{k_{\parallel c}}.
    \label{eq:isothermal_tai_maximum_xi}
\end{align}
This observation also allows us to write the peak growth rate, given by \eqref{eq:tai_growth_rate_expanded} with $k_\parallel = k_{\parallel \text{max}}$, as follows:
\begin{align}
    \gamma_\text{max} \approx \sqrt{ 2\omega_{de} \omega_{*e}  \left(1+ \taubar \right)}\left[1 -  \frac{1}{2}\left( 1+ \frac{1}{\alpha} \right)\left(\frac{k_{\parallel \text{max}}}{k_{\parallel c}} \right)^2  \right],
    \label{eq:tai_curvature_maximum_growth_rate_corrections}
\end{align}
which reduces to \eqref{eq:isothermal_tai_dispersion_relation} if we ignore the small correction due to $k_{\parallel \text{max}}$. 

For $k_\parallel > k_{\parallel c}$ and $\xi_* \ll 1$ (the isothermal KAW regime), the expansion \eqref{eq:tai_functions_expansion} with $\sigma = -1 $ gives us 
\begin{align}
    \omega_r^2 \approx \omega_\text{KAW}^2 (1+\taubar) \left[ 1 - \left( \frac{k_{\parallel c}}{k_{\parallel}} \right)^2 \right] , \quad \gamma^2 \approx \frac{\omega_{de} \omega_{*e}}{2(1+\taubar)} \left[ \left( \frac{k_{\parallel }}{k_{\parallel c}} \right)^2 -1 \right] \xi_*^2.
    \label{eq:stai_isothermal_limit_expansions}
\end{align}
At $k_\parallel \gg k_{\parallel c}$, these turn into \eqref{eq:stai_isothermal_limit}. In the collisionless limit, $\xi_* \propto k_{\parallel}^{-1}$, so $\gamma \rightarrow \text{const}$ as $k_{\parallel } \rightarrow \infty$; this constant value is \eqref{eq:tai_isothermal_collisionless_max}. In the collisional limit, $\xi_* \propto k_\parallel^{-2}$, so $\gamma \rightarrow 0$ as $k_\parallel \rightarrow \infty$, and peak growth is reached at a finite $k_{\parallel}$: from \eqref{eq:stai_isothermal_limit_expansions}, we get
\begin{align}
    \gamma^2 \propto  \left[ \left( \frac{k_{\parallel}}{k_{\parallel c}} \right)^2 -1 \right] \left( \frac{k_{\parallel c}}{k_\parallel } \right)^4,
    \label{eq:tai_growth_rate_expanded_secondary}
\end{align}
so the maximum is reached at $k_\parallel = \sqrt{2} k_{\parallel c}$. Putting this back into \eqref{eq:stai_isothermal_limit_expansions}, we find \eqref{eq:tai_isothermal_collisional_max} for $\gamma_\text{max}$.
\begin{figure}
    \centering

\begin{tabular}{cc}
     \includegraphics[width=0.5\textwidth]{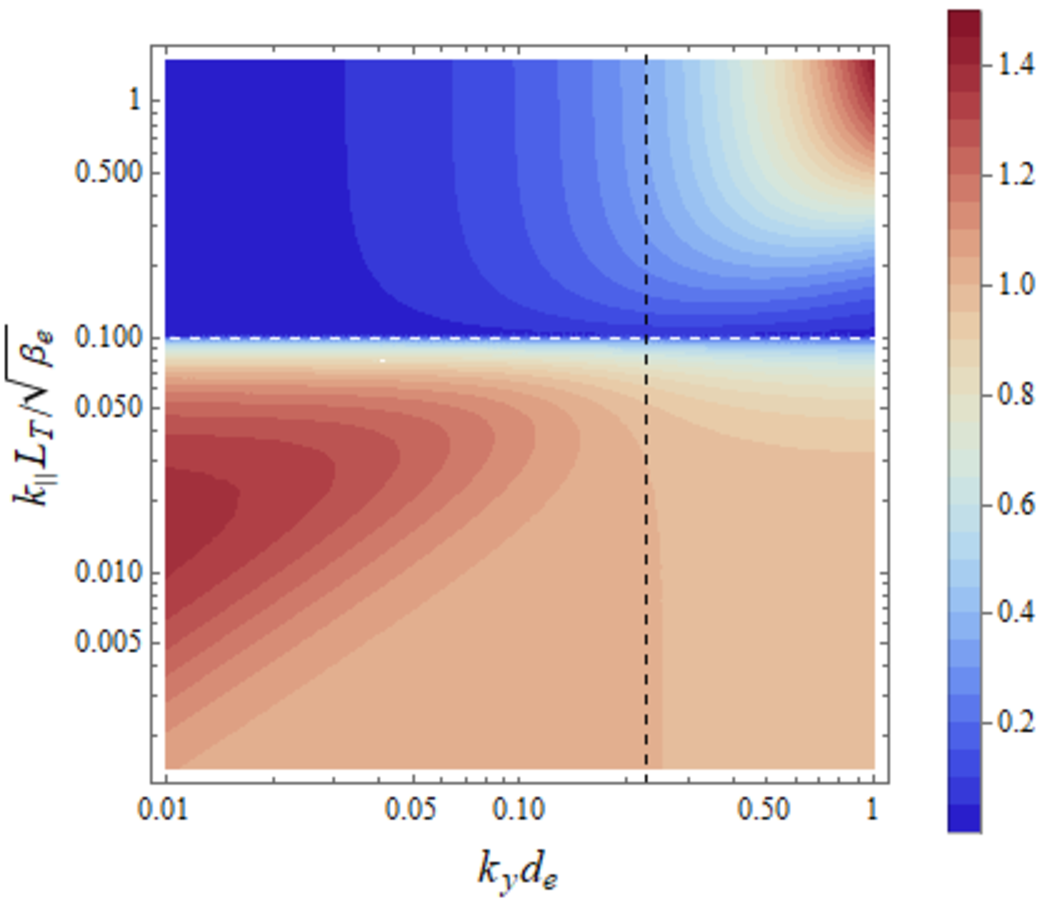}&  
     \includegraphics[width=0.5\textwidth]{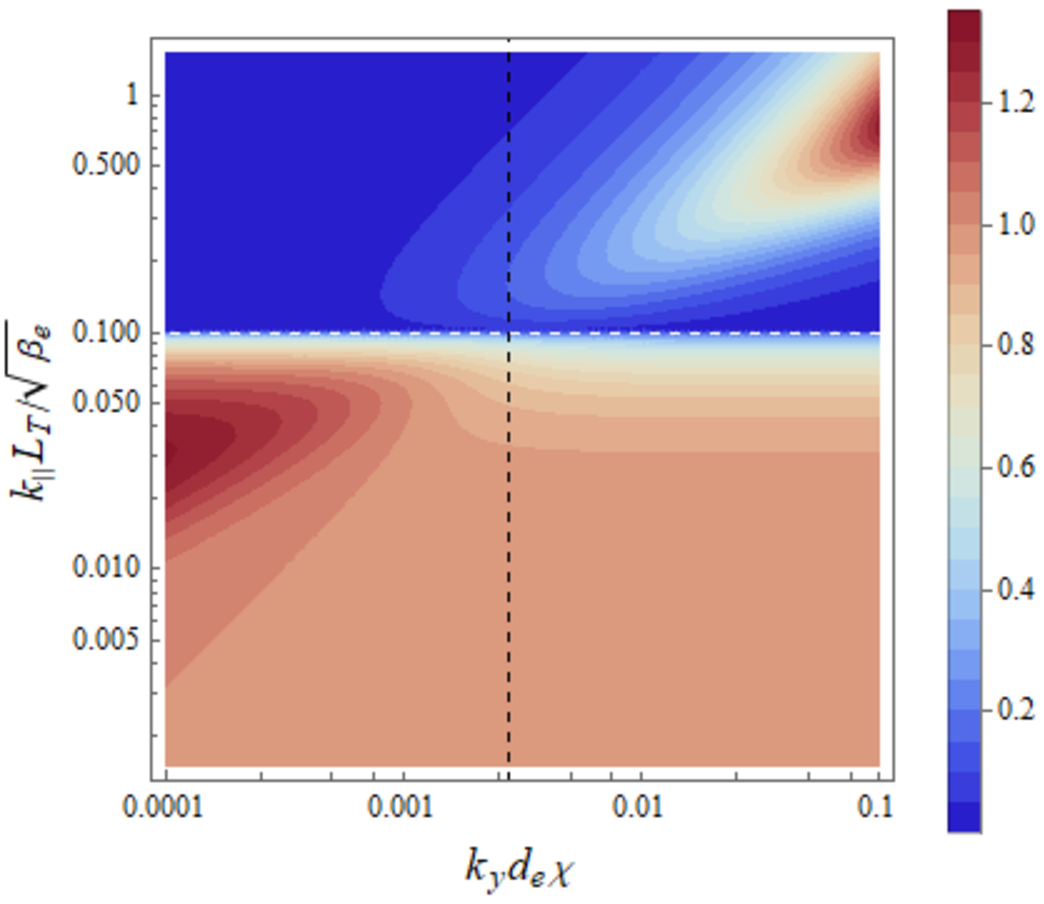} \\\\
    (a) $L_B/L_{T} =100  $, $k_x d_e = 0$ & (b) $L_B/L_{T} =100  $, $k_x d_e \chi  = 0$ \\
\end{tabular}
    \caption{Contour plots of the TAI growth rate \eqref{eq:tai_growth_rate_and_frequency} in the $(k_y,k_\parallel )$ plane, normalised to the cETG growth rate \eqref{eq:cetg_gamma}. Panels (a) and (b) show the collisionless and collisional cases, respectively. The horizontal white dashed line is $k_\parallel = k_{\parallel c}$, as defined in \eqref{eq:tai_kpar_critical}, while the vertical black dashed line is $k_\perp =  k _{\perp *}$, as defined in \eqref{eq:tai_transition_kperp}. There is clear enhancement of the cETG growth rate due to the cTAI \eqref{eq:tai_curvature_maximum_growth_rate_corrections} at $k_\perp < k_{\perp *}$ (the isothermal regime, \secref{sec:isothermal_ctai}), while there is no enhancement for $k_\perp > k_{\perp *}$ (the isobaric regime, \secref{sec:isobaric_limit}). We chose a large value of $L_B/L_T$ in order to show a clear transition between these two regimes.}
    \label{fig:tai_growth_rate_kykz}
\end{figure}

\subsection{Isobaric limit}
\label{app:isobaric_limit}
Now consider the limit $\xi_* \gg 1$, which, in \secref{sec:isobaric_limit}, we showed to be isobaric. Consider first $k_\parallel < k_{\parallel c}$ and again use \eqref{eq:tai_functions_expansion} with $\sigma =1$ to expand the growth rate, now in $\xi_* \gg 1$:
\begin{align}
    \gamma^2 & = 2 \omega_{de} \omega_{*e} \taubar \left[1 -  \left(\frac{k_\parallel}{k_{\parallel c}} \right)^2 \right] \left[1+ \frac{1+ 4\taubar}{4\taubar^2} \frac{1}{\xi_*^2} + \dots  \right] \nonumber \\
    & = 2 \omega_{de} \omega_{*e} \taubar \left[1 -  \left(\frac{k_\parallel}{k_{\parallel c}} \right)^2 + \frac{1+4\taubar}{4\taubar^2} \left( \frac{k_{\parallel c}}{k_{\parallel *}} \right)^{2\alpha} \left( \frac{k_\parallel }{k_{\parallel c}} \right)^{2\alpha} + \dots \right],
    \label{eq:tai_growth_rate_expanded_isobaric}
\end{align}
where we have used~\eqref{eq:xi_kpar_transition_definition}. When $k_{\parallel *} \gg k_{\parallel c}$, i.e., when all wavenumbers $k_{\parallel} < k_{\parallel c}$ are in the isobaric limit, the last term in \eqref{eq:tai_growth_rate_expanded_isobaric} is negligible and the resultant expression simply describes the gradual petering out of the cETG growth rate due to the stabilising effect of the KAW response --- $\gamma$ has no extrema. If $k_{\parallel *} \ll k_{\parallel c}$ and $\alpha =1$ (collisionless limit), then \eqref{eq:tai_growth_rate_expanded_isobaric} describes the increase of $\gamma$ with $k_{\parallel}$ --- it will reach the maximum \eqref{eq:isothermal_tai_maximum_solution} after is crosses over from the isobaric regime into the isothermal one around $k_{\parallel} \sim k_{\parallel *}$. In the collisional limit ($\alpha =2$), \eqref{eq:tai_growth_rate_expanded_isobaric} does have an extremum in the isobaric regime, viz., 
\begin{align}
    \frac{k_\parallel}{k_{\parallel c}} =  \sqrt{\frac{2\taubar^2}{1+4\taubar}} \left( \frac{k_{\parallel *}}{k_{\parallel c}} \right)^\alpha,
    \label{eq:tai_isobaric_minimum}
\end{align}
but this extremum is a minimum, not a maximum: the growth rate dips slightly before starting to increase again towards the isothermal maximium \eqref{eq:isothermal_tai_maximum_solution} [this is visible in \figref{fig:tai_growth_rate_and_frequency}(c)]. 

Returning to the case $k_{\parallel *} \gg k_{\parallel c}$, we must also examine the isobaric behaviour at $k_\parallel > k_{\parallel c}$, because the transition to the isothermal regime does not happen until well into this range. Using \eqref{eq:tai_functions_expansion}, we find, at $\xi_* \gg 1$, 
\begin{align}
    \gamma^2 \approx \frac{\omega_{de} \omega_{*e}}{2\taubar} \left[ \left( \frac{k_\parallel}{k_{\parallel c}} \right)^2 - 1 \right] \frac{1}{\xi_*^2} = \frac{\omega_{de} \omega_{*e}}{2\taubar}  \left[ \left( \frac{k_{\parallel *}}{k_{\parallel c}} \right)^{2} \frac{1}{\xi_*^{2/\alpha}} - 1 \right] \frac{1}{\xi_*^2}.
    \label{eq:tai_growth_rate_expanded_isobaric_secondary}
\end{align}
This increases monotonically with $1/\xi_* \propto k_\parallel^\alpha$ until the isobaric regime transitions into the isothermal one at $\xi_* \sim 1$. In the collisionless limit, $\gamma$ asymptotes to the constant~\eqref{eq:tai_isothermal_collisionless_max}, whereas in the collisional limit, it has a maximum at $\xi_* \sim 1$ before decaying at $k_\parallel \rightarrow \infty$. To find this maximum, one must extremise
\begin{align}
    \gamma^2 \approx 2 \omega_{de} \omega_{*e} \taubar \left( \frac{k_{\parallel *}}{k_{\parallel c}} \right)^{2}\frac{1}{\xi_*^{2/\alpha}} f_{+}(\xi_*)
    \label{eq:tai_growth_rate_expanded_secondary_collisional}
\end{align}
with respect to $\xi_*^2$ without further approximations --- a thankless exercise leading to a quartic equation. The answer is 
\begin{align}
    \gamma^2_\text{max} =  \omega_{de} \omega_{*e}  \left( \frac{k_{\parallel *}}{k_{\parallel c}} \right)^{2} C(\taubar) 
    \label{eq:stai_max_growth_rate}
\end{align}
where $C(\taubar)$ is a constant formally of order unity, e.g., $C(1) \approx 0.093$ (for which $\xi_* \approx 0.67$). This is the same as~\eqref{eq:stai_max_gamma}.

\section{Consequences of the choice of collision operator}
\label{app:choice_of_collision_operator}
As discussed in \apref{app:low_beta_collisional_equations}, a careful reader may have been concerned that our choice of the simplified collision operator \eqref{eq:collision_operator} would have consequences for the (collisional) physics phenomena described in this paper. In this appendix, however, we demonstrate that the collisional equations \eqref{eq:col_density_equation}-\eqref{eq:col_temperature_equation} --- and their strongly-driven counterparts \eqref{eq:density_moment_collisional}-\eqref{eq:t_moment_collisional} --- are almost unchanged if derived using the general Landau collision operator, and discuss the status of slab micro-tearing modes (MTMs) driven by higher-order collisional effects (\citealt{hassam80a,hassam80b}) within the context of this work. 

\subsection{Collisional equations with Landau operator}
\label{app:collisional_equations_with_landau_operator}
Returning to the electron drift-kinetic equation \eqref{eq:electron_drift_kinetic_equation}, let us, instead of \eqref{eq:collision_operator}, adopt the Landau collision operator (evaluated to leading order in the electron-ion mass ratio, consistently with the low-beta ordering introduced in \apref{app:low_beta_ordering}):
\begin{align}
    \left( \frac{\partial \df_e}{\partial t}\right)_c = C_{ee}\left[\df_e \right] + \mathcal{L}_{ei}\left[\df_e \right],
    \label{eq:landau_operator}
\end{align}
where $C_{ee}$ and $\mathcal{L}_{ei}$ are, respectively, the linearised electron-electron and the pitch-angle (Lorentz) collision operators (\citealt{helander05}). With these collision operators, it is no longer analytically convenient to derive the collisional equations \eqref{eq:density_moment_collisional}-\eqref{eq:t_moment_collisional} via the Hermite-Laguerre moments \eqref{eq:laguerre_hermite_moments}. Instead, we simply expand the perturbed electron distribution function as
\begin{align}
    \df_e = \df_e^{(0)} + \df_e^{(1)} + \df_e^{(2)} + \dots,
    \label{eq:electron_distribution_function}
\end{align}
where the superscripts indicate the order in the small parameter $\chi^{-1} \ll 1$ of our collisional expansion \eqref{eq:ordering_of_chi}. Since collisions are dominant to leading order, $\df_e^{(0)}$ is constrained to be a perturbed Maxwellian with no mean-flow\footnote{ In general, the Lorentz collision operator constrains the electron distribution function to be isotropic in the frame moving with the parallel ion velocity. However, the parallel ion flow is negligible within our low-beta ordering [see \eqref{eq:ion_gyrokinetic_equation} and the following discussion], meaning that the electron distribution function will have no parallel velocity moment to leading order.}, viz., 
\begin{align}
     C_{ee}\left[\df_e^{(0)} \right] + \mathcal{L}_{ei}\left[\df_e^{(0)} \right] = 0 \quad \Rightarrow \quad \df_e^{(0)} = \left[ \frac{\delta n_e}{n_{0e}} + \frac{\dTe}{T_{0e}} \left( \frac{v^2}{\vthe^2} - \frac{3}{2} \right) \right] f_{0e},
     \label{eq:collisional_expansion_leading_order}
\end{align}
where we have imposed the solvability conditions 
\begin{align}
    \int \rmd^3 \vec{v} \: \df_e^{(n)} =  \int \rmd^3 \vec{v} \: v^2 \df_e^{(n)}  = 0, \quad n \geqslant 1,
    \label{eq:solvability_density_temperature}
\end{align}
in order to determine uniquely the density and temperature moments in \eqref{eq:collisional_expansion_leading_order}.
Owing to the ordering of parallel lengthscales \eqref{eq:ordering_thermal_conduction_vs_kaw} and perpendicular magnetic-field perturbations \eqref{eq:ordering_amplitudes_collisional}, $\df_e^{(1)}$ is then determined at the next order by the balance
\begin{align}
    \vpar  \left[ \gradd_\parallel\log p_e + \left(\frac{v^2}{\vthe^2} - \frac{5}{2} \right)\gradd_\parallel \log T_e\right] f_{0e} + \vpar \frac{e E_\parallel}{ T_{0e}} f_{0e} =  C_{ee}\left[\df_e^{(1)} \right] + \mathcal{L}_{ei}\left[\df_e^{(1)} \right],
    \label{eq:collisional_expansion_next_order}
\end{align}
where we have made use of \eqref{eq:collisional_expansion_leading_order}, while $\gradd_\parallel \log T_e$ and $\gradd_\parallel \log p_e$ are defined in \eqref{eq:logt_definition_intro} and~\eqref{eq:log_p_definition}, respectively. By exploiting the fact that spherical harmonics are eigenfunctions of the Landau collision operator, \eqref{eq:collisional_expansion_next_order} can be readily inverted for $\df_e^{(1)}$ using a variational principle (see, e.g., \citealt{helander05}), which allows us to determine the parallel electron flow
\begin{align}
    u_{\parallel e} = \frac{1}{n_{0e}} \int \rmd^3 \vec{v} \: v_\parallel \df_e = \vthe d_e^2 \gradd_\perp^2 \mathcal{A},
    \label{eq:collisional_expansion_parallel_flow}
\end{align}
subject to the solvability condition
\begin{align}
    \int \rmd^3 \vec{v} \: \vpar\df_e^{(n)} = 0, \quad n \geqslant 2.
    \label{eq:solvability_velocity}
\end{align}
Finally, the evolution of $\df_e^{(0)}$ is then determined by the next-order equation
\begin{align}
    &\left( \frac{\rmd}{\rmd t} + \vec{v}_{de} \cdot \dperp \right)\df_e^{(0)} + \vpar \gradd_\parallel \df_e^{(1)} \nonumber \\
    &= (\vec{v}_{de} \cdot \gradd_\perp \varphi) f_{0e} - \frac{\rho_e \vthe}{2} \frac{\partial \varphi}{\partial y} \left[\frac{1}{L_n} + \frac{1}{L_{T_e}} \left( \frac{v^2 }{\vthe^2} - \frac{3}{2} \right) \right] f_{0e} + C_{ee}\left[\df_e^{(2)} \right] + \mathcal{L}_{ei}\left[\df_e^{(2)} \right].
    \label{eq:collisional_expansion_second_order}
\end{align}

Equations for the evolution of the density and temperature perturbations are extracted from the relevant moments of \eqref{eq:collisional_expansion_second_order}, while the electron parallel momentum equation can be derived from \eqref{eq:collisional_expansion_next_order} [using \eqref{eq:collisional_expansion_parallel_flow}], yielding 
\begin{align}
    &\frac{\rmd}{\rmd t} \frac{\dne}{n_{0e}}  + \gradd_\parallel u_{\parallel e} + \frac{\rho_e \vthe}{L_B} \frac{\partial}{\partial y}\left( \frac{\dne}{n_{0e}} -  \varphi + \frac{\delta T_{ e}}{T_{0e}} \right) = - \frac{\rho_e \vthe}{2 L_n} \frac{\partial \varphi}{\partial y}, \label{eq:col_density_equation_landau} \\
    &\frac{\rmd \mathcal{A}}{\rmd t} + \frac{\vthe}{2} \frac{\partial \varphi}{\partial z} =  \frac{\vthe}{2} \left( \gradd_\parallel \frac{\dne}{n_{0e}} - \frac{\rho_e}{L_n} \frac{\partial \mathcal{A}}{\partial y} \right) + \left(1 + \frac{c_2}{c_1} \right)\frac{\vthe}{2} \left( \gradd_\parallel \frac{\dTe}{T_{0e}} - \frac{\rho_e}{L_{T_e}} \frac{\partial \mathcal{A}}{\partial y} \right) \nonumber \\
    & \quad\quad\quad\quad\quad\quad\quad + \frac{\nu_{ei}}{c_1} d_e^2 \gradd_\perp^2 \mathcal{A}, \label{eq:col_velocity_equation_landau} \\
    & \frac{\rmd}{\rmd t} \frac{\dTe}{T_{0e}} -\kappa \gradd_\parallel^2 \log T_e + \frac{2}{3}\left(1 + \frac{c_2}{c_1} \right) \gradd_\parallel u_{\parallel e} + \frac{2}{3}\frac{\rho_e \vthe}{L_B} \frac{\partial}{\partial y} \left(  \frac{\dne}{n_{0e}} -  \varphi + \frac{7}{2} \frac{\dTe}{T_{0e}} \right) \nonumber \\
    & \quad\quad\quad\quad\quad\quad\quad =-\frac{\rho_e \vthe}{2 L_{T_e}} \frac{\partial \varphi}{\partial y} , \label{eq:col_temperature_equation_landau}
\end{align}
where $\nu_{ei}$ is defined in \eqref{eq:definition_collision_frequencies}, $\kappa = c_3 \vthe^2/3 \nu_{ei}$, and $c_1, c_2$ and $c_3$ are ion-charge-dependent coefficients
\begin{equation}
    c_1 = \frac{\frac{217}{64} + \frac{151}{8\sqrt{2}Z} + \frac{9}{2Z^2}}{1 + \frac{61}{8 \sqrt{2}Z} +  \frac{9}{2Z^2}}, \spc c_2  = \frac{\frac{5}{2}\left( \frac{33}{16} + \frac{45}{8 \sqrt{2}Z} \right)}{1 + \frac{61}{8 \sqrt{2}Z} +  \frac{9}{2Z^2}}, \spc c_3 = \frac{\frac{25}{4} \left( \frac{13}{4} + \frac{45}{8 \sqrt{2} Z} \right)}{1 + \frac{61}{8 \sqrt{2}Z}} - \frac{c_2^2}{c_1}.
    \label{eq:charge_coefficients}
\end{equation}
Rescalling the collisionality and thermal conductivty as $\nu_{ei}/c_1 \rightarrow \nu_{ei}$, $c_1\kappa \rightarrow \kappa$ respectively, it becomes clear that the only difference between \eqref{eq:col_density_equation_landau}-\eqref{eq:col_temperature_equation_landau} and \eqref{eq:col_density_equation}-\eqref{eq:col_temperature_equation} is the presence of the additive $c_2/c_1$ factors in \eqref{eq:col_velocity_equation_landau} and \eqref{eq:col_temperature_equation_landau}. These factors are due to the fact that, in the presence of a temperature gradient, the energy of a particle is dependent on the direction of its motion, with particles coming from the higher-temperature region having more energy than particles moving in the opposite direction (coming from the lower-temperature region). This gives rise to a net frictional force as the lower-temperature particles will undergo more frequent collisions than their hotter counterparts, and will thus lose more momentum than those coming from the hotter region. This effect clearly relies on the velocity dependence of the collision frequency associated with the Landau collision operator, and hence was not captured by the simplified collision operator \eqref{eq:collision_operator}, whose collision frequency was a constant. 

If one performs the same analysis with \eqref{eq:col_density_equation_landau}-\eqref{eq:col_temperature_equation_landau} as was performed with the collisional equations \eqref{eq:col_density_equation}-\eqref{eq:col_temperature_equation} used throughout this paper, one will find only finite modifications to constant factors entering into all expressions; e.g., the electromagnetic results of \secref{sec:electromagnetic_regime_tai} remain valid but with the rescaling $(1+c_2/c_1) \xi_* \rightarrow \xi_*$, with $\xi_*$ defined in \eqref{eq:xi_definition_collisional}. Thus, the results of this paper derived in the collisional limit are unaffected by our choice of the model collision operator \eqref{eq:collision_operator}. 

\subsection{Time-dependent thermal force}
\label{app:time_dependent_thermal_force}
A feature missing from our collisional model --- represented by either \eqref{eq:col_density_equation}-\eqref{eq:col_temperature_equation} or \eqref{eq:col_density_equation_landau}-\eqref{eq:col_temperature_equation_landau} --- is the so-called `time-dependent thermal force' (\citealt{hassam80a,hassam80b,drake80}), which is believed to be responsible for micro-tearing modes (MTMs) in slab geometry; in particular, it is claimed to be important for destabilising two-dimensional modes involving only perturbations of the parallel component of the vector potential $\mathcal{A}$.

In general, such a term enters as a higher-order correction to the parallel momentum equation \eqref{eq:col_velocity_equation_landau}. In a system where perturbations with a finite parallel wavenumber are allowed, any destabilising effect that this term may have will occur at sub-leading order, and so can be neglected in comparison to the more vigorous (leading-order) instabilities supported by our collisional system, such as slab or curvature-mediated TAI (see \secref{sec:electromagnetic_regime_tai}). However, in the case of two-dimensional perturbations, the perturbations of $\mathcal{A}$ are damped drift-waves \eqref{eq:lin_col_mdw_damped}, on which the time-dependent thermal force can, in theory, have a destabilising effect. In the remainder of this appendix, we demonstrate, however, that its contribution \textit{always} occurs at sub-leading order in collisional systems with $\omega \ll \nu_{ei} \sim \nu_{ee}$, and so should not be included in our analysis. 

We begin by considering the ordering adopted by \cite{hassam80a,hassam80b}:
\begin{align}
    \nu_{ee} \sim \nu_{ei} \gg \omega \sim \omega_{*\s} \sim \omega_{d\s} \sim k_\perp v_E \sim (k_\perp d_e)^2 \nu_{ei} \sim k_\parallel \vthe.
    \label{eq:hassam_ordering}
\end{align}
This is distinct to the collisional ordering \eqref{eq:ordering_collisional_frequencies} in that all of the non-collisional terms in the electron drift-kinetic equation \eqref{eq:electron_drift_kinetic_equation} are now of the same order, leading to different dominant balances at each order [compare, e.g., \eqref{eq:collisional_expansion_second_order} and \eqref{eq:collisional_expansion_second_order_hassam}]. Imposing this ordering on the electron drift-kinetic equation \eqref{eq:electron_drift_kinetic_equation} and expanding the perturbed electron distribution function as \eqref{eq:electron_distribution_function}, we once again find that the leading-order piece is given by \eqref{eq:collisional_expansion_leading_order}, while the next-order piece is determined by \eqref{eq:collisional_expansion_next_order} [after imposing the solvability conditions \eqref{eq:solvability_density_temperature}]. Then, at second order, we have, instead of \eqref{eq:collisional_expansion_second_order},
\begin{align}
    \left( \frac{\rmd}{\rmd t} + \vpar \gradd_\parallel + \vec{v}_{de} \cdot \gradd_\perp \right) \df_e^{(1)} = C_{ee}\left[\df_e^{(2)} \right] + \mathcal{L}_{ei}\left[\df_e^{(2)} \right].
    \label{eq:collisional_expansion_second_order_hassam}
\end{align}
Solving this equation for $\df_e^{(2)}$ yields the contribution of the time-dependent thermal force to the electron parallel momentum equation. 

Following \cite{hassam80a}, we assume that the right-hand sides of \eqref{eq:collisional_expansion_next_order} and \eqref{eq:collisional_expansion_second_order_hassam} can be approximated solely by the pitch-angle scattering operator (see \citealt{helander05, abel08}): 
\begin{align}
    \mathcal{L}_{ei} \left[ \df_e \right] =\frac{1}{2} \nu_D\frac{\partial}{\partial \xi} \left[ (1-\xi^2) \frac{\partial \df_e}{\partial \xi} \right], \quad \nu_D(v) = \frac{3\sqrt{\pi}}{4} \nu_{ei} \left( \frac{\vthe}{v} \right)^3,
    \label{eq:pitch_angle_scattering}
\end{align}
where $\xi = \vpar /v $ is the pitch-angle coordinate, and $\nu_{ei}$ is as defined in \eqref{eq:definition_collision_frequencies}. Ignoring, for brevity, the contributions from the magnetic drifts, both \eqref{eq:collisional_expansion_next_order} and \eqref{eq:collisional_expansion_second_order_hassam} can be straightforwardly solved for the perturbed distribution functions:
\begin{align}
    \df_e^{(1)} & = - \frac{\vpar}{\nu_D} \left[ \frac{e E_\parallel}{T_{0e}} + \gradd_\parallel \log p_e + \left(\frac{v^2}{\vthe^2} - \frac{5}{2} \right)\gradd_\parallel \log T_e\right] f_{0e}, \label{eq:hassam_df1}\\
    \df_e^{(2)} & = - \frac{1}{\nu_D} \left(\frac{\rmd}{\rmd t} + \frac{1}{3} \vpar \gradd_\parallel \right)  \df_e^{(1)} \label{eq:hassam_df2}. 
\end{align}
Then, the electron parallel momentum equation can be derived from \eqref{eq:collisional_expansion_next_order} [using \eqref{eq:collisional_expansion_parallel_flow}], yielding
\begin{align}
    &\frac{\rmd \mathcal{A}}{\rmd t} + \frac{\vthe}{2} \frac{\partial \varphi}{\partial z} =  \frac{\vthe}{2} \left( \gradd_\parallel \frac{\dne}{n_{0e}} - \frac{\rho_e}{L_n} \frac{\partial \mathcal{A}}{\partial y} \right) + \left(1 + \frac{c_2}{c_1} \right)\frac{\vthe}{2} \left( \gradd_\parallel \frac{\dTe}{T_{0e}} - \frac{\rho_e}{L_{T_e}} \frac{\partial \mathcal{A}}{\partial y} \right) \nonumber \\
    & + \frac{\nu_{ei}}{c_1} d_e^2 \gradd_\perp^2 \mathcal{A} + \frac{c_4}{\nu_{ei}} \frac{\rmd}{\rmd  t} \left[\frac{\rmd \mathcal{A}}{\rmd t} + \frac{\vthe}{2} \frac{\partial \varphi}{\partial z} - \frac{\vthe}{2} \left( \gradd_\parallel \frac{\dne}{n_{0e}} - \frac{\rho_e}{L_n} \frac{\partial \mathcal{A}}{\partial y} \right) \right. \nonumber \\
    & \quad\quad\quad\quad\quad\quad\quad\quad\quad\quad\quad - \left. \left(1 + \frac{2c_2}{c_1} \right)\frac{\vthe}{2} \left( \gradd_\parallel \frac{\dTe}{T_{0e}} - \frac{\rho_e}{L_{T_e}} \frac{\partial \mathcal{A}}{\partial y} \right) \right], \label{eq:hassam_velocity_equation} 
\end{align}
where now $c_1 = 32/3\pi$, $c_2 = 16/\pi$ and $c_4 = 105/16$ (these coefficients are the same as those in \citealt{hassam80a}). We note that \eqref{eq:hassam_velocity_equation} is the same as \eqref{eq:col_velocity_equation_landau}, up to the terms known as the time-dependent thermal force (proportional to $\nu_{ei}^{-1}$) on the right-hand side.

Given that we argued above that these terms can only ever matter for two-dimensional perturbations, we set $k_\parallel = 0$ in \eqref{eq:hassam_velocity_equation}. Linearising and Fourier-transforming the resultant expression, we find the dispersion relation 
\begin{align}
     \omega - \left( 1 + \frac{1}{\eta_e} + \frac{c_2}{c_1} \right)\omega_{*e} + i \frac{(k_\perp d_e)^2\nu_{ei}}{c_1} + i c_4 \left[ \omega - \left( 1 + \frac{1}{\eta_e} + \frac{2c_2}{c_1} \right)\omega_{*e}\right] \frac{\omega}{\nu_{ei}} = 0 .
    \label{eq:hassam_dispersion_relation}
\end{align}
This has a perturbative solution [neglecting small modifications of the real frequency, cf. equation (6) in \citealt{hassam80b}]:
\begin{align}
    \omega = \left( 1 + \frac{1}{\eta_e} + \frac{c_2}{c_1} \right)\omega_{*e} -  i \frac{(k_\perp d_e)^2\nu_{ei}}{c_1} + i \frac{c_2 c_4}{c_1}\left( 1 + \frac{1}{\eta_e} + \frac{c_2}{c_1} \right) \frac{\omega_{*e}^2}{\nu_{ei}}.
    \label{eq:hassam_solution}
\end{align}
The last, destabilising term only arises for collision operators with a velocity-dependent collision frequency [being proportional to $c_2$; see discussion following \eqref{eq:charge_coefficients}], and so is not captured in our simplified collision operator~\eqref{eq:collision_operator}; in particular, if it were not for the velocity dependence of $\nu_D$ [see \eqref{eq:pitch_angle_scattering}] and the resultant factor of two multiplying $c_2/c_1$ in the last term on the right-hand side of \eqref{eq:hassam_velocity_equation}, this destabilising term would vanish. However, this term can never actually cause \eqref{eq:hassam_solution} to have a positive growth rate within the ordering~\eqref{eq:hassam_ordering}, as it is a higher-order correction to the resistive damping manifest in the second term. This statement remains true in the non-zero $k_\parallel$ case; the last term on the right-hand side of~\eqref{eq:hassam_velocity_equation} is clearly higher-order in $\omega/\nu_{ei} \ll 1$ --- arising from $\df_e^{(2)}$, rather than $\df_e^{(1)}$ --- and so can be neglected in comparison to the remaining contributions. This means that the time-dependent thermal force cannot be responsible for driving MTMs in a collisional ($\omega \ll \nu_{ei}$) system, and so we are justified in neglecting it in our analysis. The effect is also absent in the collisionless limit ($\omega \gg \nu_{ei}$). There is, of course, an intermediate regime $\omega \sim \nu_{ei}$ in which physics typified by such a term can be important (see, e.g., \citealt{larakers20}), but this regime is outside the scope of this paper (and indeed of any perturbative collisional expansion).

\end{appendix}

\bibliography{bibliography.bib}{}
\bibliographystyle{jpp}

\end{document}